\documentclass[aps,pre,twocolumn,showpacs]{revtex4}
\usepackage{graphicx}
\usepackage{amssymb}
\usepackage{subfigure}
\usepackage{amsmath}
\usepackage{captcont}

\begin{document}

%
%
%
%
%
%
%
%
\title{\bf Dynamics and Pattern Formation in Large Systems of 
       Spatially-Coupled Oscillators with Finite Response Times}

\author{Wai Shing Lee$^1$, Juan G. Restrepo$^2$, Edward Ott$^1$, Thomas M. Antonsen$^1$\\
       }

\affiliation{{}$^1$ Institute for Research in Electronics and Applied Physics, 
                  University of Maryland, College Park, Maryland 20742, USA, \\
             {}$^2$ Department of Applied Mathematics, University of Colorado, 
                  Boulder, Colorado 80309, USA }


\begin{abstract}
We consider systems of many spatially distributed phase oscillators that
interact with their neighbors. Each oscillator is allowed to have a 
different natural frequency, as well as a different response time
to the signals it receives from other oscillators in its neighborhood. 
Using the ansatz of Ott and Antonsen (Ref. \cite{OA1}) and adopting 
a strategy similar to that employed in the
recent work of Laing (Ref. \cite{Laing2}), we reduce the 
microscopic dynamics of these
systems to a macroscopic partial-differential-equation description.
Using this macroscopic formulation, we 
numerically find that finite oscillator
response time leads to interesting spatio-temporal dynamical
behaviors including propagating fronts, spots, target patterns, chimerae,
spiral waves, etc., and we
study interactions and evolutionary behaviors of these spatio-temporal
patterns.
\end{abstract}

\maketitle
 
{\bf Many physical systems can be thought of as consisting of
a large number of oscillating units that are distributed in space
and coupled to neighboring units that are within some limited distance.
The individual coupled units of such systems,
moreover,
can have non-negligible response times,
and it is well known that delays
can give rise to a set of possible behaviors that is significantly
richer than would be the case without delays.
Our work addresses two issues:
(1) derivation of a macroscopic description for such systems, and
(2) the possible characteristic behaviors that may be 
revealed through study of such macroscopic descriptions.}

\section{Introduction}
Systems of large coupled oscillator networks 
appear in many physical and 
engineering systems \cite{PRK}-\cite{Winfree}.
Examples include synchronous flashing of fireflies \cite{Buck}, 
pedestrian induced
oscillations of the Millennium Bridge \cite{MillB},
cardiac pace-maker cells \cite{Glass}, 
alpha rhythms in the brain \cite{Brain}, 
glycolytic oscillations in yeast populations \cite{Yeast}, 
cellular clocks governing circadian rhythm in mammals \cite{Yamaguchi},
oscillatory chemical reactions \cite{Kuramoto1}-\cite{Taylor}, etc.

Many previous studies of oscillator networks developed in
the setting of network couplings on graphs
with different topological characterizations, such as 
small-world, Erd\"{o}s-Renyi, and scale-free (e.g., Refs. \cite{Hong}-\cite{Restrepo}).
Here we consider applications in which
the oscillators are distributed spatially,
for example, when there is
a row of trees each occupied with a large number of fireflies. 
Indeed,
in the past decade studies of spatially distributed 
coupled oscillators
have aroused much interest. An example is
the chimera states (e.g., Refs. \cite{Kura_Batto}-\cite{Laing2}),
in which there is a stable coexistence of
both coherent and incoherent states distributed in space. 

Another important aspect of the dynamics of oscillator networks
is that
physical oscillators may have significant delays in their response
to signals and these signals may also take a significant
time to propagate. Studies
of time-delay effects in the 
context of all-to-all coupled networks with a homogeneous 
distribution of time delays (\cite{YS}, \cite{CKK})
show that interesting features such as
bistable behaviors and multiple coherent states
are induced in the presence of time delays.
Reference \cite{Lee}, building on the machinery developed in 
Refs. \cite{OA1}-\cite{OA3}, extends this line of study to
a heterogeneous nodal response time distribution. In addition,
Ref. \cite{Sethia} studies the dynamics of
a one dimensional ring of spatially distributed and non-locally coupled
oscillator network when the time delays are
due to signal propagation between interacting oscillators. 

The problem studied 
in this paper is that of uncovering the spatio-temporal dynamics 
of a system of coupled oscillators with
heterogeneous oscillator response times.
We first give a microscopic description of the individual oscillators
and their couplings. 
We then spatially coarse-grain this description and use the methods
developed in Refs. \cite{Laing2} and \cite{Lee} to
derive a set of partial differential equations giving
a macroscopic description of the system dynamics.
Using our derived macroscopic equations, we then numerically
explore the spatio-temporal dynamics and resulting pattern
formation in both one- and two-dimensions.
We find that a rich variety of behaviors are induced by the presence
of time delay in the oscillator response. These include
hysteresis, propagating fronts, spots, target patterns,
chimerae, spiral waves, etc.

\section{Formulation}
We consider a system of $N$ spatially distributed 
interacting phase oscillators with
time delays between the response of an oscillator and the signal it receives.
The evolution equation of oscillator $m$ is
\begin{align}
\frac{d}{dt}\theta_m(t) &= \omega_m +
   \sum_{n \neq m}^{N}\hat{K}_{mn} 
   \left\{ \sin[\theta_{n}(t-\tau_{mn}) - \theta_m(t)] \right\} \notag \\
&= \omega_m + \sum_{n\neq m}^{N} \hat{K}_{mn} 
   \frac{1}{2i} \{ e^{-i [\theta_m(t) -  \theta_{n}(t-\tau_{mn})]}- c.c.\}, 
\label{eq:Fund}
\end{align}

\noindent
where $\hat{K}_{mn}$ is the interaction strength
between oscillators $m$ and $n$, which is assumed to be spatial in 
character (i.e., $\hat{K}_{mn}$ becomes small or zero if the distance
between oscillator $m$ and oscillator $n$ is large),
$\tau_{mn}$ is the interaction time delay in the effect of oscillator
$n$ on oscillator $m$, and $c.c.$ denotes complex conjugate. 

Assuming a separation in the scales of the macroscopic and microscopic
system dynamics,
we follow a path similar to that employed by kinetic theory
to reduce the study of a 
gas of many interacting molecules to a fluid description.
We begin by partitioning the continuous space into 
discrete regions $I_{\bar{x}}$ centered at the discrete set of spatial points $\bar{x}$,
such that
the domain of interest is $\cup_{\bar{x}} I_{\bar{x}}$, and
$I_{\bar{x}} \cap I_{\bar{x}'} = \emptyset$ for $\bar{x} \neq \bar{x}'$.
The diameter of each region is $|I_{\bar{x}}| \sim w$, and the 
volume of each region is $w^d$ where 
$d$ denotes the dimension of space.

These regions are assumed to be
small enough that $\hat{K}_{mn} \approx \hat{K}_{ml}$
if oscillators $n$ and $l$ are in the same region $I_{\bar{x}'}$, yet
large enough that many oscillators ($N_{I_{\bar{x}'}} \gg 1$) are contained within
each $I_{\bar{x}'}$. Thus we can meaningfully define
\begin{align}
\rho(\bar{x}') &\equiv \frac{N_{I_{\bar{x}'}}}{w^d}, \notag \\
r(\bar{x}',t) &\equiv \frac{1}{N_{I_{\bar{x}'}}} 
        \sum_{n \in I_{\bar{x}'}} e^{i \theta_{n}(t)},
\label{eq:local_para}
\end{align}

\noindent
respectively
as the local density and the local order
parameter in $I_{\bar{x}'}$.
In addition,
for all $m \in I_{\bar{x}}$ and
$n \in I_{\bar{x}'}$, we approximate $\hat{K}_{mn} \approx K_{\bar{x}\bar{x}'}$.
The summation in (\ref{eq:Fund}) can thus first be
approximated as
\begin{equation}
\frac{1}{2i} \left[ 
   \sum_{I_{\bar{x}'}} K_{\bar{x},\bar{x}'} 
   N_{I_{\bar{x}'}} e^{-i \theta_m(t)}
   \frac{1}{N_{I_{\bar{x}'}}} \sum_{n \in I_{\bar{x}'}} 
     e^{i \theta_{n}(t - \tau_{mn})}
   - c.c. \right].
\label{eq:Interact}
\end{equation}

\noindent
In all of what follows, we consider only the simple illustrative
case that $\tau_{mn}=\tau_m$, i.e., we suppose that the delay
in the effect of oscillator $n$ upon oscillator $m$ is 
independent of $n$. This would, e.g., apply if the signal
propagation time from $n$ to $m$ was very fast, but each oscillator had a finite
reaction time. Together with Eq.(\ref{eq:local_para}), Eq. (\ref{eq:Interact})
can then be written as
\begin{equation}
\sum_{I_{\bar{x}'}}  w^d K_{\bar{x}\bar{x}'}
\rho(\bar{x}') \text{Im} \{e^{-i \theta_m(t)} r(\bar{x}',t-\tau_m)\}.
\label{eq:Interact2}
\end{equation}

\noindent
Since we assume $N_{I_{\bar{x}}} \gg 1$ for all $\bar{x}$, it is 
appropriate to introduce a distribution function $F(\theta,\omega,\bar{x},\tau,t)$
proportional to
the fraction of oscillators in $I_{\bar{x}}$ with 
$\theta \in [\theta,\theta+d\theta]$, $\omega \in [\omega,\omega+d\omega]$
and $\tau \in [\tau,\tau+d\tau]$ at time $t$. 
We furthermore pass to the limit of continuous space
by replacing the discrete variable
$\bar{x}$ by a new variable $x$ which we now regard as 
continuous. In terms of this distribution, we introduce the marginal distribution
$\hat{g}(\omega,\tau,x)$,
\begin{equation}
\hat{g}(\omega,\tau,x) =
\int_0^{2\pi} F(\theta,\omega,\tau,x,t) d\theta.
\label{eq:Dist_fun}
\end{equation}

\noindent
Here, note that since $\omega,\tau$ and $x$ for any oscillator 
are assumed to be constant in time,
the $\theta-$integral of $F$ is time independent even though $F$ itself
depends on time.
With Eq. (\ref{eq:Dist_fun}), the quantity $r$
in Eq. (\ref{eq:local_para}) becomes

\begin{align}
r(x,t)&=\frac{\int_0^{\infty} \int_{-\infty}^{\infty} \int_0^{2\pi} 
       F(\theta,\omega,\tau,x,t) e^{i\theta} d\theta d\omega d\tau}
      {\int_0^{\infty} \int_{-\infty}^{\infty} \int_0^{2\pi} 
      F(\theta,\omega,\tau,x,t) d\theta d\omega d\tau} \notag \\
&=\frac{1}{\rho(x)} \int_0^{\infty}\int_{-\infty}^{\infty} \int_0^{2\pi} 
     F(\theta,\omega,\tau,x,t) e^{i\theta} d\theta d\omega d\tau
\label{eq:order_par}
\end{align}

\noindent
The overall system dynamics can be studied in terms of the evolution 
equation for $F(\theta,\omega,\tau,x,t)$,
\begin{equation}
\frac{\partial F}{\partial t} + \frac{\partial}{\partial \theta}
\left( F \{ \omega + \text{Im}[\eta(x,t-\tau) e^{-i \theta}] \} \right) = 0,
\label{eq:Density_evol}
\end{equation}

\noindent
where 
\begin{equation}
\eta(x,t)=\int \rho(x') K(x,x') r(x',t) dx'
\label{eq:eta0}
\end{equation}

\noindent
is Eq.(\ref{eq:Interact2}) in the continuum limit, and the integration
in (\ref{eq:eta0}) is over the $d$-dimensional spatial
domain. Referring back to our
previous analogy to kinetic theory of a gas, we think of 
Eqs.(\ref{eq:Density_evol}) and (\ref{eq:eta0}) as a kinetic description
roughly analogous to the Boltzmann equation.

To proceed we wish to reduce our kinetic description 
(\ref{eq:Density_evol}) and (\ref{eq:eta0}) to a PDE (partial differential
equation) system analogous to the fluid equations of gas dynamics. We do
this using the recent work of Ott and Antonsen (Refs. \cite{OA1}-\cite{OA2}).
We expand $F$ in a Fourier series of the form
\begin{equation}
F(\theta,\omega,\tau,x,t) = \frac{\hat{g}(\omega,\tau,x)}{2\pi} \left\{
 1 + \left[ \sum_{n=1}^{\infty} f_n(\omega,\tau,x,t) e^{i n \theta} + c.c. 
     \right] \right\}.
\label{eq:Density_Four}
\end{equation}

\noindent
As discussed and justified in Refs. \cite{OA1} and \cite{OA2}, we seek a solution
in the form
\begin{equation}
f_n(\omega,\tau,x,t) = \hat{\alpha}(\omega,x,t-\tau)^n.
\label{eq:Lowdim_mani}
\end{equation}

\noindent
Equations (\ref{eq:order_par}) to (\ref{eq:eta0}) then yield
\begin{align}
\begin{split}
\frac{\partial}{\partial t} \hat{\alpha}(\omega,x,t-\tau) &+ 
i\omega \hat{\alpha}(\omega,x,t-\tau)  \\
+ \frac{1}{2}  & \left[ \eta(x, t-\tau) 
\hat{\alpha}^2(\omega,x,t-\tau) - \eta^*(x,t-\tau) \right] = 0, 
\end{split}   
\label{eq:alpha_1} \\
\eta(x,t-\tau) =  \int & \rho(x') K(x,x') r(x',t-\tau) dx', \label{eq:eta_1} \\
r(x,t) =  \int \frac{1}{\rho(x)} & \int_{-\infty}^{\infty} 
    \hat{g}(\omega,\tau',x) \hat{\alpha}^*(\omega,x,t-\tau') d\omega d\tau',
  \label{eq:r_1}
\end{align}

\noindent
where the star ${^*}$ denotes complex conjugate, and 
$\tau'$ is written inside Eq. (\ref{eq:r_1}) to emphasize its role
as a dummy integration
variable as compared with $\tau$'s in the other equations.

In what follows, we study an illustrative case
corresponding to
\begin{align}
\hat{g}(\omega,\tau, x) = g(\omega) h(\tau) \rho_0, 
    \label{eq:ghat} \\
K(x,x') = k q(x-x'), 
    \label{eq:Kdcre}
\end{align}

\noindent
where $\int_{-\infty}^{\infty} g(\omega) d\omega 
= \int_0^{\infty} h(\tau) d\tau = 1$. Equation (\ref{eq:ghat}) implies
that the oscillator frequencies, locations, and delay distributions are
uncorrelated, and that the oscillator density $\rho_0$ is uniform.
Equation (\ref{eq:Kdcre}) states that the strength of the coupling between 
oscillators at two points depends uniformly
on their spatial separation.
Further, in (\ref{eq:Kdcre}) we take $q(x)$ to be suitably normalized, so 
that the constant $k$ may be regarded as an overall coupling strength.
With these assumptions, together with the transformation $t \rightarrow t+\tau$
in Eqs. (\ref{eq:alpha_1}) and (\ref{eq:eta_1}), and 
rewriting $\tau'$ as $\tau$ in Eq. (\ref{eq:r_1}), we obtain
\begin{align}
\begin{split}
\frac{\partial}{\partial t} \hat{\alpha}(\omega,x,t) 
 &+ i\omega \hat{\alpha}(\omega,x,t) \\
+ \frac{k}{2} & \left[ \eta(x,t) \hat{\alpha}^2(\omega,x,t) - 
   \eta^*(x,t) \right] = 0, 
\end{split}
\label{eq:alpha_2} \\
\eta(x,t) &= \int \rho_0 q(x-x') r(x',t) dx', \label{eq:eta_2} \\
r(x,t) = \int & \left[  \int_{-\infty}^{\infty} 
    g(\omega) \hat{\alpha}^*(\omega,x,t-\tau) d\omega \right] h(\tau) d\tau.
\label{eq:r_2}
\end{align}

\noindent
In order to reveal generic expected behavior, we now further specify 
particular convenient choices for the frequency distribution, $g(\omega)$,
the response time distribution, $h(\tau)$, and the spatial interaction
kernel, $q(x)$. 

We assume a Lorentzian form for $g(\omega)$,
\begin{align}
g(\omega) &= \frac{\Delta/\pi}{(\omega-\omega_0)^2 + \Delta^2} \notag \\
       &= \frac{1}{2 \pi i} 
          \left\{ \frac{1}{\omega-\omega_0-i\Delta} -
                 \frac{1}{\omega-\omega_0+i\Delta} \right\}.
\label{eq:Lorentz}
\end{align}

\noindent
Assuming $\hat{\alpha}$ is analytic in $\omega$, we close
the $\omega-$integration path in (\ref{eq:r_2}) with a large
semi-circle of radius $R \rightarrow \infty$ in the lower half complex
$\omega-$plane. Thus we obtain from the pole of $g(\omega)$ at 
$\omega=\omega_0-i\Delta$ [see Eq. (\ref{eq:Lorentz})],
\begin{equation}
r(x,t) = \int \alpha^*(x,t-\tau) h(\tau) d\tau,
\label{eq:r_2.5}
\end{equation}

\noindent
where $\alpha(x,t)=\hat{\alpha}(\omega_0-i\Delta,x,t)$, and we have assumed
(Ref. \cite{OA1}) that, as $\mbox{Im} (\omega) \rightarrow -\infty$, 
$\hat{\alpha}(\omega,x,t)$ is sufficiently well-behaved that the 
contribution from the integration over the large semicircle 
approaches zero as $R \rightarrow \infty$.
Setting $\omega=\omega_0-i\Delta$ in Eq. (\ref{eq:alpha_2})
we obtain the following equation for the time evolution of $\alpha(x,t)$,
\begin{equation}
\frac{\partial}{\partial t} \alpha(x,t) + (\Delta+i\omega_0) 
  \alpha(x,t) 
+ \frac{k}{2} \left[ \eta(x,t) \alpha^2(x,t) - 
   \eta^*(x,t) \right] = 0. 
\label{eq:alpha_3}
\end{equation}

Our assumed form for the response time distribution
$h(\tau)$ is given by \cite{Lee},
\begin{equation}
h_n(\tau) = A_n \tau^n e^{-\beta_n \tau},
\label{eq:delay_dist}
\end{equation}

\noindent
where $A_n$ and $\beta_n$ are defined by $\int_0^\infty h(\tau) d\tau=1$
and $\int_0^\infty \tau h(\tau) d\tau = T$. Noting the 
convolution form of Eq. (\ref{eq:r_2.5}), we can re-express 
(\ref{eq:r_2.5}) as 

\begin{equation}
\left( \frac{T}{n+1}\frac{\partial}{\partial t} +1 \right)^{n+1} 
   r(x,t) = \alpha^*(x,t). 
\label{eq:r_3}
\end{equation}

For the interaction kernel, we choose $q(x)$ to be the solution to the
problem,

\begin{equation}
\left(\nabla^2 - \frac{1}{L^2}\right) q(x) = - \frac{1}{L^2} \delta(x).
\label{eq:q_def}
\end{equation}

\noindent
For example, for an unbounded domain with
boundary conditions $q(x) \rightarrow 0$ as $|x| \rightarrow \infty$,
we obtain
\begin{align}
q(x) = \left\{ \begin{array}{ll}
\frac{1}{2L} \exp \left( -\frac{|x|}{L} \right) & \mbox{for} \hspace{1mm} d=1,  
                   \\ \vspace{3mm}
\frac{1}{2\pi L^2} K_0 \left(\frac{|x|}{L} \right)   & \mbox{for} 
\hspace{1mm} d=2, \\ \vspace{2mm}
\frac{1}{4 \pi |x| L^2} \exp \left( -\frac{|x|}{L}  \right) & \mbox{for} 
\hspace{1mm} d=3, 
\label{eq:q_exp}
\end{array} \right.
\end{align}

\noindent
where $K_0(|x|/L)$ is a zero order Bessel function of imaginary argument.
Using Eq. (\ref{eq:q_def}), Eq. (\ref{eq:eta_2}) can be rewritten by acting
on it with the operator
$(\nabla^2 - \frac{1}{L^2})$, giving
\begin{equation}
\nabla^2 \eta(x,t) - \frac{1}{L^2}\eta(x,t) = 
-\frac{1}{L^2} \rho_0 r(x,t)
\label{eq:eta_3}
\end{equation}

Thus we obtain a closed system of three PDE's in the independent variables
$x$ and $t$ given by Eq. (\ref{eq:alpha_3}) for $\alpha(x,t)$, Eq. (\ref{eq:r_3})
for $r(x,t)$, and Eq. (\ref{eq:eta_3}) for $\eta(x,t)$.
In the rest of this paper we study solutions of these equations
in one- and
two- dimensional domains of size $D$ with periodic boundary conditions. 
The parameters of this system are
\begin{displaymath}
\Delta,\omega_0,k,L,T,D,\rho_0,n.
\end{displaymath}

\noindent
By suitable normalization we can remove three of these parameters. We will do
this by redefining $\eta$ and $k$ to absorb $\rho_0$ and by
normalizing time to $\Delta^{-1}$ and distance to $L$. This can 
also be viewed as using our original parameter set with the 
choices $\Delta=1, L=1, \rho_0=1$. In either case, our
normalized PDE description becomes
\begin{align}
\begin{split}
\frac{\partial}{\partial t} \alpha(x,t) +  (1+i\omega_0) &\alpha(x,t)   \\
+ \frac{k}{2}  & \left[ \eta(x,t) \alpha^2(x,t) - 
    \eta^*(x,t) \right] =  0, 
\end{split}
\label{eq:alpha_4} \\
\left( \frac{T}{n+1}\frac{\partial}{\partial t} +1 \right)^{n+1} 
    & r(x,t) = \alpha^*(x,t), \label{eq:r_4} \\
(\nabla^2  - 1)  & \eta(x,t) =  - r(x,t).
\label{eq:eta_4}
\end{align}

\noindent
In addition, in what follows we will only consider $n=0$ corresponding
to $h(\tau) = T^{-1} \exp(-\tau/T)$. Thus, our reduced parameter set is
\begin{equation}
\omega_0,k,T,D. \label{eq:para_set}
\end{equation}

Before turning to the study of Eqs. (\ref{eq:alpha_4})-(\ref{eq:eta_4}),
we briefly comment on the analogy 
of the derivation of our evolution equations 
(\ref{eq:alpha_4})-(\ref{eq:eta_4}) to the derivation of the equations
of gas dynamics from Boltzmann's equation. 
Substituting (\ref{eq:Lowdim_mani}) into (\ref{eq:Density_Four})
and summing the geometric series ($|\hat{\alpha}|<1$ is assumed for convergence),
we obtain 
\begin{equation}
F(\theta,\omega,x,\tau,t) = \frac{\hat{g}(\omega,x,\tau)}{2\pi}
\left(
\frac{1-|\hat{\alpha}|^2}{1+|\hat{\alpha}|^2-2
|\hat{\alpha}|\cos(\theta-\psi)} \right),
\label{eq:Poisson_ker}
\end{equation}

\noindent
where $\hat{\alpha}=|\hat{\alpha}| \exp(-i \psi)$.
It is shown in Refs. \cite{OA1} and \cite{OA2} that, 
under very general conditions, the solution
to our Eq. (\ref{eq:Density_evol}) relaxes to this form. In gas dynamics, the solution
to Boltzmann's equation, via the Chapman-Enskog expansion 
(Ref. \cite{Chapman}), is assumed to
approximately relax to a local Maxwellian distribution whose velocity-space
width is controlled by the temperature, and whose velocity-space
maximum is located at the fluid velocity. In analogy with this situation,
Eq. (\ref{eq:Poisson_ker}) is peaked in $\theta$ (analogous to velocity space)
at the location $\theta=\psi$ (analogous to the fluid velocity), and the
width of this peak is controlled by $|\hat{\alpha}|$ (analogous to temperature)
with $F$ becoming a delta function in $\theta$ as $|\hat{\alpha}|\rightarrow 1$
(analogous to temperature $\rightarrow 0$). In contrast to the derivation 
of gas dynamics from the Boltzmann equation, our relaxation to (\ref{eq:Poisson_ker})
is due to the phase mixing of many oscillators with different natural frequencies,
whereas relaxation to a local Maxwellian in gas dynamics is due to chaos
in the collisional dynamics of interacting particles. Another difference is that
(\ref{eq:Poisson_ker}) is an exact, rigorous result (as shown in Refs. \cite{OA2} 
and \cite{OA3}), while relaxation to a local Maxwellian in the derivation
of gas dynamics is an asymptotic result in the ratio of
the mean free path (and mean free time) to the macroscopic length (and time) scale.



\section{Numerical studies and discussions}

\subsection{1D propagating fronts, ``bridge'' and ``hole'' patterns}
\begin{figure}[h] 
  \begin{center}
    \includegraphics[width=6cm]{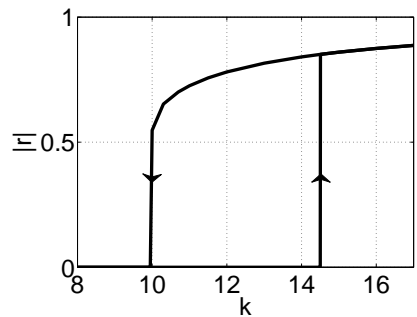}
    \caption{Hysteresis loop for
             $\omega_0=5,T=1$. The upper and lower branches correspond
             to stable coherent and incoherent states.}
    \label{fig:hys_ome05}
  \end{center}
\end{figure}

The simplest solutions of our system, 
Eqs. (\ref{eq:alpha_4})-(\ref{eq:eta_4}),
are the homogeneous incoherent state solution ($r=0$ everywhere) 
and the homogeneous coherent state solution
($r = r_0 e^{i \Omega t}$ where
$r_0$ and $\Omega$ are real constants). As
shown in Refs. \cite{YS}-\cite{Lee} 
for the case of globally coupled oscillators [corresponding
to $\nabla^2 \rightarrow 0$ in Eq. (\ref{eq:eta_4})], a distribution of
interaction time-delays induces bistability and hysteretic behaviors.
Figure \ref{fig:hys_ome05} shows an example of the hysteresis loop 
in the $|r|-k$ plane for spatially
homogeneous states with $\omega_0=5,T=1$, which is obtained  
by solving Eqs. (\ref{eq:alpha_4}) to (\ref{eq:eta_4}) with 
$\eta = r$ for the coherent solution $r = r_0 e^{i \Omega t}$.

\begin{figure}[h] 
  \begin{flushleft}
    \subfigure[]
    {\includegraphics[width=4.1cm]{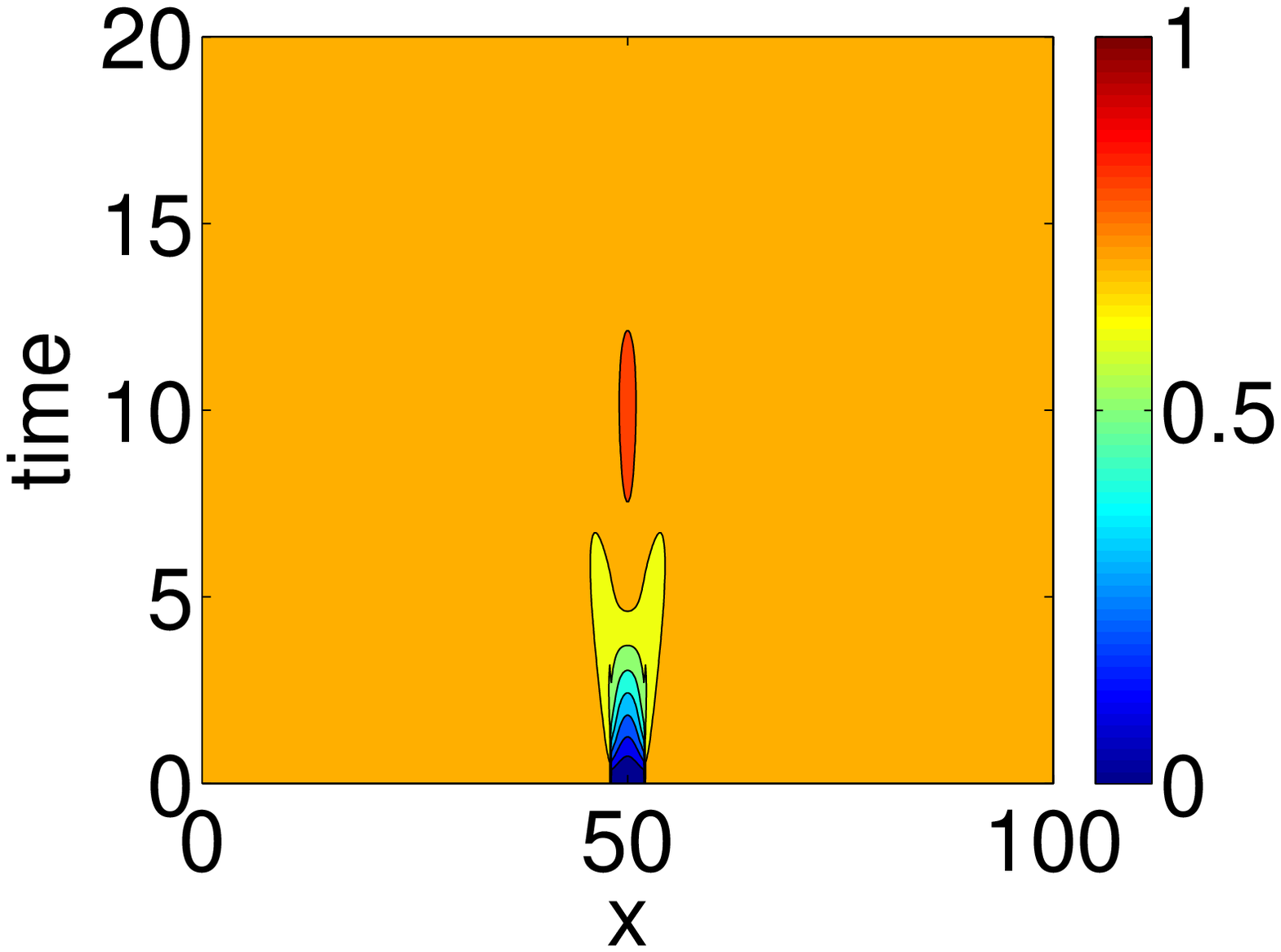}}
    \subfigure[]
    {\includegraphics[width=4.1cm]{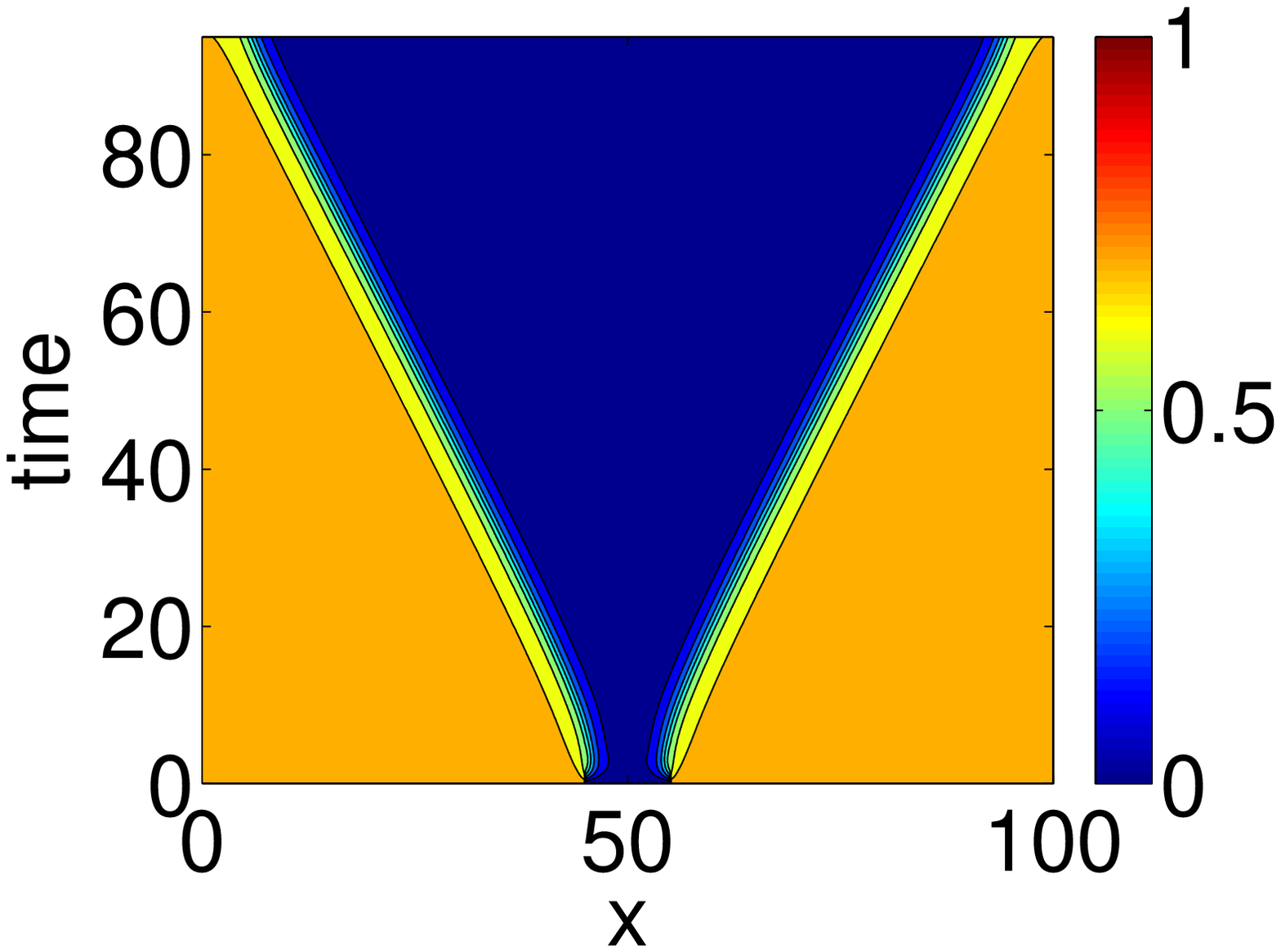}}
    \subfigure[]
    {\includegraphics[width=4.1cm]{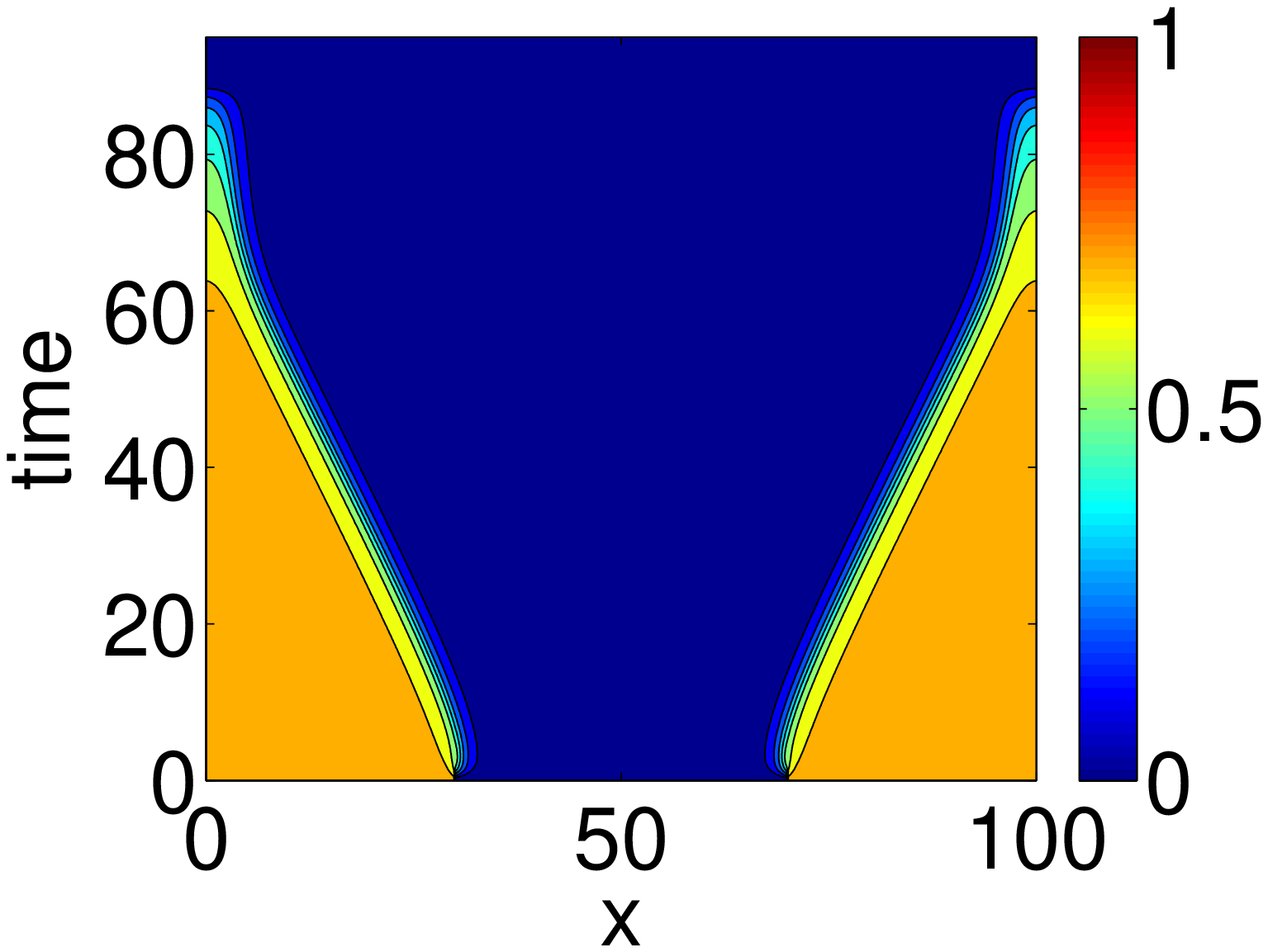}}
    \caption{
             $|r(x,t)|$ for, (a) an initial
             configuration with a small part of the one-dimensional spatial domain 
             in the 
             incoherent state (blue) and a large part in the
             coherent state (orange), (b) a larger part of the spatial domain is
             initially in the incoherent
             state than that in (a), and (c) a still larger initial
             incoherent region. 
             ($\omega_0=5, T=1, D=100, k=12$).}
    \label{fig:Inhomostab}
  \end{flushleft}
\end{figure}

We first consider a one-dimensional
version of our system, Eqs. (\ref{eq:alpha_4})-(\ref{eq:eta_4}), for
a $k$ value within
the bistable region, $k=12$, and examine the evolution resulting from
several initialized configurations with different
spatial regions in the 
homogeneous incoherent and coherent state solutions. Results
are shown in Fig. \ref{fig:Inhomostab}.
Note that the final state
is either coherent or incoherent depending on how large the initial
incoherent region is. Thus, there appears to be a critical initial
size of the incoherent region beyond which the incoherent region takes over.
Furthermore, from Fig. \ref{fig:Inhomostab}, we see that
the evolutionary process leading to this final state is by 
propagation of fronts separating coherent and incoherent regions,
and that these fronts propagate at an approximately 
constant speed. In addition to this initial example, 
we find a variety of other one-dimensional spatio-temporal behaviors
to be reported in the following.

\begin{figure*}[h] 
  \begin{flushleft}
    \subfigure [$k=10$]
    {\includegraphics[width=4.1cm]{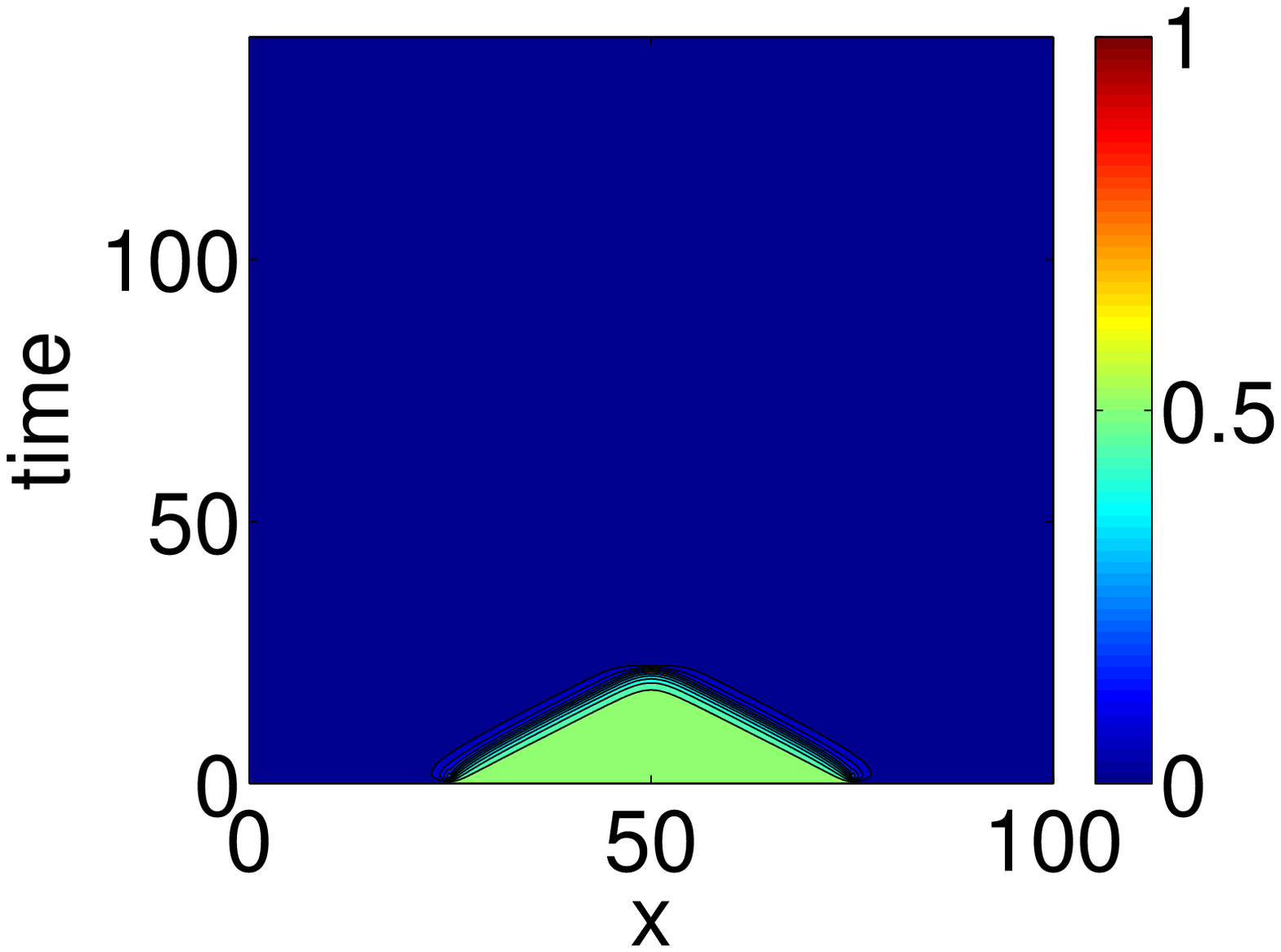} \label{fig:1dpf_k10}}
    \subfigure [$k=11$]
    {\includegraphics[width=4.1cm]{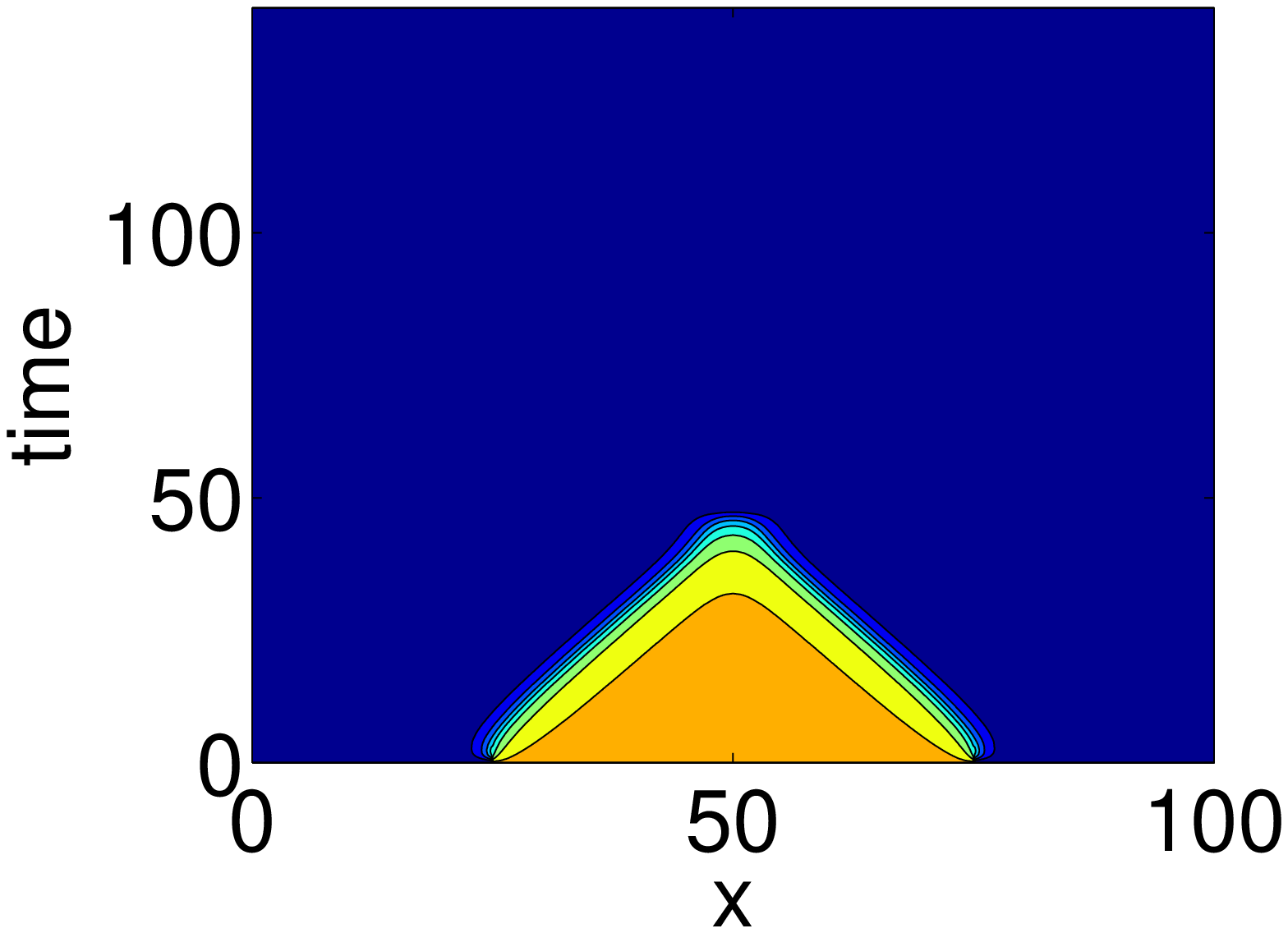} \label{fig:1dpf_k11}}
    \subfigure [$k=12$]
    {\includegraphics[width=4.1cm]{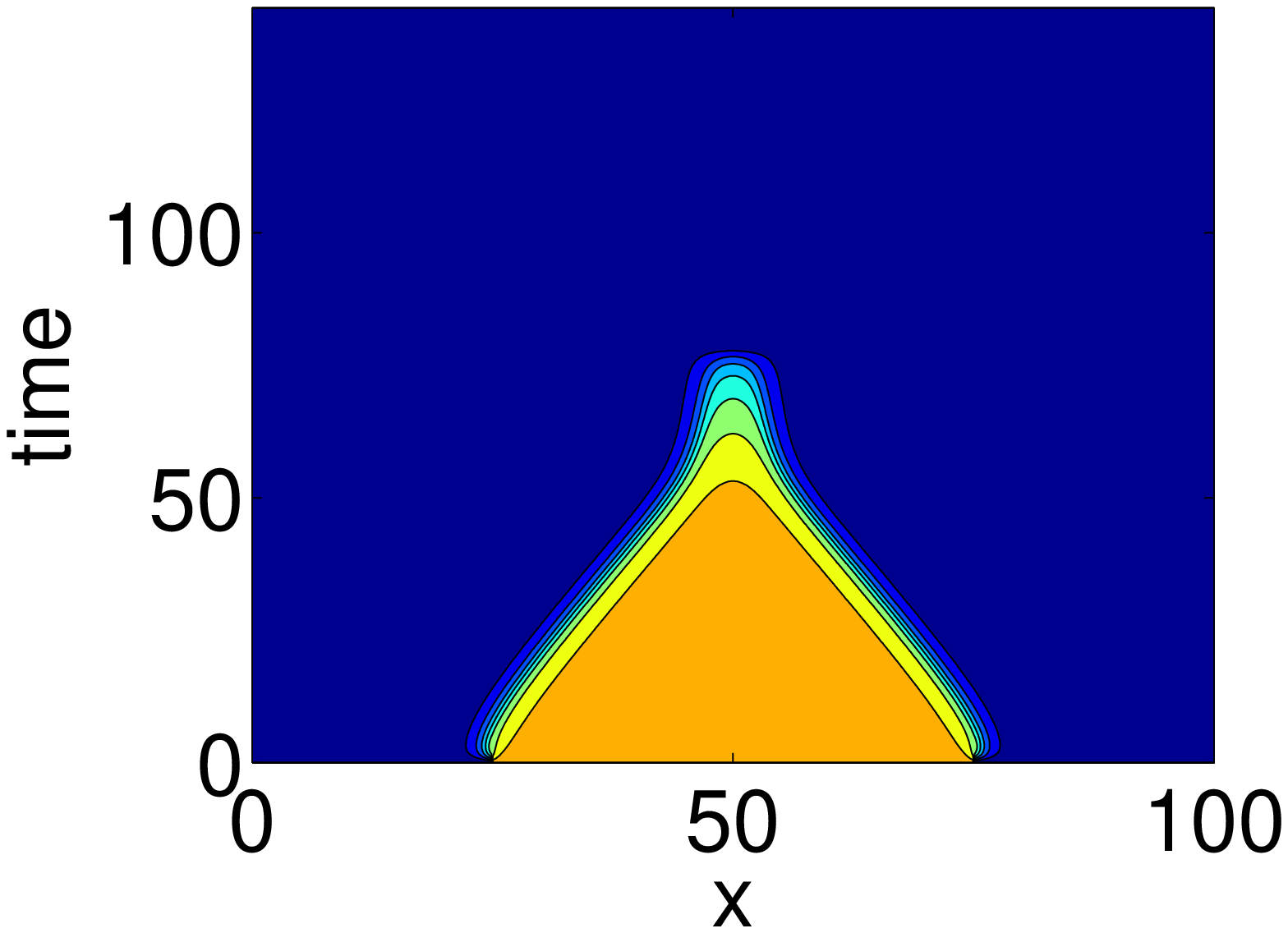}\label{fig:1dpf_k12}}
    \subfigure [$k=13$]
    {\includegraphics[width=4.1cm]{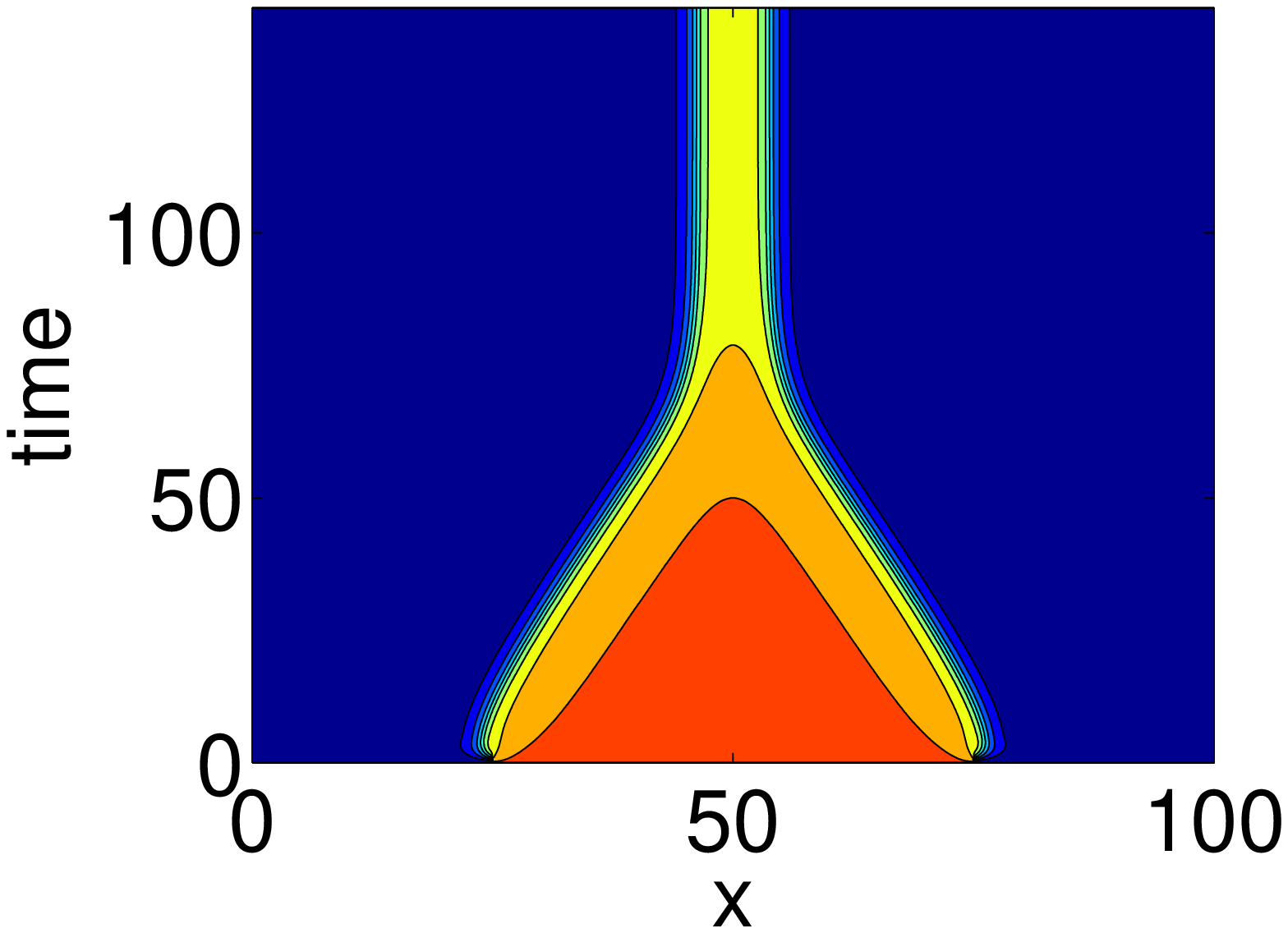}\label{fig:1dpf_k13}}
    \subfigure [$k=14$]
    {\includegraphics[width=4.1cm]{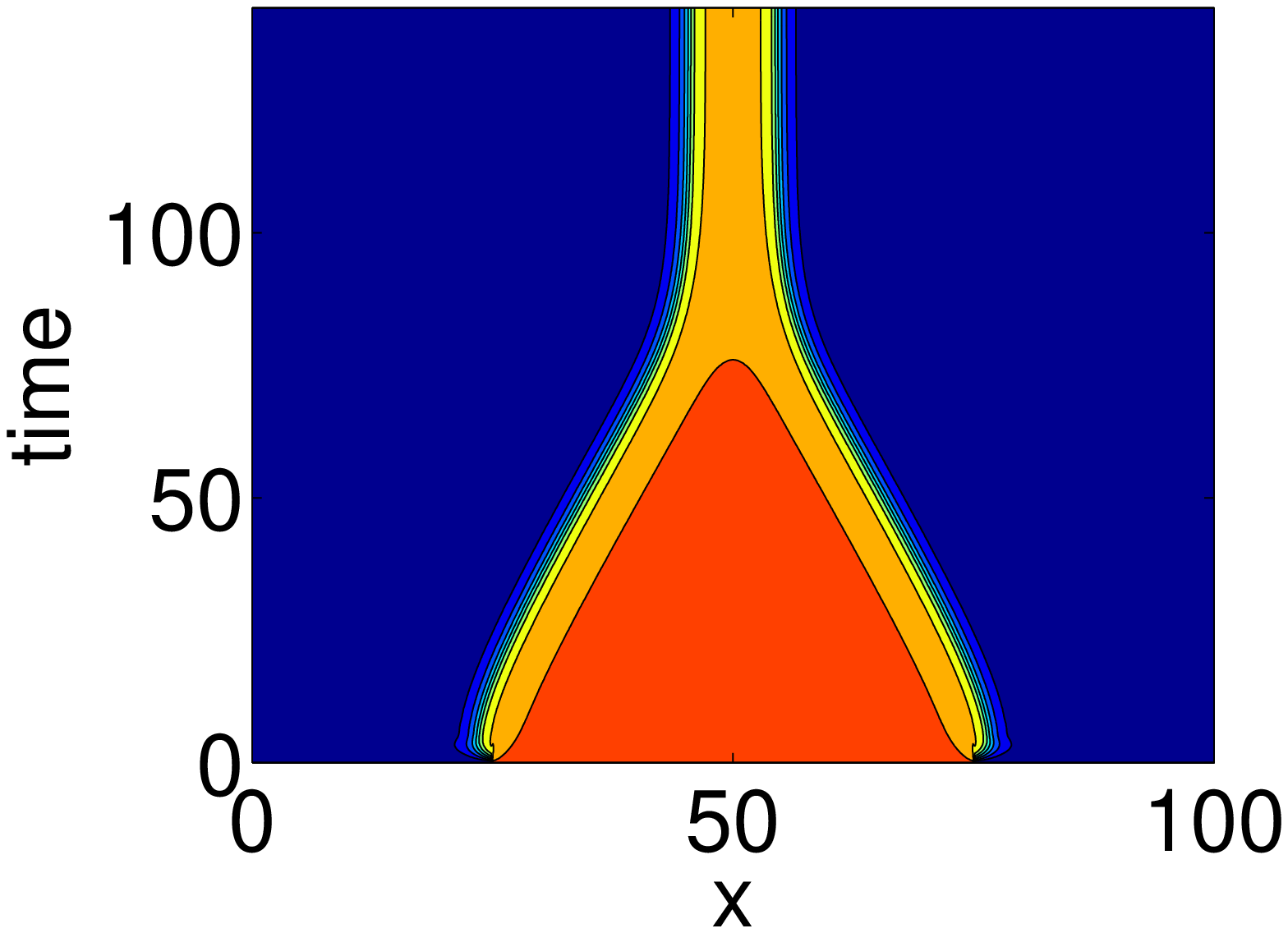}\label{fig:1dpf_k14}}
    \subfigure [$k=14.5$]
    {\includegraphics[width=4.1cm]{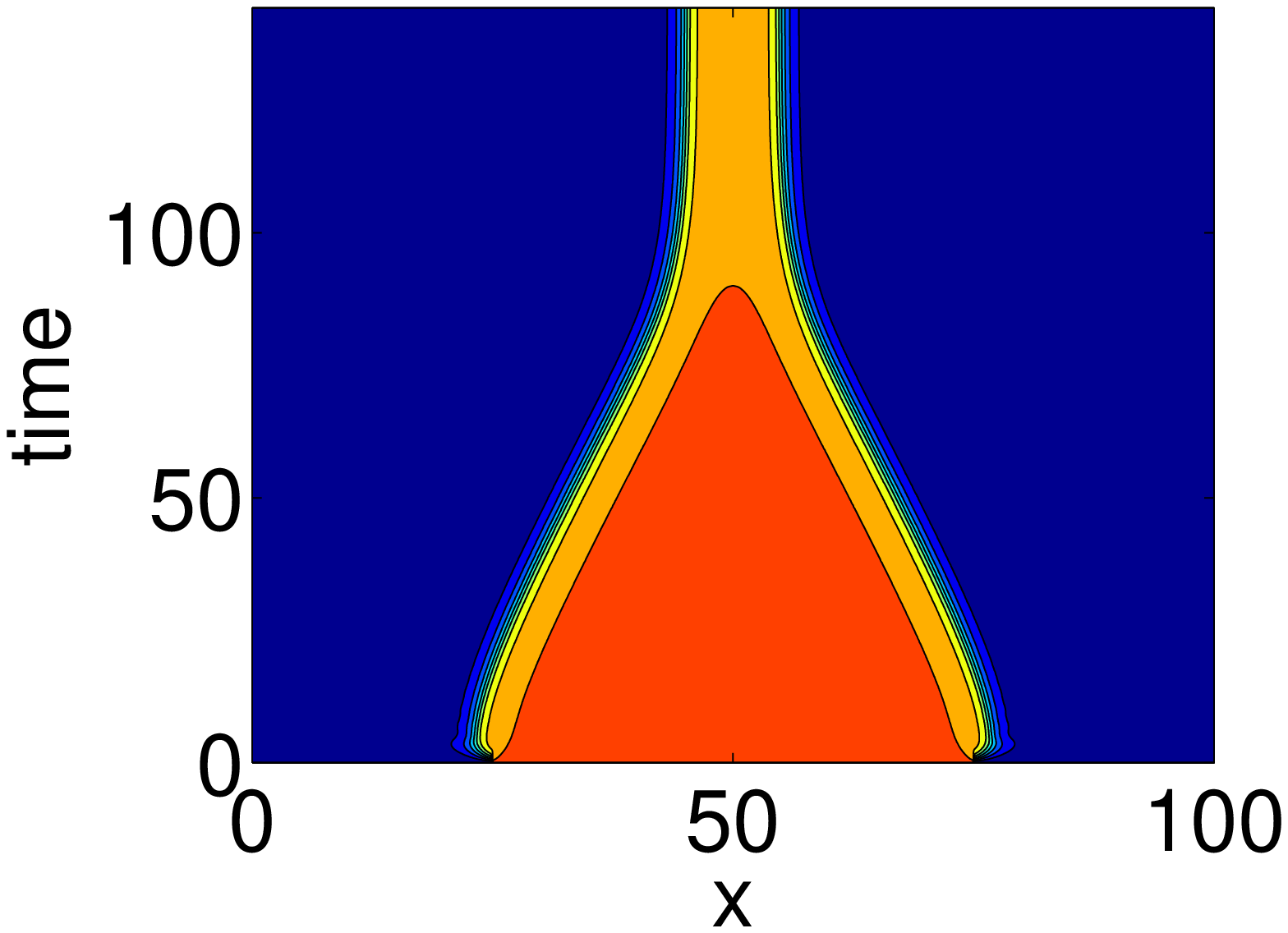}\label{fig:1dpf_k14_5}}
    \subfigure[$k=14.6$]
    {\includegraphics[width=4.1cm]{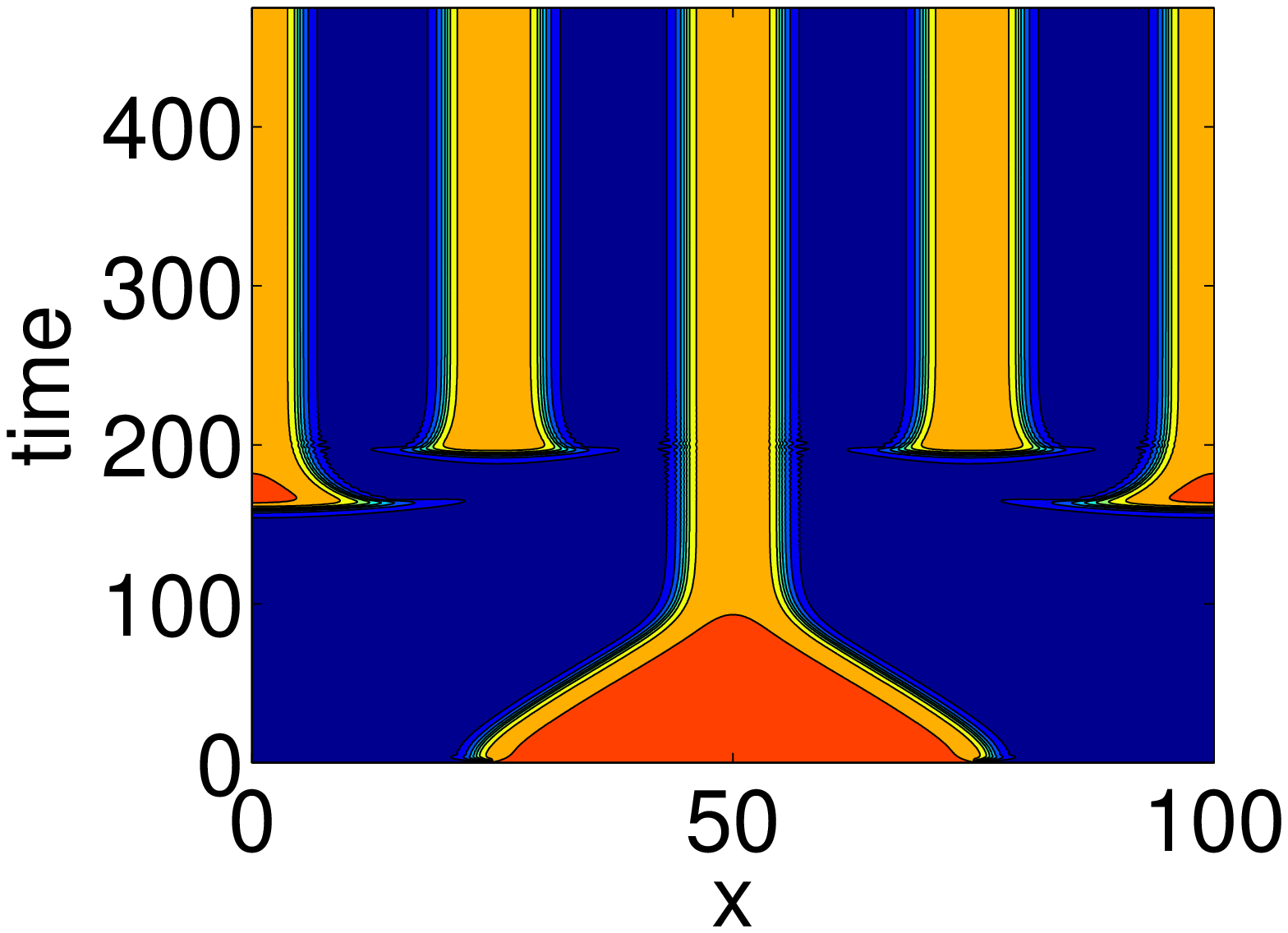} \label{fig:1dpf_k14_6}}  
    \subfigure[$k=14.8$]
    {\includegraphics[width=4.1cm]{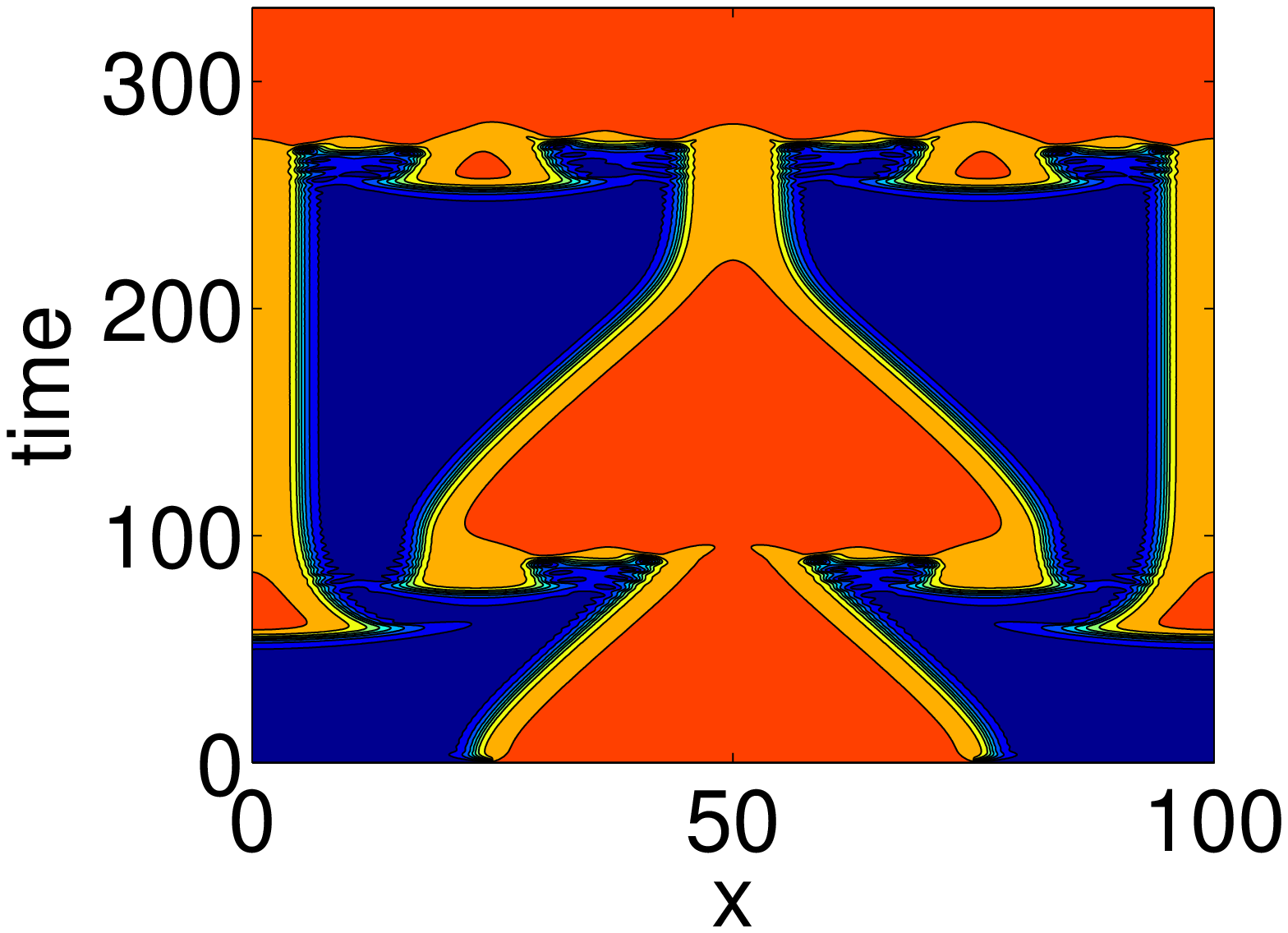} \label{fig:1dpf_k14_8}}  
    \subfigure [$k=15$]
    {\includegraphics[width=4.1cm]{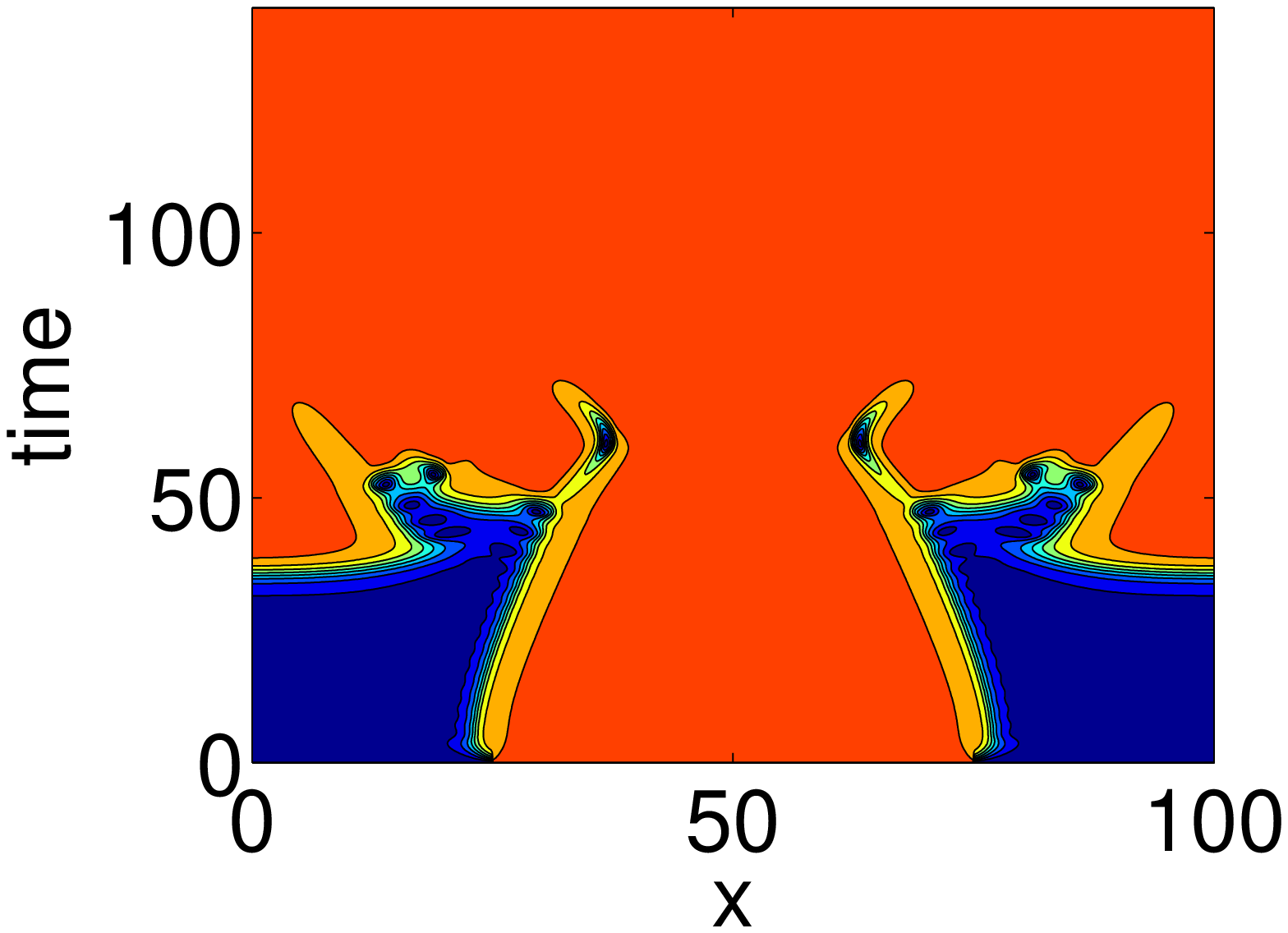}\label{fig:1dpf_k15}}
    \subfigure [$k=16$]
    {\includegraphics[width=4.1cm]{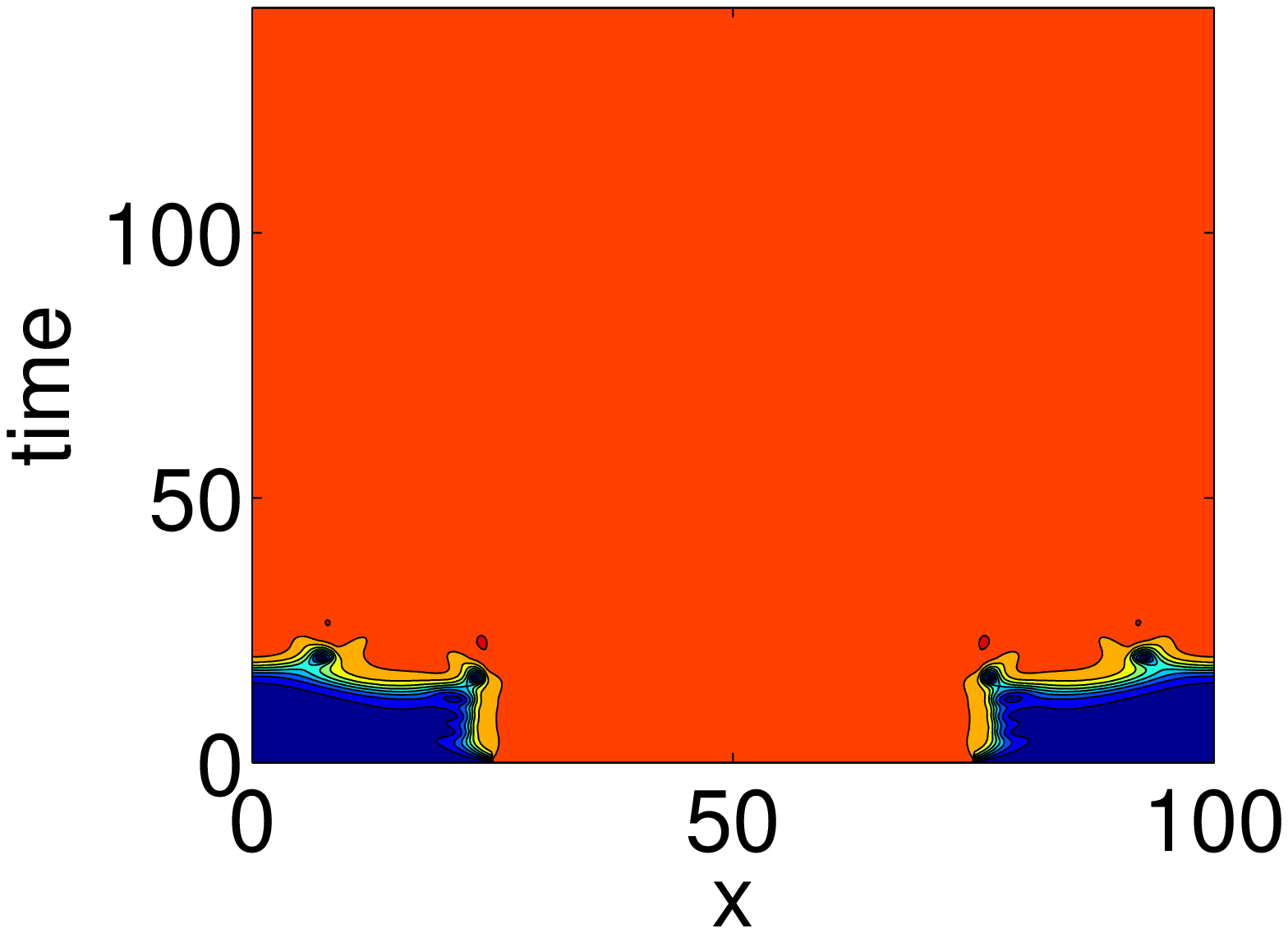}\label{fig:1dpf_k16}}
    \subfigure [$k=17$]
    {\includegraphics[width=4.1cm]{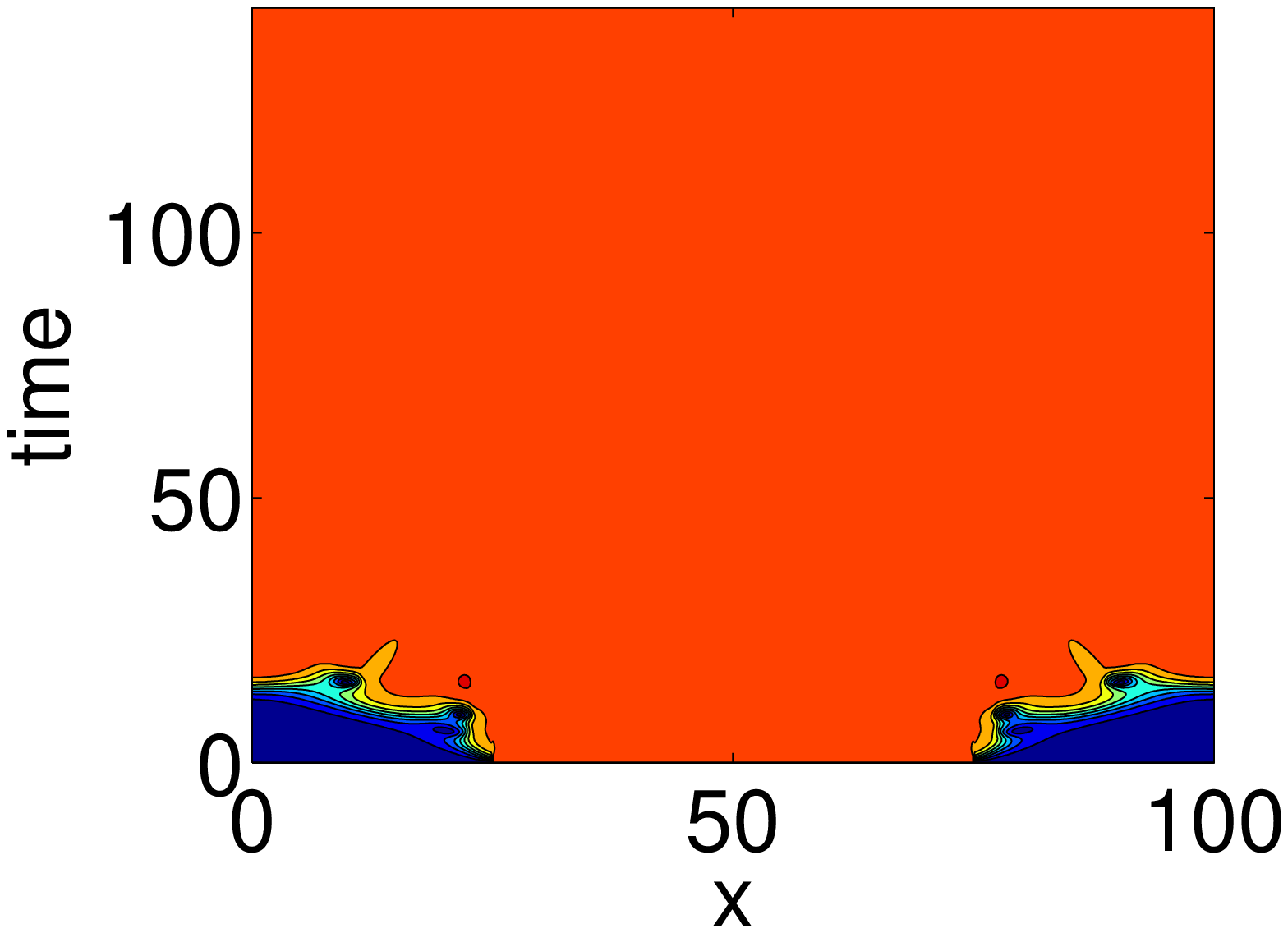}\label{fig:1dpf_k17}}
    \caption{
             A comparison of the time evolutions of $|r(x,t)|$ for 
             different values of $k$ where $r$ is initialized with 
             half of the interval
             at the coherent state ($25 \leq x \leq 75$)
             and half at the incoherent state. 
             Notice the difference
             in time scales of Fig. \ref{fig:1dpf_k14_6} and 
             Fig. \ref{fig:1dpf_k14_8} from other figures
             ($\omega_0=5, T=1, D=100$; periodic boundary conditions are imposed).}
    \label{fig:1dpf}
  \end{flushleft}
\end{figure*}

\begin{figure*}[h] 
  \begin{flushleft}
    \subfigure [$|r(x,t)|$]
    {\includegraphics[width=4.1cm]{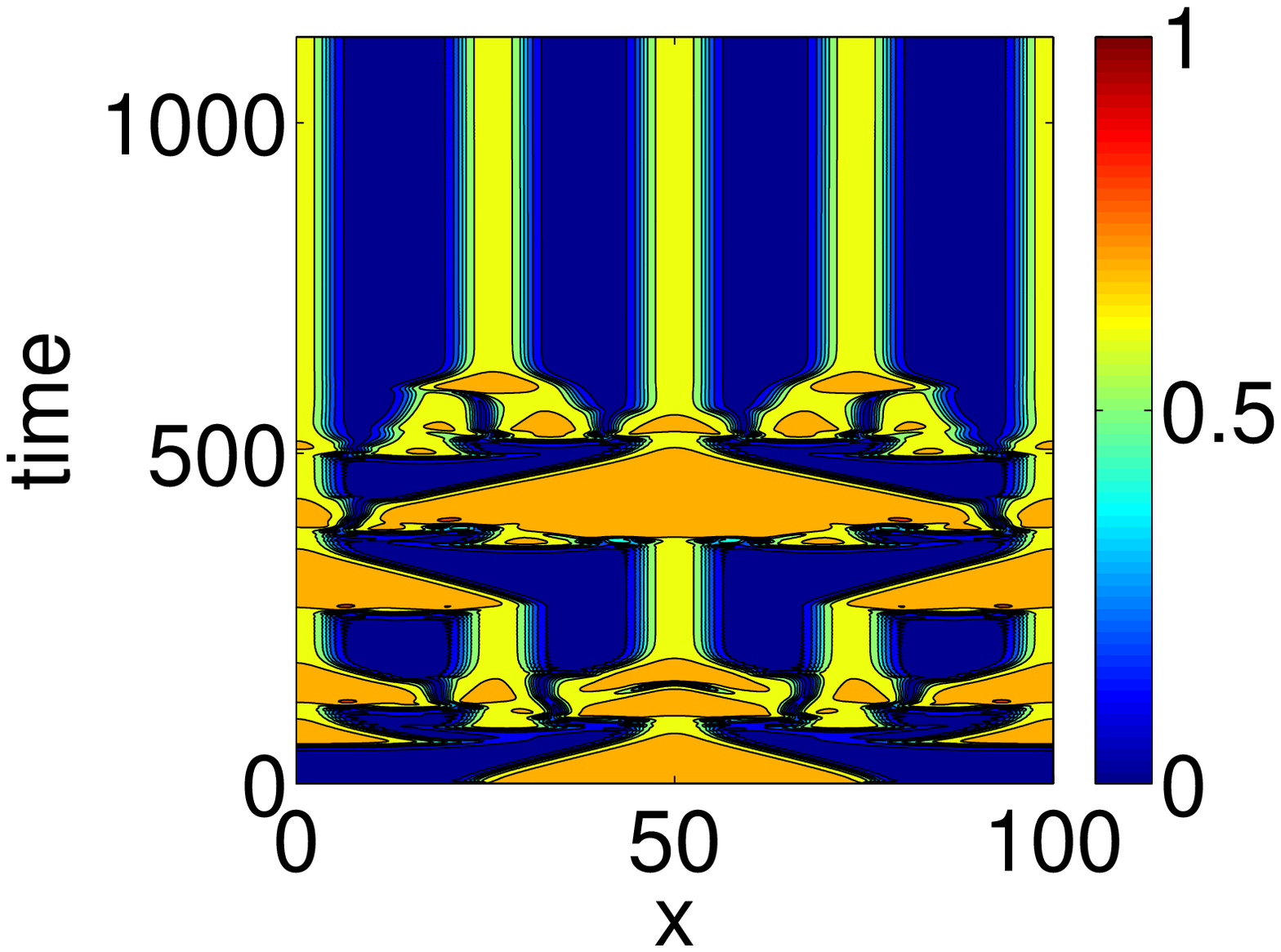} \label{fig:glassy_abs}}
    \subfigure [$|r(x,t)|$]
    {\includegraphics[width=4.1cm]{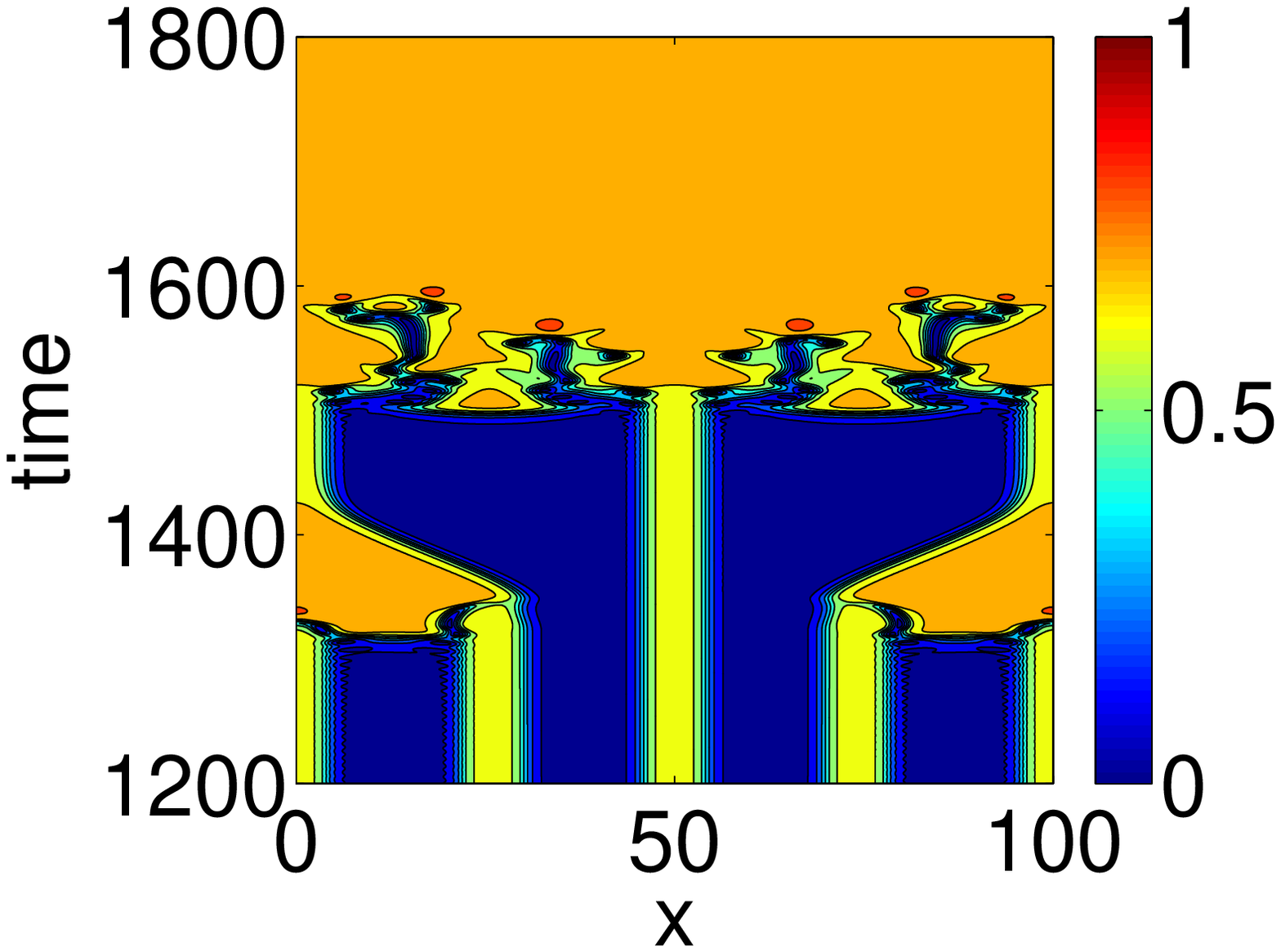} \label{fig:glassy_abs_p2}}
    \subfigure [$\theta(x,t)$ in the plateau centered at $x=50$ and $t=420$.]
    {\includegraphics[width=4.1cm]{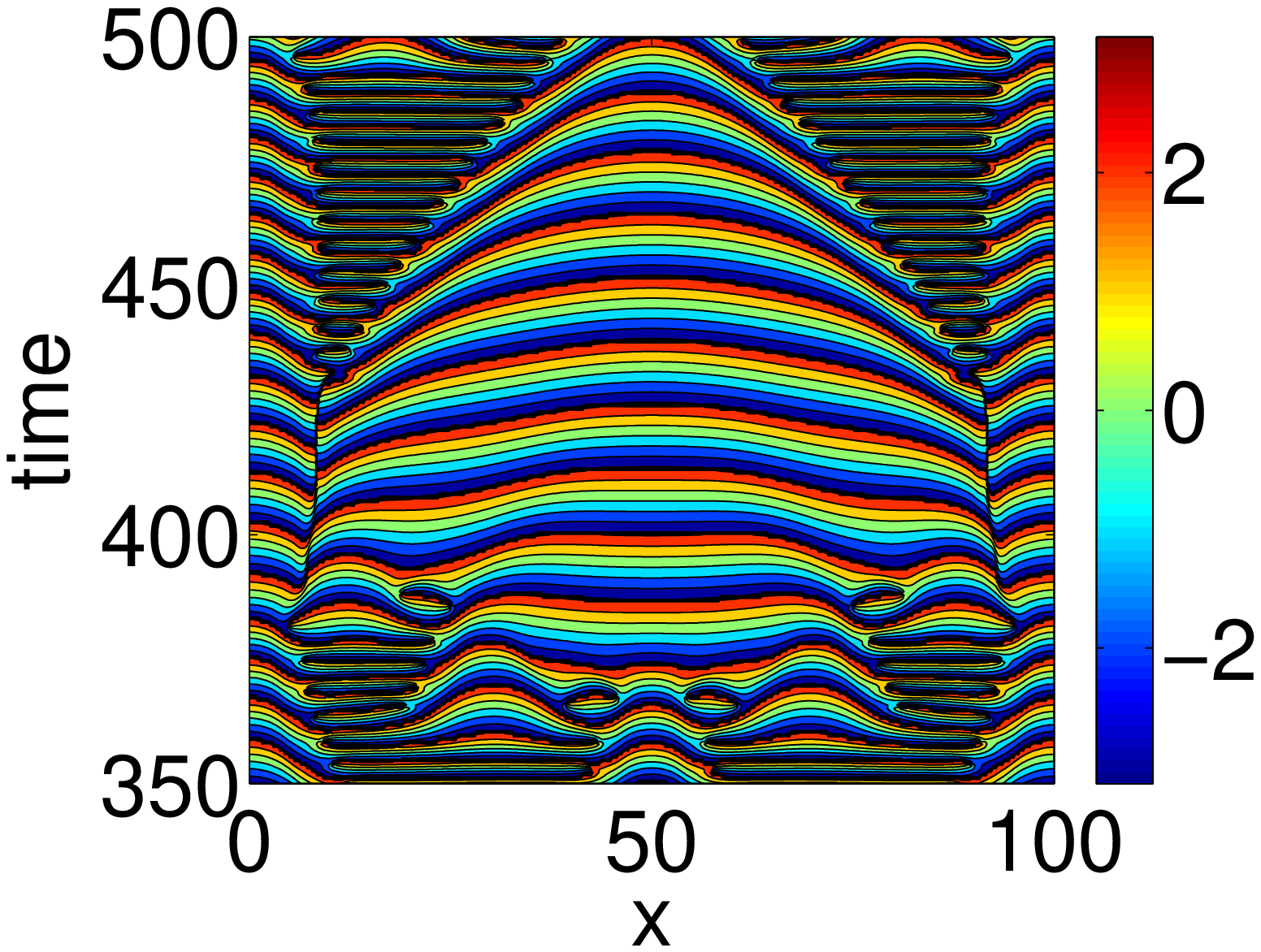}\label{fig:glassy_plateau}}
    \subfigure [$\theta(x,t)$ in part of the 
                 multiple-bridge pattern for $700 \leq  t \leq 1300$.]
    {\includegraphics[width=4.1cm]{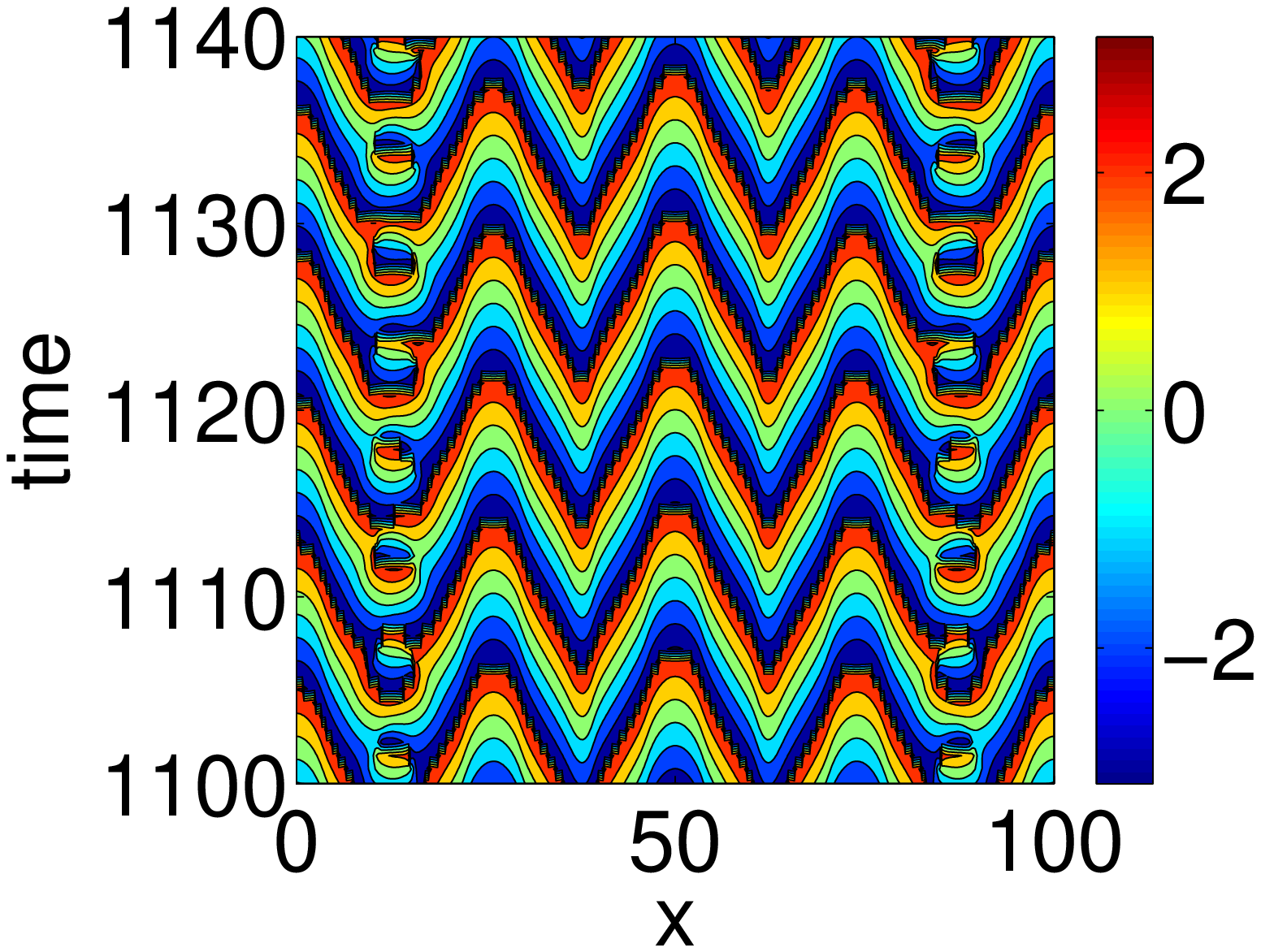}\label{fig:glassy_bridge}}
    \subfigure [$|r(x,148)|$]
    {\includegraphics[width=4.1cm]{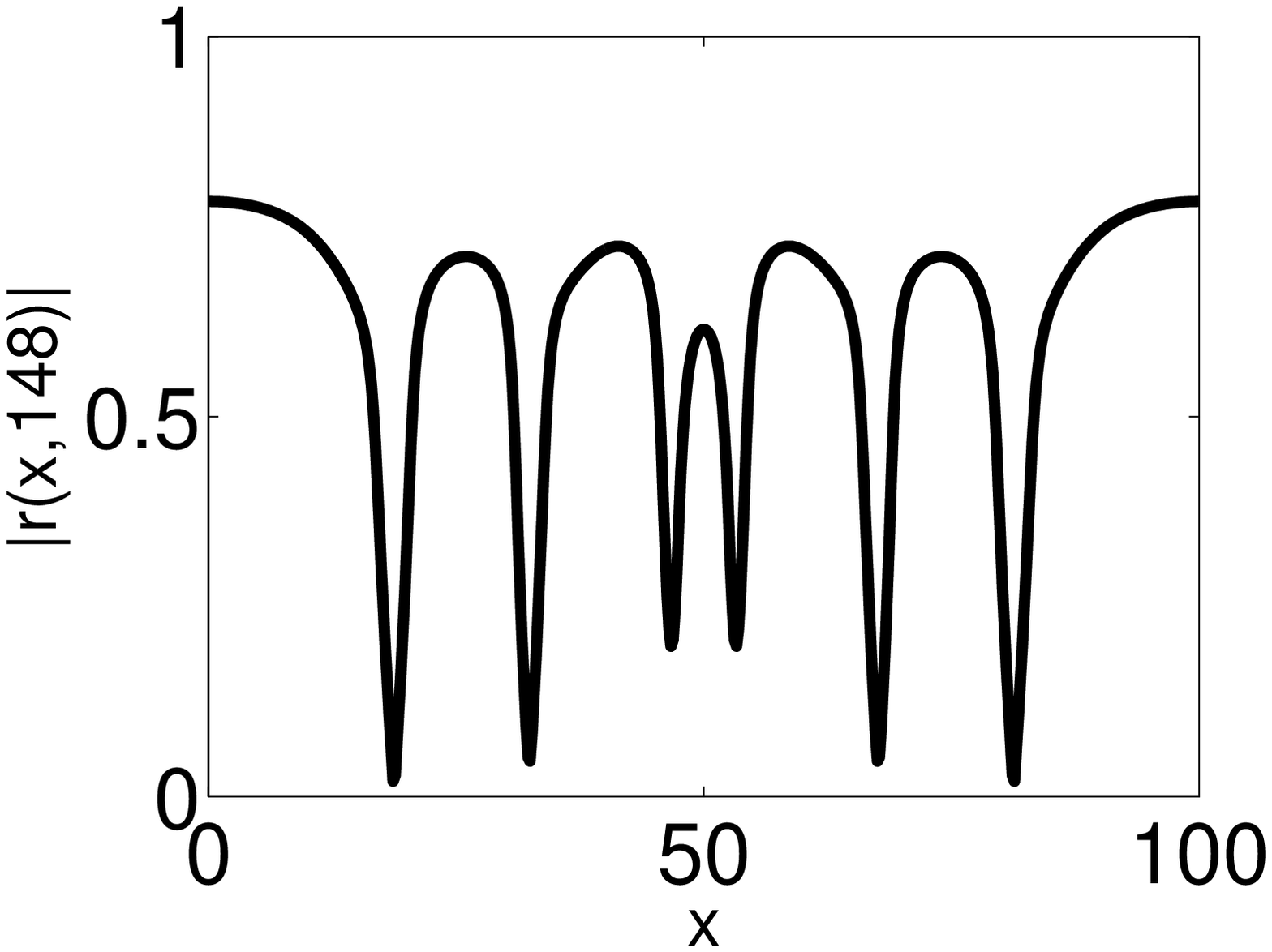}\label{fig:glassy_abs_t148}}
    \subfigure [$\theta(x,148)$]
    {\includegraphics[width=4.1cm]{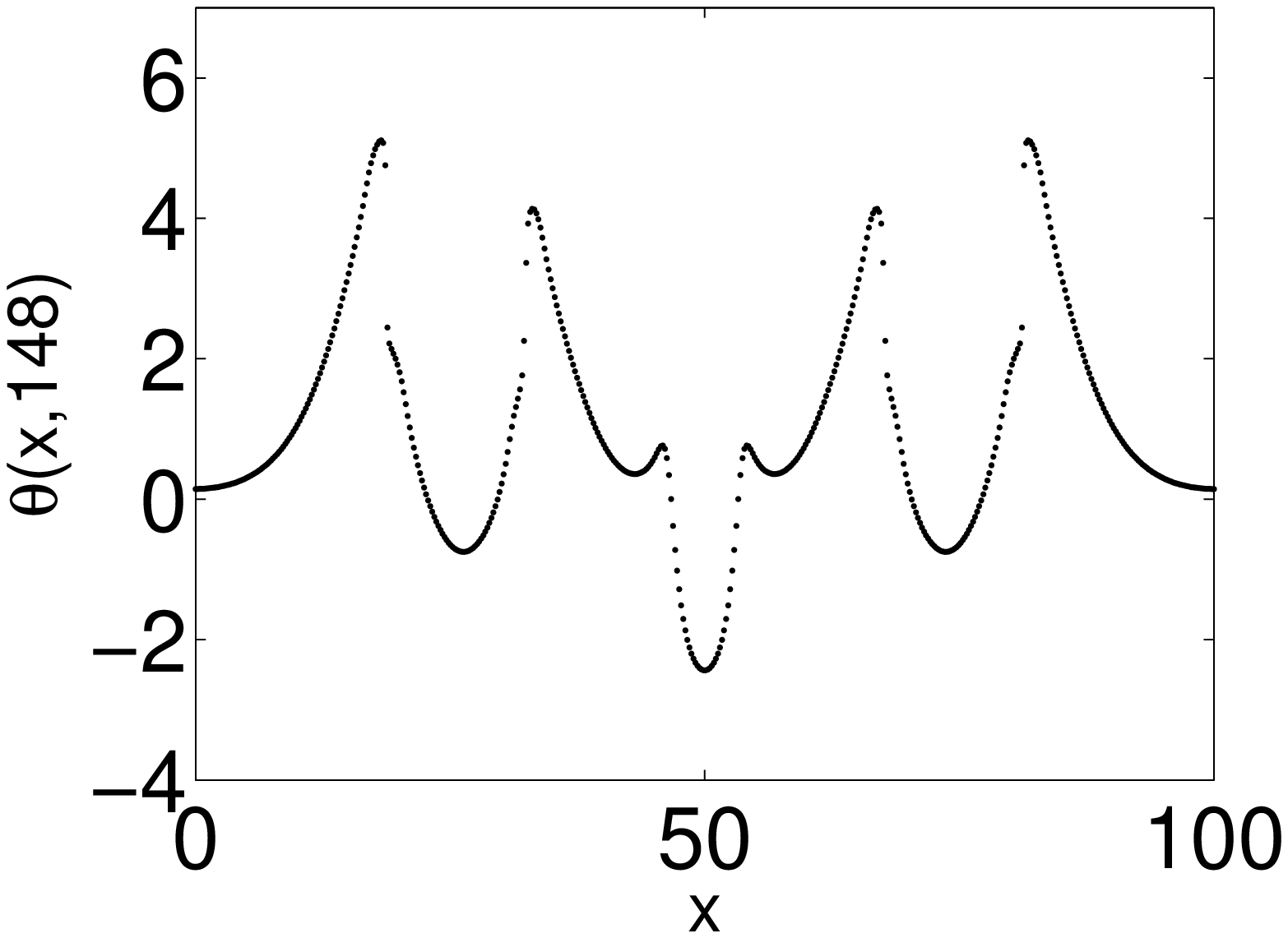}\label{fig:glassy_ph_t148}}
    \subfigure [$|r(x,1200)|$]
    {\includegraphics[width=4.1cm]{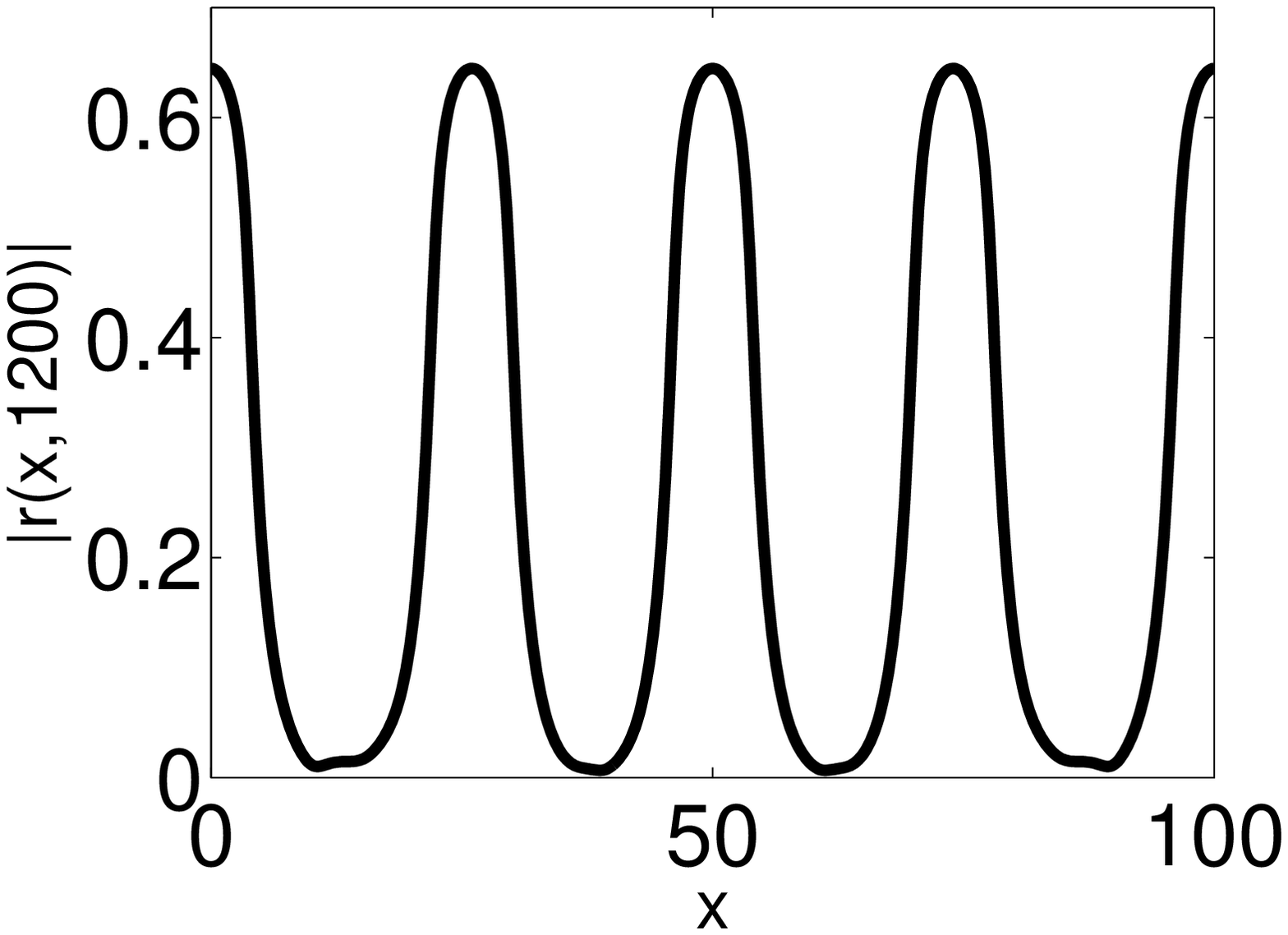}\label{fig:glassy_abs_t1200}}
    \subfigure [$\theta(x,1200)$]
    {\includegraphics[width=4.1cm]{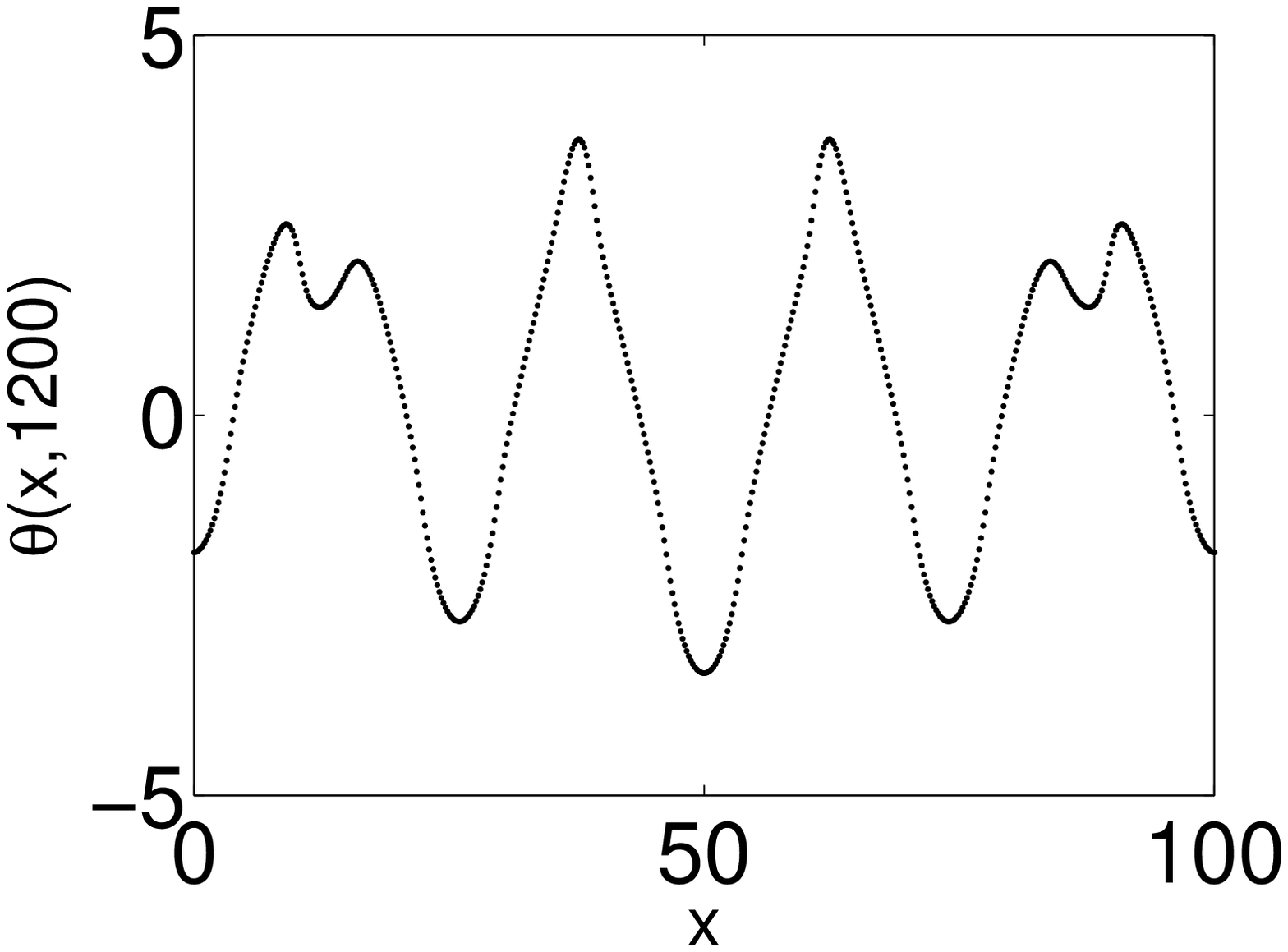}\label{fig:glassy_ph_t1200}}
  \end{flushleft}
  \caption{
           (a,b) Glassy state of transition, formation of plateaus of coherent regions 
           and hole patterns, and
           final evolution into the homogeneous coherent state.
           (c,d) Phase evolution in the plateau and multiple-bridge regions.
           (e,f) $|r(x,148)|$ and $\theta(x,148)$. (g,h) 
            $|r(x,1200)|$ and $\theta(x,1200)$  
           ($\omega_0=4, T=1, D=100, k=10.3 \hspace{1mm} (> k_c=10)$; initial condition:
           $r$ is given by the homogeneous coherent state solutions for 
           $25 \leq x \leq 75$, and $r = 0$ otherwise; periodic
           boundary conditions are imposed).}
 \label{fig:glassy}
\end{figure*}

Next, we consider the dynamics as a function of the coupling strength $k$. 
Recall from Fig. \ref{fig:hys_ome05} that there is a 
hysteretic region of coexisting coherent and incoherent states for the region
$k'_c < k < k_c$ where $k'_c=10$ and $k_c=14.5$.
Figure \ref{fig:1dpf} shows the 
time evolution of
$|r(x,t)|$ as a function of $k$. When the state is initialized
with half ($25 \leq x \leq 75$) the domain in the homogeneous incoherent state
and the remaining half in the homogeneous coherent state,
it is seen that 
if $k$ is sufficiently close to $k'_c$, 
then the incoherent region engulfs the 
coherent region, while if $k$ is sufficiently greater than $k_c$, the 
homogeneous coherent solution
takes over, and by
comparing Figs. \ref{fig:1dpf_k10} to \ref{fig:1dpf_k14},
we find that the propagation velocity decreases as $k$ is increased
toward $k_c$. As $k$ increases past $k \sim 12$, the simple propagating front phenomenon
seen in Figs. \ref{fig:Inhomostab} and \ref{fig:1dpf_k10}-\ref{fig:1dpf_k12}
is replaced by more complex behavior.
For example, in Fig. \ref{fig:1dpf_k13} we observe
the formation of a ``bridge'' at $k=13$ ($<k_c$),
i.e., a narrow stable coherent region sandwiched between 
two broad incoherent regions. This solution is apparently a long-time stable
state. It develops as the two propagating fronts collapsing the coherent
regions slow to a halt as they approach each other.
We note further that the bridge has an amplitude
which is smaller than that of the stable homogeneous solution, and the
oscillation frequency is different as well (graphs not shown).
Further, this bridge type
solution persists for $k > k_c$, and can give
rise to further intriguing dynamics like multiple bridges,
as shown in 
Figs. \ref{fig:1dpf_k14_6}, and even more vigorous behaviors of 
merging and re-creation of plateaus of coherent regions
and bridges, as seen in Fig. \ref{fig:1dpf_k14_8}.
Comparing Fig. \ref{fig:1dpf_k14_5} to \ref{fig:1dpf_k14_8},
it is notable that a wide variety of evolutionary behaviors occurs within
a relatively small range in $k$, including the formation of single and
multiple bridges, as well as collapse and re-creation
of plateaus. 
Figure \ref{fig:glassy} studies the glassy-like
behavior related to that seen in Fig. \ref{fig:1dpf_k14_8}
at a slightly different set of system parameters.
The figure shows
plateaus of coherent
regions (orange triangles in Figs. \ref{fig:glassy_abs} and \ref{fig:glassy_abs_p2})
and bridge-like patterns (yellow stripes), connected through dynamical 
creation, merging 
and re-creation of such structures
until the system eventually evolves
into the homogeneous coherent state. Figure \ref{fig:glassy_plateau}
shows the phase evolution inside the plateau region (orange triangle) of
Fig. \ref{fig:glassy_abs} centered at $t \approx 420, x \approx 50$.
Figure \ref{fig:glassy_bridge} shows the phase evolution corresponding
to the four-bridge-structure between the top of Fig. \ref{fig:glassy_abs} and
the bottom of Fig. \ref{fig:glassy_abs_p2} ($700 \leq t \leq 1300$).
We note that within a plateau, the whole
region oscillates roughly homogeneously 
(see the nearly parallel evolving fronts in Fig. \ref{fig:glassy_plateau}), and
each bridge pattern functions as
a sink of incoming waves (see the zig-zag-like pattern in Fig. \ref{fig:glassy_bridge}). 
Further important dynamical characteristics
during this vigorous glassy-like transition state are revealed in
Figs. \ref{fig:glassy_abs_t148} and
\ref{fig:glassy_ph_t148}, which show 
$|r|$ and $\theta$ (where $r= |r|e^{i\theta}$)
respectively at $t=148$. We see that
there are multiple hole-like patterns (deep dips in
$|r|$ in Fig. \ref{fig:glassy_abs_t148}), 
at which the phase changes 
sharply,
(see Fig. \ref{fig:glassy_ph_t148}, and 
note that the changes in phase for the outer two holes
appear to be virtually discontinuous, as discussed 
in more detail
shortly). In comparison,
for the multiple-bridge region at $t=1200$, 
Figs. \ref{fig:glassy_abs_t1200} and \ref{fig:glassy_ph_t1200} show that both 
$|r|$ and $\theta$ change smoothly in space. 

\begin{figure*}[h] 
  \begin{flushleft}
    \subfigure [$|r(x,t)|$]
    {\includegraphics[width=4.1cm]{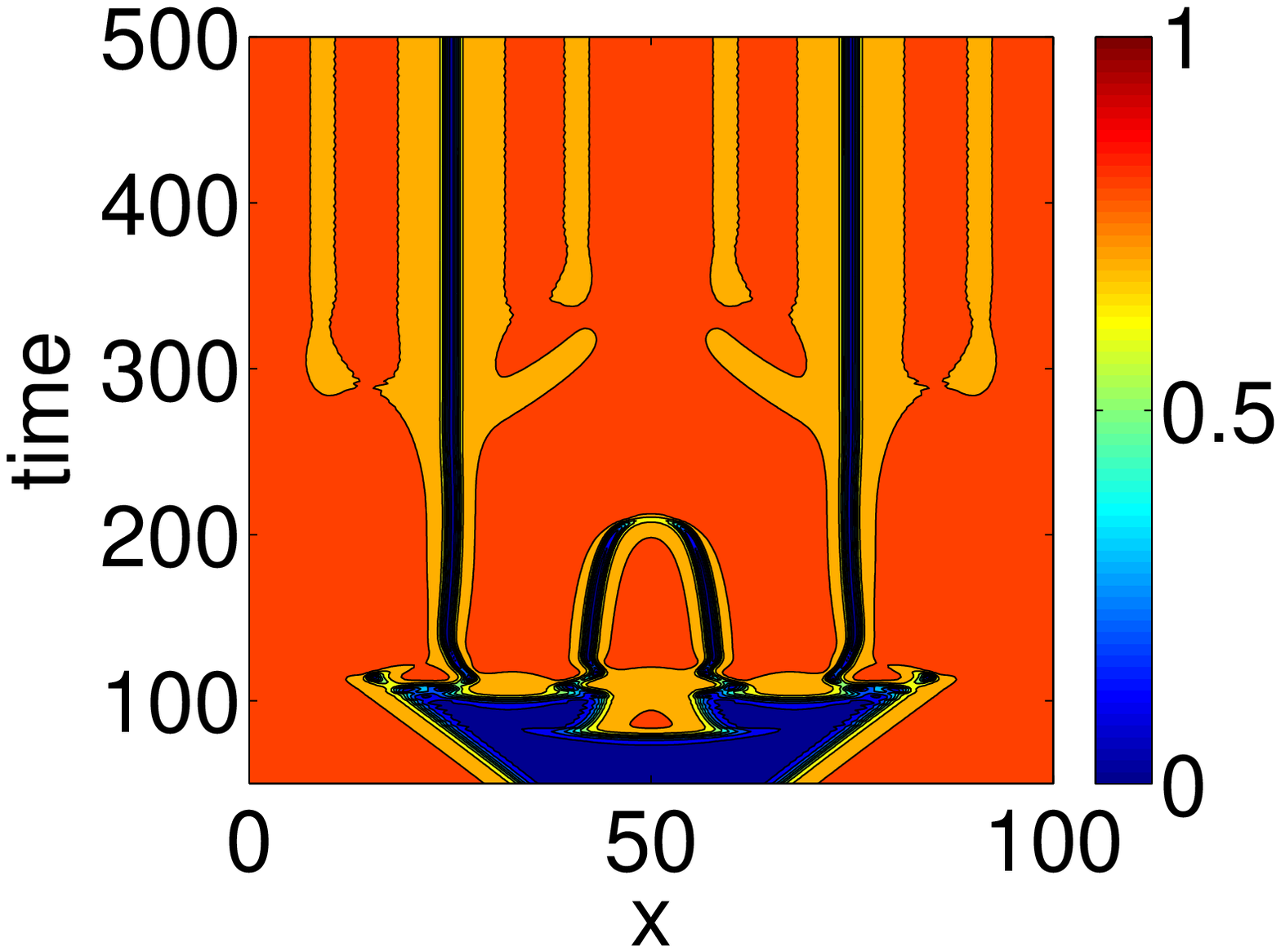} \label{fig:four_hole_abs}}
    \subfigure [$\theta(x,t)$]
    {\includegraphics[width=4.1cm]{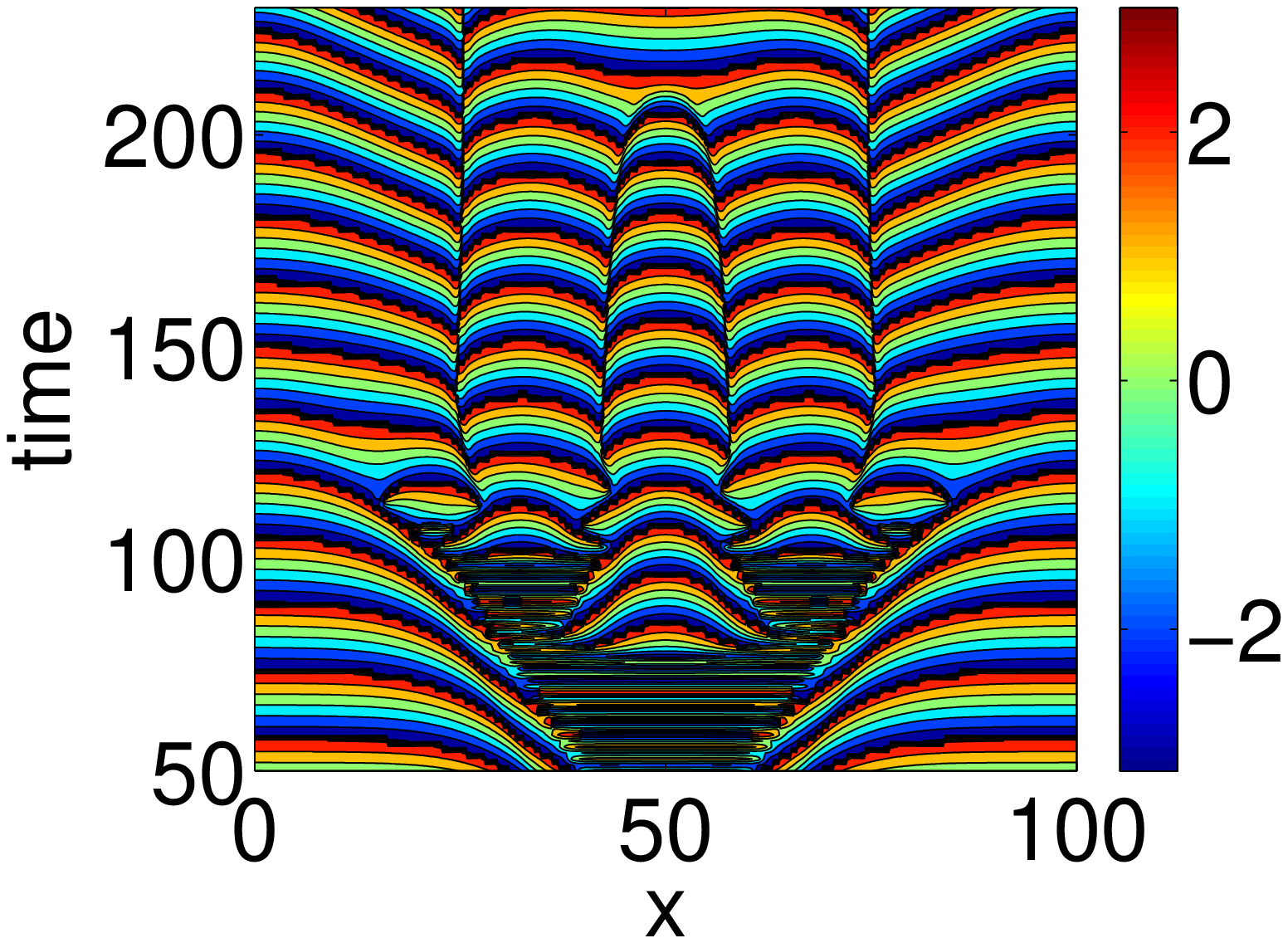} \label{fig:four_hole_ph_p1}}
    \subfigure [$|r(x,192)|$]
    {\includegraphics[width=4.1cm]{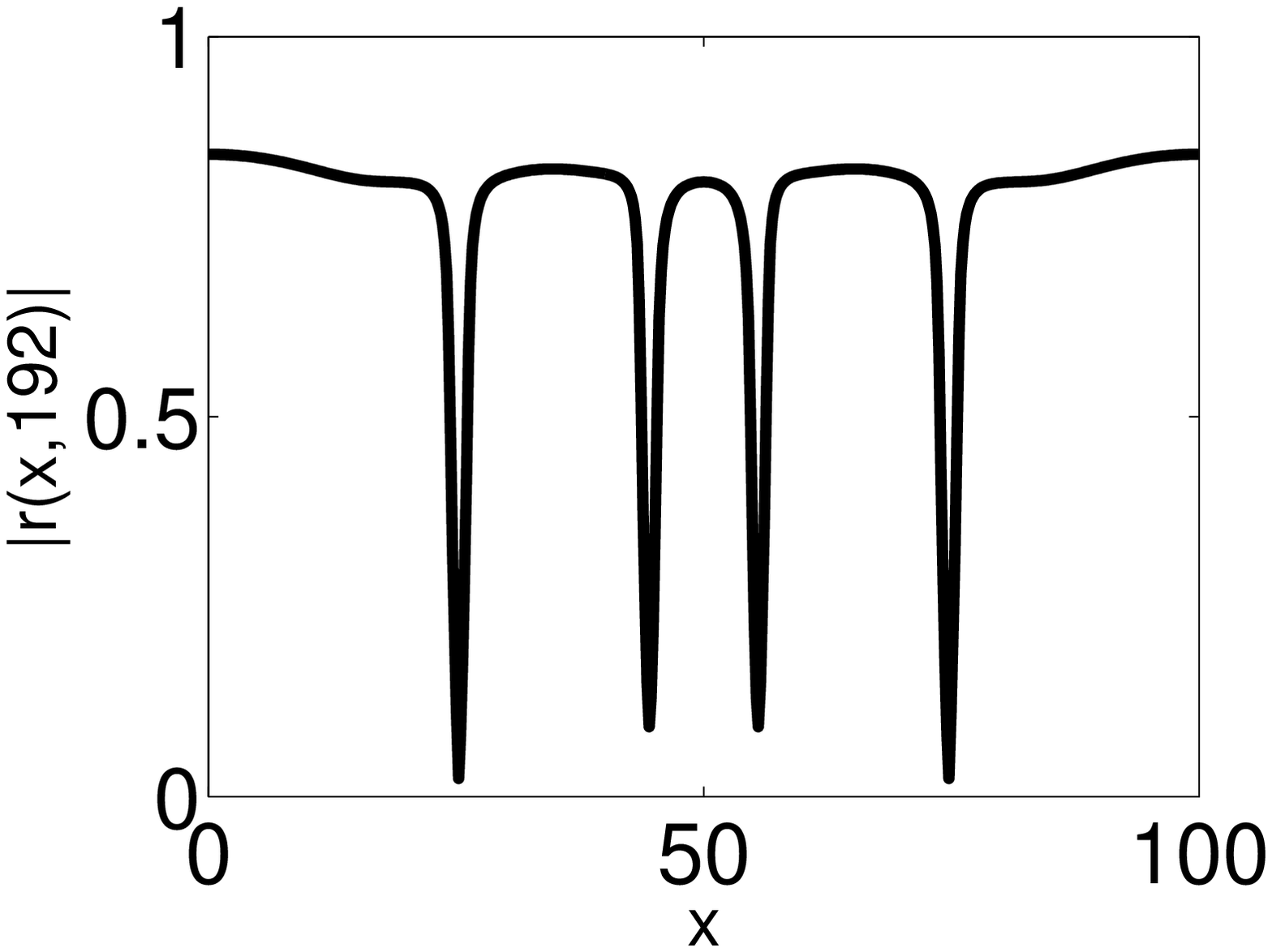}\label{fig:four_hole_abs_t192}}
    \subfigure [$\theta(x,192)$]
    {\includegraphics[width=4.1cm]{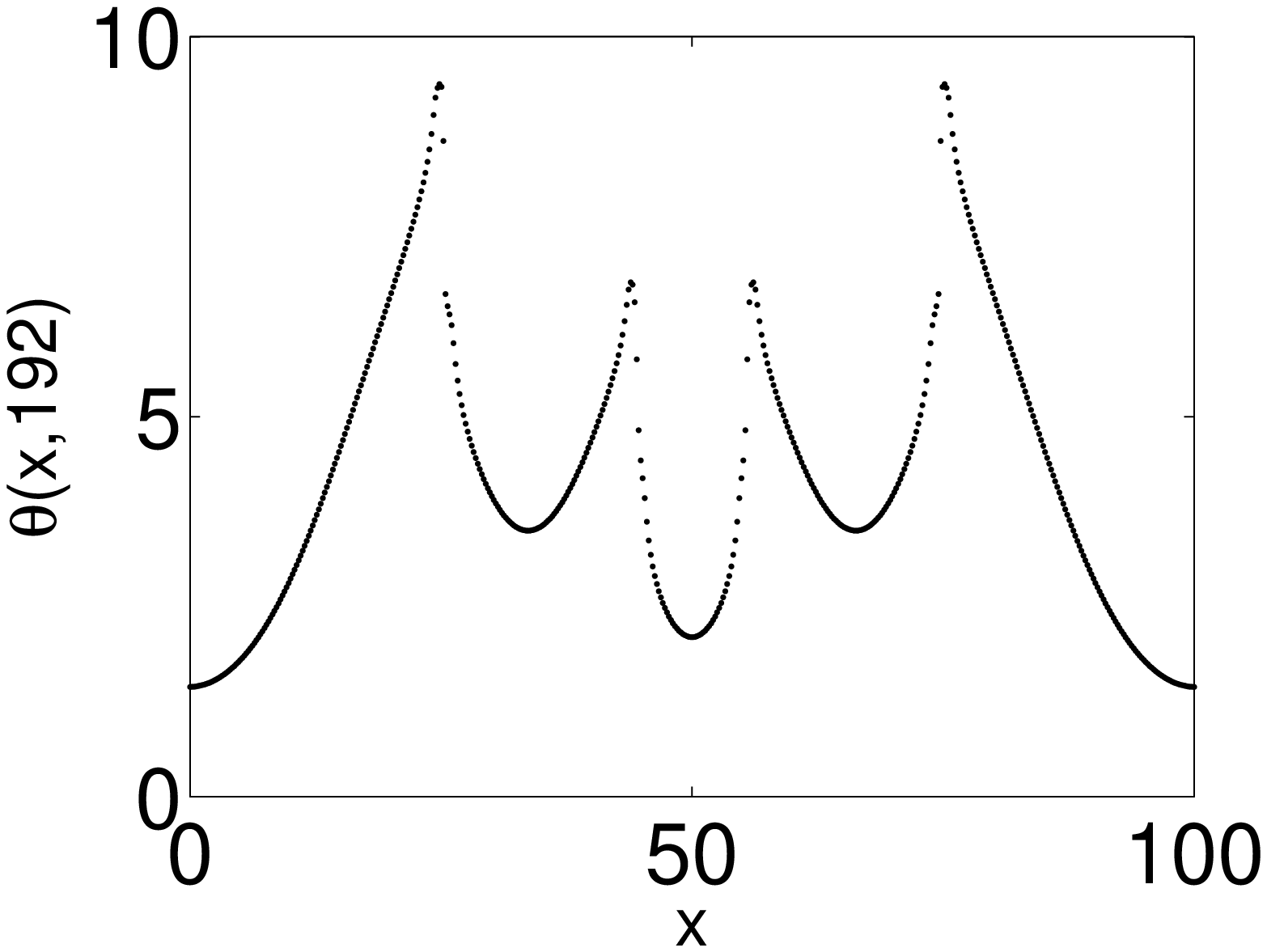}\label{fig:four_hole_ph_t192}}
    \subfigure [$\theta(x,t)$]
    {\includegraphics[width=4.1cm]{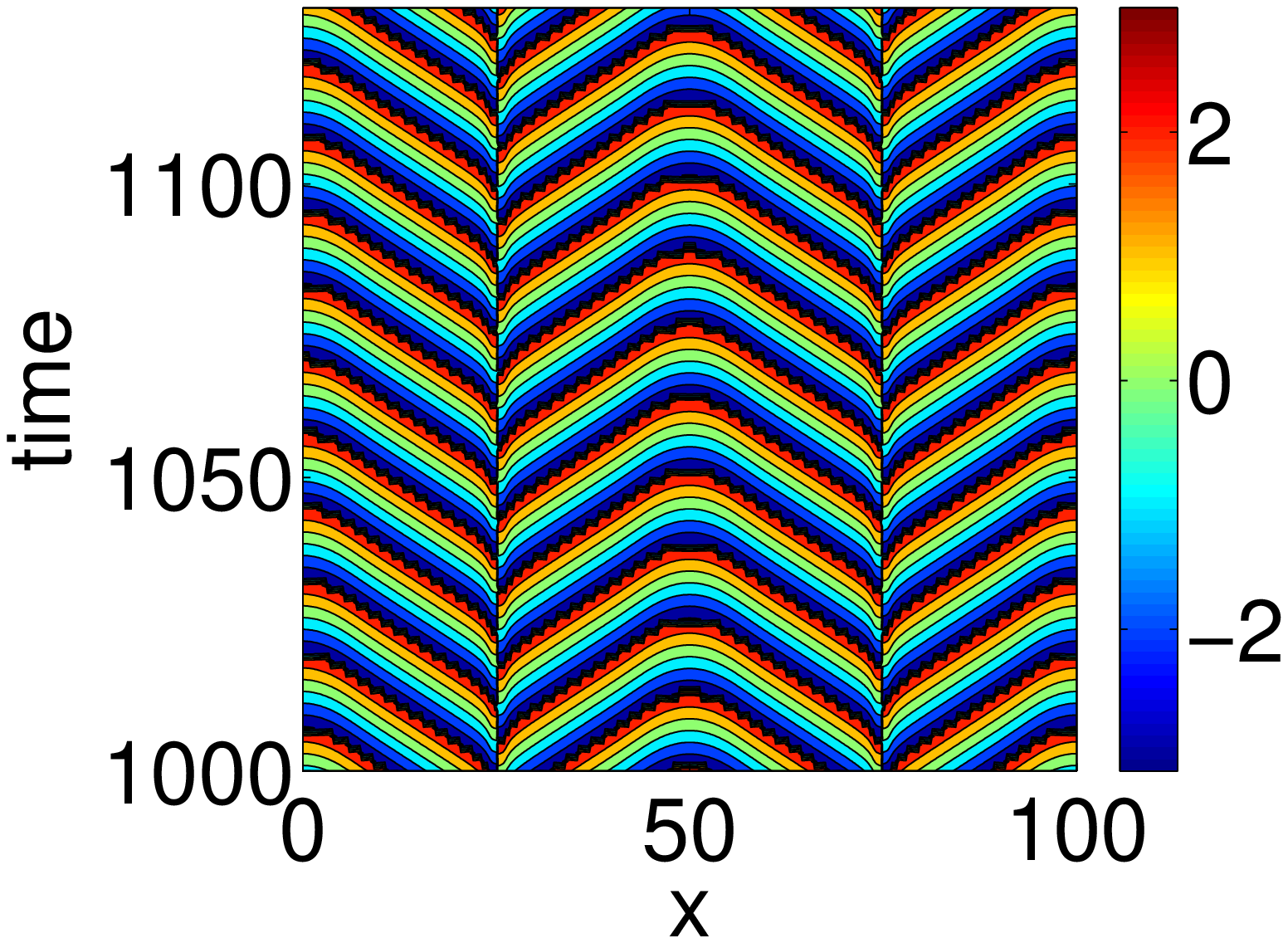} \label{fig:four_hole_ph_p2}}
    \subfigure [$r(x,1200)$]
    {\includegraphics[width=4.1cm]{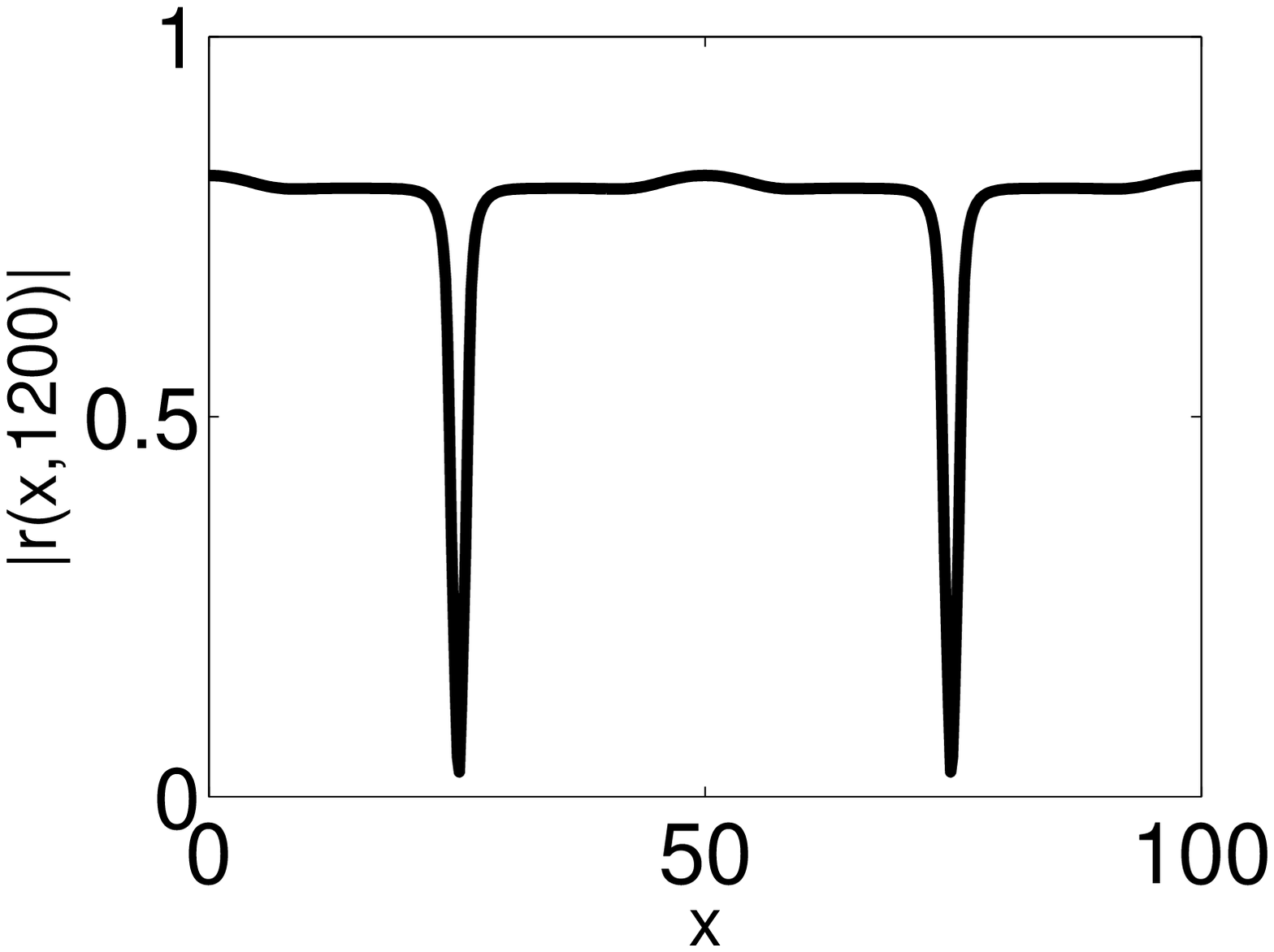}\label{fig:four_hole_abs_t1200}}
    \subfigure [$\theta(x,1200)$]
    {\includegraphics[width=4.1cm]{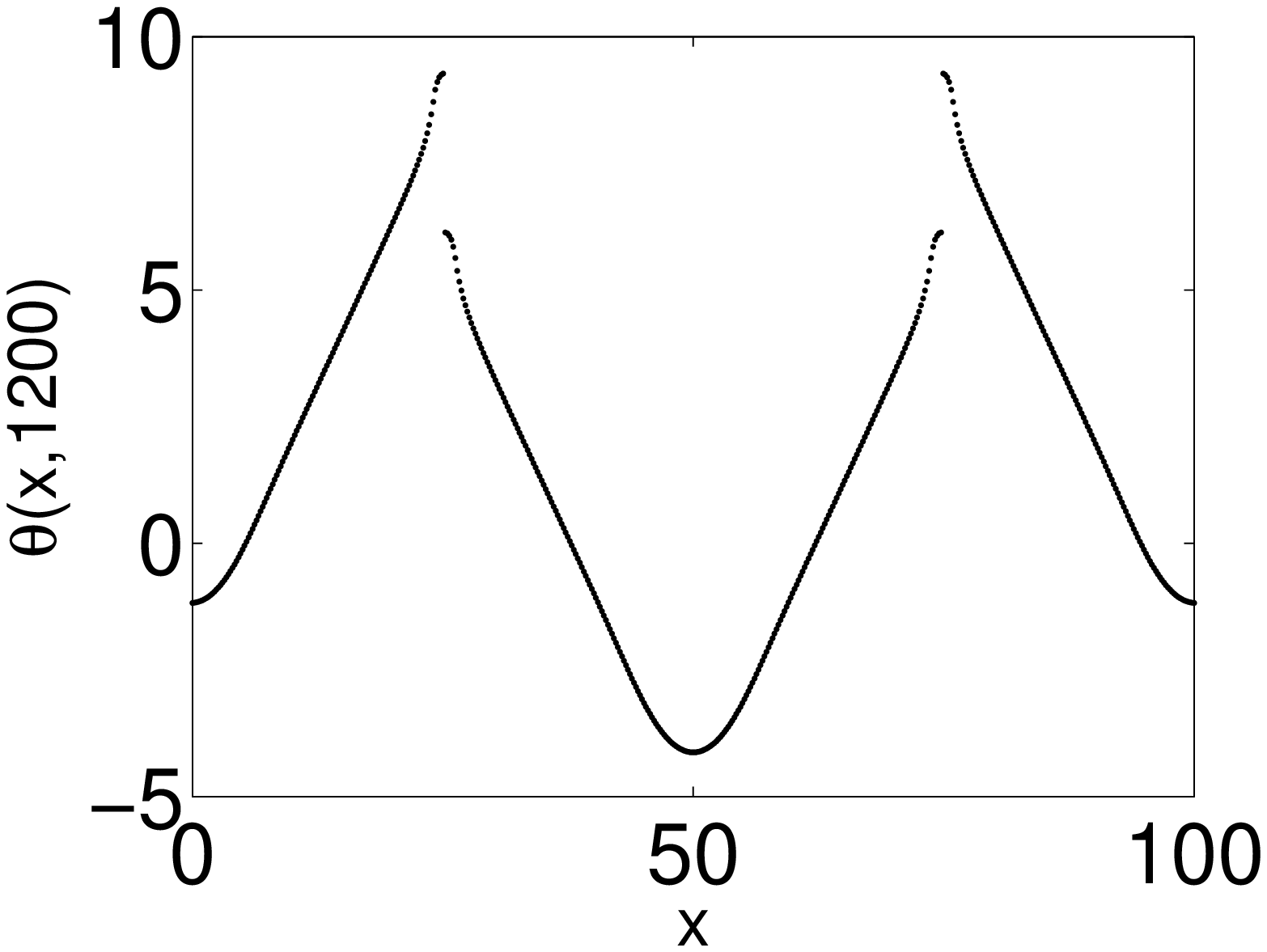}\label{fig:four_hole_ph_t1200}}
  \end{flushleft}
  \caption{
           Formation and dynamical evolution of hole patterns. 
           (a) $|r(x,t)|$. (b-d) Close-up
           views of four hole patterns with two inner traveling holes. 
           (e-g) Close-up views of two stationary
           hole patterns.
           ($\omega_0=5, T=1, D=100, k=14.8$; initial condition:
           $r$ is given by the homogeneous coherent state solutions for 
           $0 \leq x \leq 41$ and $59 \leq x \leq 100$, and $r=0$ otherwise; periodic
           boundary conditions are imposed).}
  \label{fig:four_hole}
\end{figure*}

Figure \ref{fig:four_hole} shows the dynamical
characteristics associated with the hole-like patterns
in another setting where these patterns dominate and are not
interspersed with other spatial features (like bridges and plateaus).
The figure corresponds to
the same parameters as those in
Fig. \ref{fig:1dpf_k14_8}, but initialized with
different incoherent and coherent regions. 
Compared
with Fig. \ref{fig:1dpf_k14_8}, there is a
relatively short time for the system to stay in the plateau-like
regions, and instead of settling in the homogeneous 
coherent state solution as in Fig. \ref{fig:1dpf_k14_8}, four distinct
hole-like patterns emerge (black lines starting at $t \approx 130$ in
Fig. \ref{fig:four_hole_abs}).
As time evolves, the two inner holes 
move toward each other and
annihilate, while the outer two continue to
evolve, apparently becoming stationary.
Note also that for the two merging holes, they approach each other
at a faster speed when they are closer to each other.
Examination of the phase evolution of the system
(Figs. \ref{fig:four_hole_ph_p1} and \ref{fig:four_hole_ph_p2})
suggests the center of each hole act as a source of plane waves,
in contrast with the bridge solution which acts as a sink
(see Fig. \ref{fig:glassy_bridge}).
For the inner two moving holes, while each is characterized by a dip
in magnitude (see Fig. \ref{fig:four_hole_abs_t192} at $t=192$), the dips decrease 
in magnitude as the two holes 
approach each other,
with the relative phase difference 
on the two sides of the hole center 
close to being continuous (see Fig.\ref{fig:four_hole_ph_t192}). However, if
the holes are stationary, e.g., the outer two holes in 
Figs. \ref{fig:four_hole_abs_t192} and \ref{fig:four_hole_abs_t1200},
each dip in $|r|$ is close to zero, with
the relative phase
difference on the two sides being an essentially discontinuous slip of $\pm 3\pi$. 
A further observation in the case of two stationary holes
is that there is a bump in $|r|$
half-way between them corresponding to the location at which
incoming waves emitted from the holes converge
(see $x = 50$ in Fig. \ref{fig:four_hole_abs_t1200}).

\begin{figure*}[h] 
  \begin{flushleft}
    \subfigure [$|r(x,t)|$]
    {\includegraphics[width=4.1cm]{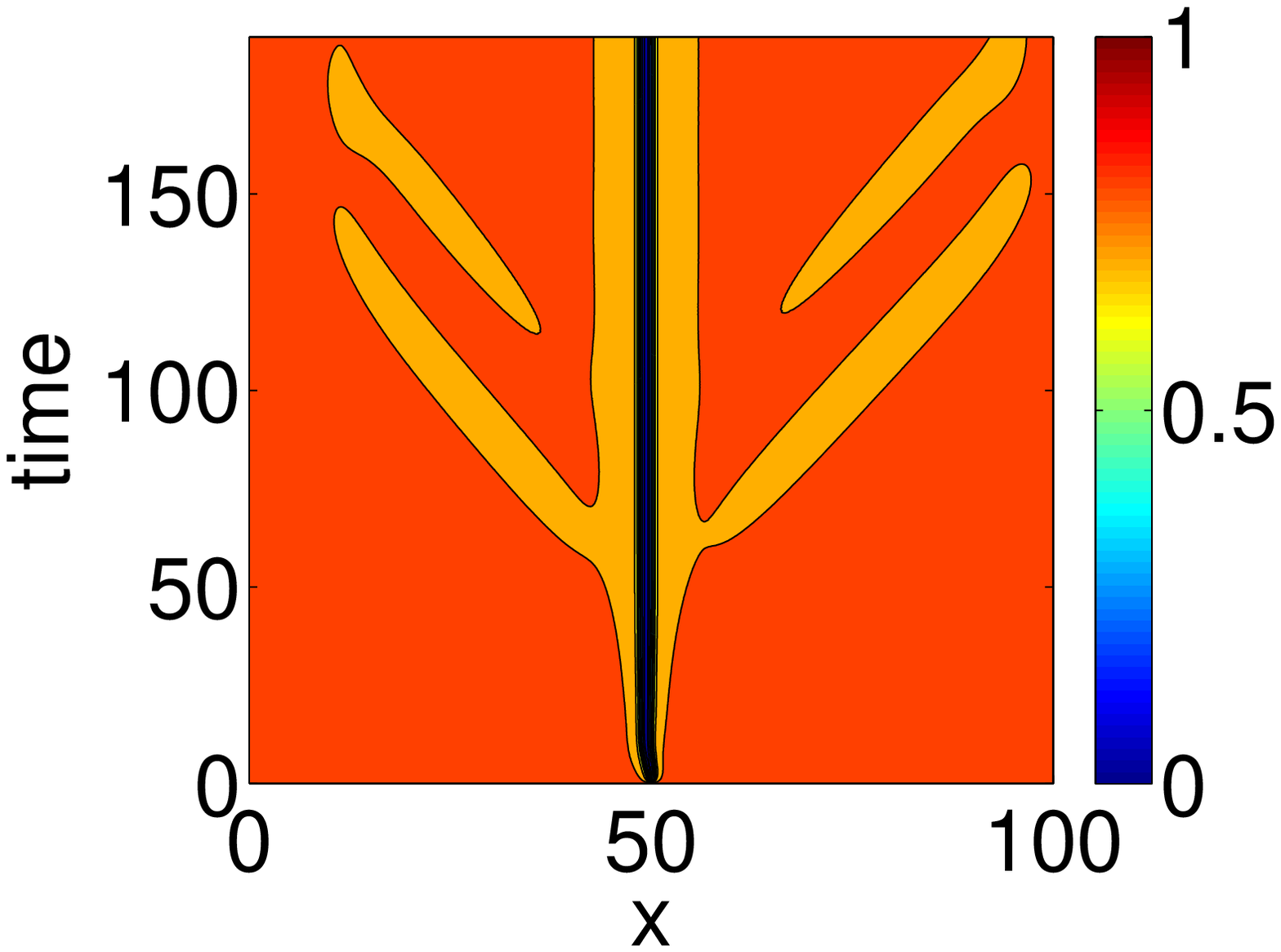} \label{fig:coll_wav_abs}}
    \subfigure [$\theta(x,t)$]
    {\includegraphics[width=4.1cm]{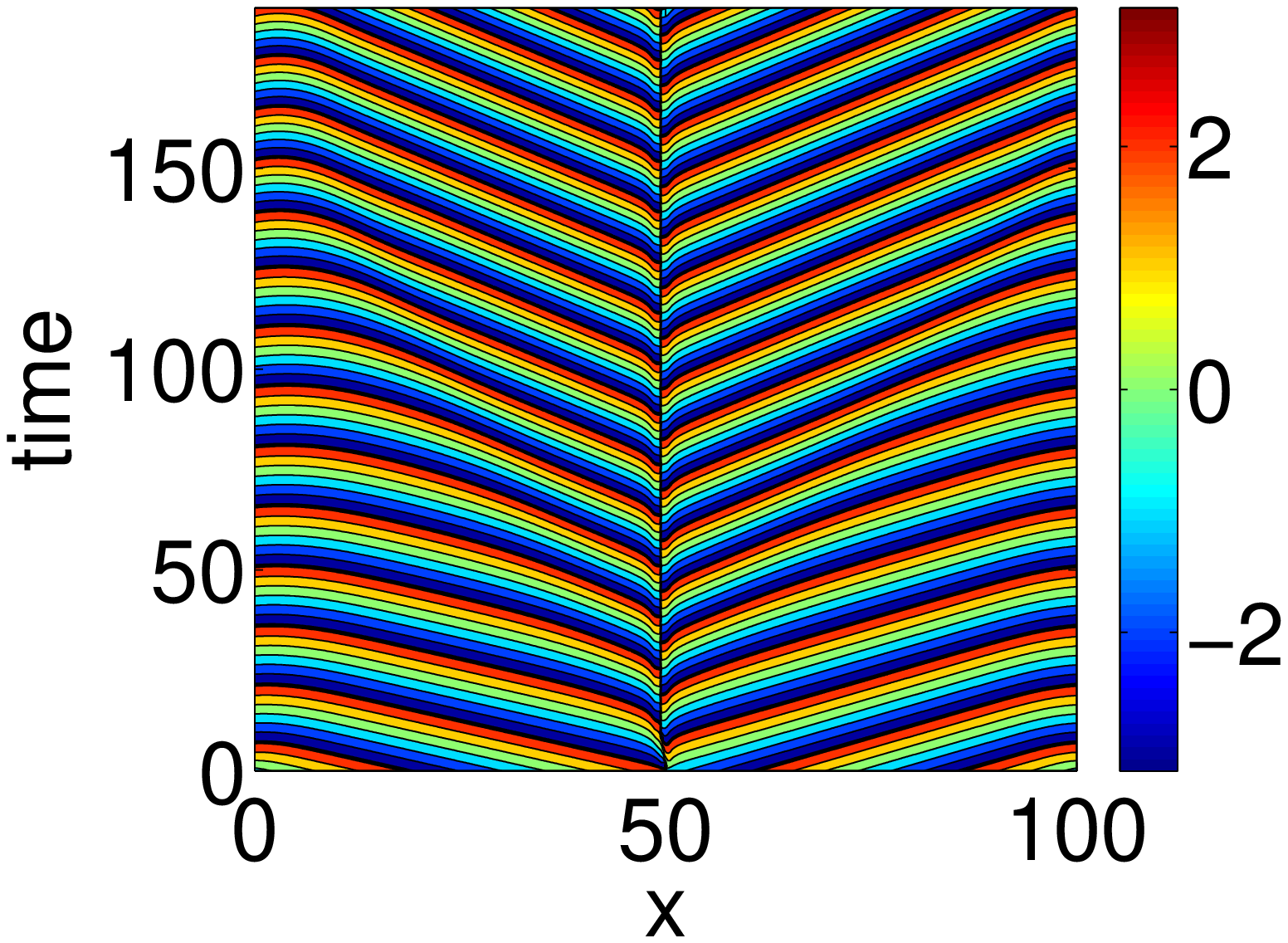} \label{fig:coll_wav_ph}}
    \subfigure [$|r(x,200)|$]
    {\includegraphics[width=4.1cm]{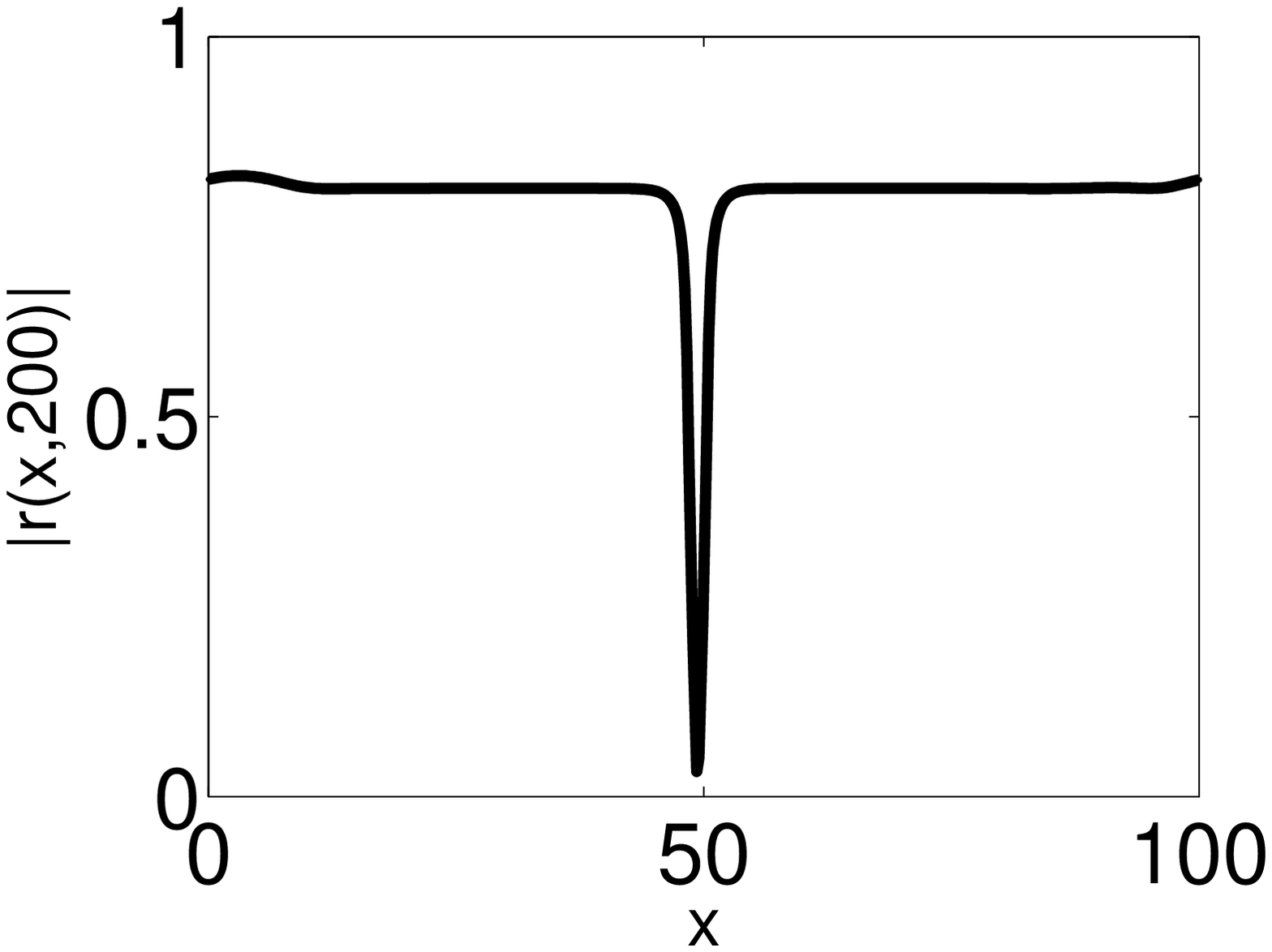}
                     \label{fig:coll_wav_abs_t200}}
    \subfigure [$\theta(x,200)$]
    {\includegraphics[width=4.1cm]{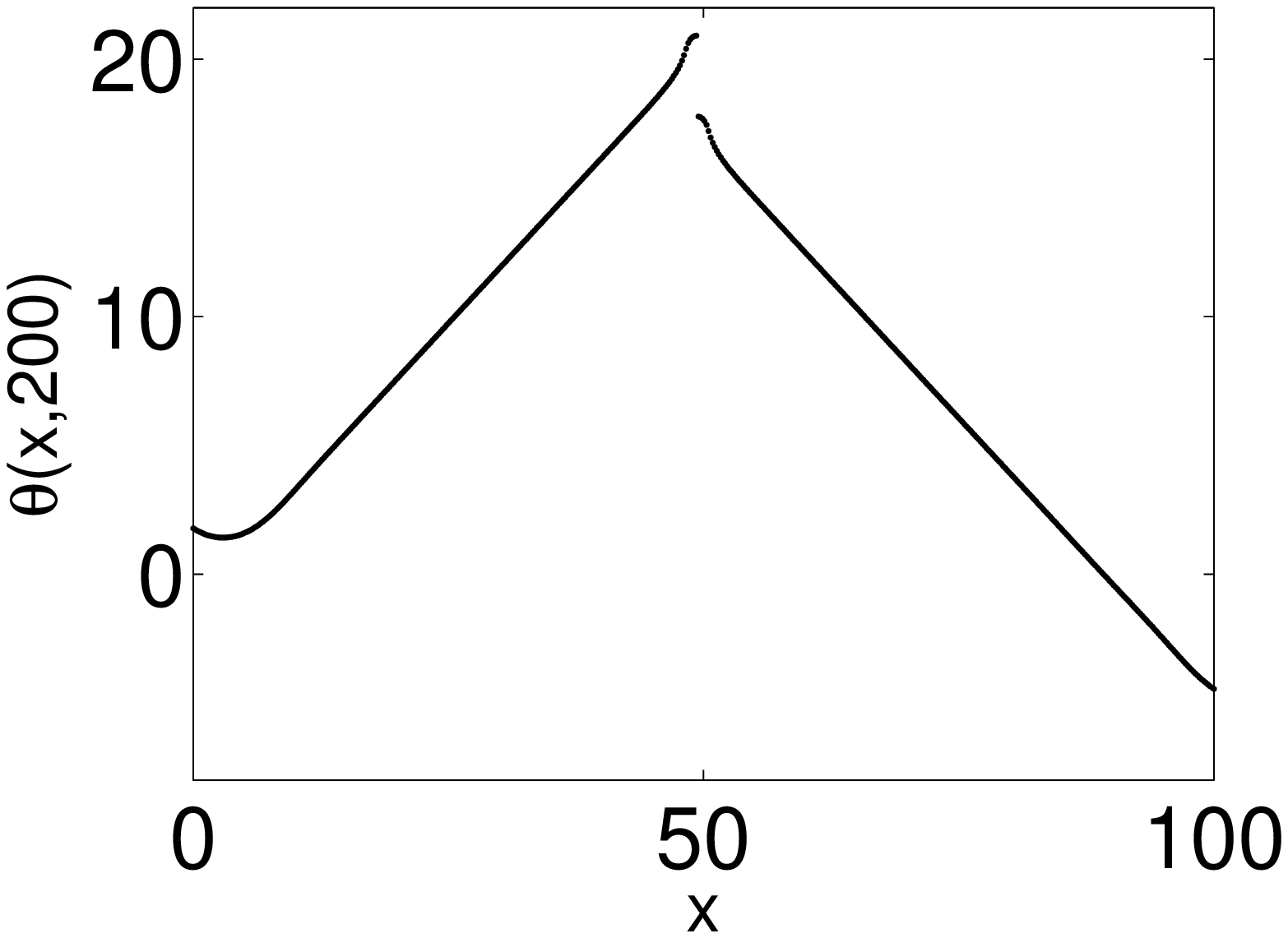}
                     \label{fig:coll_wav_ph_t200}}
  \end{flushleft}
  \caption{
           An example of the hole solution by collision of two 
           plane wave solutions.
           The two waves meet at $x=50$ with a $\pi$ phase difference.
           ($\omega_0=5, T=1, D=100, k=14.8$ and periodic boundary conditions). The 
           initial condition corresponds to a discontinuous $r$
           given by a right traveling plane wave solution with $m=3$ 
           (where the wave number is $2m\pi/D$) for 
           $0 \leq x \leq 50$ and a left traveling plane wave solution with 
           $m=4$ for $50 < x \leq 100$. Correspondingly, we observe from (d) that 
           $[\theta(0,200)-\theta(100,200)] = 2\pi$.
           }
\label{fig:coll_wav}
\end{figure*}

\begin{figure*}[h] 
  \begin{center}
  \subfigure[$|r(x)|$]
  {\includegraphics[width=4.1cm]{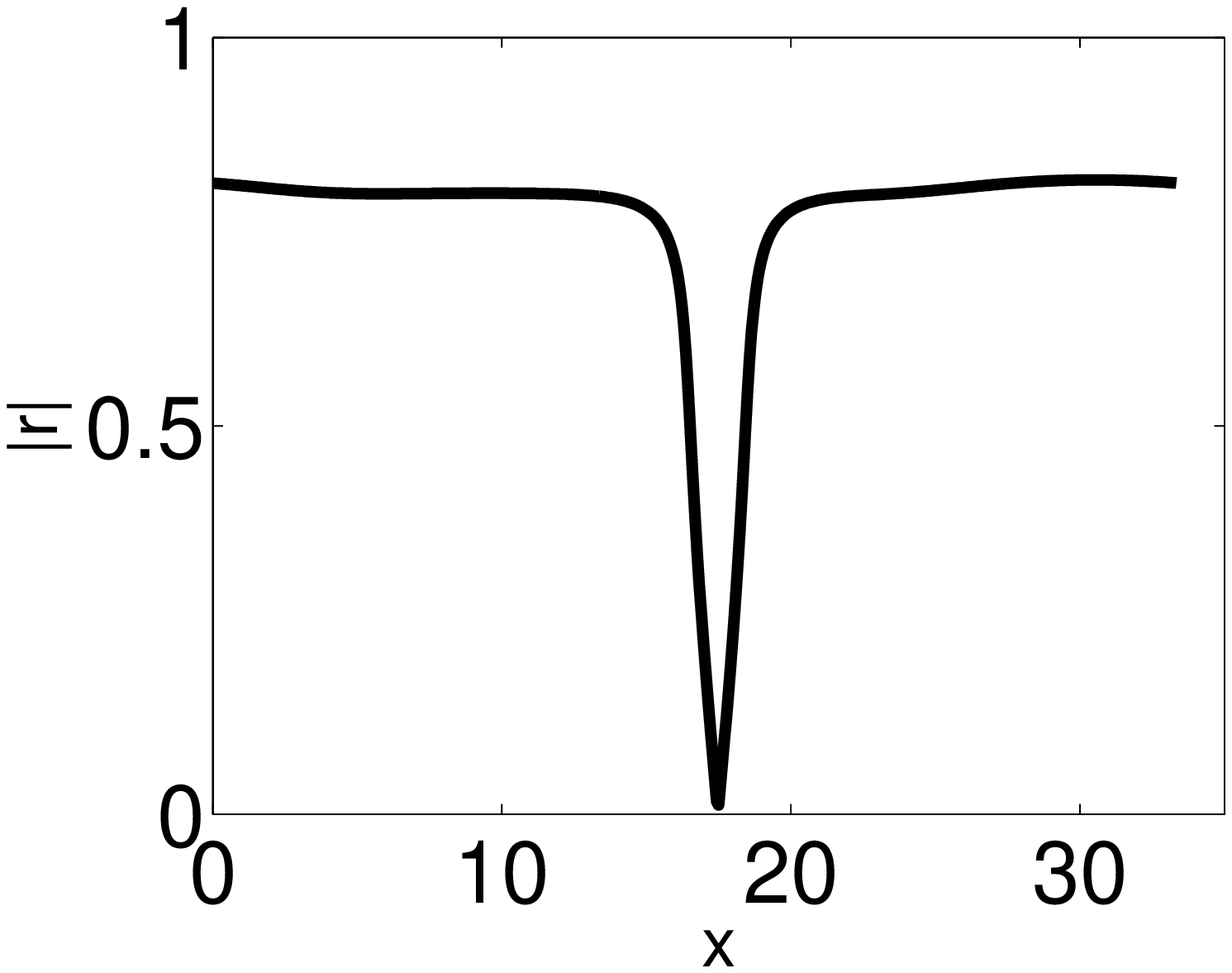}\label{fig:hol_cor__abs}}
  \subfigure[$\theta(x)$]
  {\includegraphics[width=4.1cm]{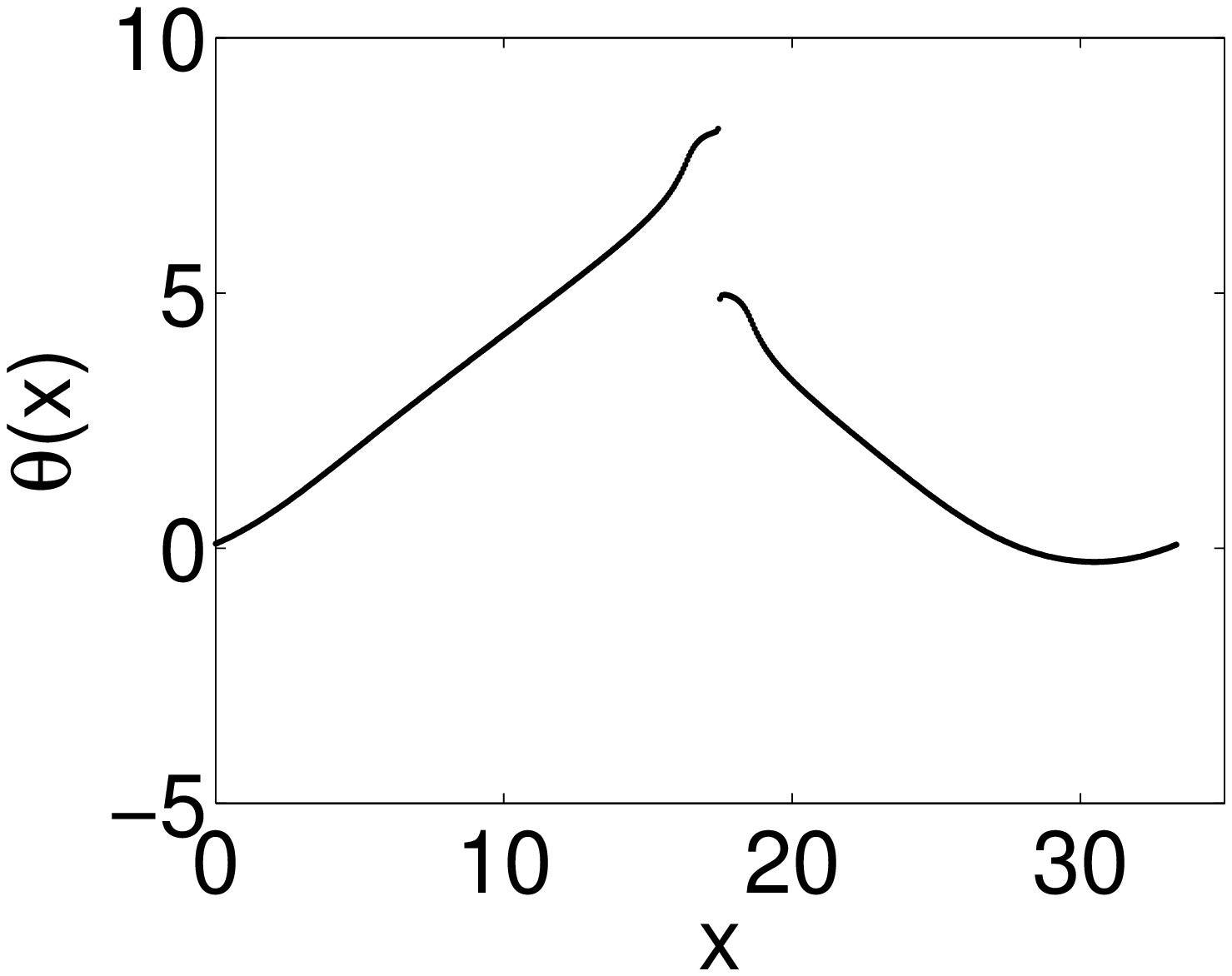} \label{fig:hol_cor_ph_t200}}
  \end{center}
\caption{
         Finite width of (one-dimensional) hole core
         ($\omega_0=5, T=1, D=33.3, k=14.8$).}
\label{fig:hol_core}
\end{figure*}

In fact, when $k \approx k_c$,
the hole-like pattern is a feature that shows up 
readily when two
plane waves with a relative phase difference of  $\pm \pi$
(or odd-multiples of them) collide. 
An example is studied in Fig. \ref{fig:coll_wav}
where two waves of relative phase difference $\pi$ 
collide giving rise to a hole pattern. 
This observation is consistent with the relative phase differences 
observed at
the two outer holes studied in Fig. \ref{fig:four_hole_ph_t1200}. Furthermore,
although the hole pattern seems to arise only under relatively specific conditions,
it is found to be pretty stable with respect to changes in parameters
or small perturbations once it is formed. 
Finally, as shown in
Fig. \ref{fig:hol_core}, we note that the hole core
occupies a finite width and so is not a point singularity when
$T \neq 0$.
This will be shown to have a close correspondence with
the spiral wave in our two-dimensional study (sec. \ref{sec:target_spiral}).

It is further interesting to note some similarity between
our observations in the region
$k \approx k_c$ and the intermittency regime of the complex Ginzburg-Landau equation
(CGLE) (See for example, section III of
Ref. \cite{Aranson}, and section 2.5 of Ref. \cite{Mori_Kura}).
There, the CGLE 
displays similar glassy-like transition patterns characterized by
large plateaus of coherent regions with hole-like patterns being continuously
created and destroyed.
However, there are
also differences between the two systems. For example, 
the CGLE does not
seem to have a close counterpart to the bridge pattern observed
in our system, while more intricate dynamics of hole creation
and destruction leading to zigzagging holes and
defect chaos have not been observed in our study.

\subsection{2D propagating fronts and ``bridge'' patterns}
\begin{figure*}[h] 
  \begin{flushleft}
  \subfigure [$t=0$]
  {\includegraphics[width=4.4cm]{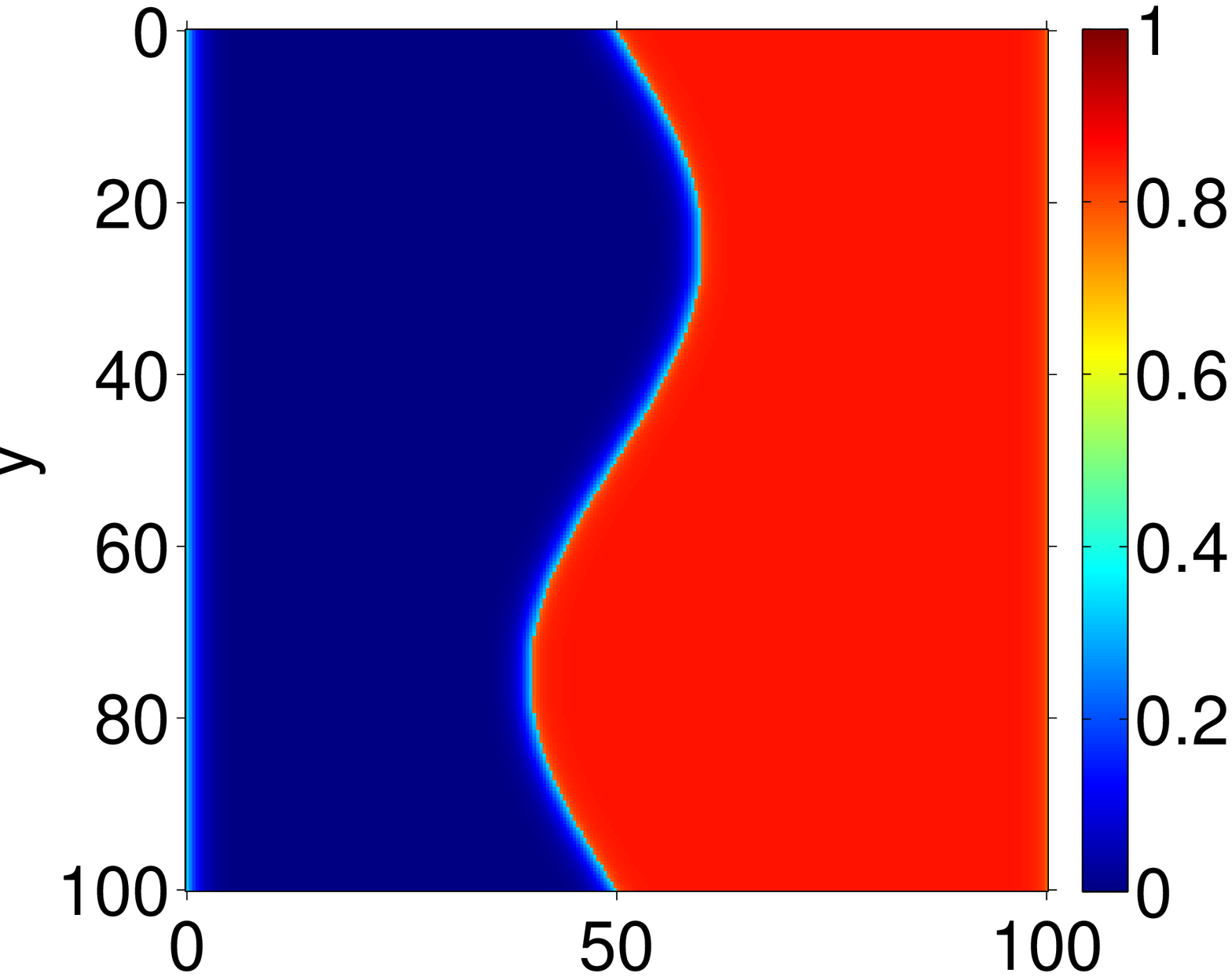} \label{fig:sineb_klkc_0}}
  \subfigure [$t=46$]
  {\includegraphics[width=4.1cm]{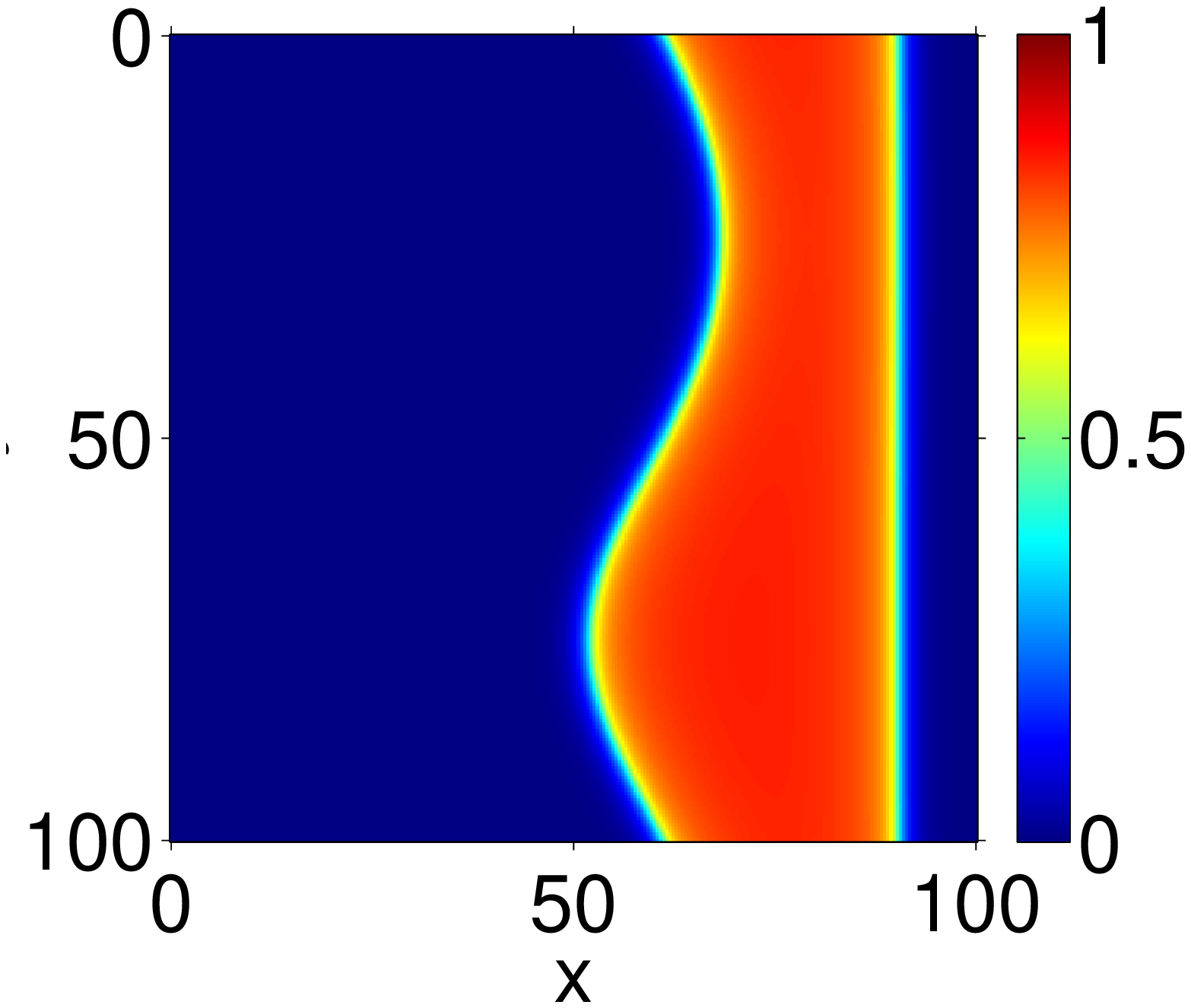} \label{fig:sineb_klkc_58}}
  \subfigure [$t=72$]
  {\includegraphics[width=4.1cm]{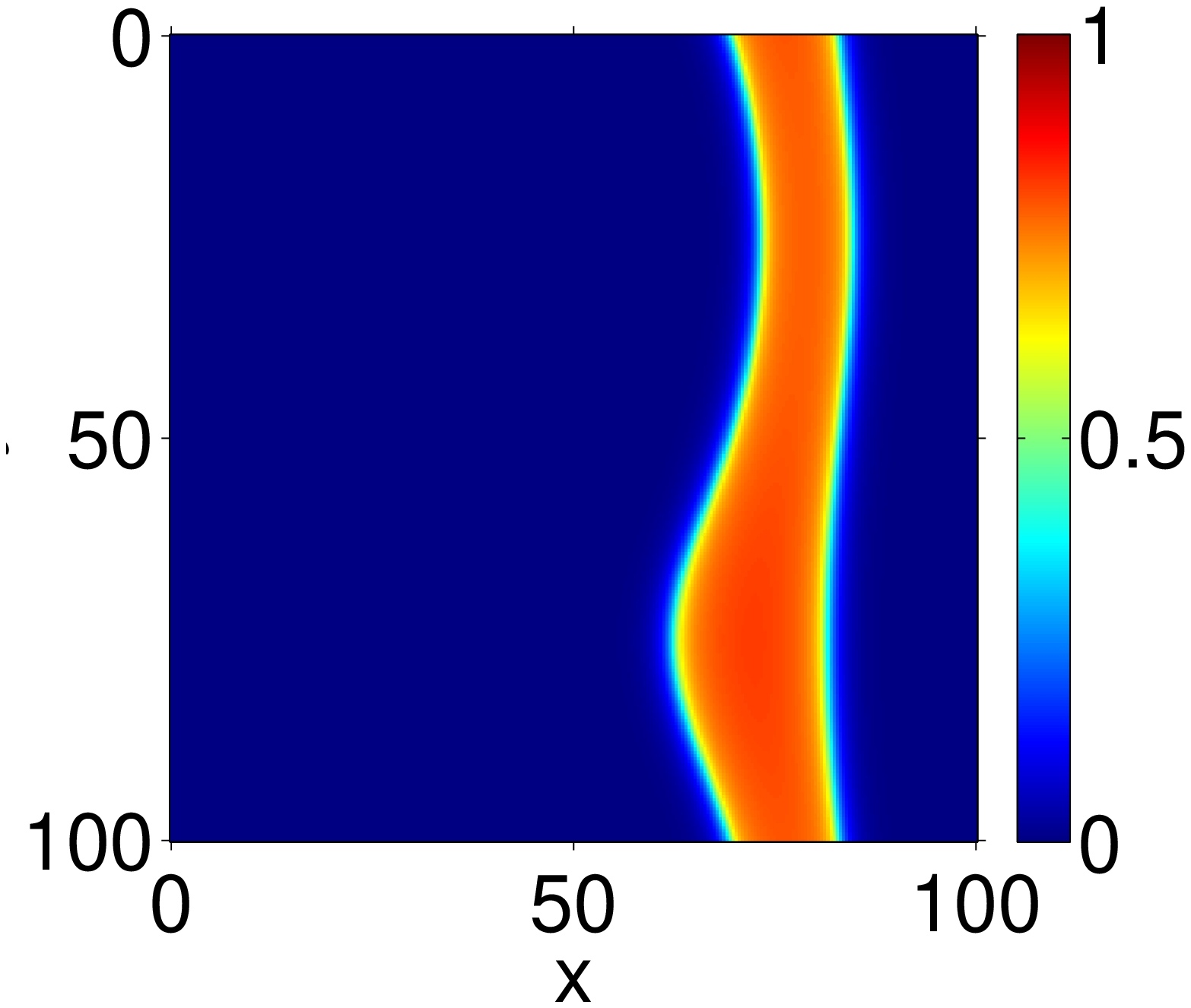} \label{fig:sineb_klkc_90}}
  \subfigure [$t=216$]
  {\includegraphics[width=4.1cm]{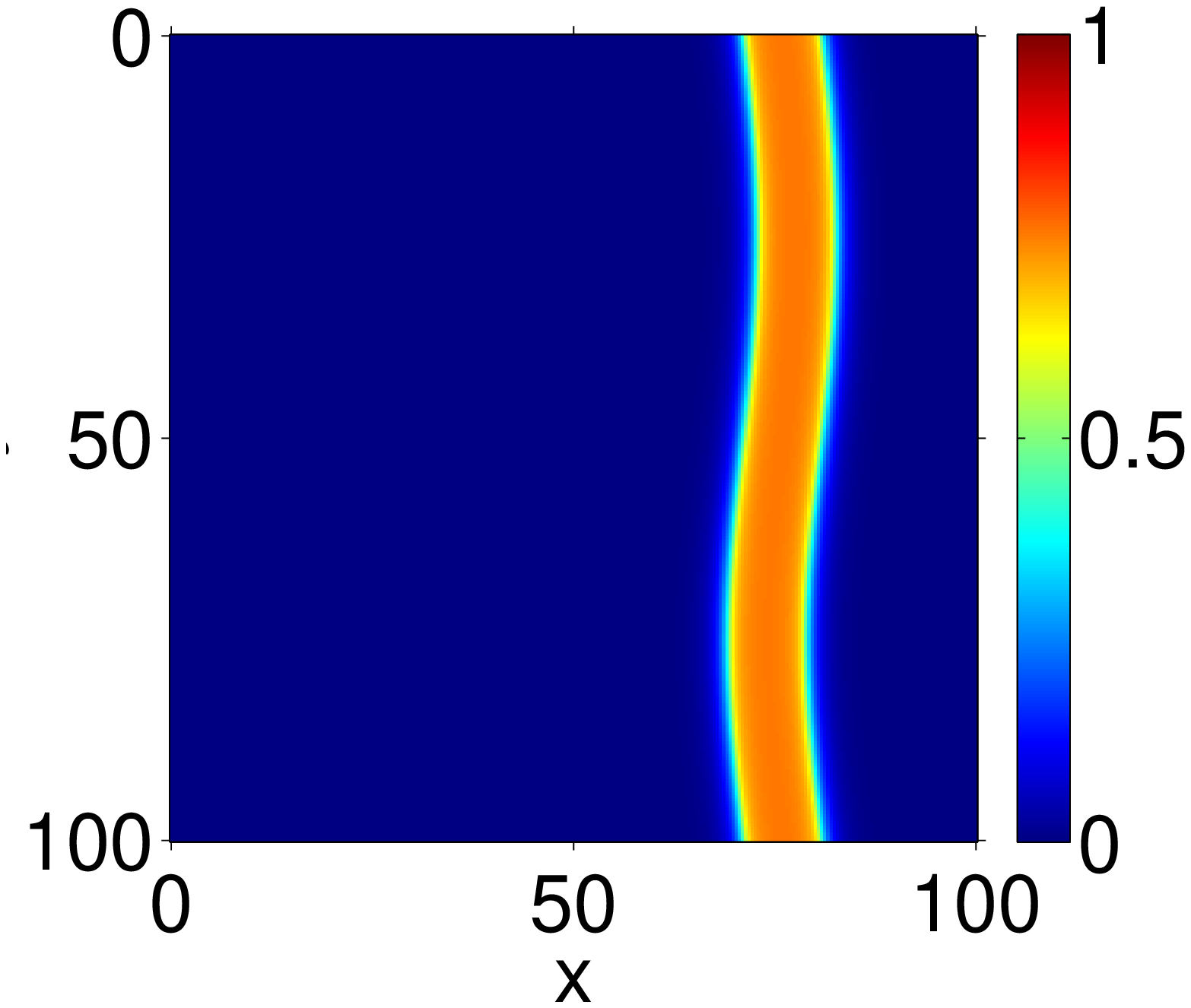} \label{fig:sineb_klkc_270}}
  \end{flushleft}
  \caption{
           Time evolution of $|r({\bf x},t)|$ 
           of a $d=2$ configuration  initialized
           with half of the region at the incoherent state and half at the coherent
           state divided by a wiggled boundary with $k=14.4$ ($<k_c=14.5$)
           ($\omega_0=5, T=1, D=100$; 
           periodic boundary conditions are imposed).
           }
  \label{fig:sineb_klkc}
\end{figure*}

\begin{figure*}[h] 
  \begin{flushleft}
    \subfigure [$t=0$]
    {\includegraphics[width=4.1cm]{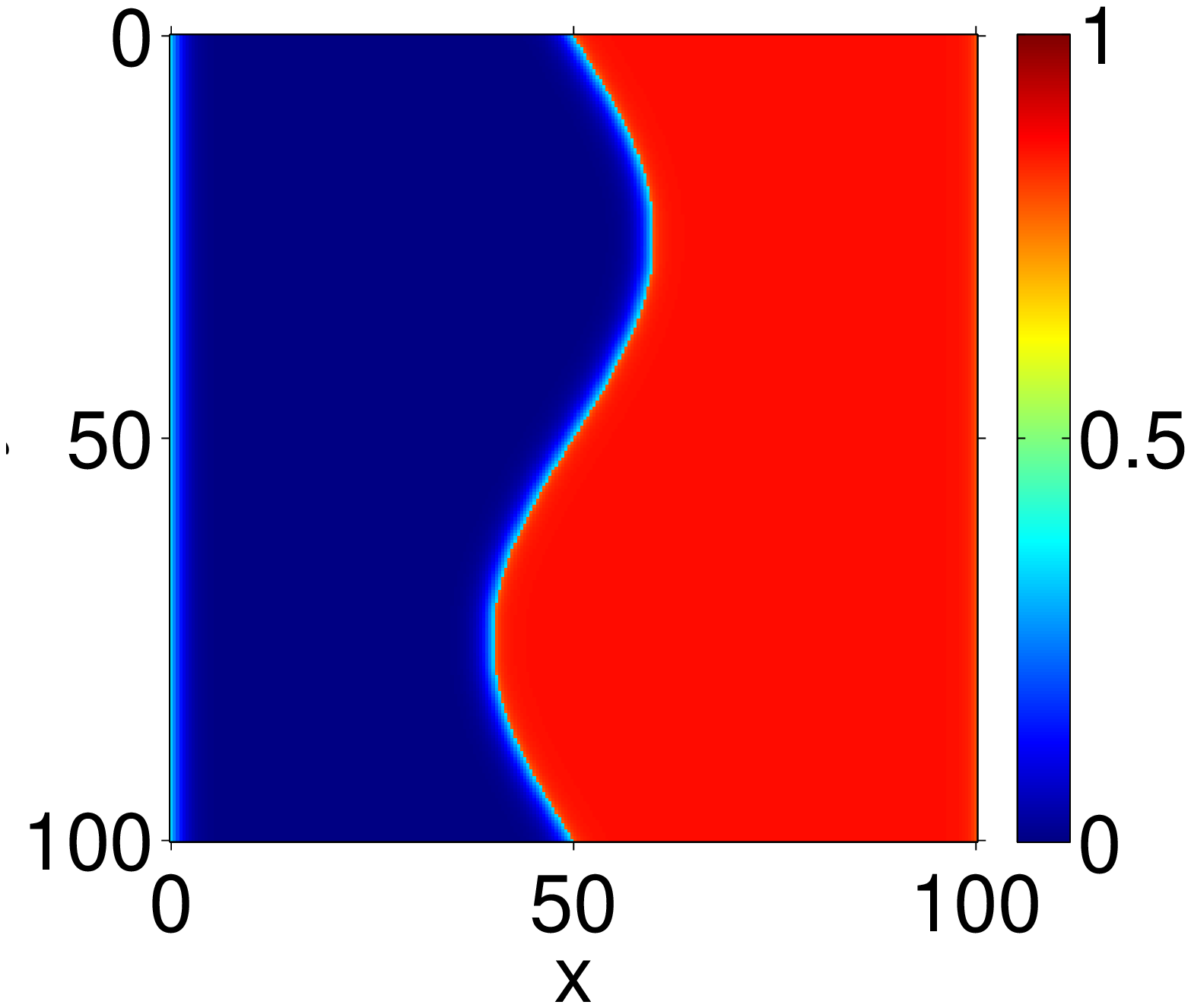} \label{fig:sineb_kgkc_0}}
    \subfigure [$t=55$]
    {\includegraphics[width=4.1cm]{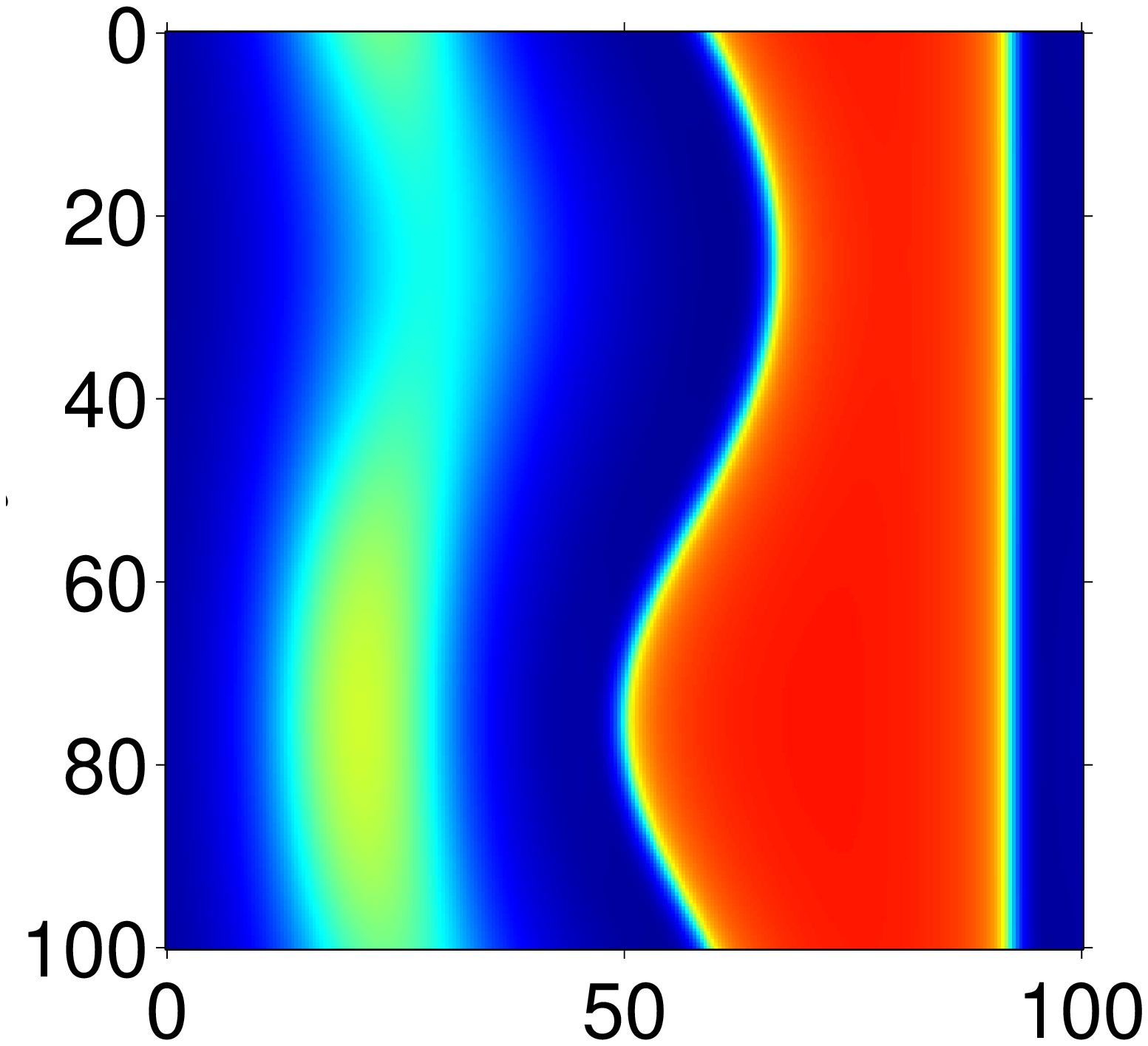} \label{fig:sineb_kgkc_69}}
    \subfigure [$t=83$]
    {\includegraphics[width=4.1cm]{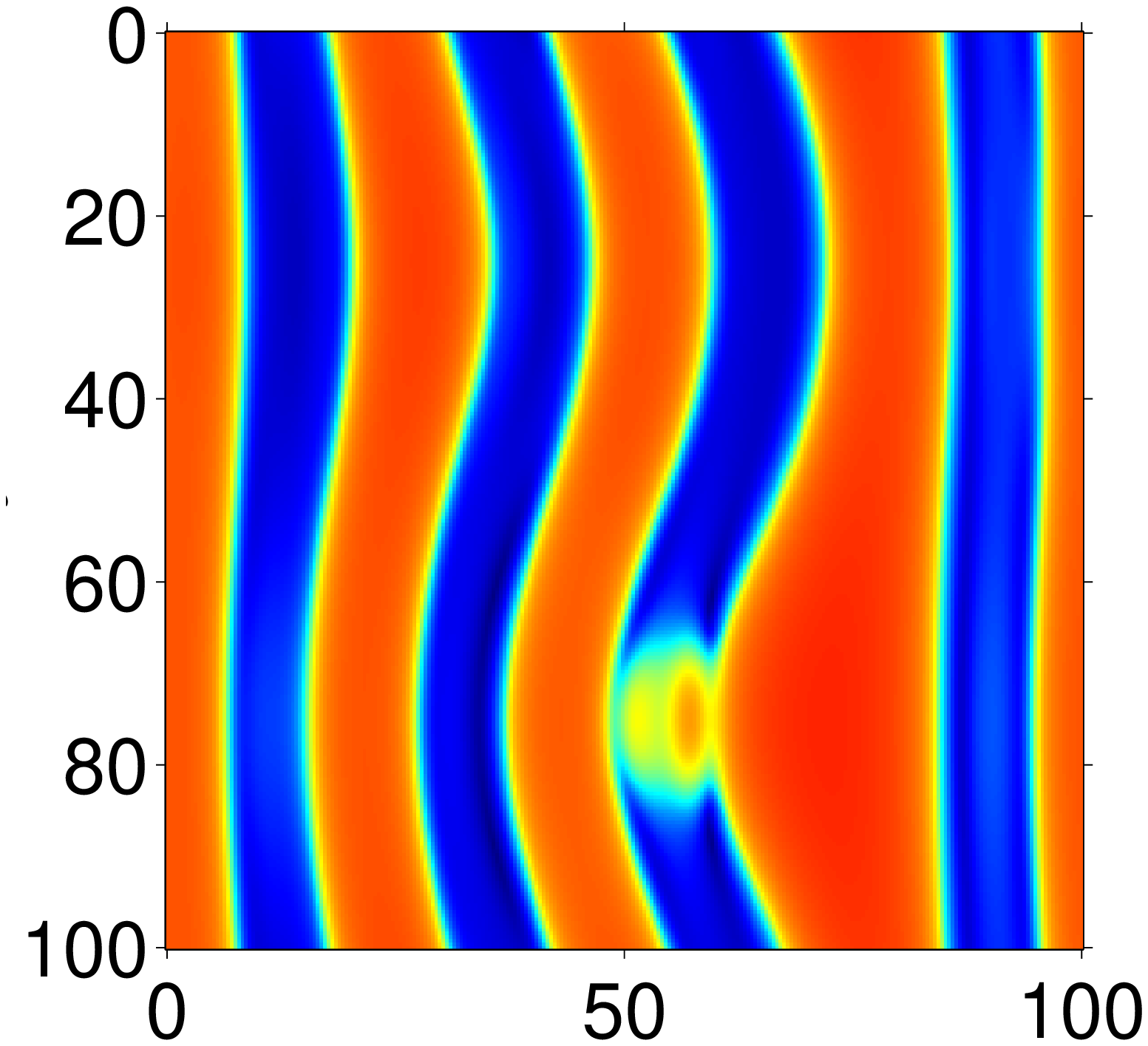}\label{fig:sineb_kgkc_104}}
    \subfigure [$t=98$]
    {\includegraphics[width=4.1cm]{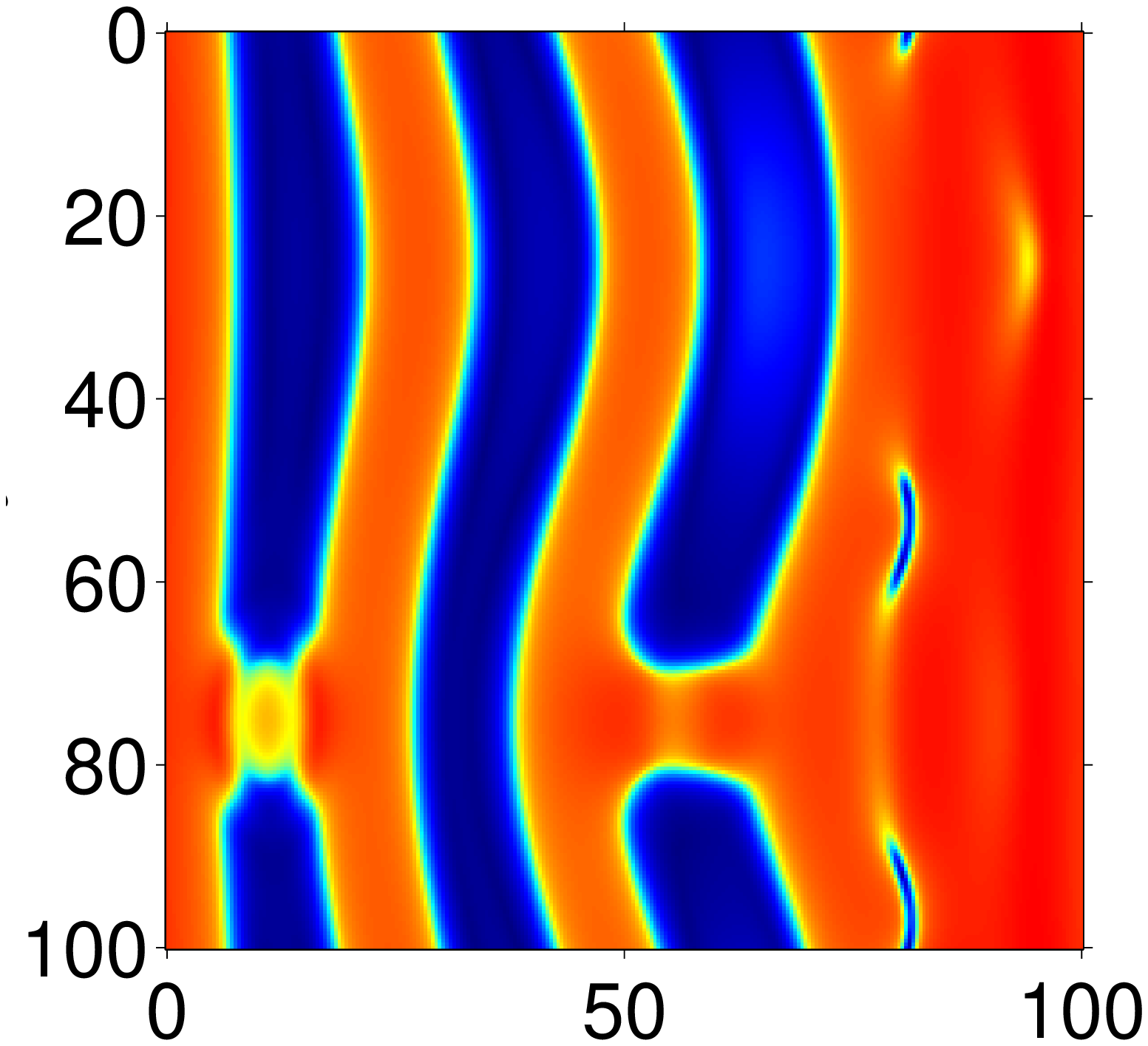}\label{fig:sineb_kgkc_123}}
    \subfigure [$t=138$]
    {\includegraphics[width=4.1cm]{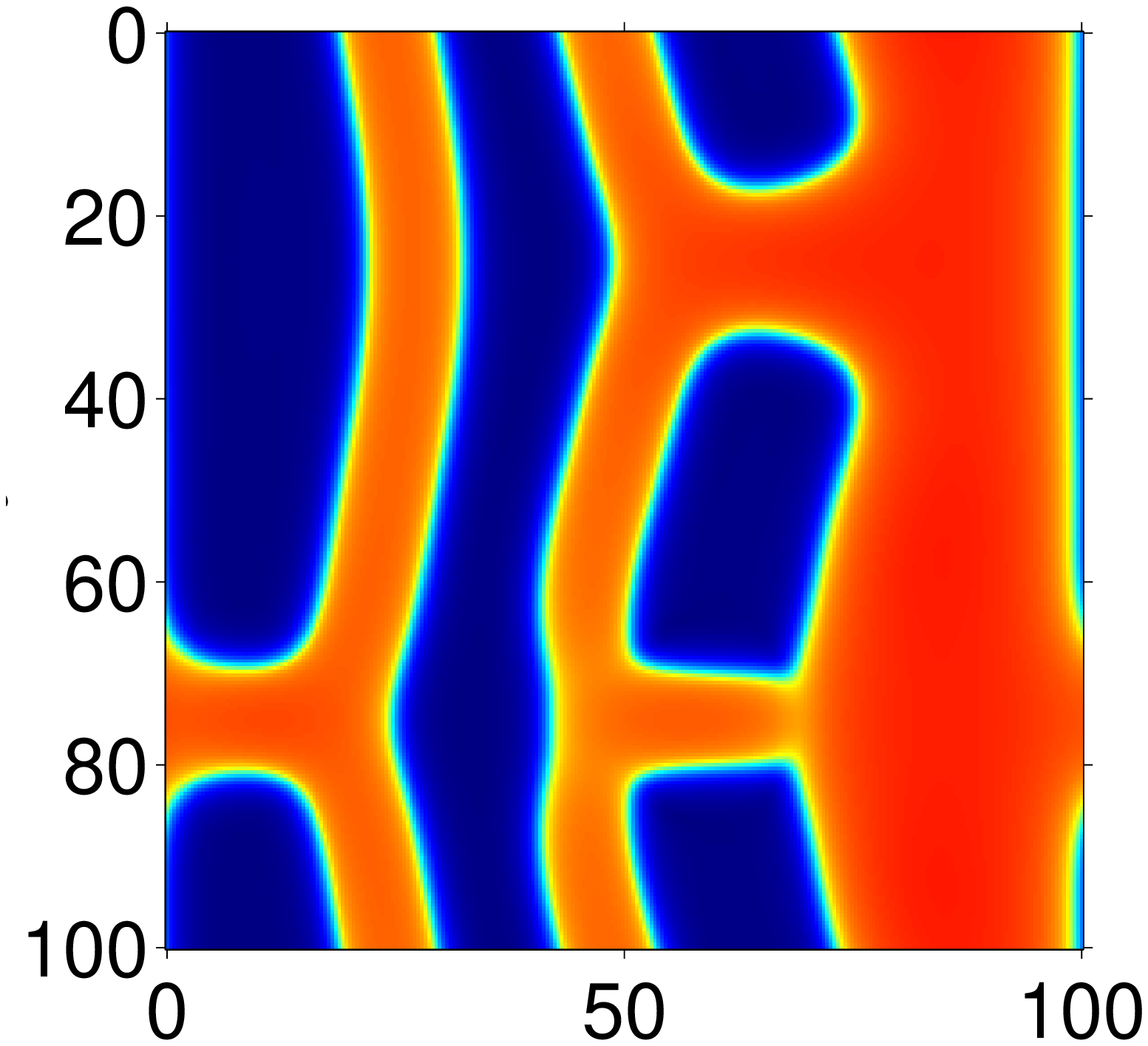}\label{fig:sineb_kgkc_172}}
    \subfigure [$t=150$]
    {\includegraphics[width=4.1cm]{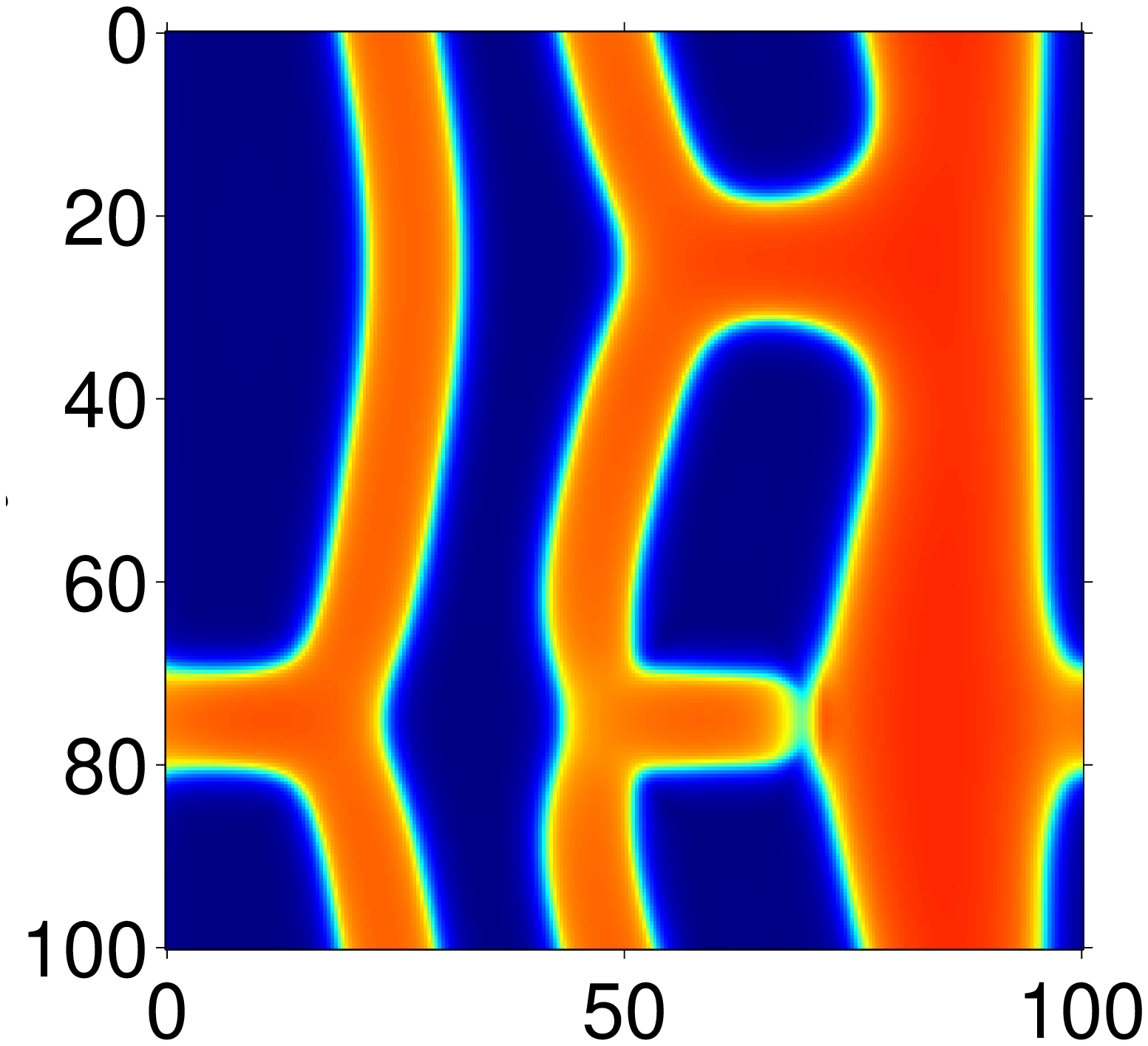}\label{fig:sineb_kgkc_188}}
    \subfigure [$t=170$]
    {\includegraphics[width=4.1cm]{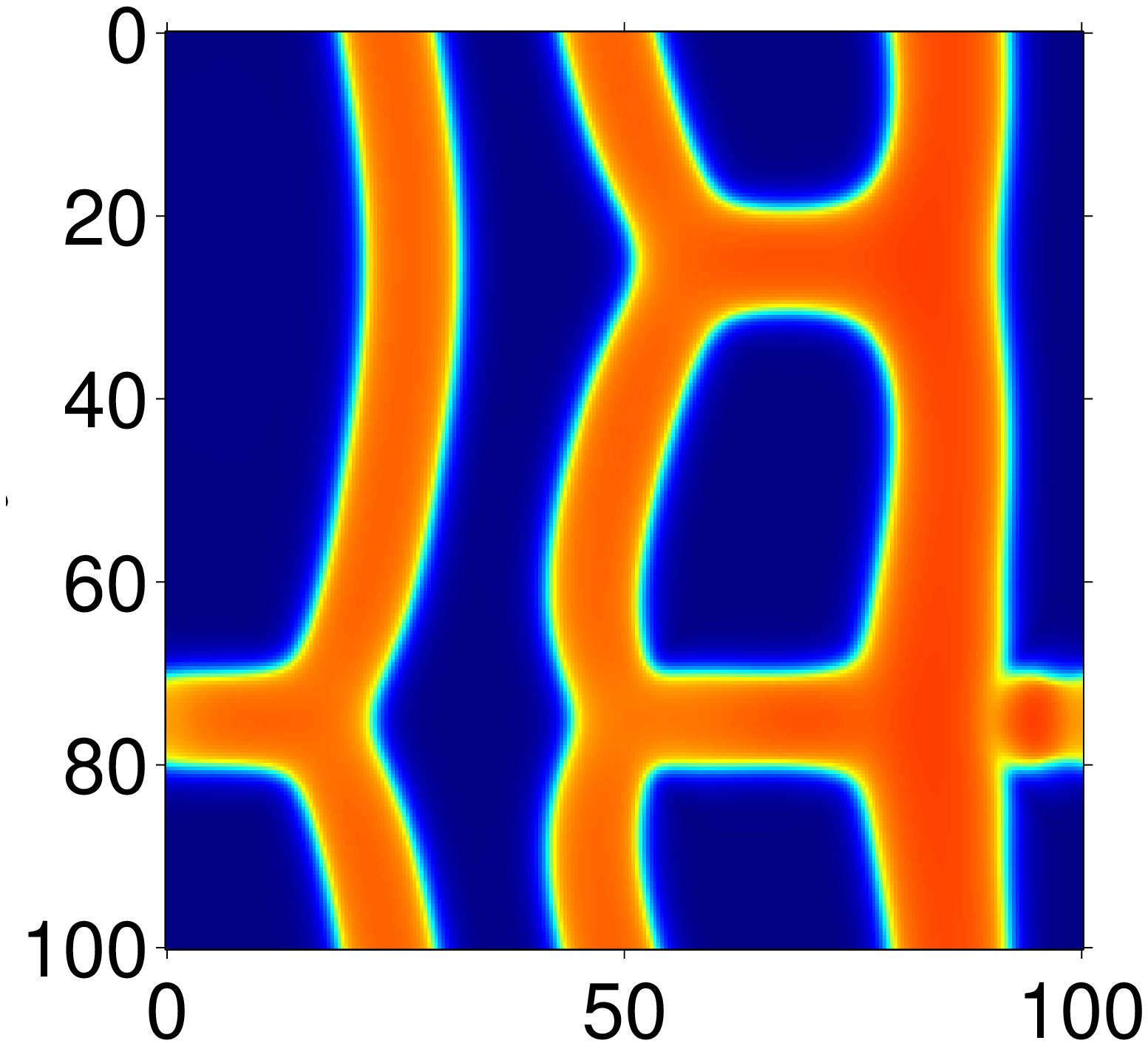}\label{fig:sineb_kgkc_213}}
    \subfigure [$t=245$]
    {\includegraphics[width=4.1cm]{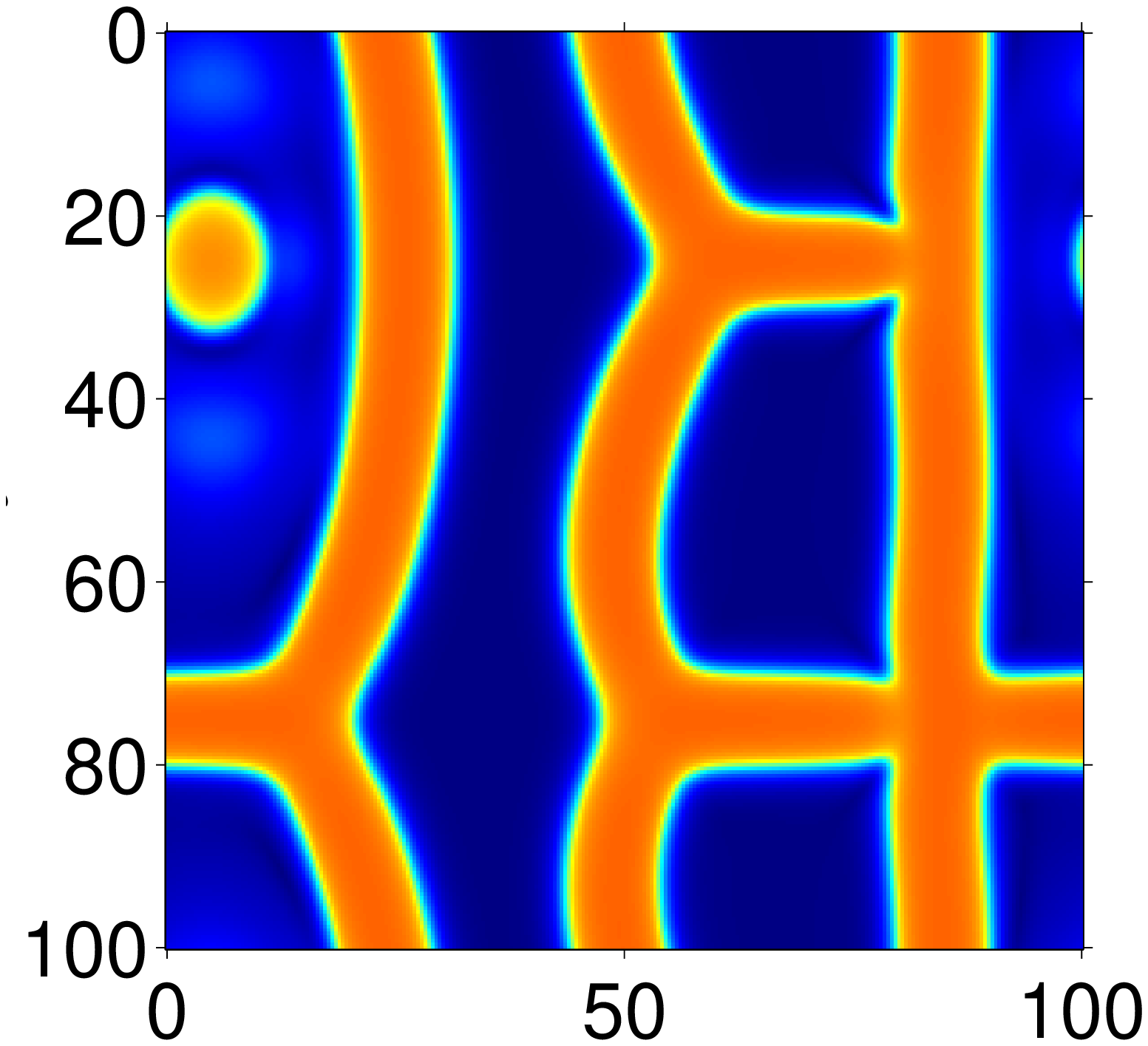}\label{fig:sineb_kgkc_311}}
    \subfigure [$t=282$]
    {\includegraphics[width=4.1cm]{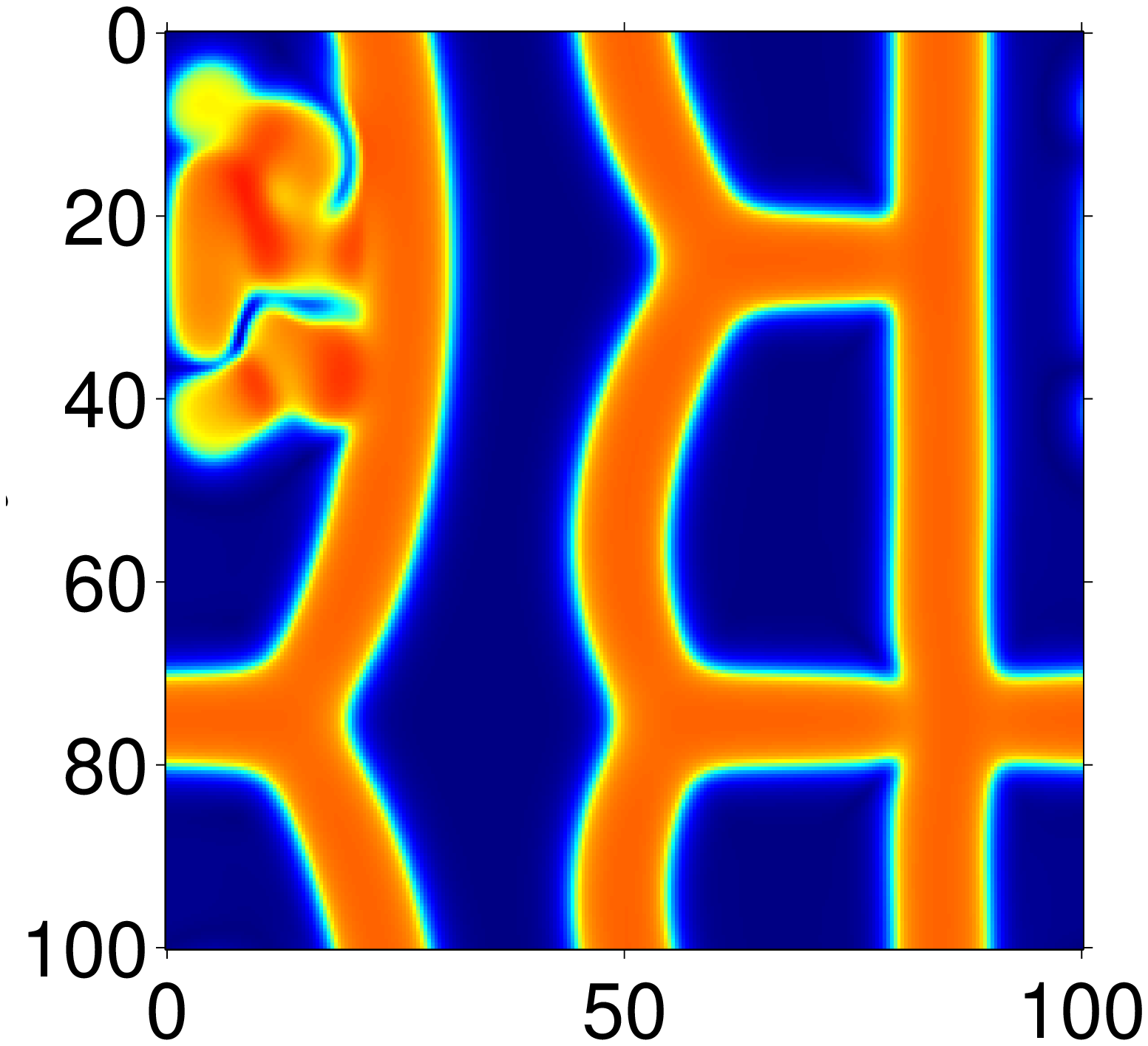}\label{fig:sineb_kgkc_353}}
    \subfigure [$t=303$]
    {\includegraphics[width=4.1cm]{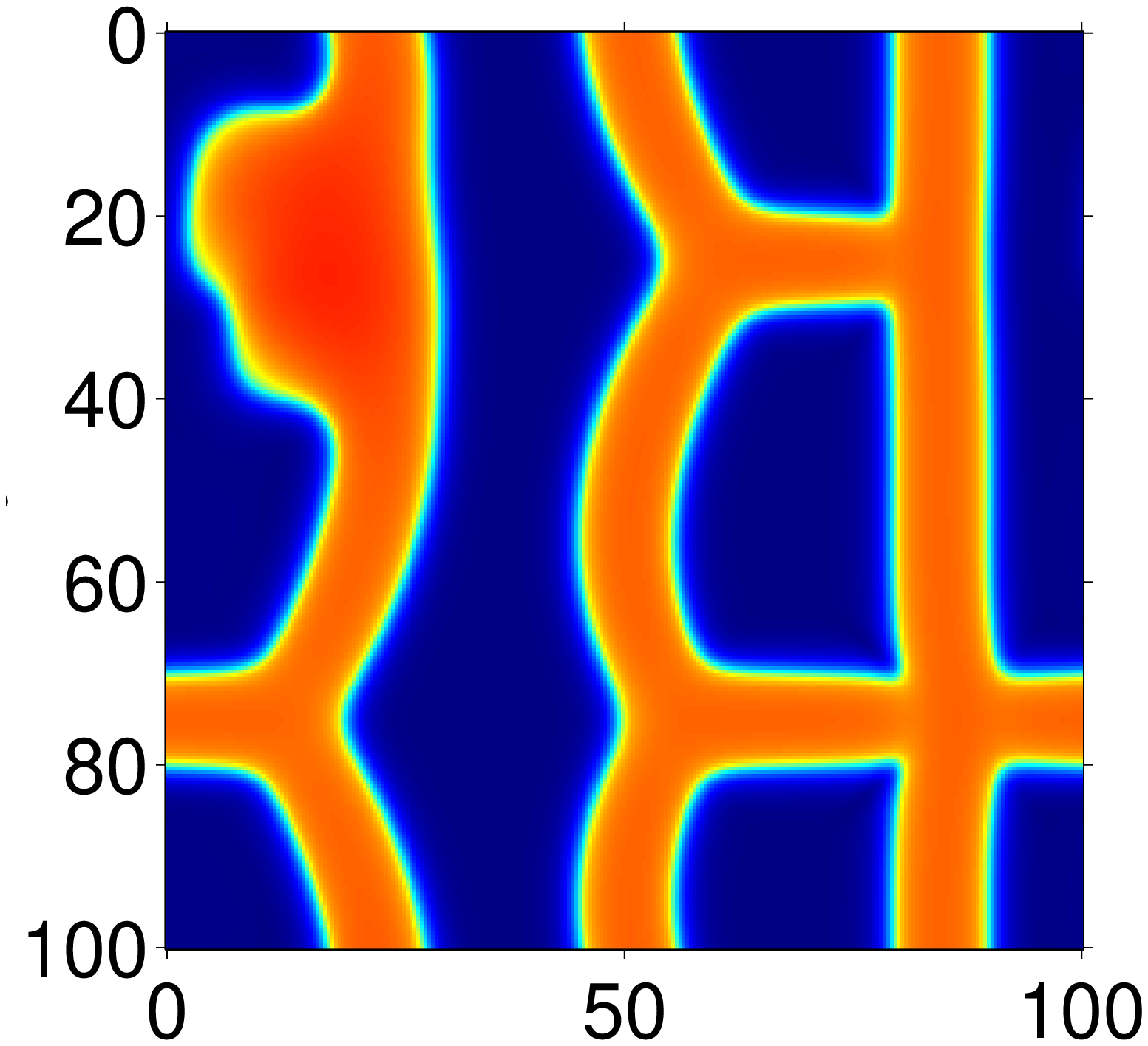}\label{fig:sineb_kgkc_385}}
    \subfigure [$t=400$]
    {\includegraphics[width=4.1cm]{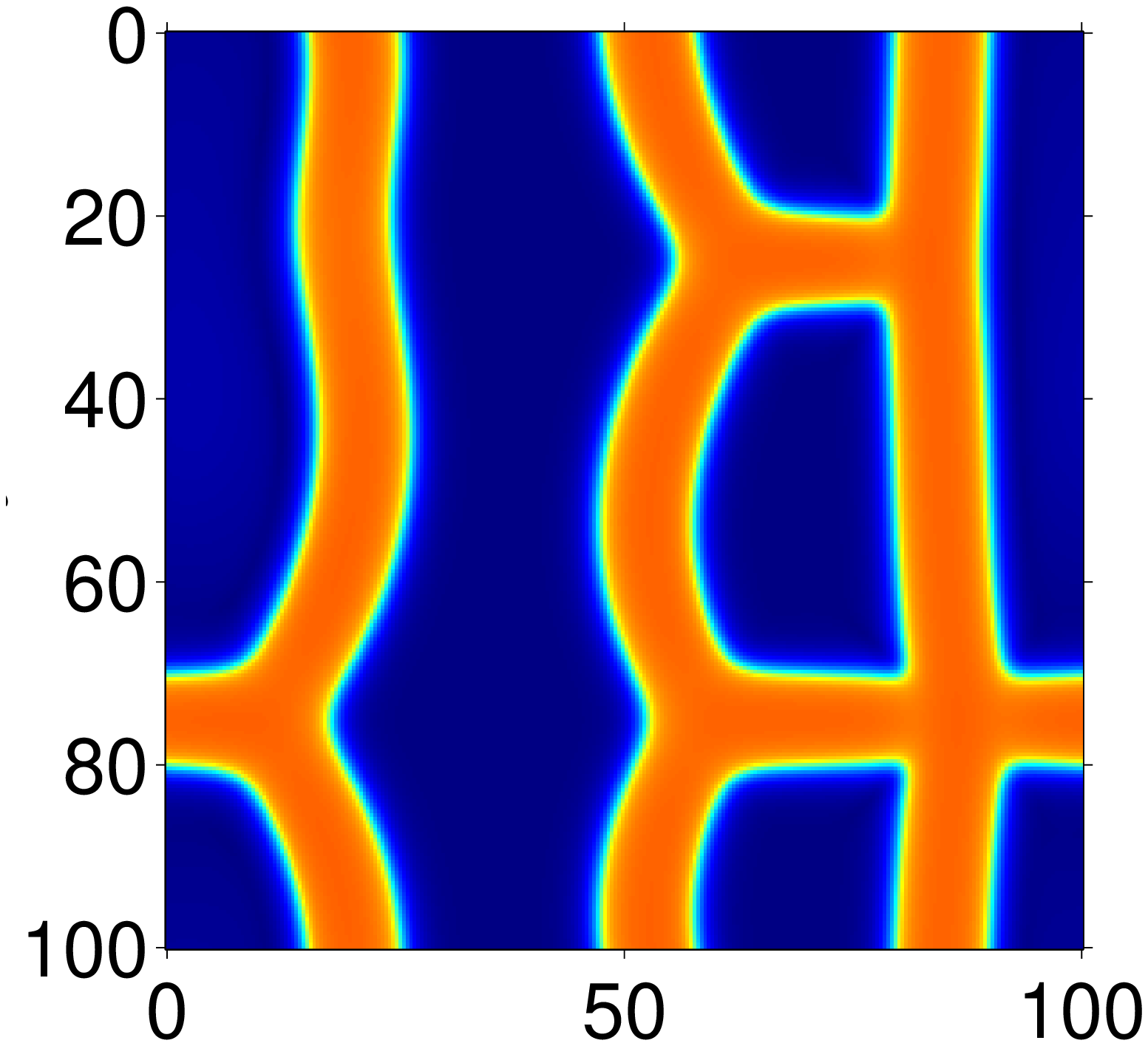}\label{fig:sineb_kgkc_500}}
    \subfigure [$t=431$]
    {\includegraphics[width=4.1cm]{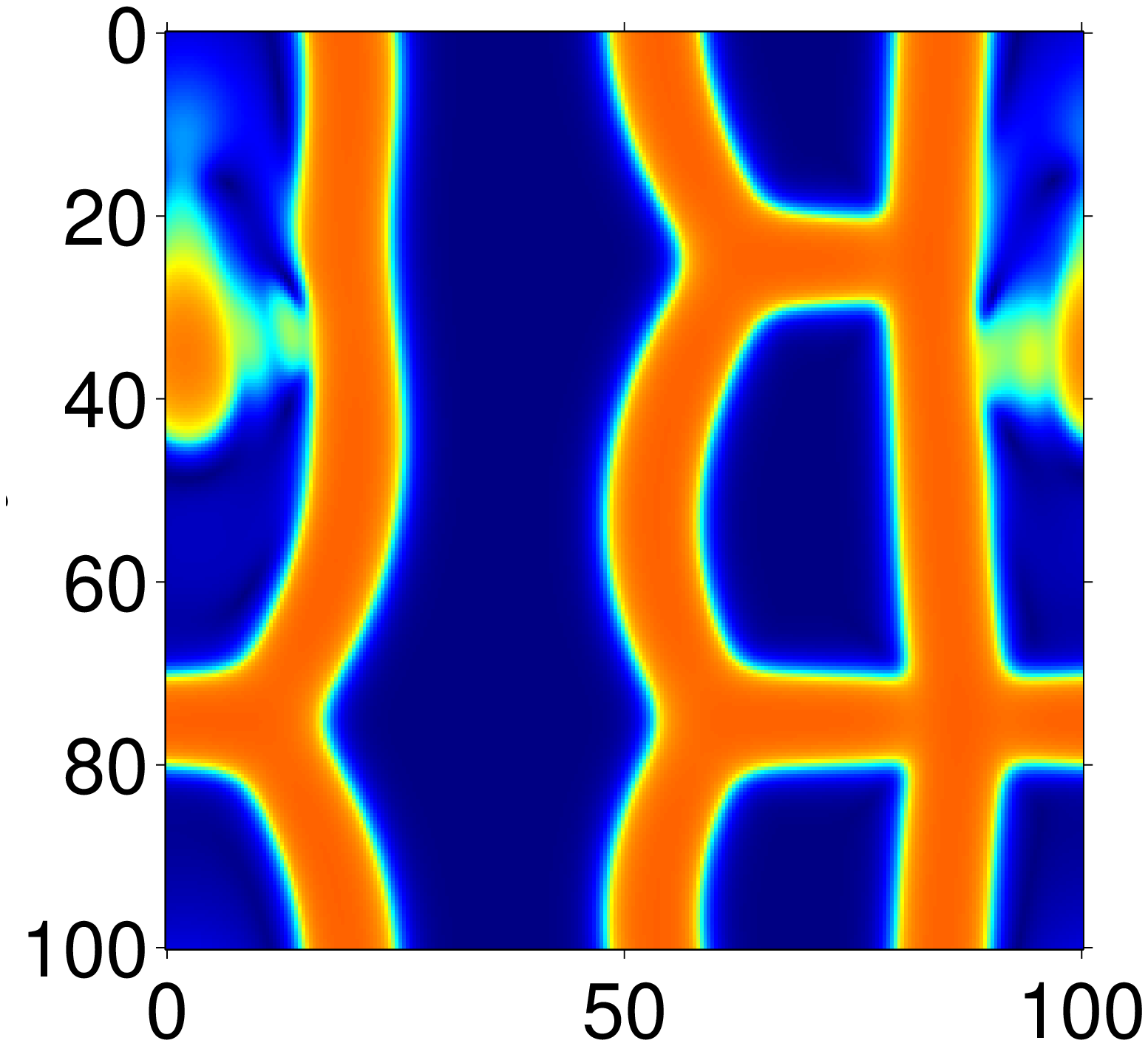} \label{fig:sineb_kgkc_t700_308}}
    \subfigure [$t=561$]
    {\includegraphics[width=4.1cm]{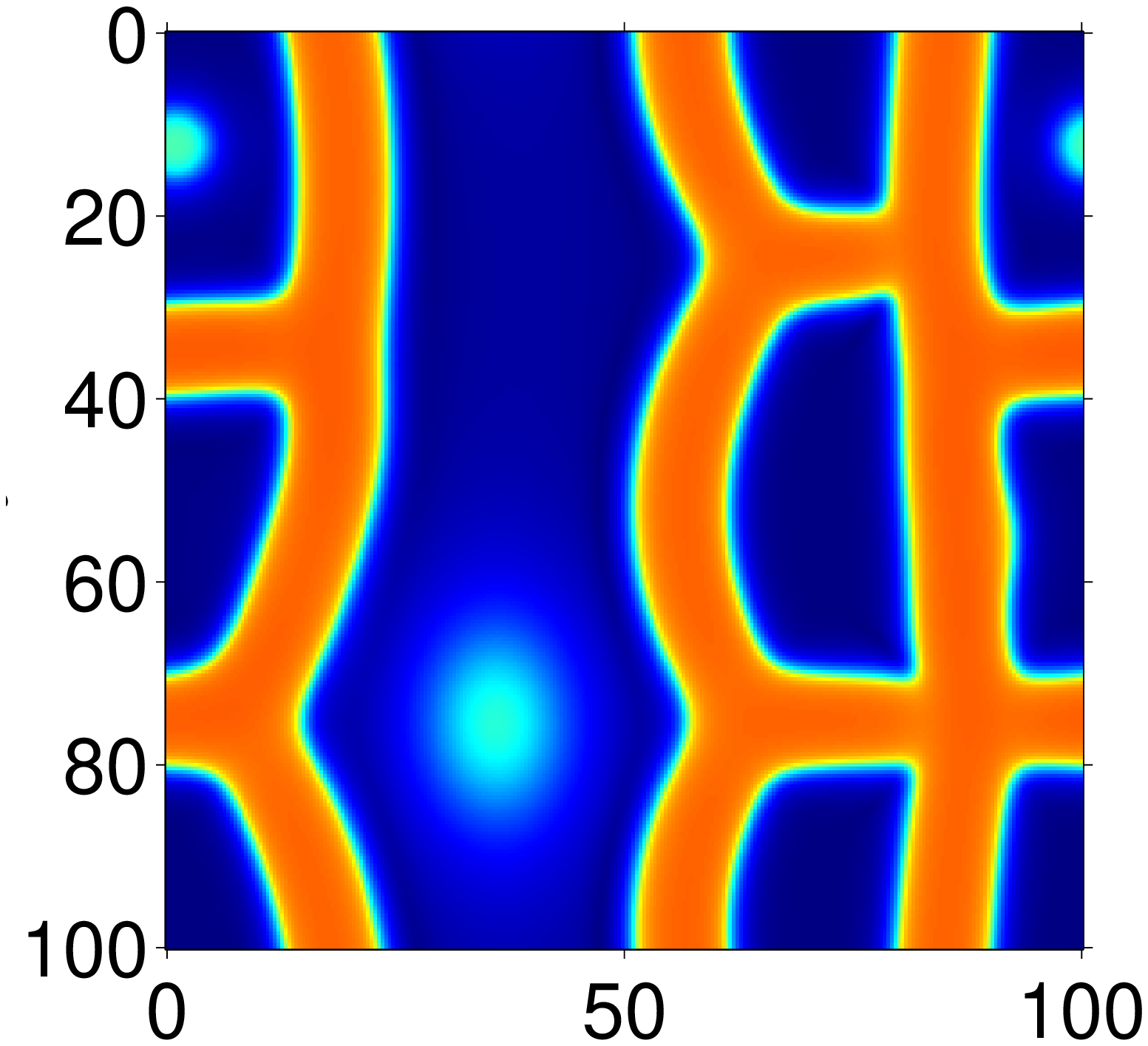} \label{fig:sineb_kgkc_t700_401}}
    \subfigure [$t=611$]
    {\includegraphics[width=4.1cm]{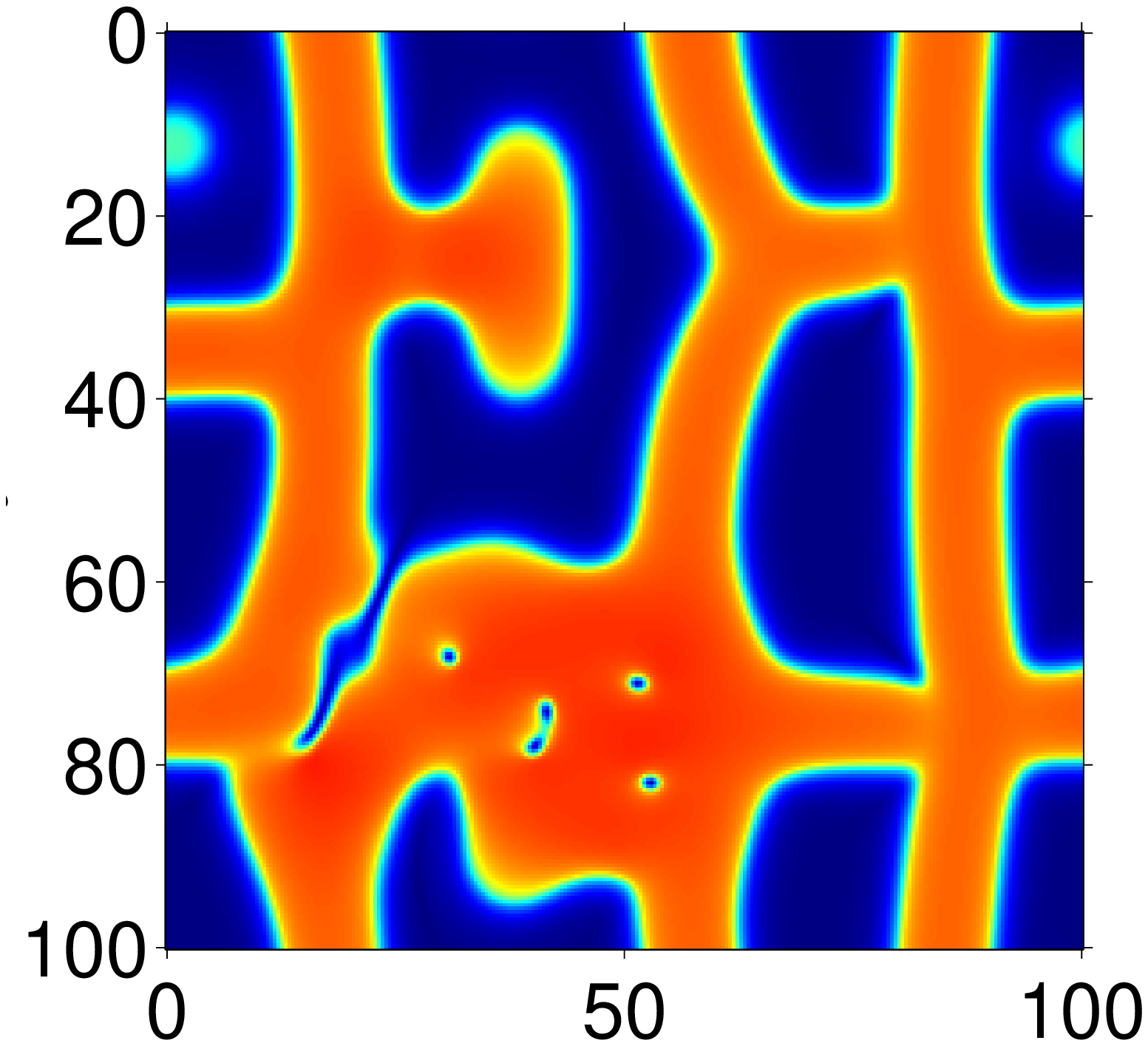}\label{fig:sineb_kgkc_t700_437}}
    \subfigure [$t=669$]
    {\includegraphics[width=4.1cm]{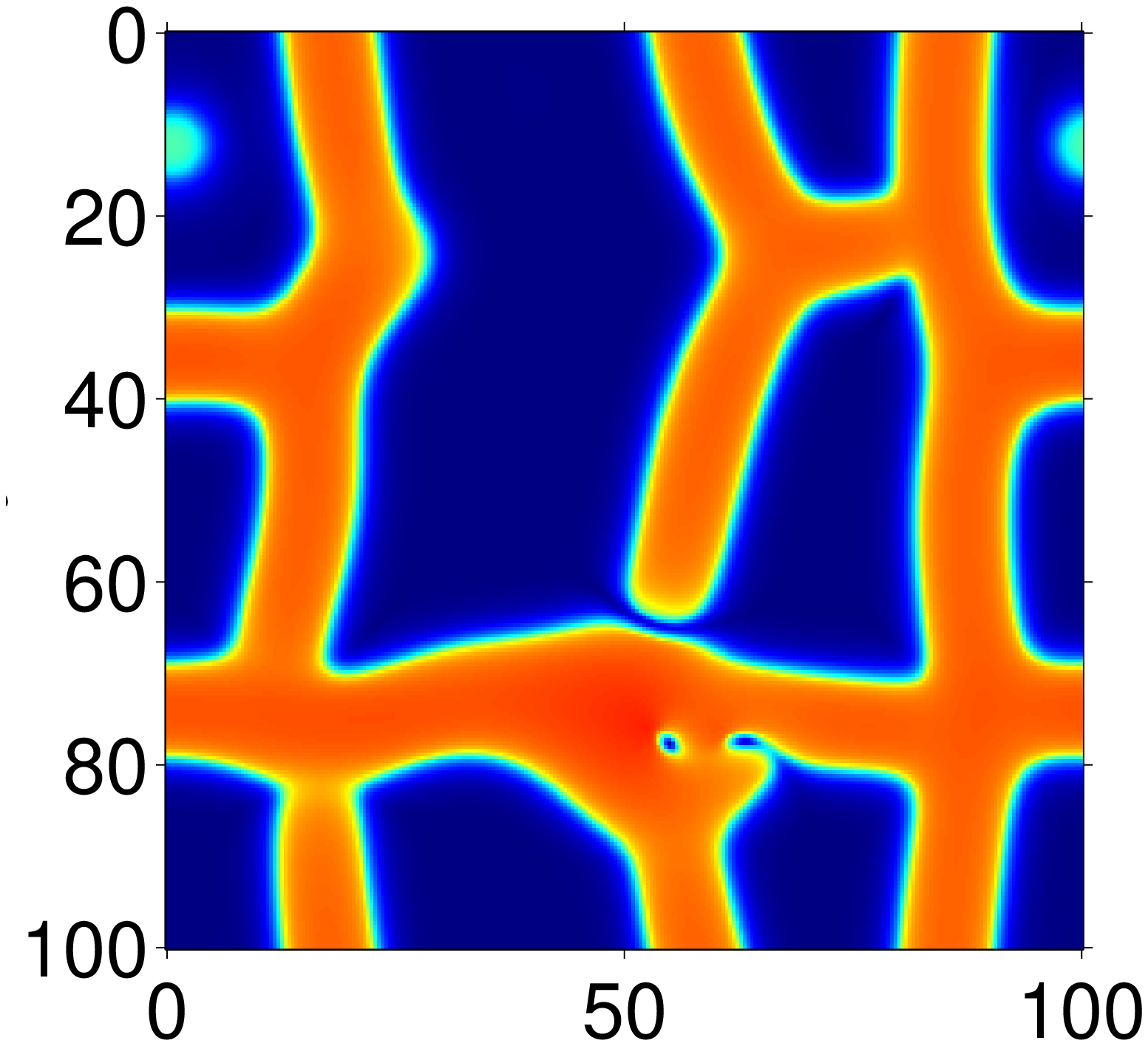}\label{fig:sineb_kgkc_t700_478}}
    \subfigure [$t=700$]
    {\includegraphics[width=4.1cm]{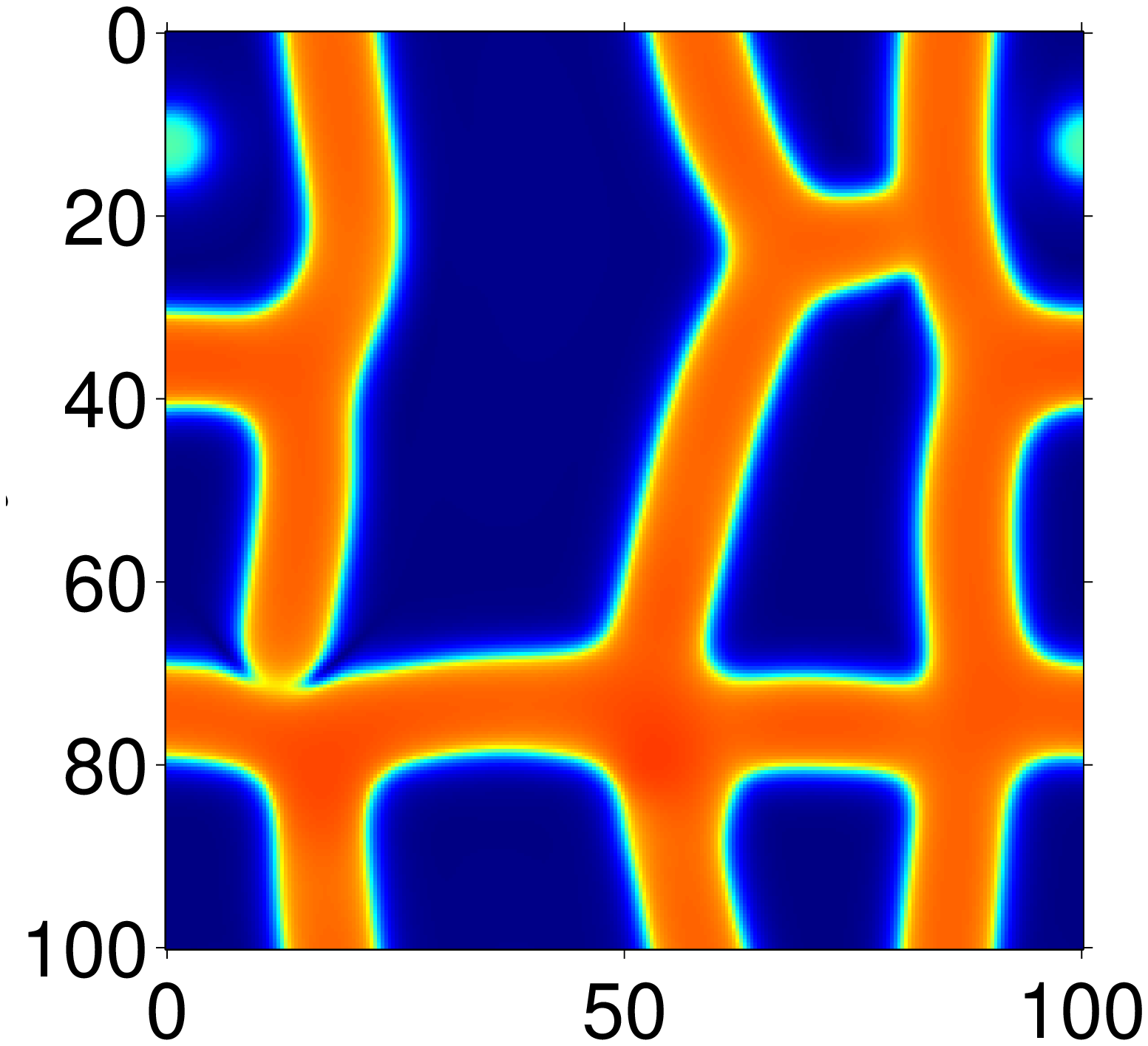}\label{fig:sineb_kgkc_t700_500}}
    \end{flushleft}
    \caption{
           Time evolution of $|r({\bf x},t)|$ 
           from an initial configuration (a)
           with half of the region at the incoherent state and half at the coherent
           state divided by a wiggled boundary with $k=14.8$ ($>k_c=14.5$). A comparison
           with Fig. \ref{fig:sineb_klkc} shows a much richer spatio-temporal dynamical pattern
           ($\omega_0=5, T=1, D=100$; 
           periodic boundary conditions are imposed).
           }
    \label{fig:sineb_kgkc}
\end{figure*}

Figures \ref{fig:sineb_klkc} and \ref{fig:sineb_kgkc} show the
$d=2$ counterparts to the $d=1$
propagating fronts and associated features.
Similar to what was previously done for $d=1$, half of
the system is initialized
in the homogeneous incoherent state and
the remaining half in the homogeneous coherent state, and they
are divided by a sinusoidally wiggling
boundary (Figs. \ref{fig:sineb_klkc_0} and \ref{fig:sineb_kgkc_0}).
Analogous to the $d=1$ case, for $d=2$,
the homogeneous
incoherent state and homogeneous coherent state
take over when $k$ is sufficiently small or large compared to 
$k_c$ respectively. 
The most interesting behaviors again take place when $k \approx k_c$.
With $k=14.4 < k_c$, Fig. \ref{fig:sineb_klkc} shows the
development of a stable
bridge solution. In contrast,
with $k=14.8 > k_c$, Fig. \ref{fig:sineb_kgkc} shows a surprisingly rich
spatio-temporal pattern evolution. As in Fig. \ref{fig:sineb_klkc},
the originally coherent half apparently starts to shrink into a bridge
(see Fig. \ref{fig:sineb_kgkc_69}); however, as time progresses further,
we see that coherent regions arise out of the 
originally incoherent regions to form new features 
(see also Figs. \ref{fig:1dpf_k14_6} and \ref{fig:1dpf_k14_8} in the 
$d=1$ case), and these new features interact in a nontrivial two dimensional manner.
For example,
when two neighboring coherent regions get close to each other, they 
can form bonds and merge into each other: see the connections formed between
bridge-like structures from $t=83$
to $t=98$; also see the coherent spot formed at the upper left hand corner
at $t=245$ and see how it merges into the bridge on its right 
as time progresses to $t=400$. 
We also observe that, during the process of merger, bridge-like structures 
may also temporarily separate and then re-connect: see the
connecting bridge at the bottom right hand corner from $t=138$ to $t=170$.
A further notable feature is the coherent spot on the top left hand 
corner at $t=561$
(a target pattern in the phase plot as shown in the next section),
which survives from $t=561$ to the end of the numerical run.
In the above reported numerical 
experiments we observe that both incoherent and coherent regions coexist
for a long time. We do not know, however, whether a coherent or incoherent state
ultimately will take over the whole domain at longer time.

\subsection{2D Spots, spiral waves and target patterns} 
\label{sec:target_spiral}
Figure \ref{fig:int_chg} shows the time evolution 
of both
$|r({\bf x},t)|$ and
$\sin[\theta({\bf x},t)]$ 
$[$ where $r({\bf x},t) = |r({\bf x},t)| \exp[i\theta({\bf x},t)]$, and ${\bf x}=(x,y)$ in
$2D$ $]$ when
the system is initialized with a small random
initial condition at each grid point, 
and the coupling strength is $k=15.5$ ($>k_c=14.5$). 
As expected from our previous studies, when $k>k_c$, 
coherent regions ($|r| \approx 1$) emerge
from the initial incoherent state. Further, the phase plots show
some distinct target-like patterns of nested closed surfaces of constant
phase (see $t=40$ and $t=217$).
As time progresses, coherent regions (red in the $|r|$ plots)
become dominant and only small
islands of incoherent regions remain (blue in the $|r|$ plots).
Similar to our previous observation of propagating fronts when
$k>k_c$ (compare Figs. \ref{fig:sineb_kgkc_213} and \ref{fig:sineb_kgkc_311}),
coherent regions can form in an originally incoherent region ($|r| \ll 1$).
For example, see the figures from $t=139$ to $t=161$, and especially
from $t=195$ to $t=225$, where we see coherent regions (red/yellow) appearing
and growing in the interior of incoherent (blue) blob, eventually 
destroying it.
As can be inferred by comparing the $|r|$ and $\sin(\theta)$ plots,
small blue, dot-like features in the $|r|$ plots represent phase defects
in the complex amplitude (i.e., counter clockwise encirclement 
of such a feature leads to a phase change of either $+2\pi$ or
$-2\pi$), and these blue dot features are commonly seen as spiral wave type patterns
in the phase plots. When, as in the previously noted plots from $t=195$ to $225$,
coherent regions take over from an incoherent patch, we also note that
a number of phase defects result
(which must be formed in opposite-spiral-parity pairs 
due to the conservation of topological charge); see $t=250$.
The isolated phase defects subsequently wander about,
and some of them are seen to 
annihilate with others of opposite parity
(see the two defects closest to the bottom of the picture
at $t=267$ and their evolution up to $t=293$), or sometimes 
they are absorbed into 
an incoherent region (e.g., compare the $|r|$ plots at $t=195$ and $t=217$).
Lastly, regarding the speed of motion of spiral patterns, we note that
similar to the observation in Fig. \ref{fig:four_hole_abs},
when oppositely charged spirals get close enough to each other, 
their speed of
approach becomes distinctively faster till they annihilate 
each other.

\begin{figure*}[h] 
  \begin{center}
  \subfigure[$|r({\bf x})|$]
  {\includegraphics[width=4.1cm]{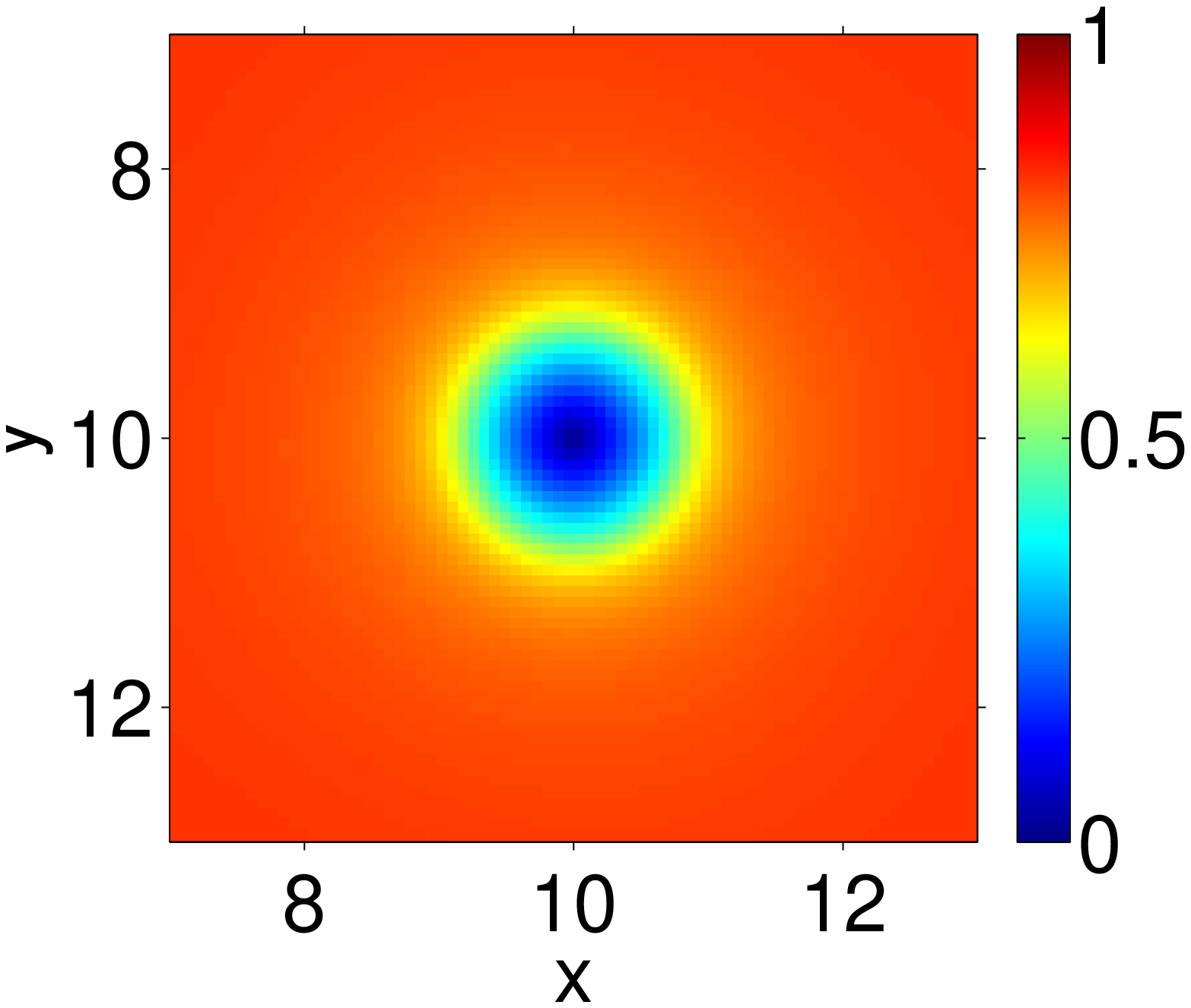} \label{fig:spi_core_abs}}
  \subfigure[$\sin (\theta({\bf x}))$]
  {\includegraphics[width=4.1cm]{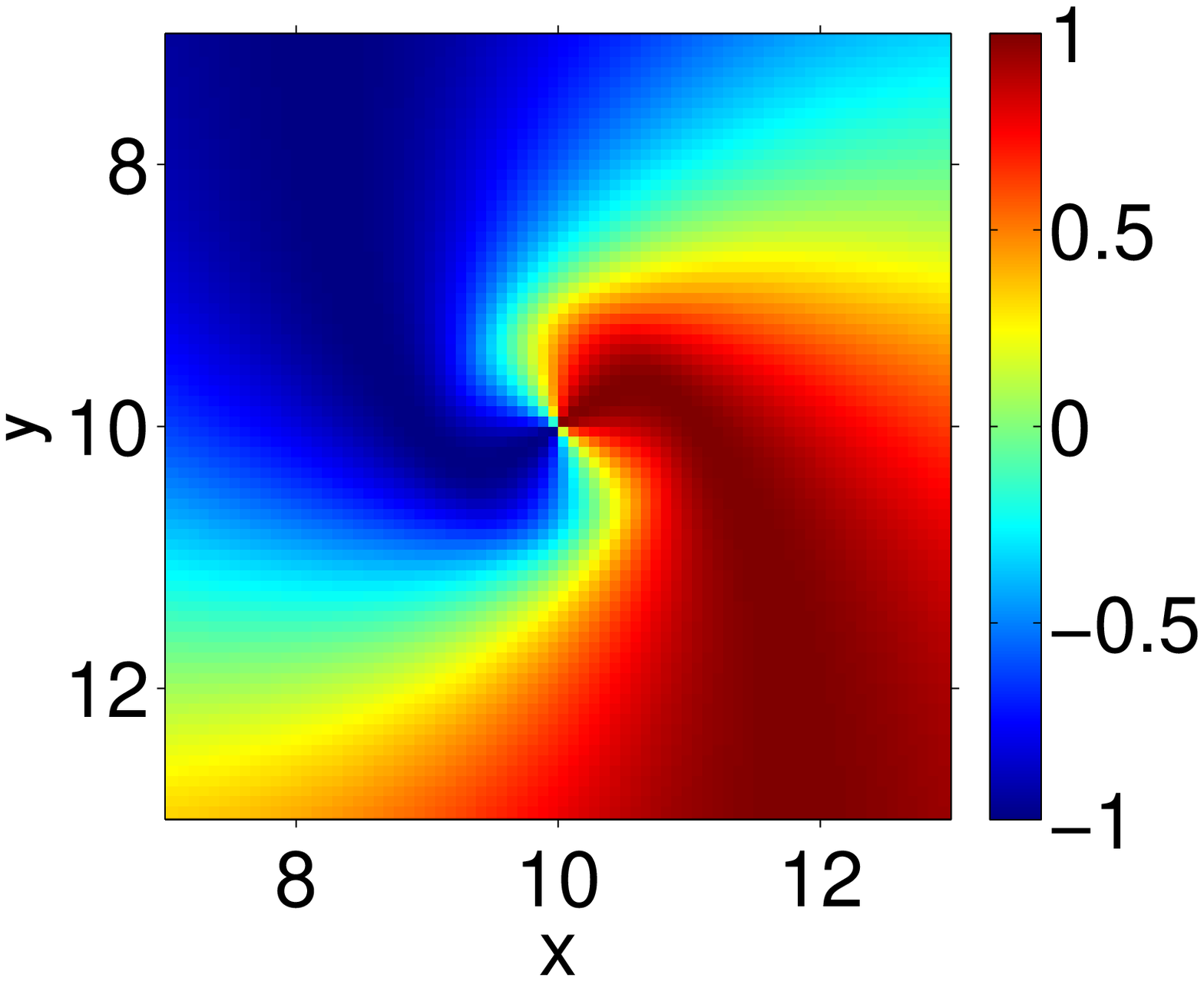} \label{fig:spi_core_ph}}
  \end{center}
\caption{Finite area of (two-dimensional) spiral cores
         ($\omega_0=5, T=1, D=20, k=15$).}
\label{fig:spi_core}
\end{figure*}

In studies of the CGLE, the hole pattern and
spiral wave pattern are analogous phenomena occurring in $d=1$ and $d=2$,
respectively. Indeed, the hole pattern and spiral wave pattern exhibit
similar characteristics in our study. 
Both features are stable with respect to small changes in parameters, and exhibit
similar dynamical characteristics of approach and
annihilation as described above. In addition, Fig. \ref{fig:spi_core} shows, in
parallel with Fig. \ref{fig:hol_core}, that
the central core of the spiral wave pattern occupies 
a finite area when $T \neq 0$. This is similar to the chimera-centered spirals
noted in Refs. \cite{Martens,KSB}.

 
\subsection{2D pulsating pattern}
\begin{figure*}[h]  
  \begin{center}
  \subfigure[$|r({\bf x},129)|$]
  {\includegraphics[width=4.1cm]{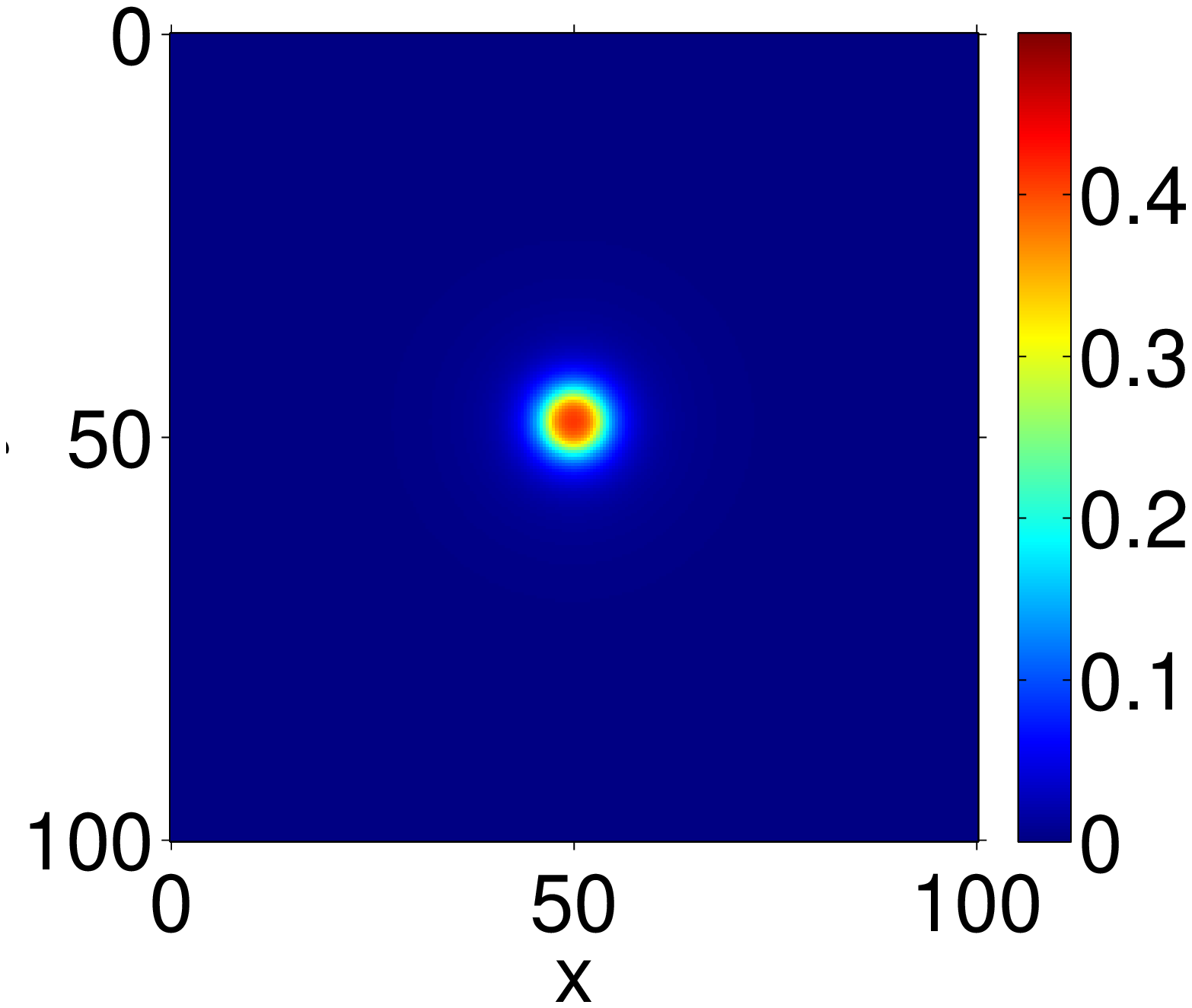} \label{fig:pulse_abs_t129}}
  \subfigure[$|r({\bf x},148)|$]
  {\includegraphics[width=4.1cm]{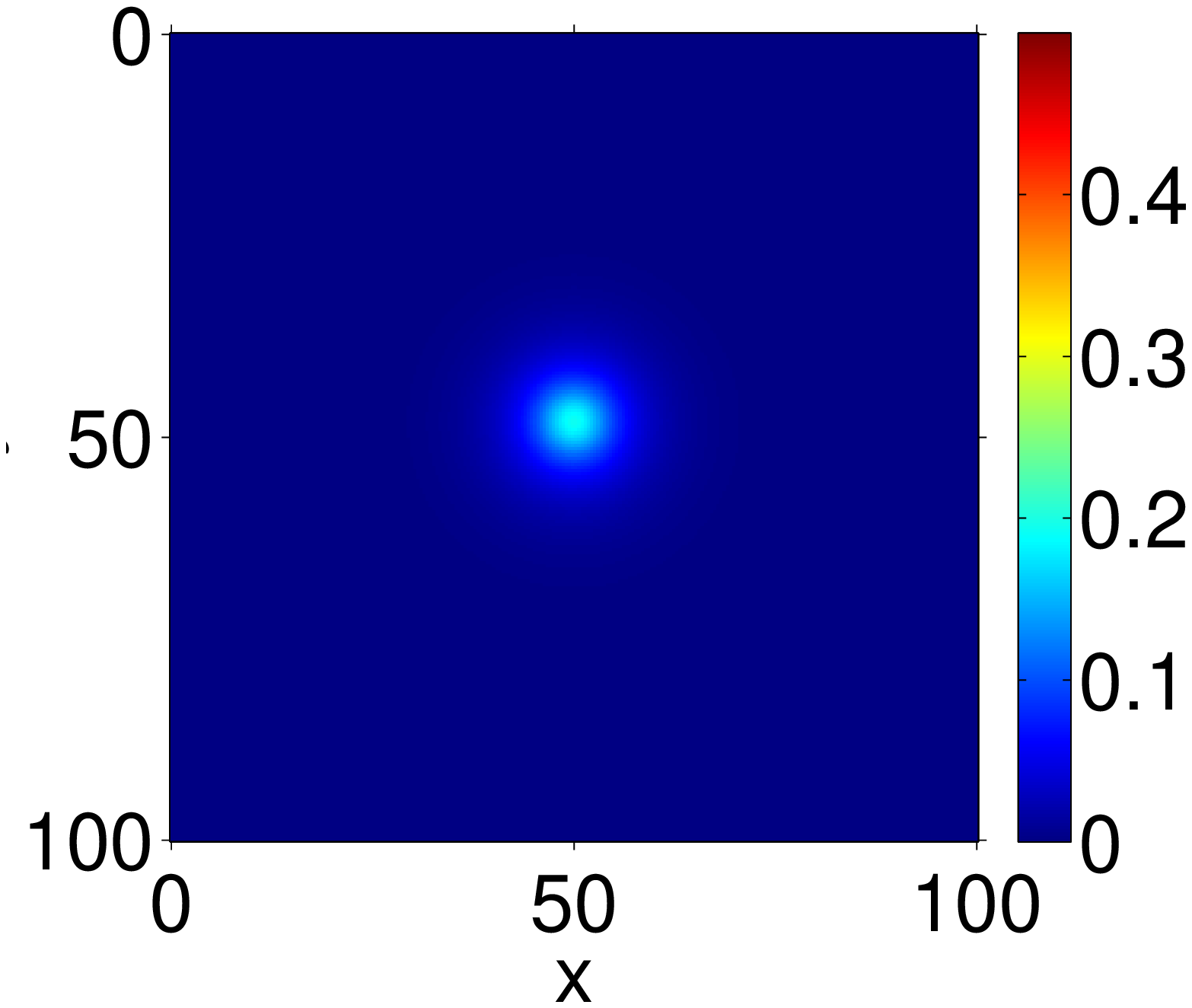} \label{fig:pulse_abs_t148}}
  \subfigure[$|r({\bf x},158)|$]
  {\includegraphics[width=4.1cm]{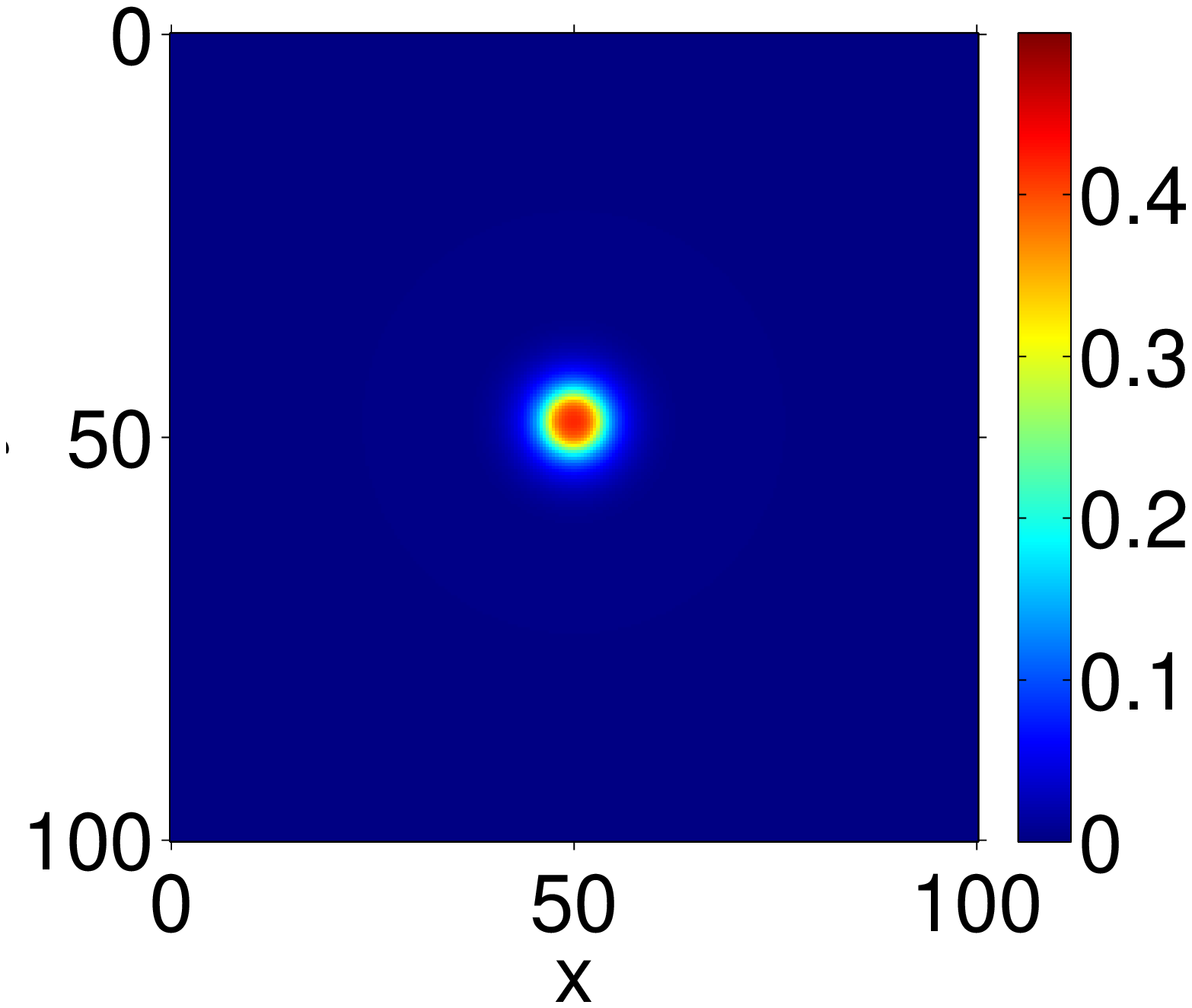} \label{fig:pulse_abs_t158}}
  \subfigure[$|r(50,50,t)|$]
  {\includegraphics[width=4.1cm]{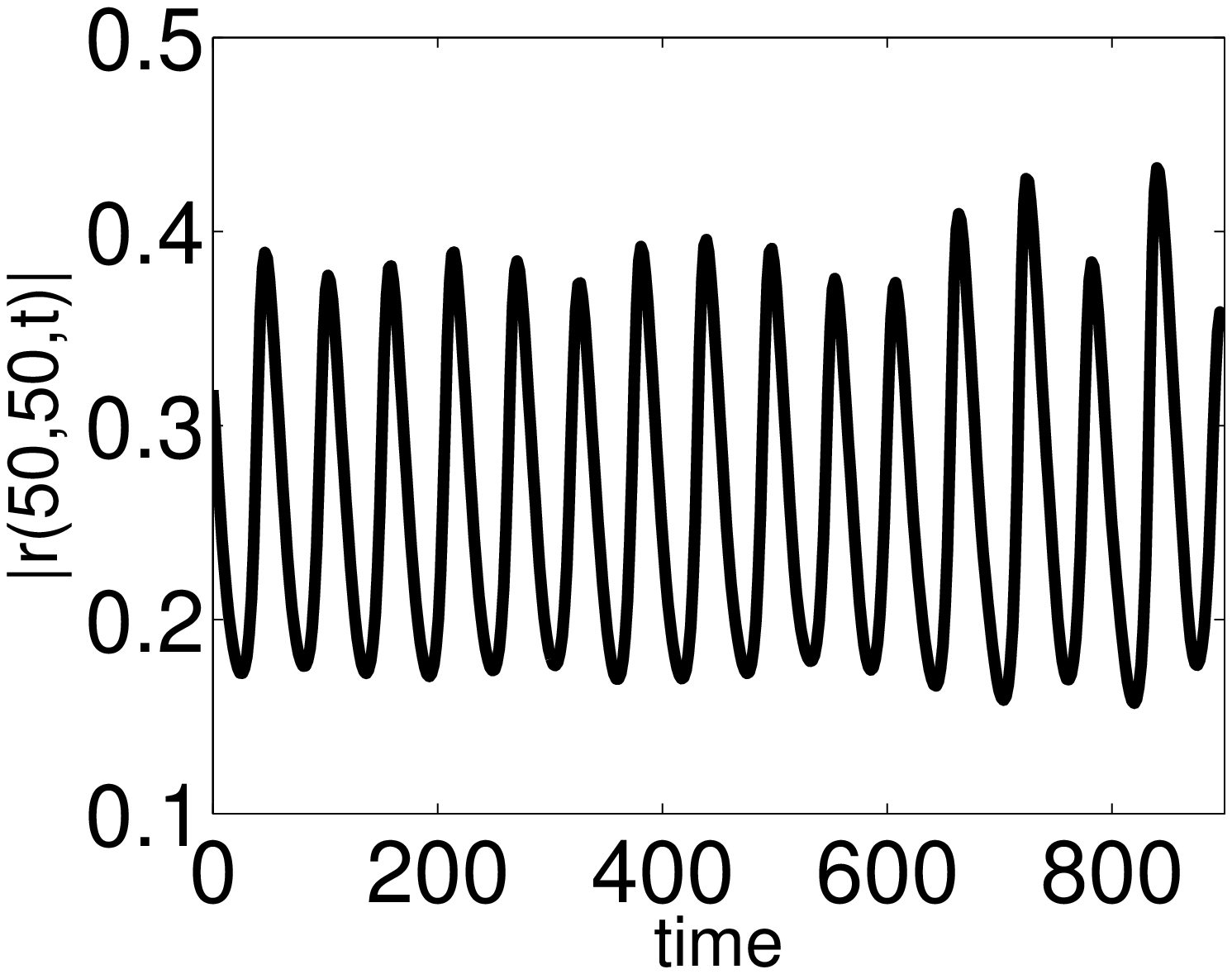} \label{fig:pulse_abs_tvariation}}
  \end{center}
\caption{
         Pulsating pattern: amplitude variation. 
         Figures (a) to (c) show approximately one ``period''
         of oscillation in $|r|$. Figure (d) shows the time variation of $|r|$ at the center
         of the pulse;
         compare with Fig. \ref{fig:pulse_ph} for
         oscillations in phase
         ($\omega_0=5, T=1, D=100, k=14.52$; periodic boundary conditions are imposed).     
         }
\label{fig:pulse_abs}
\end{figure*}

\begin{figure*}[h]
  \begin{center}
  \subfigure[$\sin(\theta), t=129$]
  {\includegraphics[width=4.1cm]{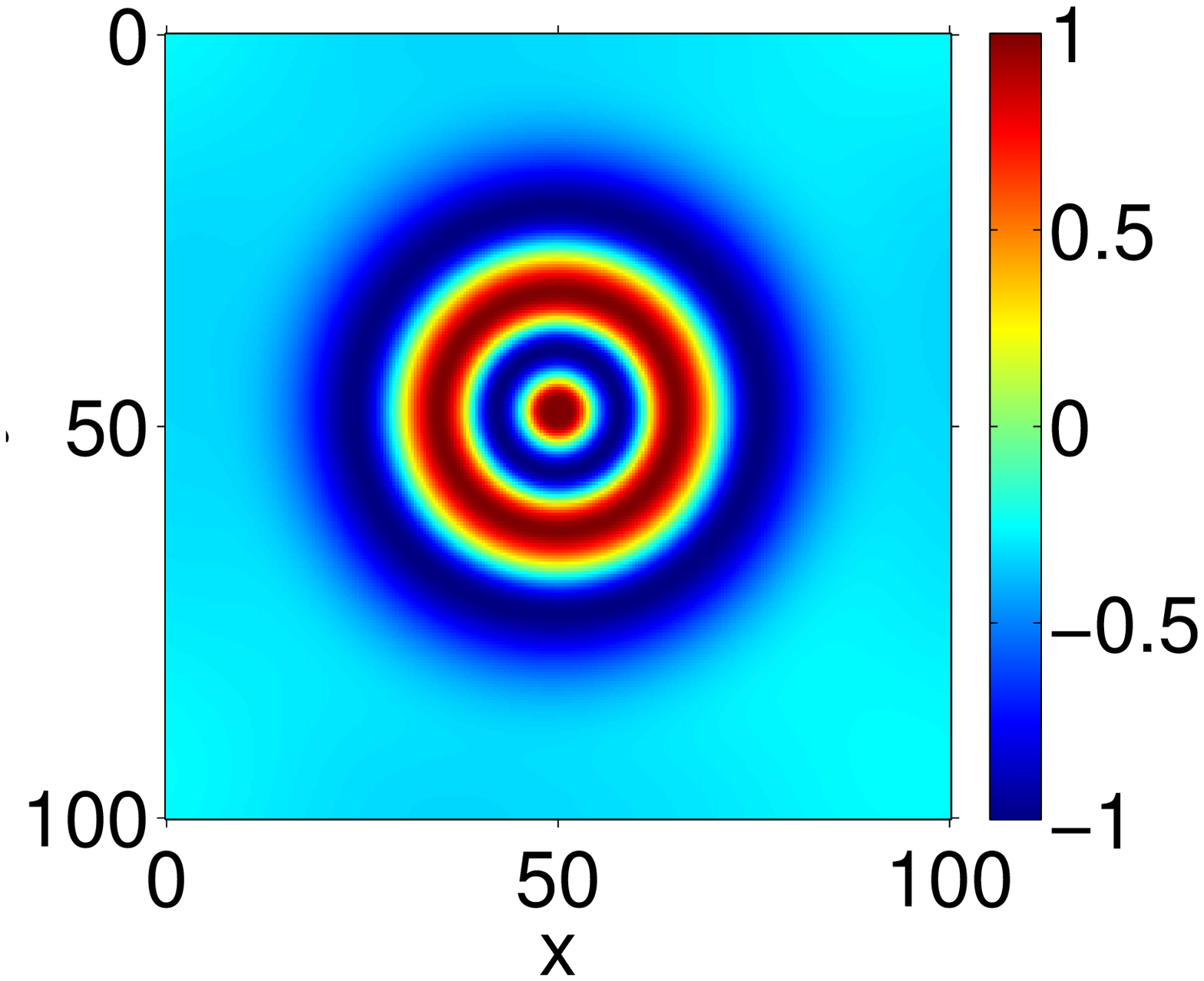} \label{fig:pulse_ph_t129}}
  \subfigure[$\sin(\theta), t=130$]
  {\includegraphics[width=4.1cm]{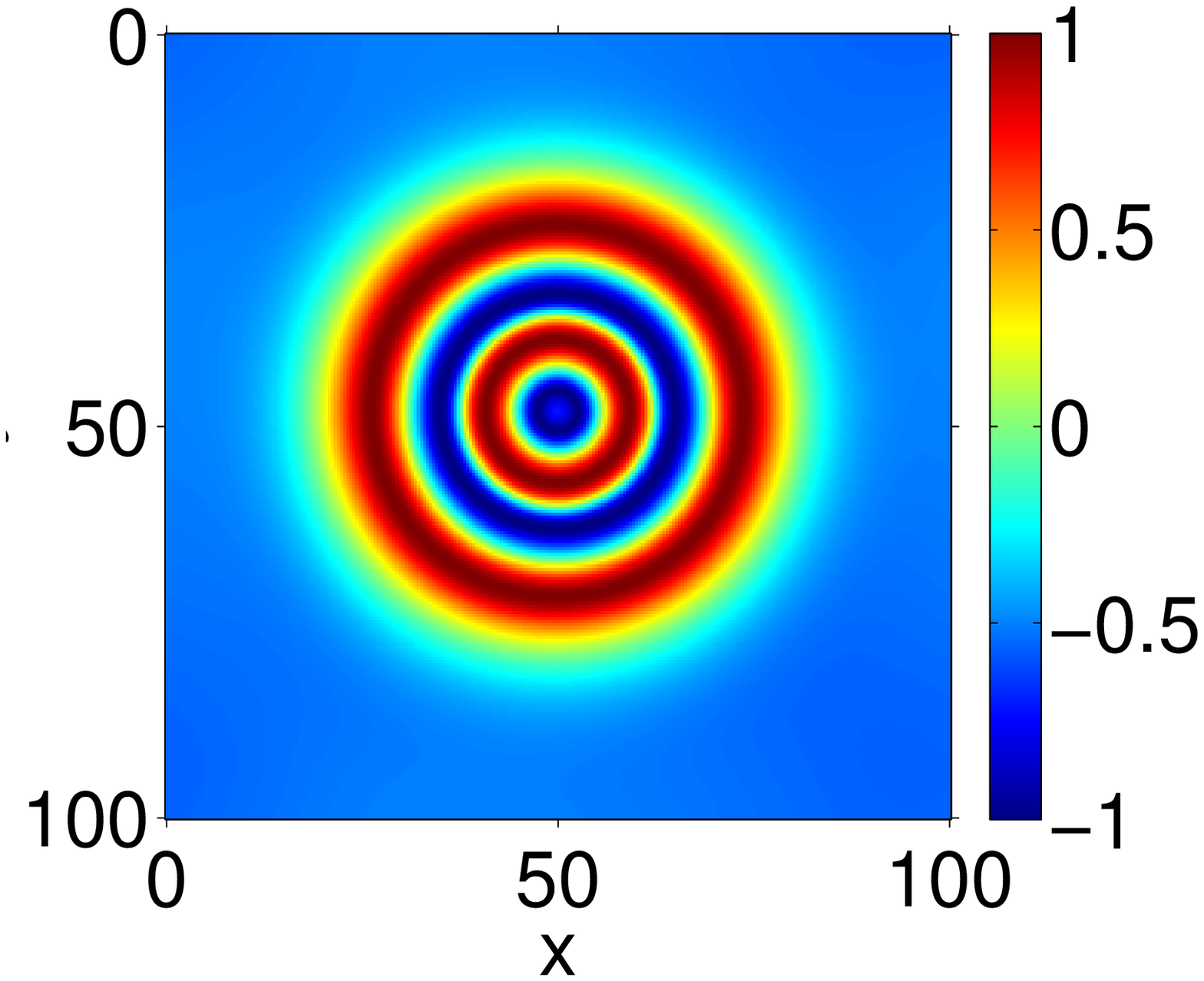} \label{fig:pulse_ph_t130}}
  \subfigure[$\sin(\theta), t=131$]
  {\includegraphics[width=4.1cm]{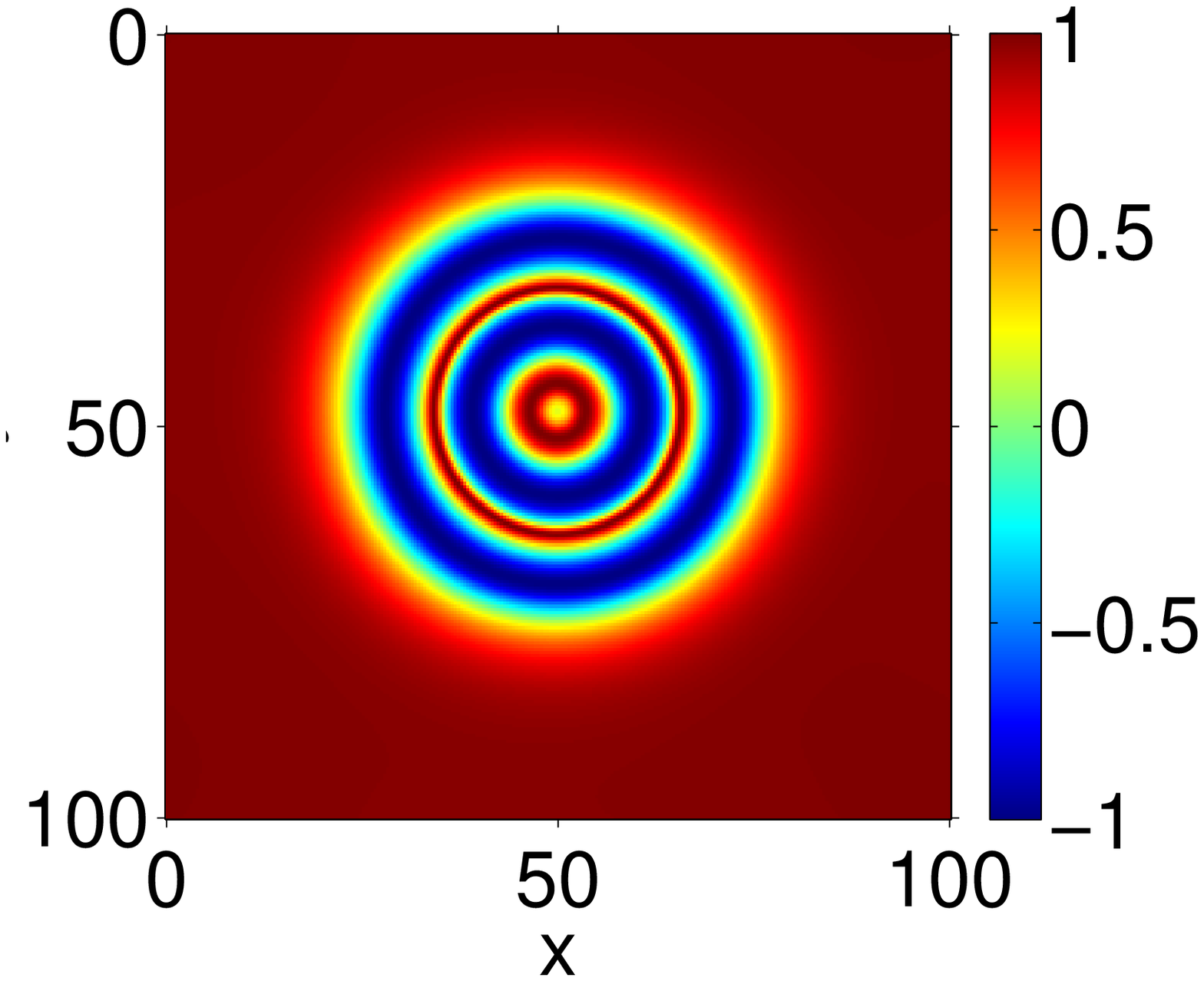} \label{fig:pulse_ph_t131}}
  \subfigure[$\sin(\theta), t=132$]
  {\includegraphics[width=4.1cm]{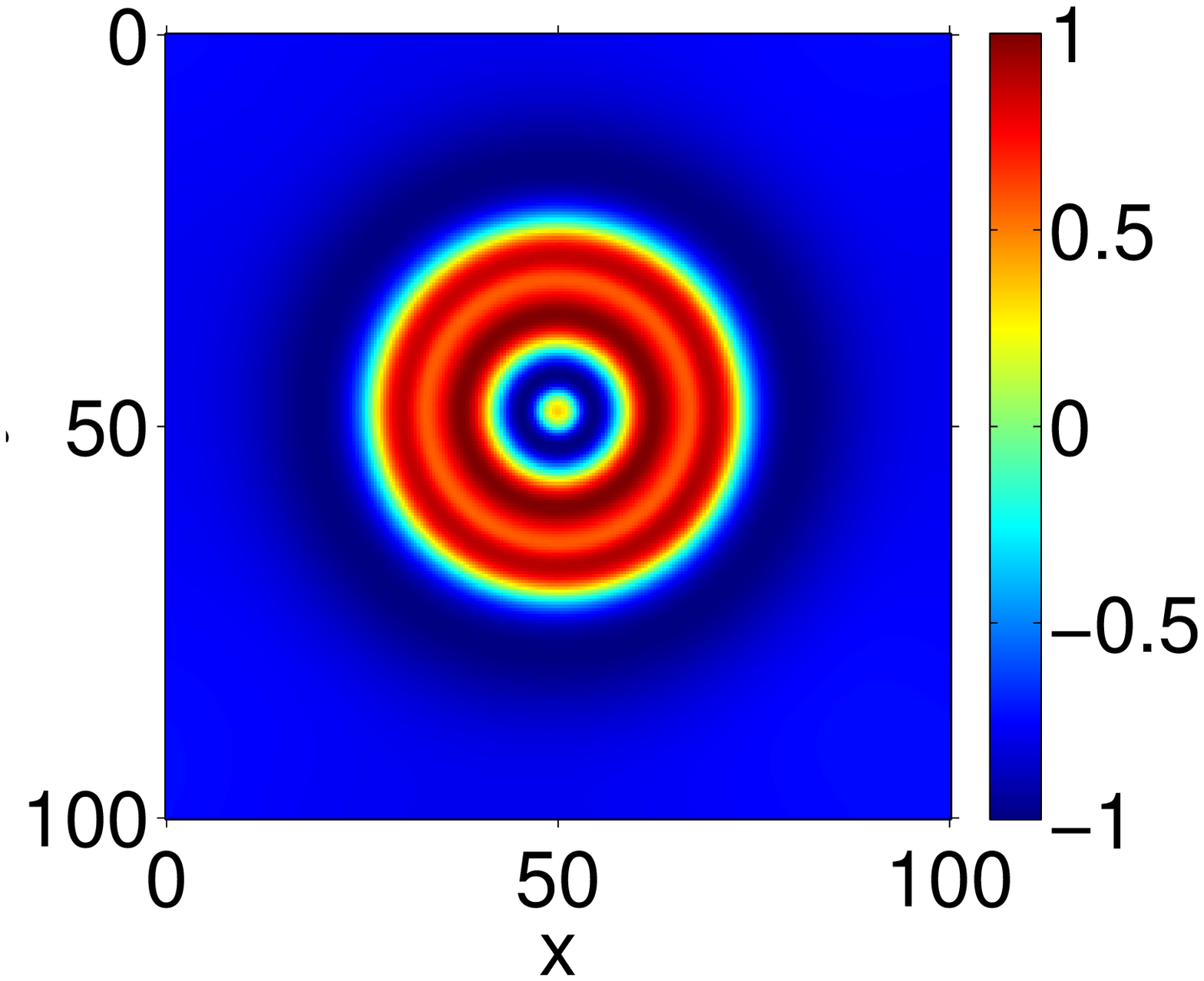} \label{fig:pulse_ph_t132}}
  \subfigure[$\sin(\theta), t=133$]
  {\includegraphics[width=4.1cm]{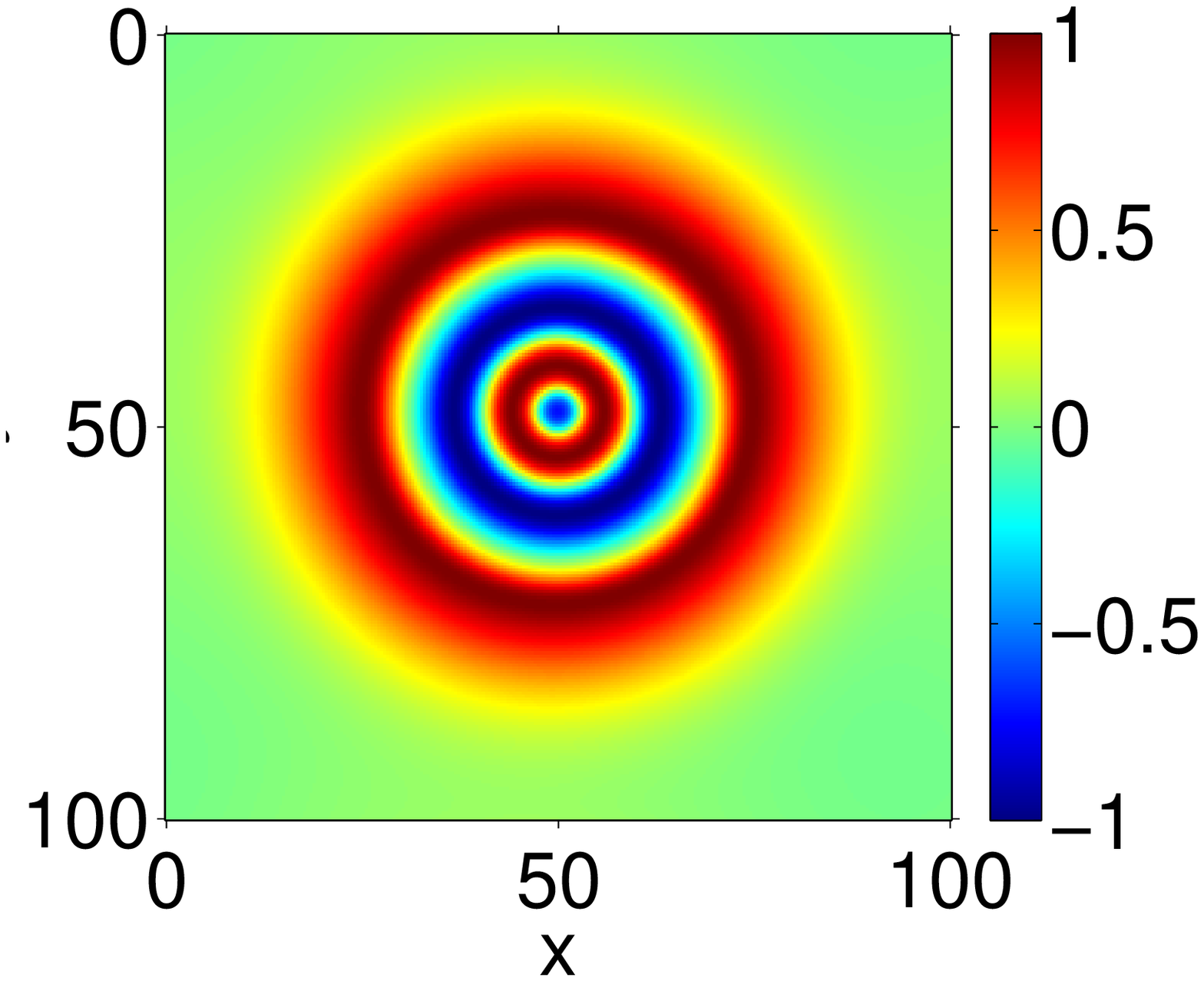} \label{fig:pulse_ph_t133}}
  \subfigure[$\sin(\theta), t=134$]
  {\includegraphics[width=4.1cm]{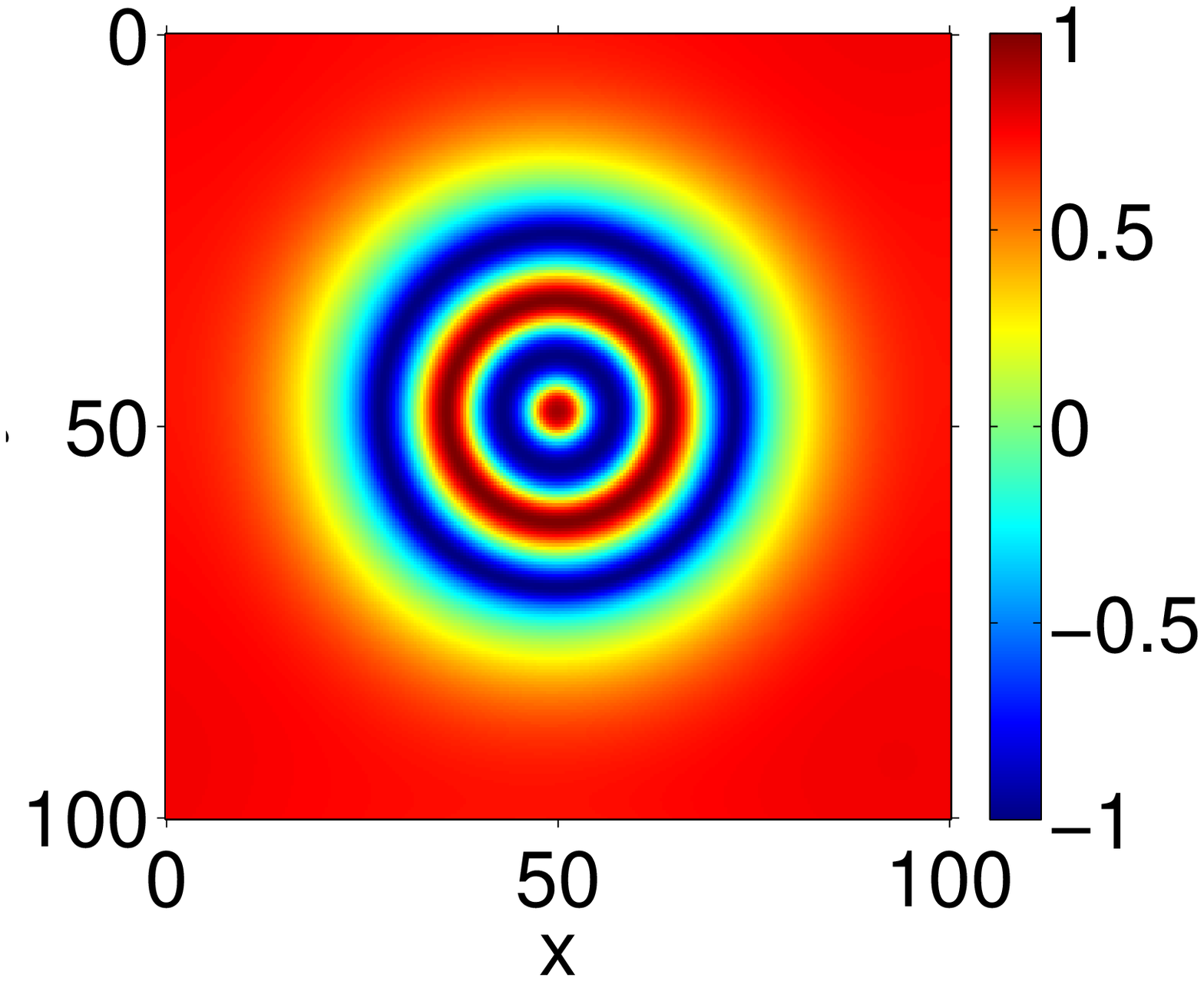} \label{fig:pulse_ph_t134}}
  \subfigure[$\sin(\theta(50,50,t))$]
  {\includegraphics[width=4.1cm]{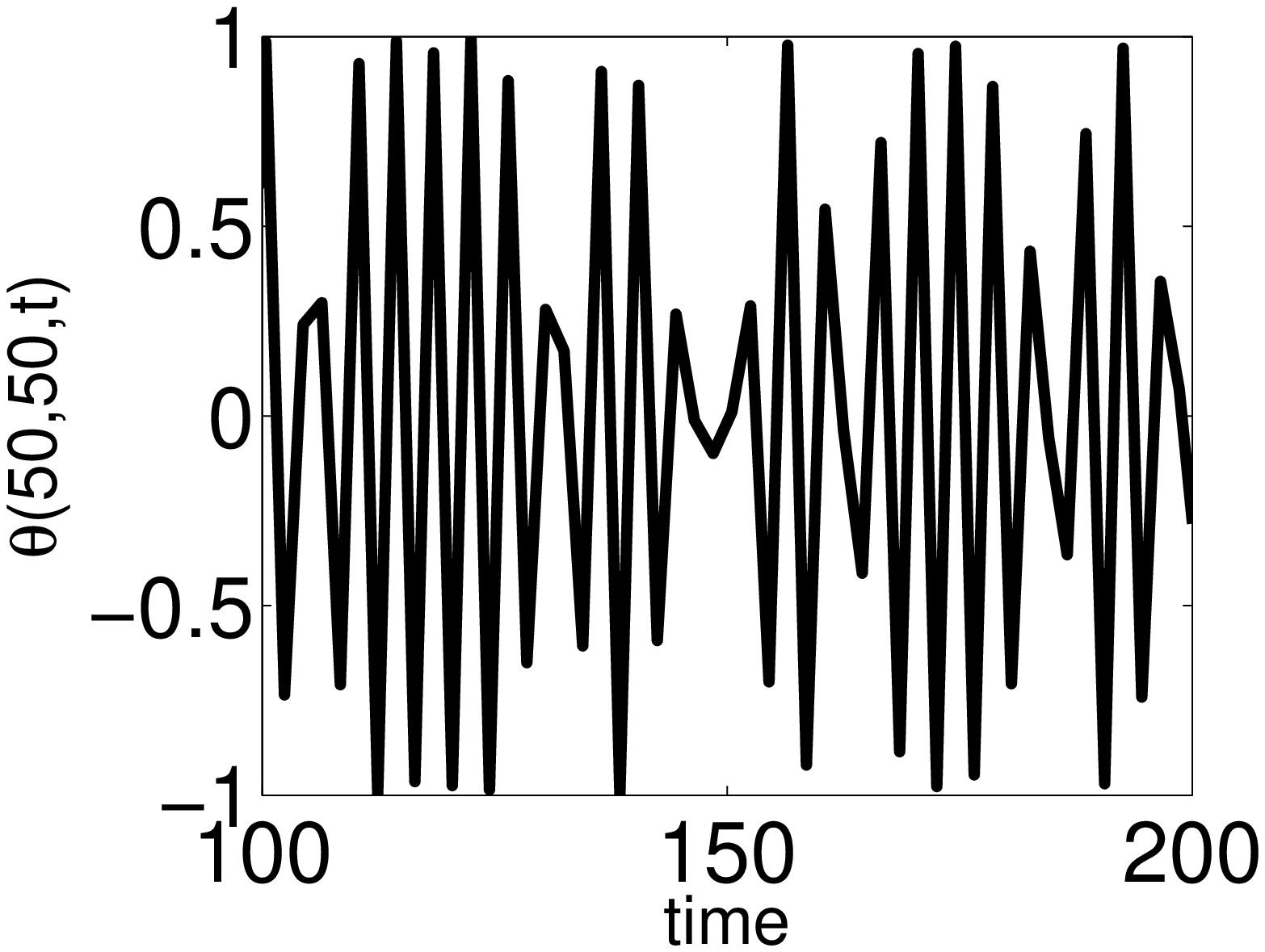}
                                                   \label{fig:pulse_ph_tvariation}}
  \end{center}
\caption{
         Pulsating pattern: phase variation. Figures (a) to (f) show the rapid
         time variations of the phase. Figure (g) shows the time variation of $sin(\theta)$
         at the center of the pulse
         (Parameters are as indicated in Fig. \ref{fig:pulse_abs}).
        }
\label{fig:pulse_ph}
\end{figure*}

Another class of local coherent structures supported in the
$d=2$ case is shown in Figs. \ref{fig:pulse_abs} and \ref{fig:pulse_ph},
which shows a localized pulsating spot in an incoherent background. 
It is interesting to notice that 
oscillations of the magnitude and phase (which show up in the form
of target patterns) of $r(x,t)$
are not the same, with that of the phase oscillation being 
more irregular and more than an order of
magnitude
faster than the amplitude oscillation 
(Figs. \ref{fig:pulse_abs_tvariation}
and \ref{fig:pulse_ph_tvariation}). It 
is interesting to note that for the
CGLE, stable pulsating patterns come only with the addition
of a quintic term (see Ref. \cite{Deissler},
and the later work
Ref. \cite{Akh} and references therein).



\section{Summary and Concluding Remarks}
In this paper, we have studied the spatio-temporal dynamics of 
spatially coupled oscillator systems where the oscillators have
a heterogeneous distribution of response times. Using the results of
Refs. \cite{OA1}-\cite{OA3}, we have derived a macroscopic PDE
description for this situation [Eqs. (\ref{eq:alpha_4})-(\ref{eq:eta_4})].
The resulting macroscopic dynamics 
are found to exhibit a wide variety of pattern 
formation behaviors. We characterized the possible behaviors 
roughly according to the hysteresis loop corresponding to bistable homogeneous 
incoherent and homogeneous coherent state solutions. Numerical 
studies show that the system behaviors
for $k$ sufficiently below/above the bistable $k$-range
are simple in that
the homogeneous incoherent/coherent state eventually takes over the
entire domain.
In contrast, for $k$ in or near the bistable range the system can exhibit
a variety of interesting spatio-temporal phenomena.
These include
propagating fronts, 
bridge patterns, hole patterns ($d=1$), spiral waves ($d=2$),
spots, target patterns, pulsating patterns, etc. 

Finally, it is interesting to consider the role of time delay in
contributing to the features that we observed.
If there is no time delay (i.e., $T= 0$),
there is no homogeneous bistable behavior as observed in Fig. \ref{fig:hys_ome05}, and
the transition from the homogeneous incoherent state 
to the homogeneous coherent state is supercritical and takes place
at $k_c = 2\Delta$. In this case, many of the interesting
spatio-temporal phenomena that 
we have found for $T > 0$ are absent.
For example, when $T=0$,
the intricate 1D glassy state transitions 
were not observed, and the system typically evolves relatively rapidly into
either
homogeneous incoherent or homogeneous coherent state solutions.
The 2D waves arisen from topological defects are still present; 
however,
for $T=0$ the system will be similar to the case
of zero nonlinear dispersion
in Ref. \cite{Shima_Kura}, where the incoherent core
remains a point defect but not a finite area as observed when 
$T \neq 0$. Thus
finite response time introduces additional
dynamics, leading to the large variety of 
behaviors observed.

\vspace{1cm}
Acknowledgments. Work at the University of Maryland was supported by a MURI 
grant (ONR N00014-07-1-0734). Work at the University of Colorado was supported
by the NSF (DMS 1030586).

\begin{figure*}[htbp] 
  \begin{center}
    \subfigure [$t=0$]
    {\includegraphics[width=4.5cm]{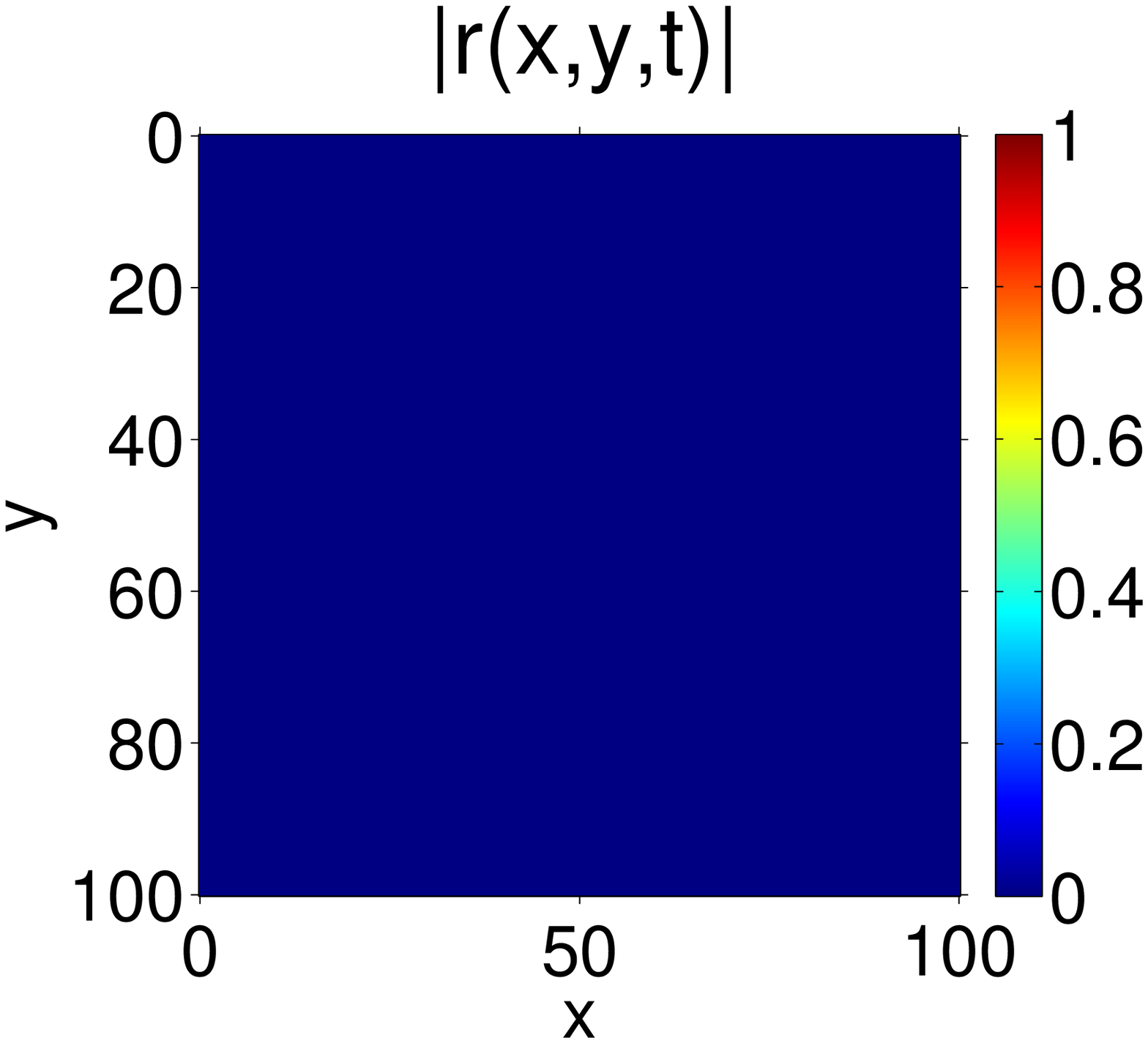}\label{fig:dyn_int_chg0a}}
    \subfigure [$t=0$]
    {\includegraphics[width=4.5cm]{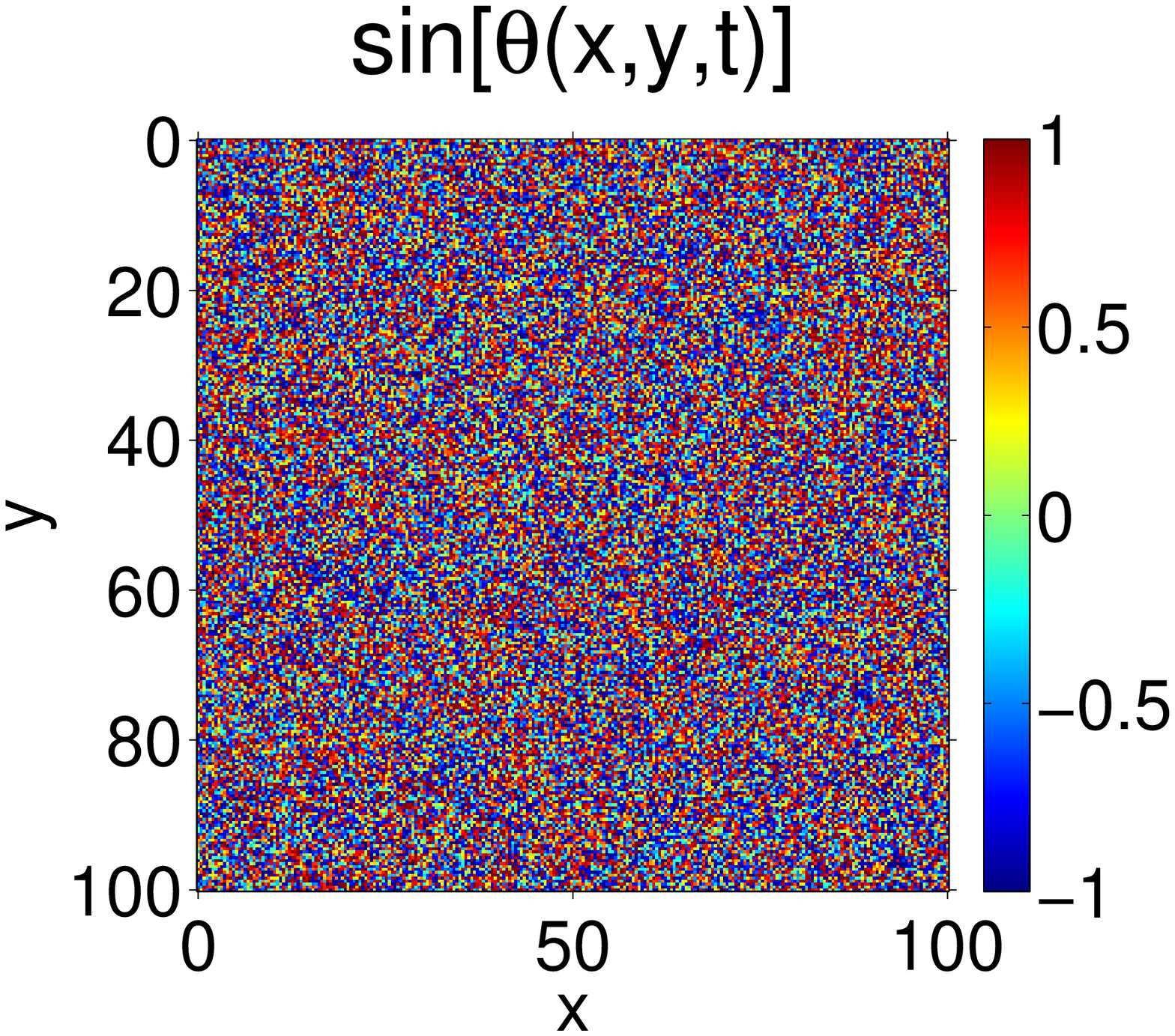}\label{fig:dyn_int_chg0b}}\\
    \subfigure [$t=40$]
    {\includegraphics[width=4.7cm]{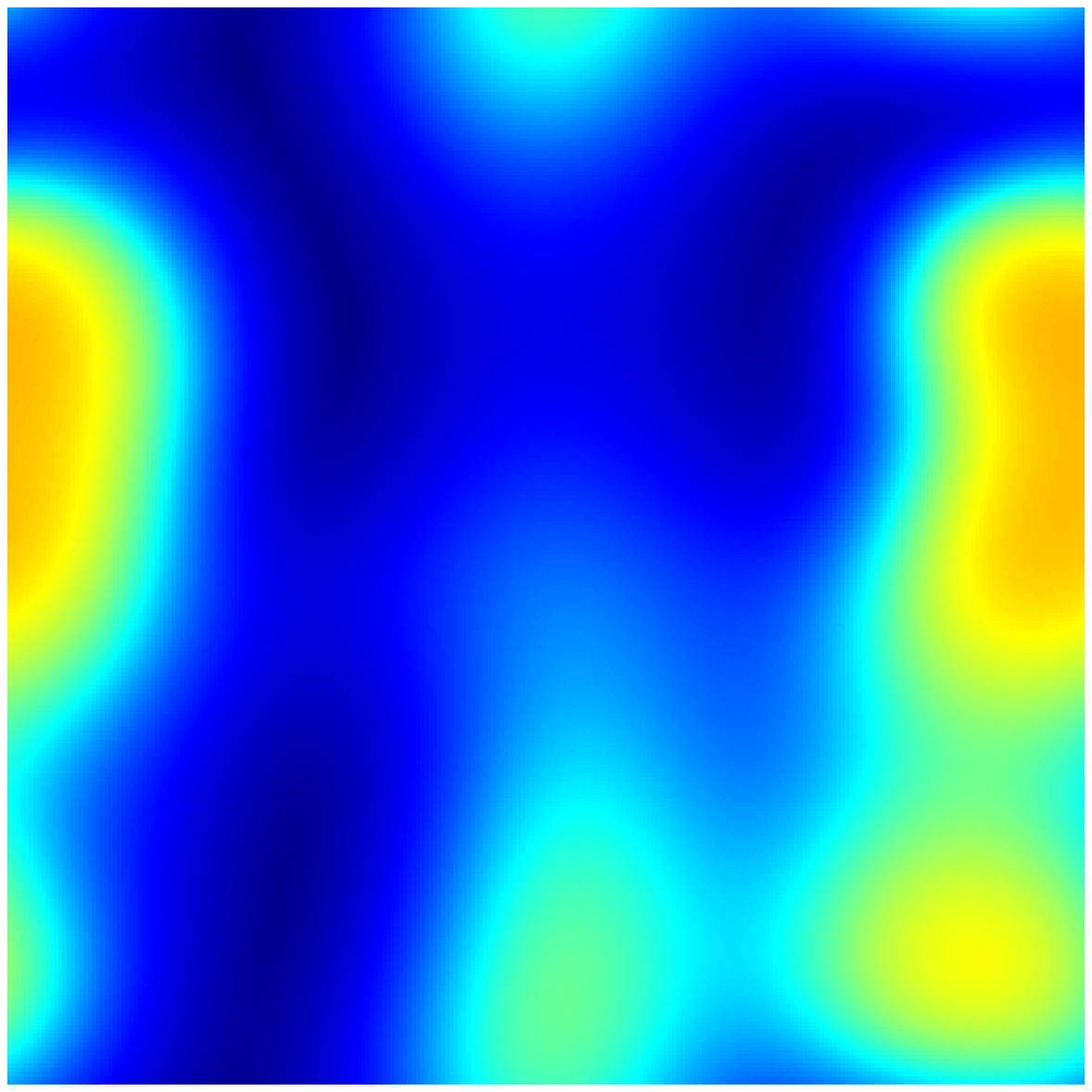} \label{fig:dyn_int_chg1a}}
    \subfigure [$t=40$]
    {\includegraphics[width=4.7cm]{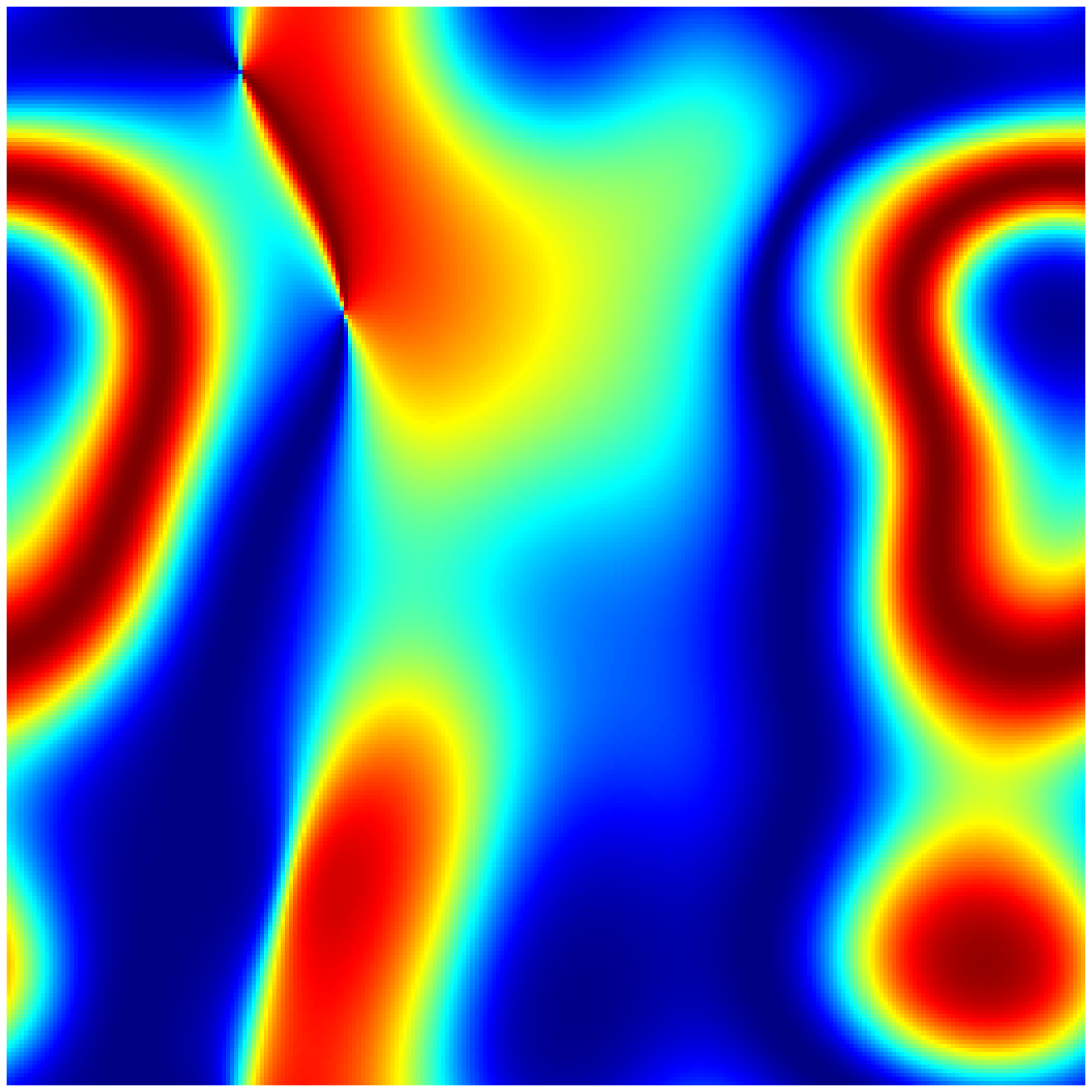} \label{fig:dyn_int_chg1b}}\\
    \subfigure [$t=57$]
    {\includegraphics[width=4.1cm]{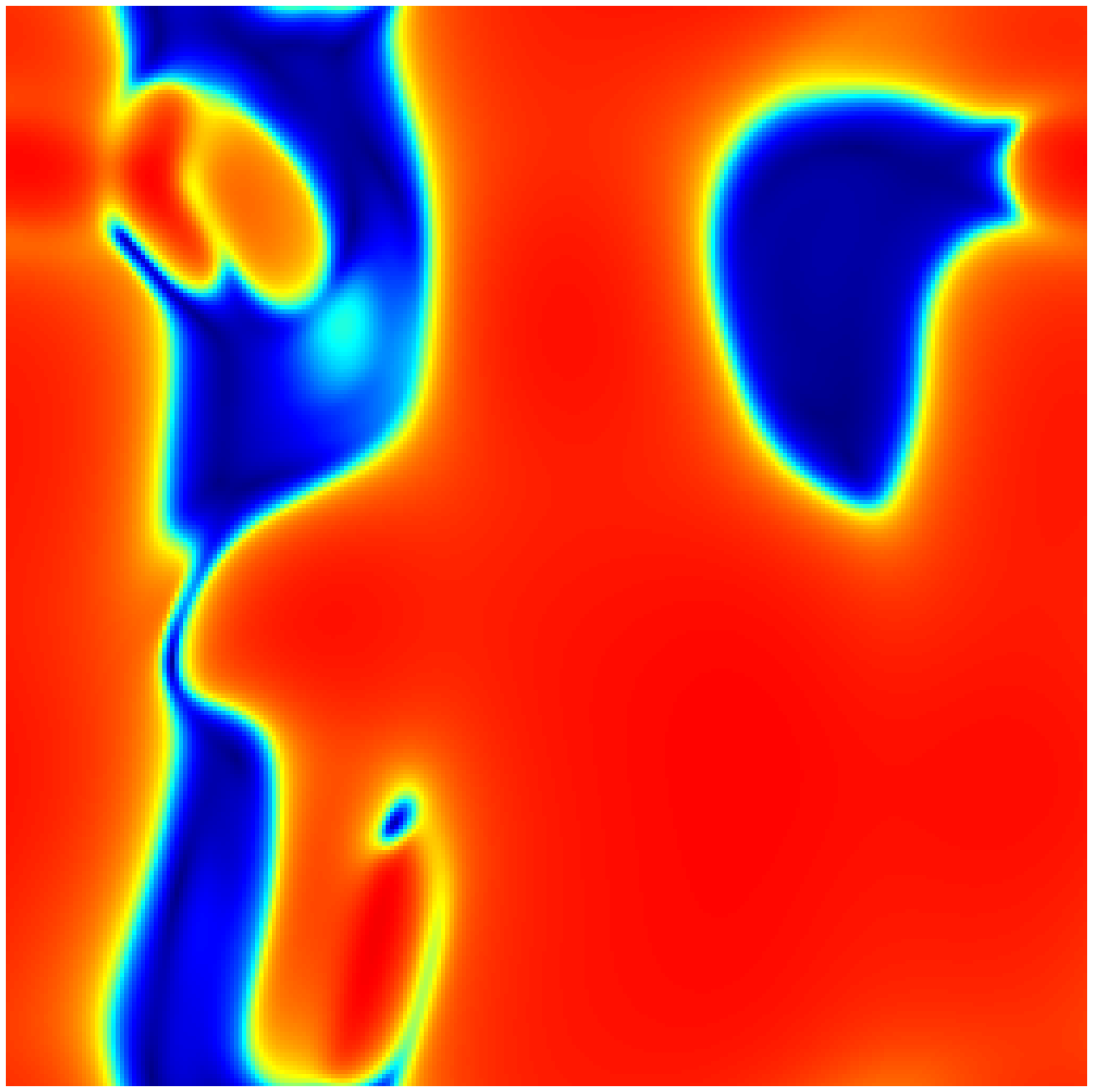}\label{fig:dyn_int_chg2a}}
    \subfigure [$t=57$]
    {\includegraphics[width=4.1cm]{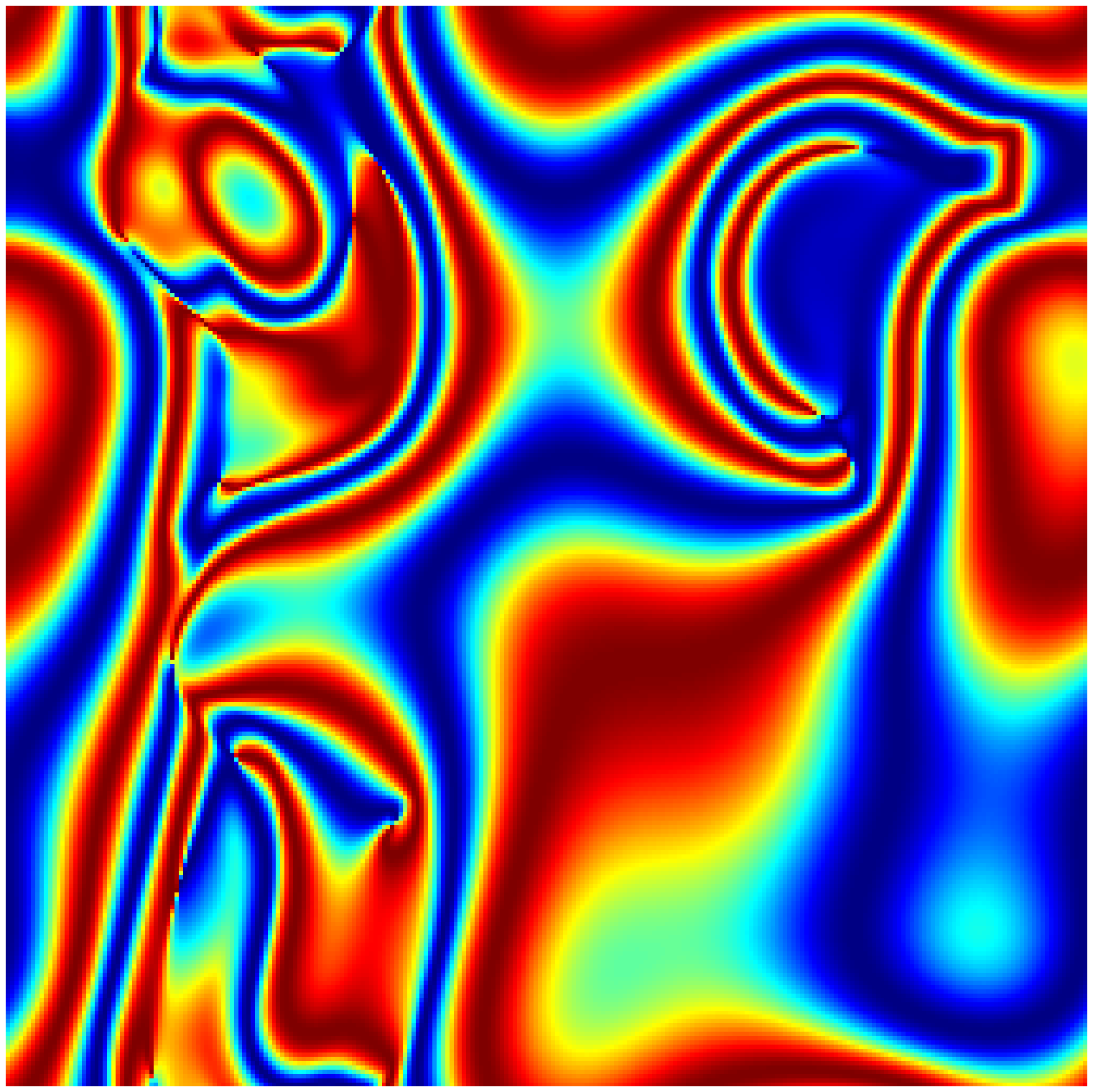}\label{fig:dyn_int_chg2b}}\\
    \subfigure [$t=87$]
    {\includegraphics[width=4.1cm]{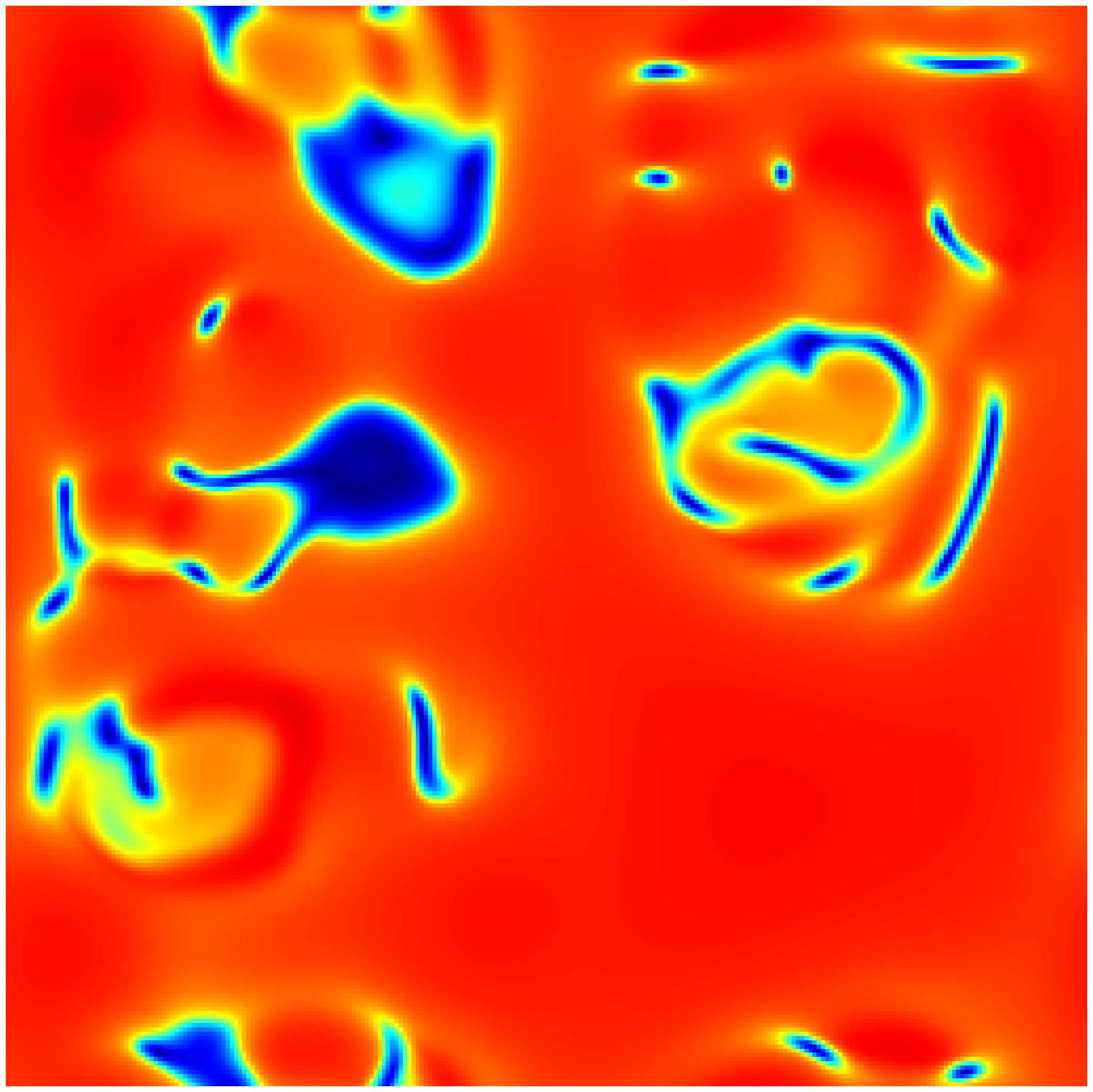}\label{fig:dyn_int_chg3a}}
    \subfigure [$t=87$]
    {\includegraphics[width=4.1cm]{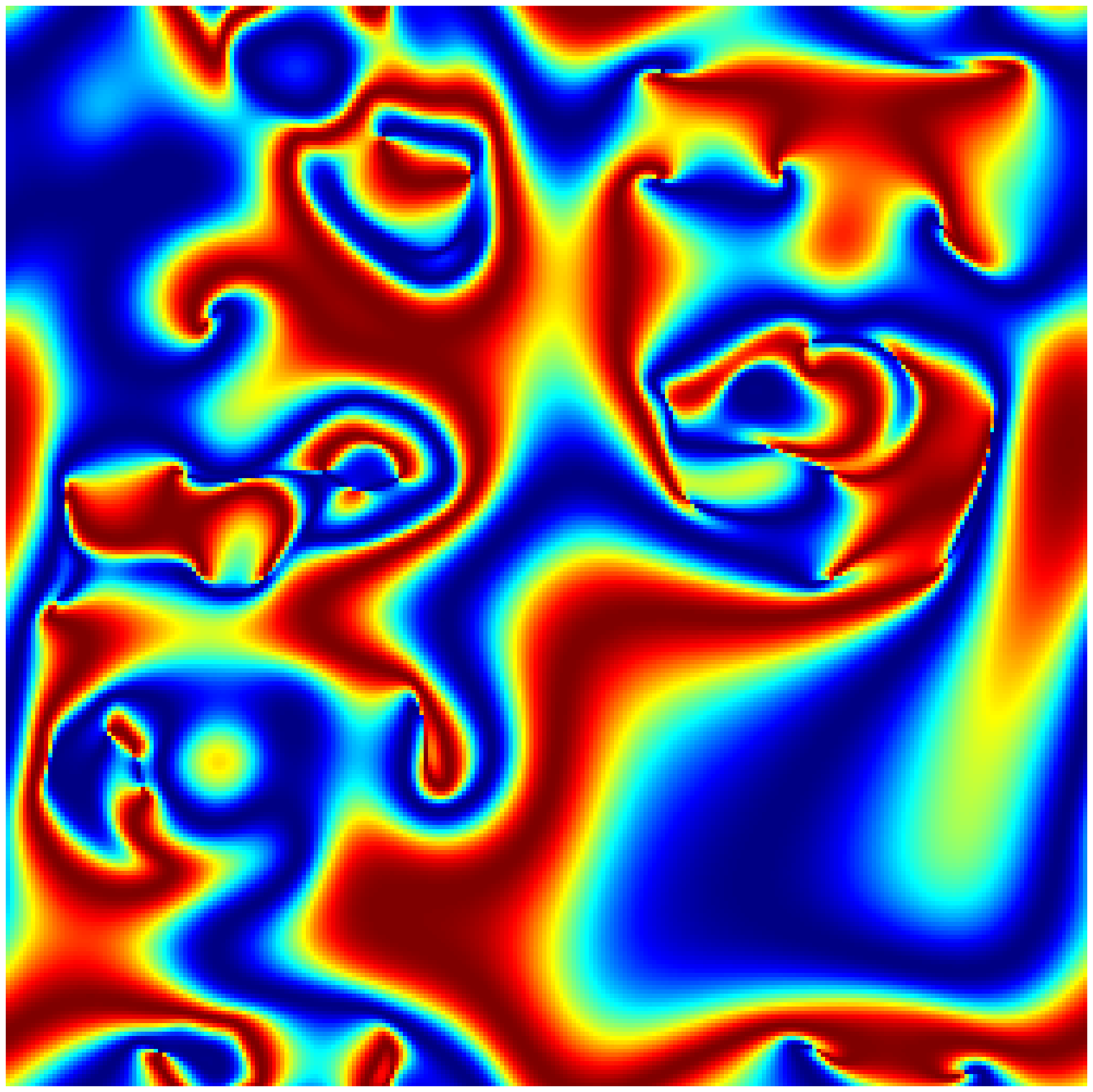}\label{fig:dyn_int_chg3b}}\\
    \subfigure [$t=139$]   
    {\includegraphics[width=4.1cm]{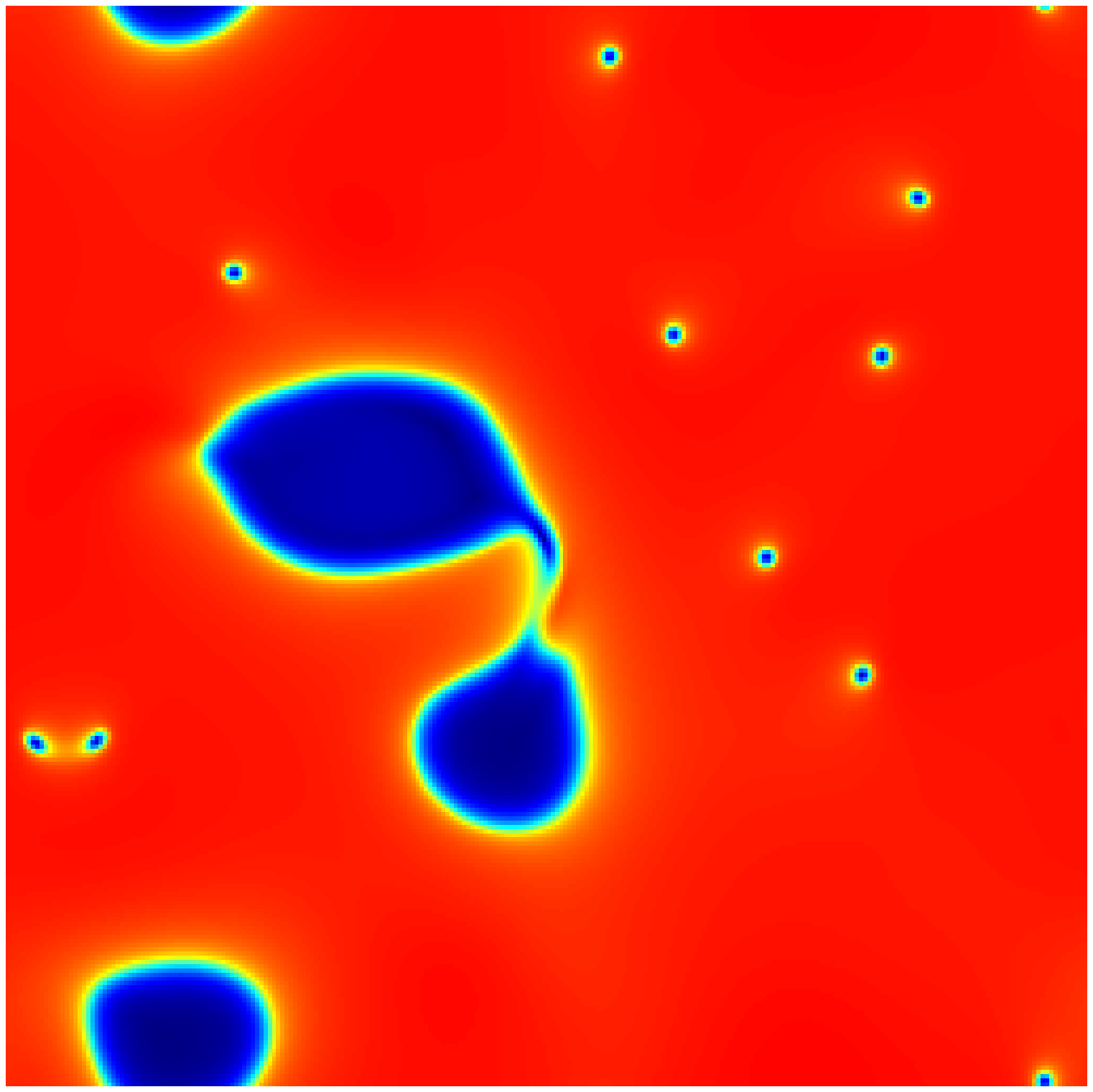}\label{fig:dyn_int_chg4a}}
    \subfigure [$t=139$]
    {\includegraphics[width=4.1cm]{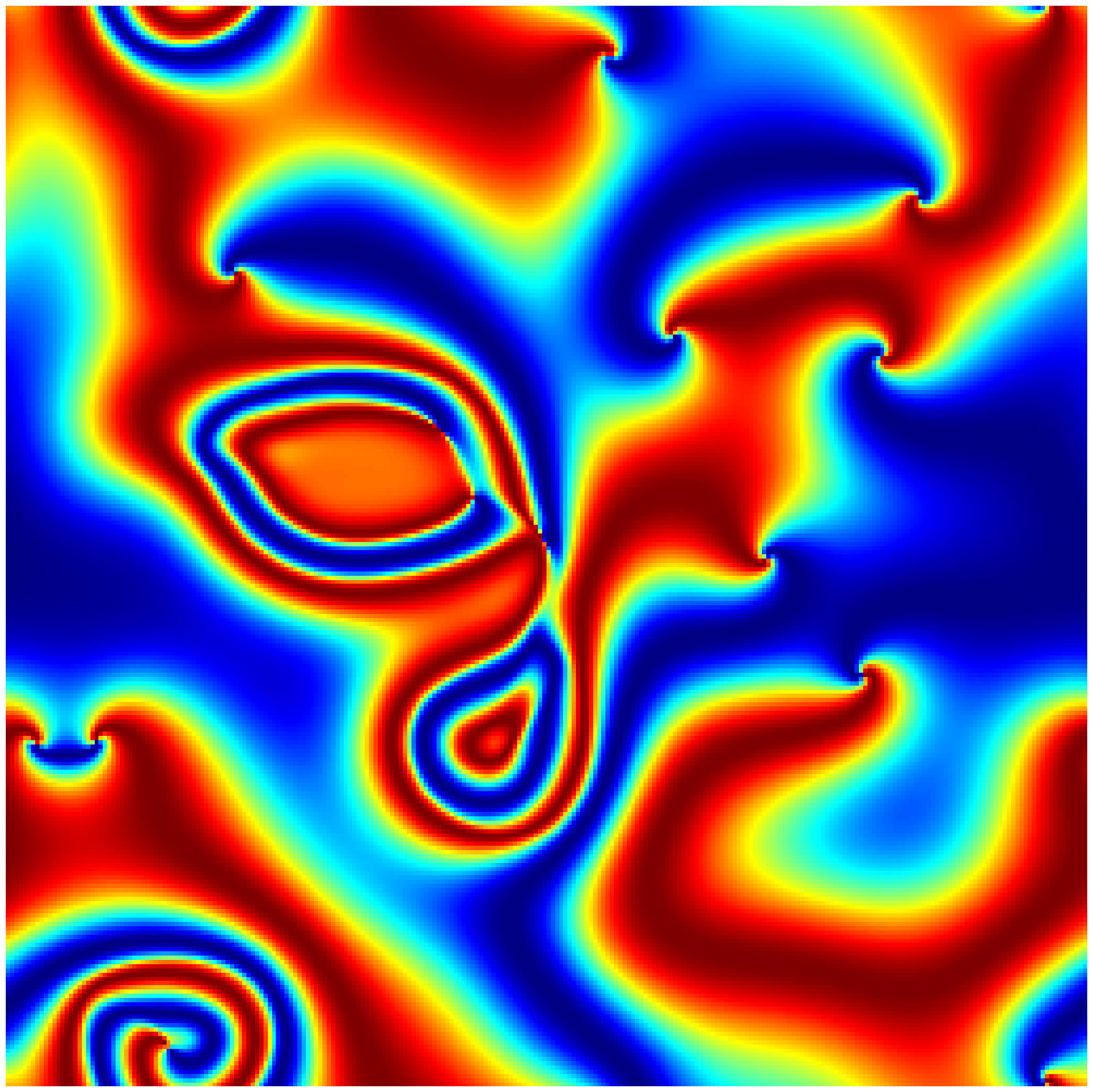}\label{fig:dyn_int_chg4b}}\\

    \captcont{
             Time evolution of 
             $|r({\bf x},t)|$ and $\sin[\theta({\bf x},t)]$ 
             (where $r({\bf x},t) = |r({\bf x},t)| \exp[i \theta({\bf x},t)]$) 
             from random initial condition, (a) and (b)
             ($\omega_0=5, T=1, D=100, k=15.5 \hspace{1mm} (> k_c=14.5)$; 
             periodic boundary conditions are imposed). 
             }
    \label{fig:int_chg}
  \end{center}
\end{figure*}

\begin{figure*}[htbp]
  \begin{center}
    \subfigure [$t=153$]
    {\includegraphics[width=4.1cm]{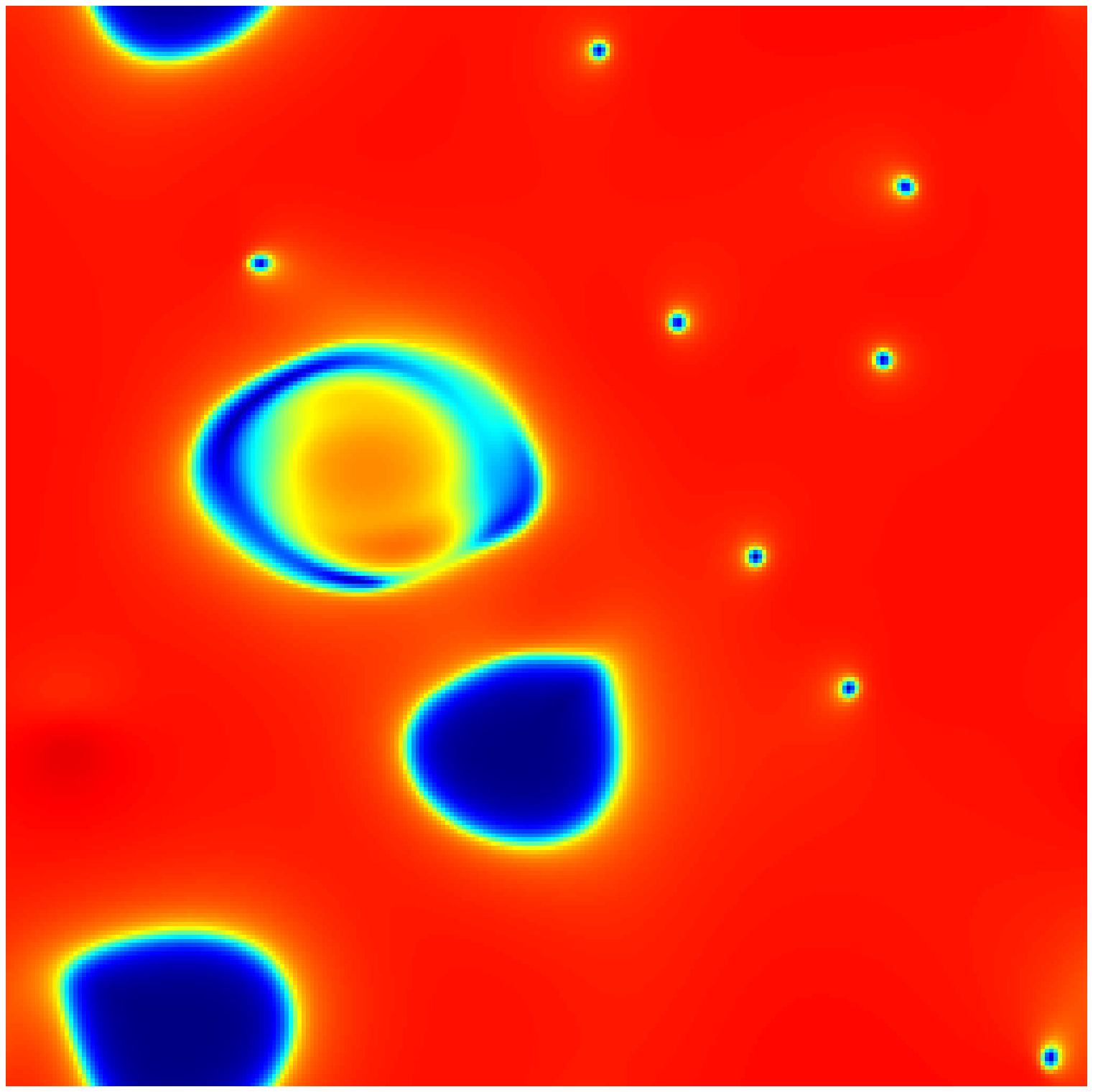}\label{fig:dyn_int_chg5a}}
    \subfigure [$t=153$]
    {\includegraphics[width=4.1cm]{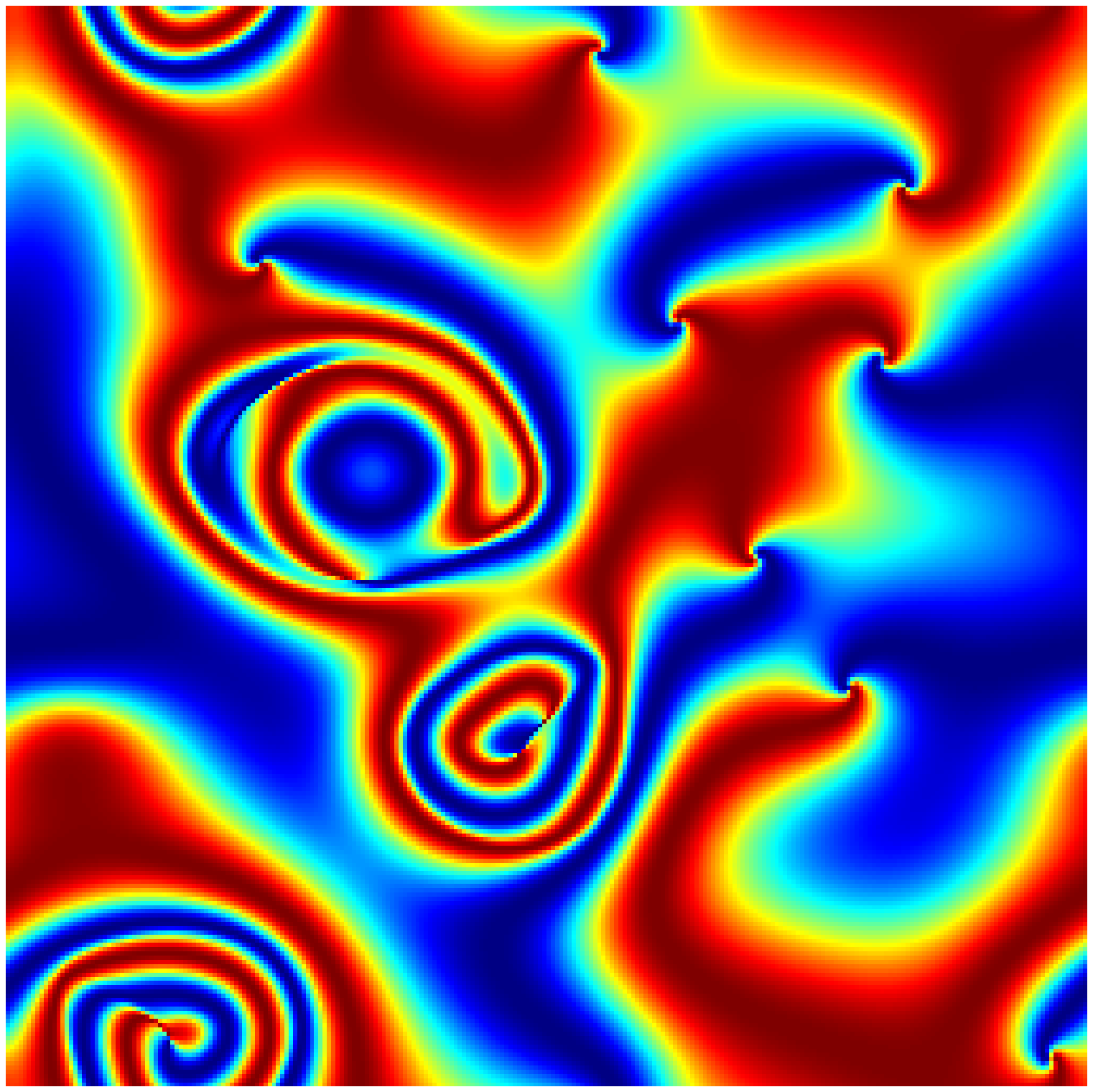}\label{fig:dyn_int_chg5b}}\\
    \subfigure [$t=161$]
    {\includegraphics[width=4.1cm]{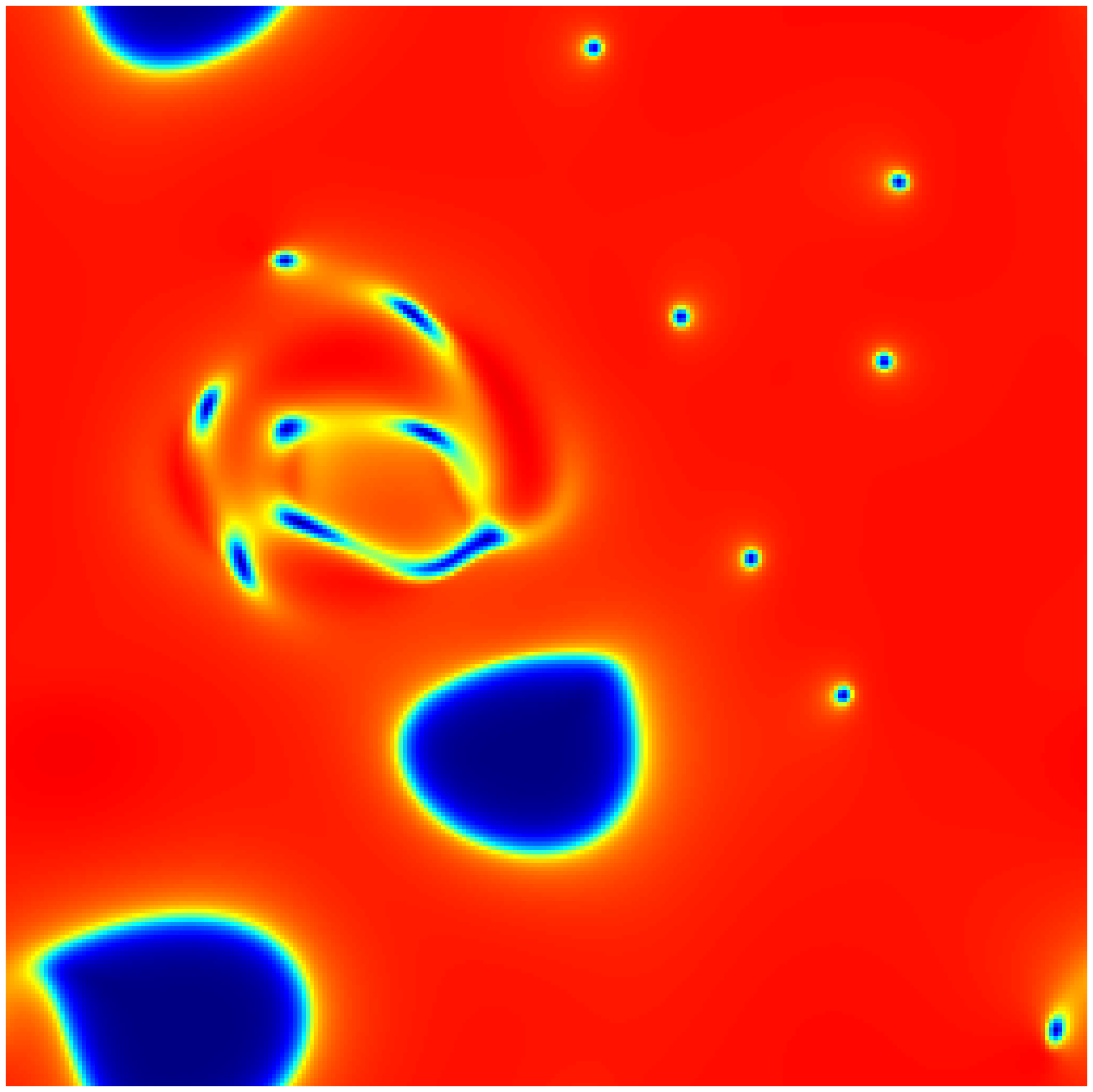}\label{fig:dyn_int_chg6a}}
    \subfigure [$t=161$]
    {\includegraphics[width=4.1cm]{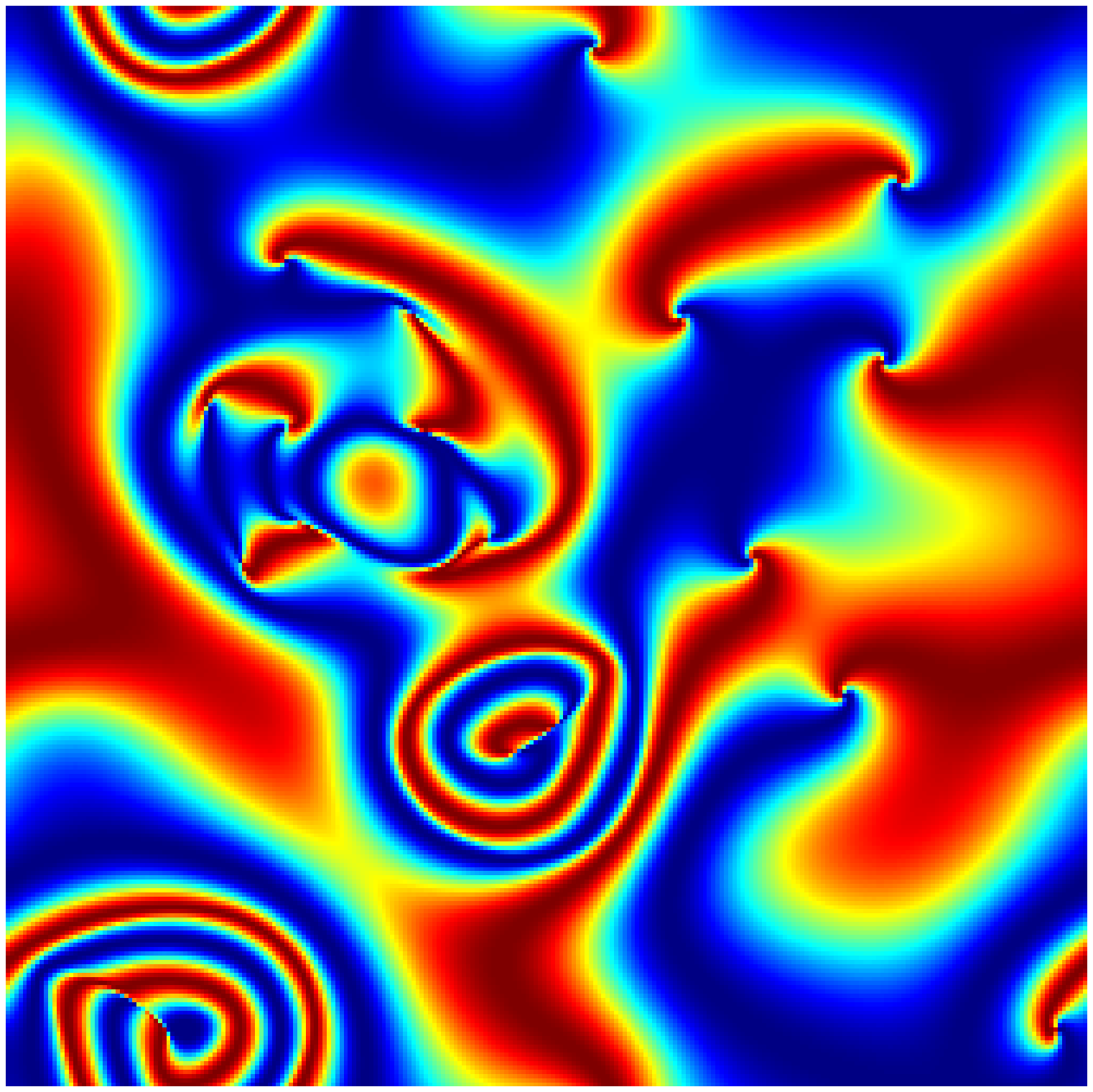}\label{fig:dyn_int_chg6b}}\\
    \subfigure [$t=195$]
    {\includegraphics[width=4.1cm]{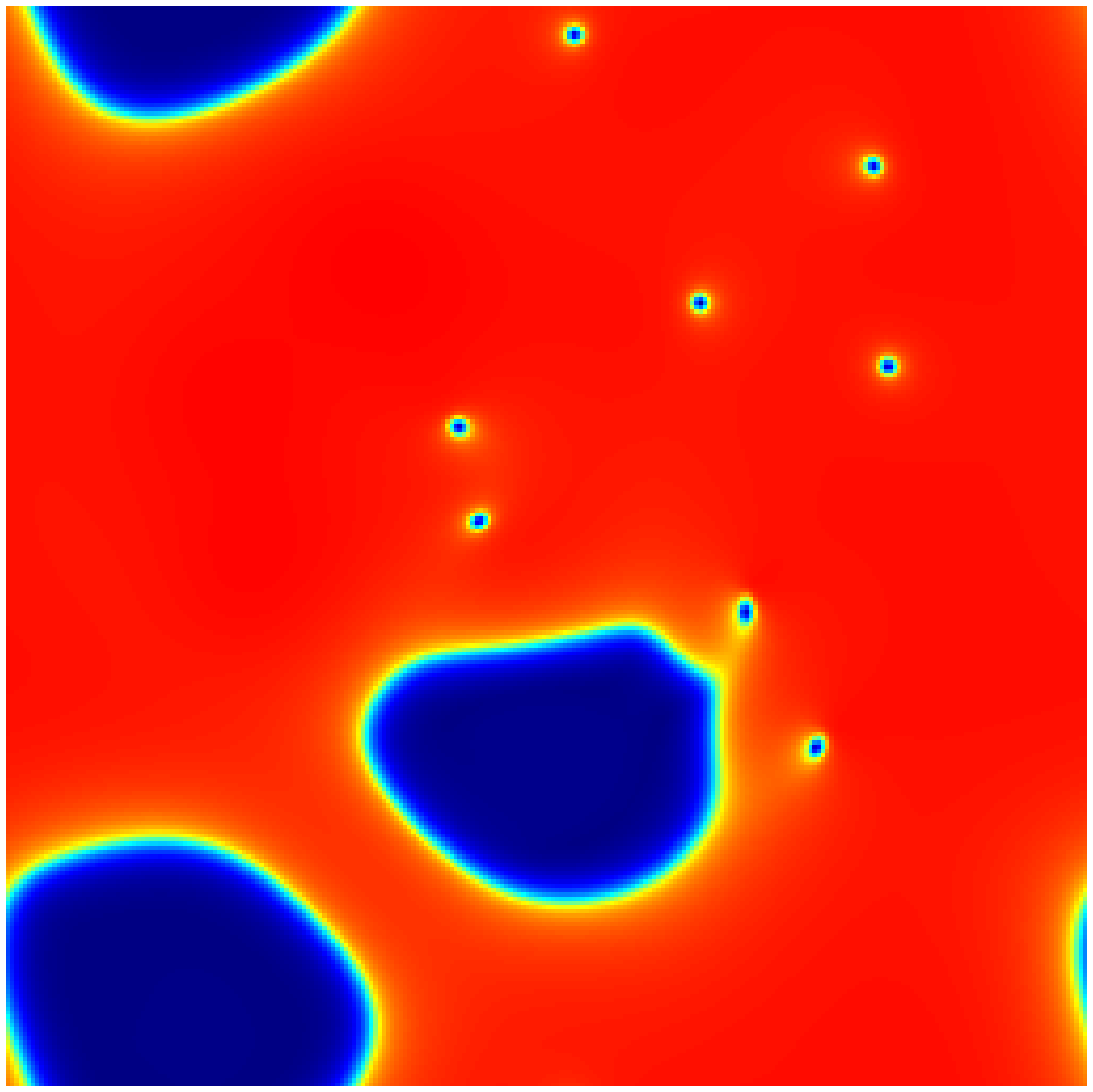}\label{fig:dyn_int_chg7a}}
    \subfigure [$t=195$]
    {\includegraphics[width=4.1cm]{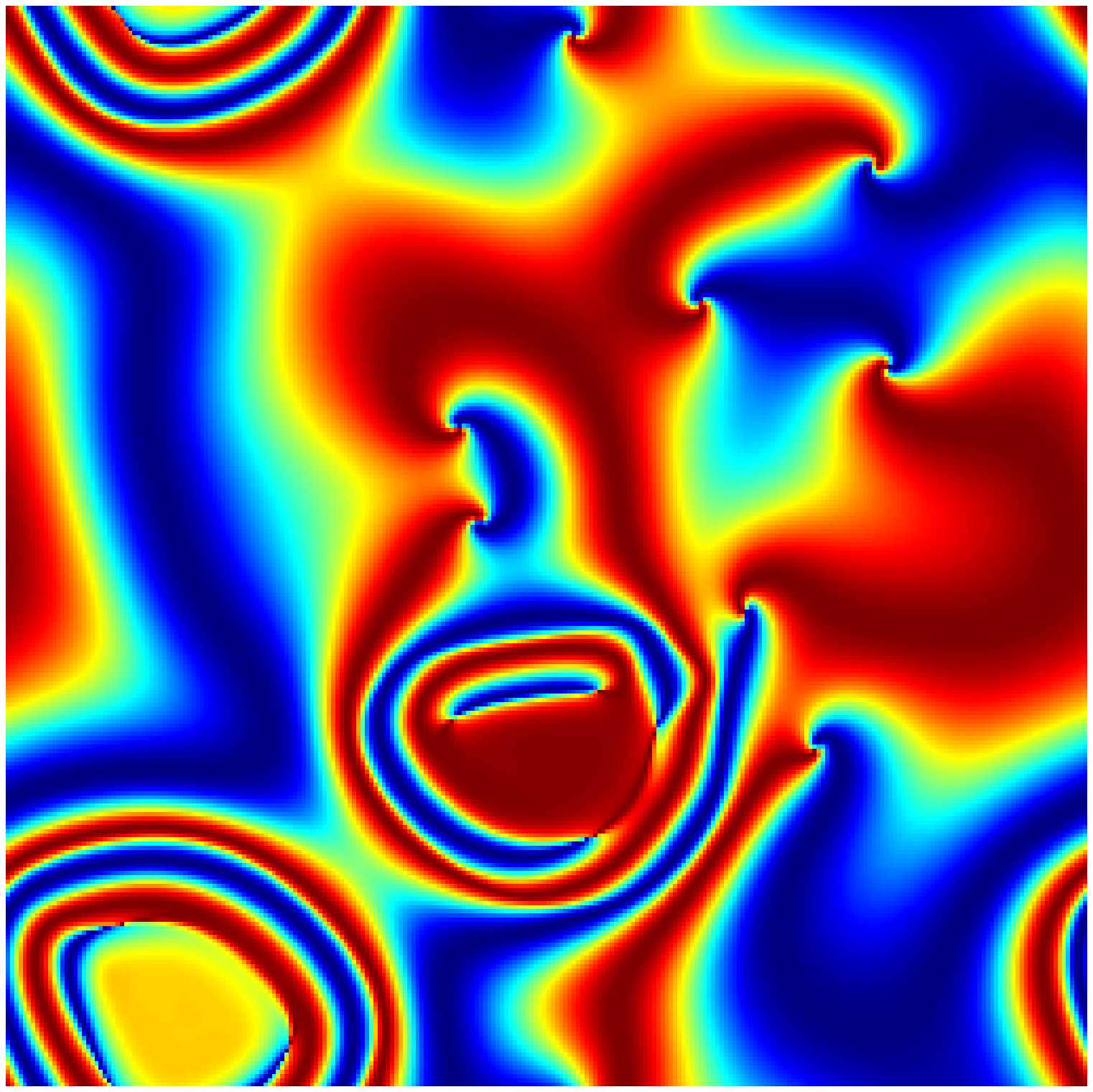}\label{fig:dyn_int_chg7b}}\\
    \subfigure [$t=217$]
    {\includegraphics[width=4.1cm]{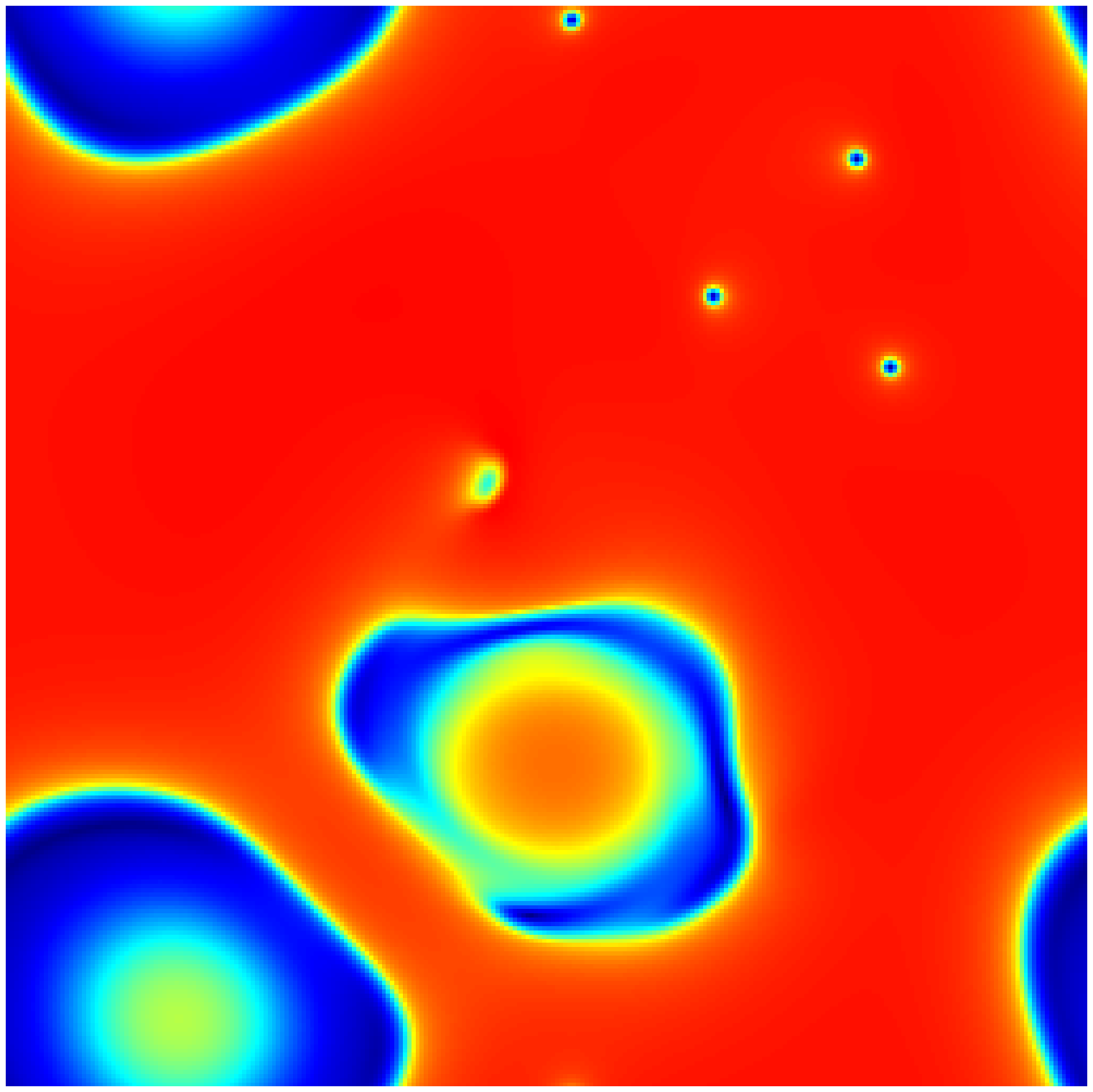}\label{fig:dyn_int_chg8a}}
    \subfigure [$t=217$]
    {\includegraphics[width=4.1cm]{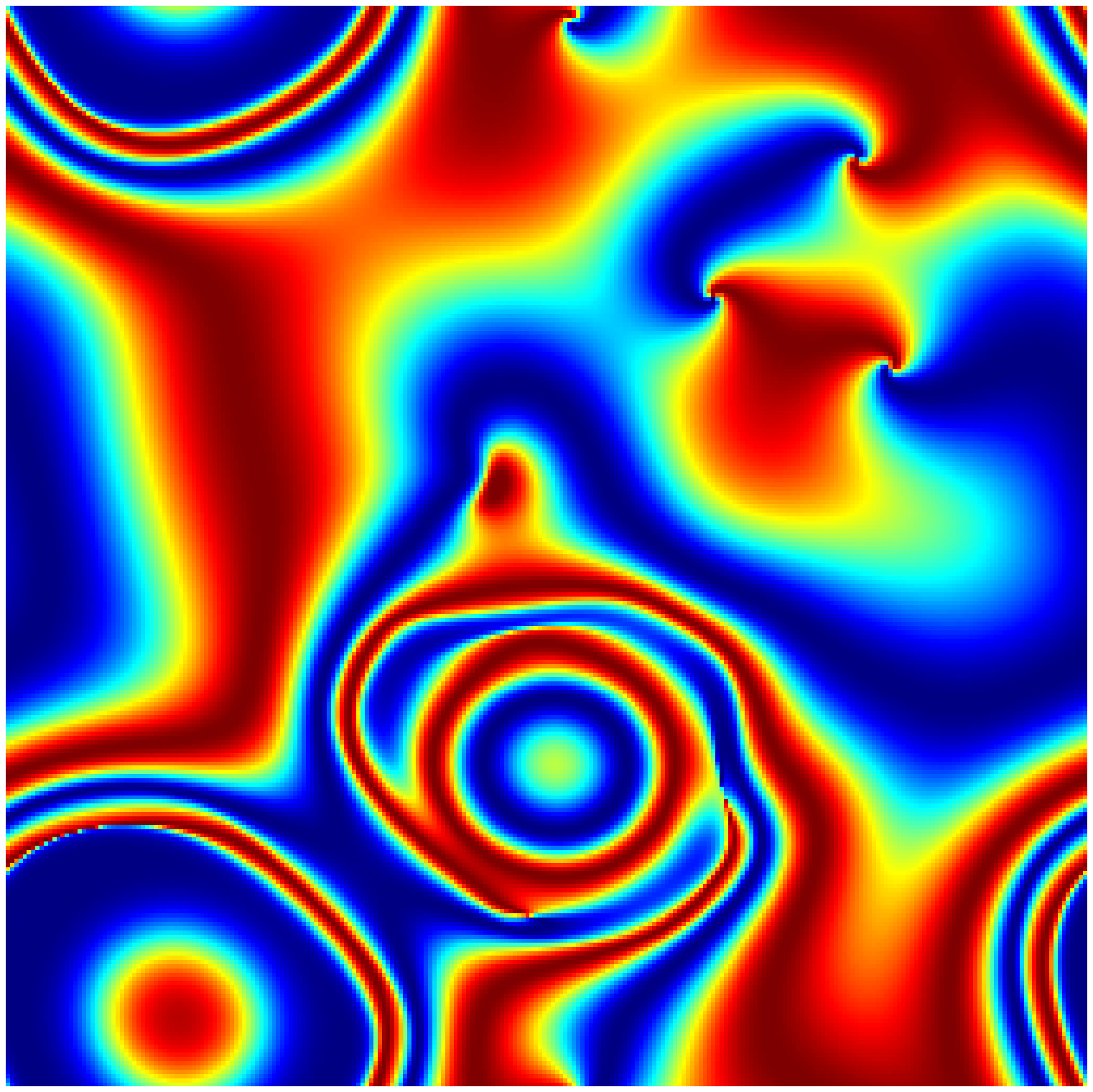}\label{fig:dyn_int_chg8b}}\\
    \subfigure [$t=225$]
    {\includegraphics[width=4.1cm]{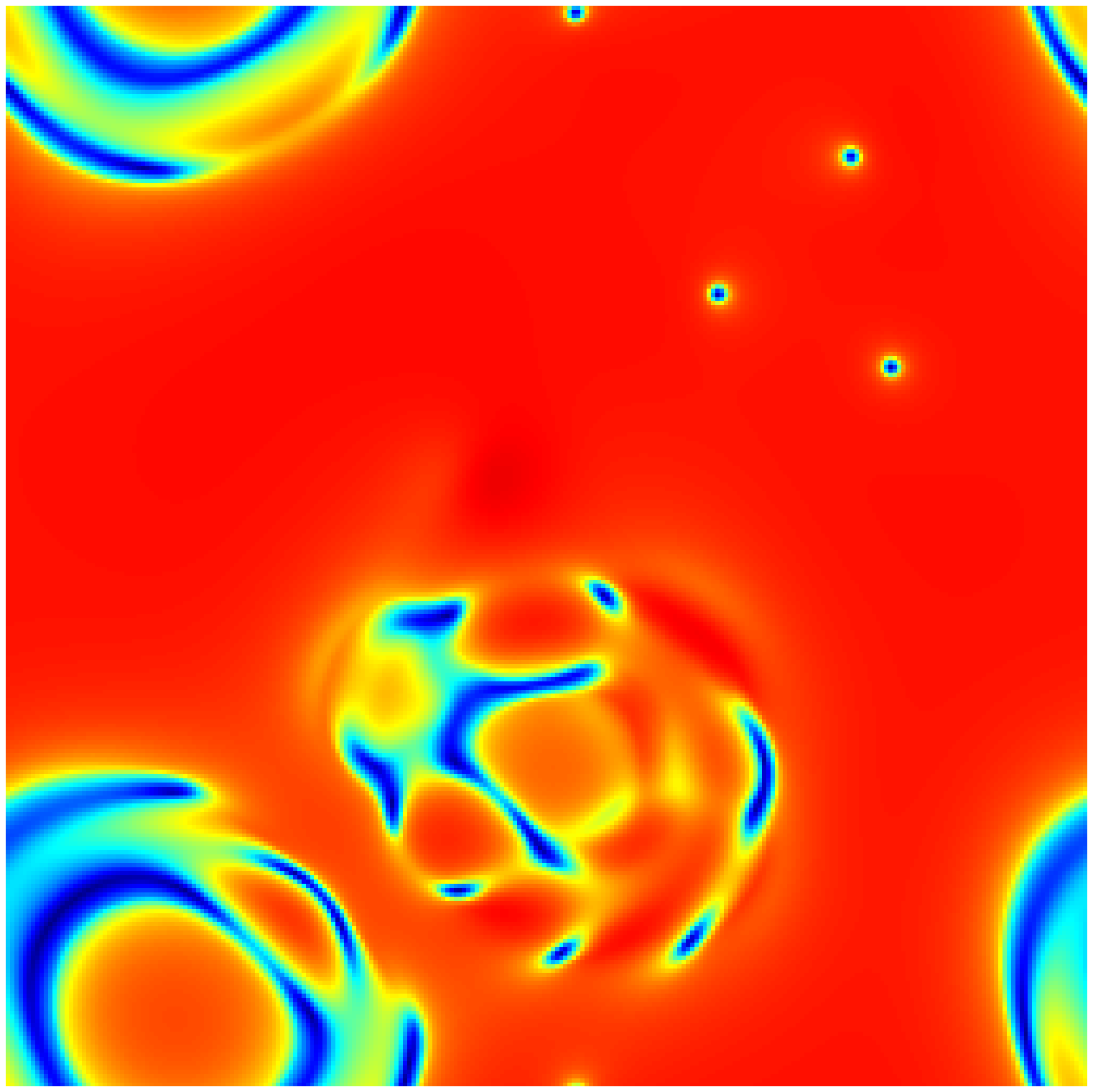}\label{fig:dyn_int_chg9a}}
    \subfigure [$t=225$]
    {\includegraphics[width=4.1cm]{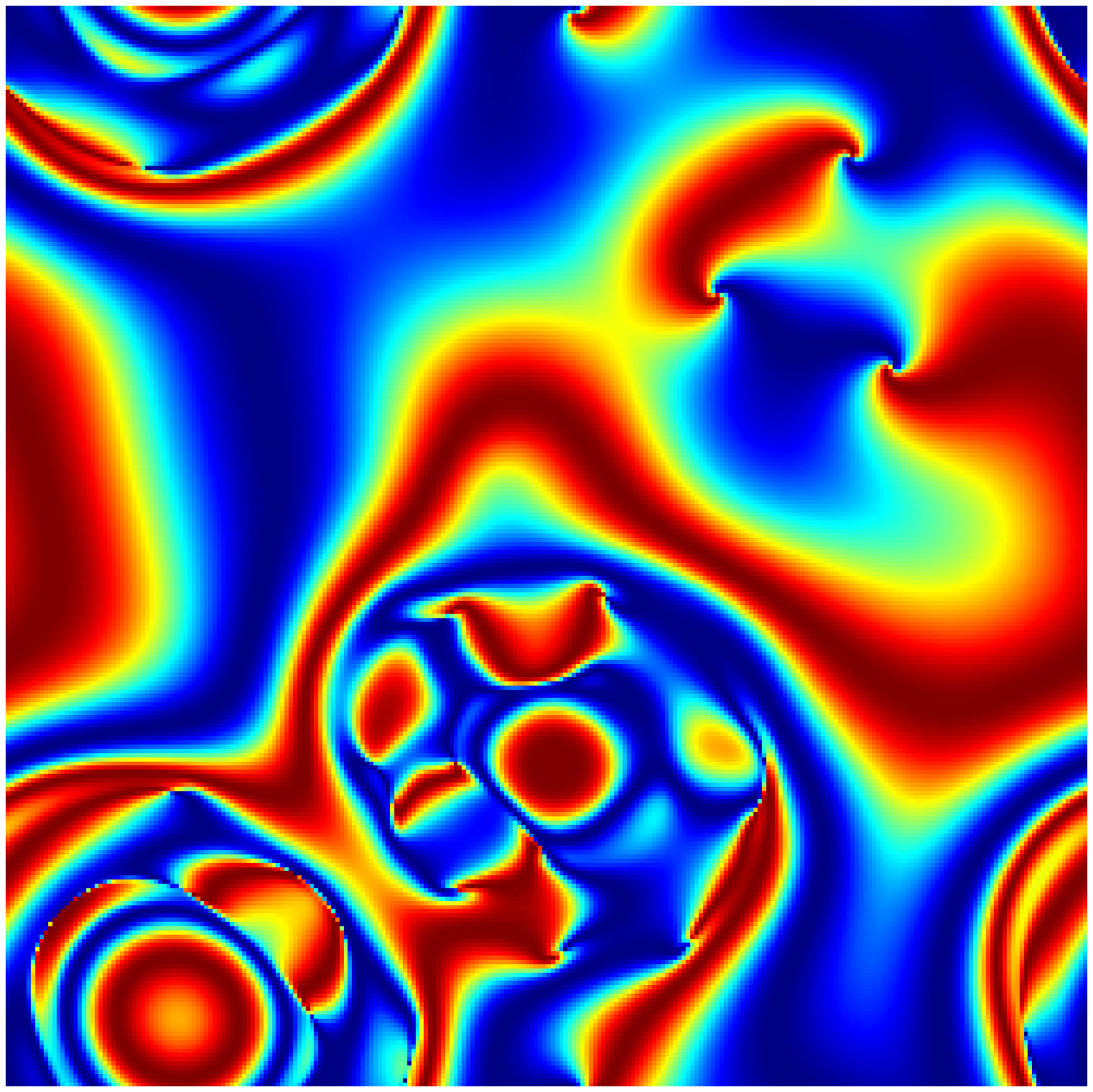}\label{fig:dyn_int_chg9b}}\\
    \captcont*{Cont'd}
    \label{fig:int_chg}
  \end{center}
\end{figure*}

\begin{figure*}[htbp]
  \begin{center}
    \subfigure [$t=238$]
    {\includegraphics[width=4.1cm]{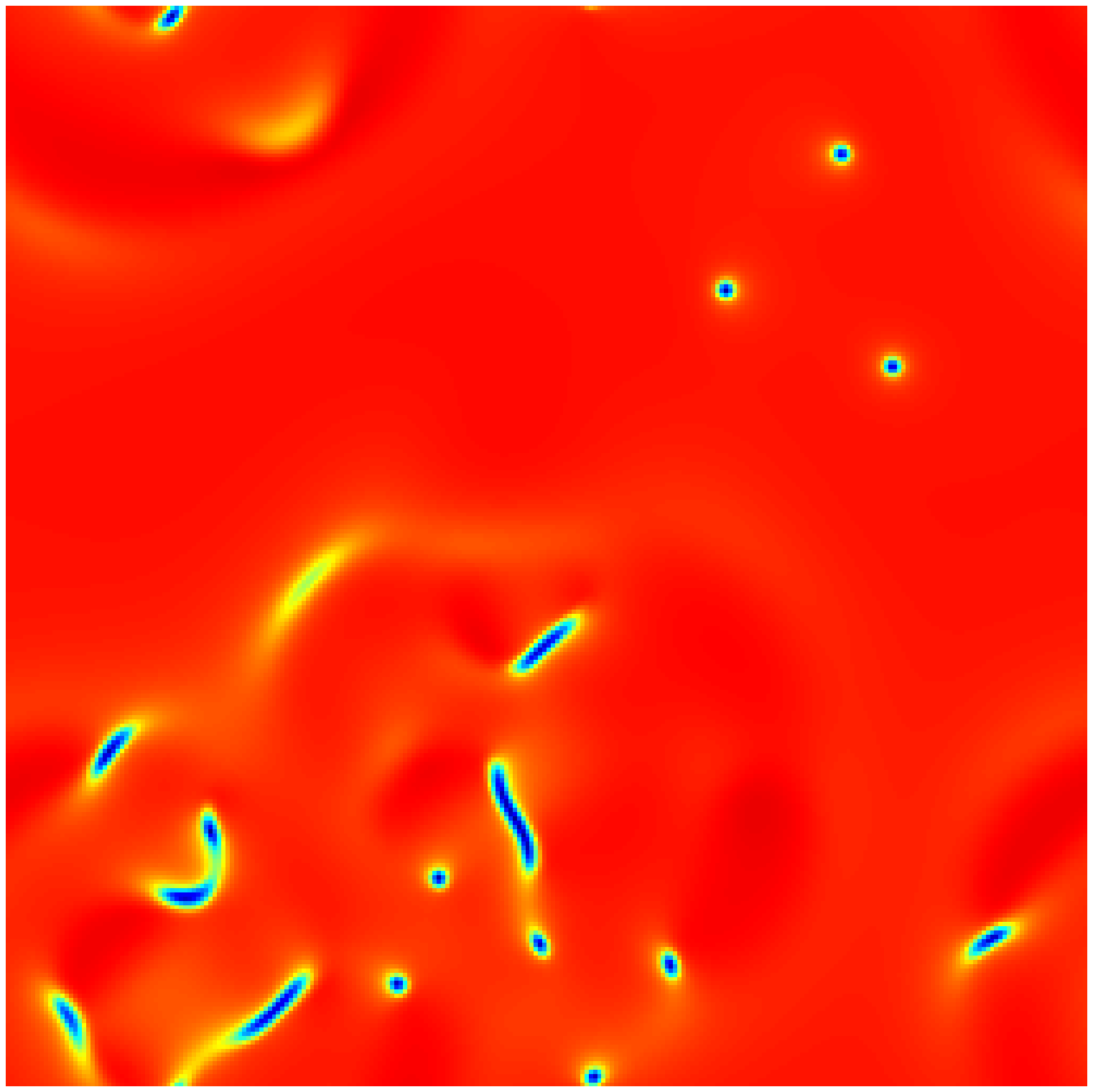}\label{fig:dyn_int_chg10a}}
    \subfigure [$t=238$]
    {\includegraphics[width=4.1cm]{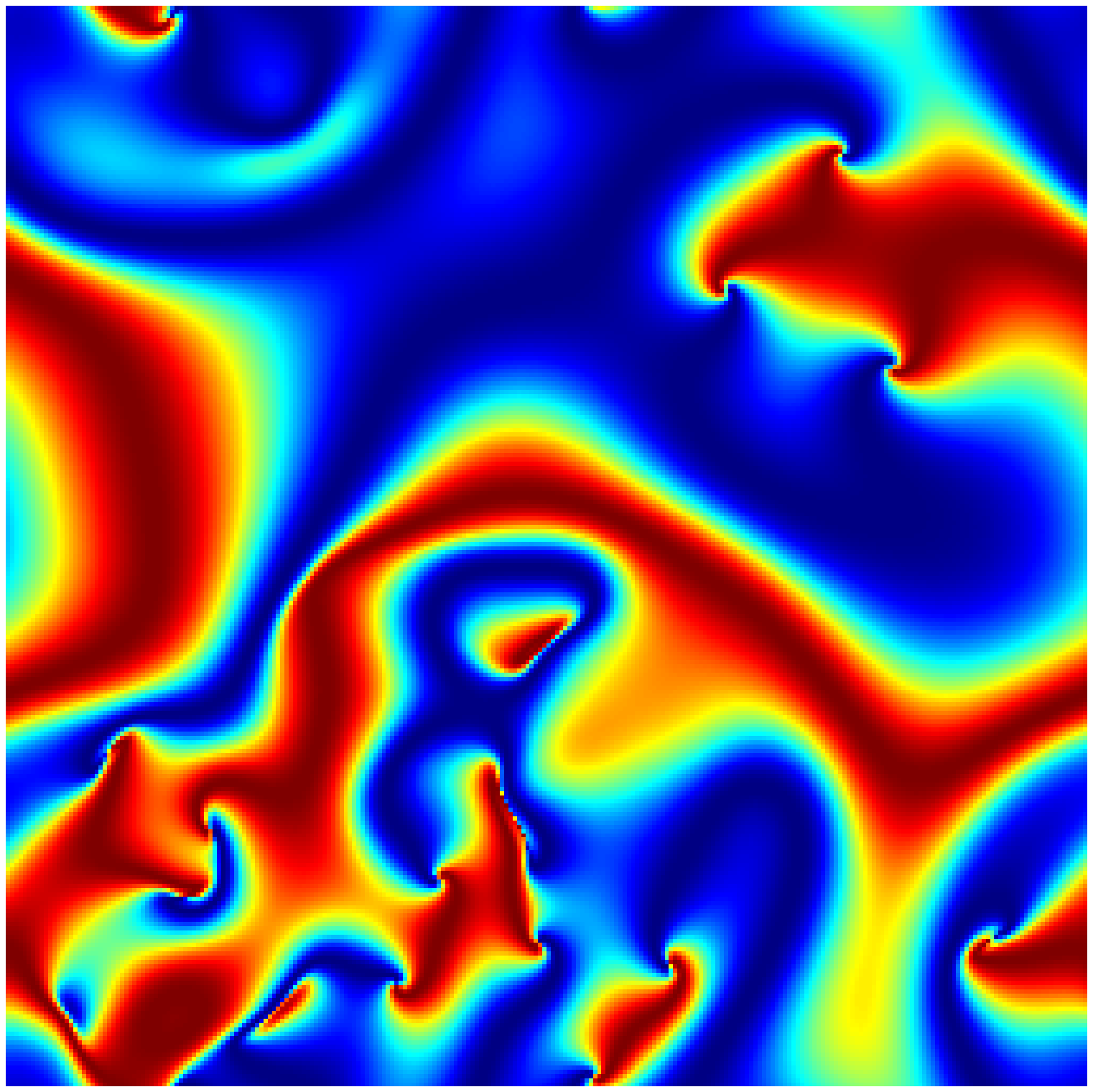}\label{fig:dyn_int_chg10b}}\\
    \subfigure [$t=250$]
    {\includegraphics[width=4.1cm]{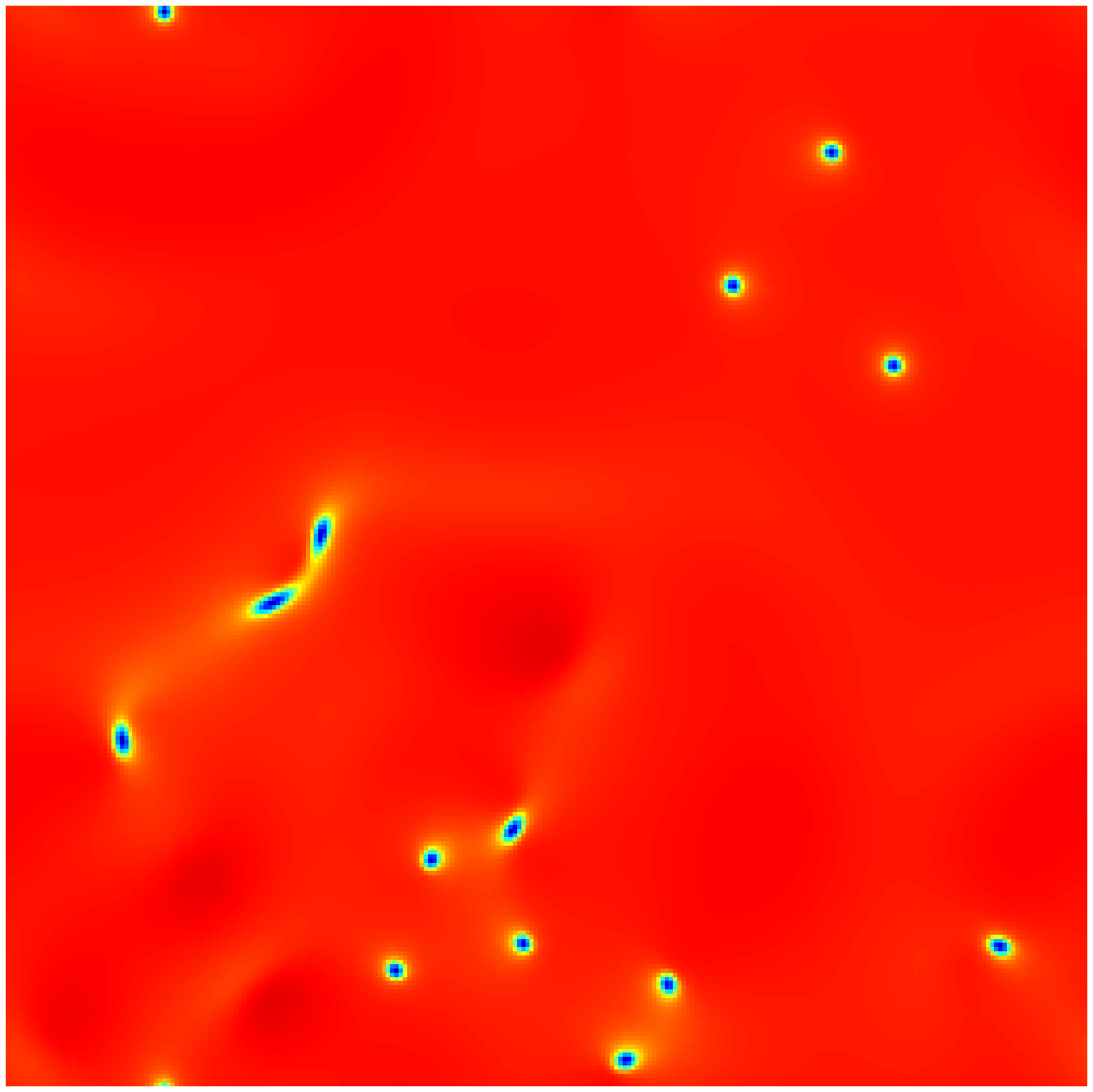}\label{fig:dyn_int_chg11a}}
    \subfigure [$t=250$]
    {\includegraphics[width=4.1cm]{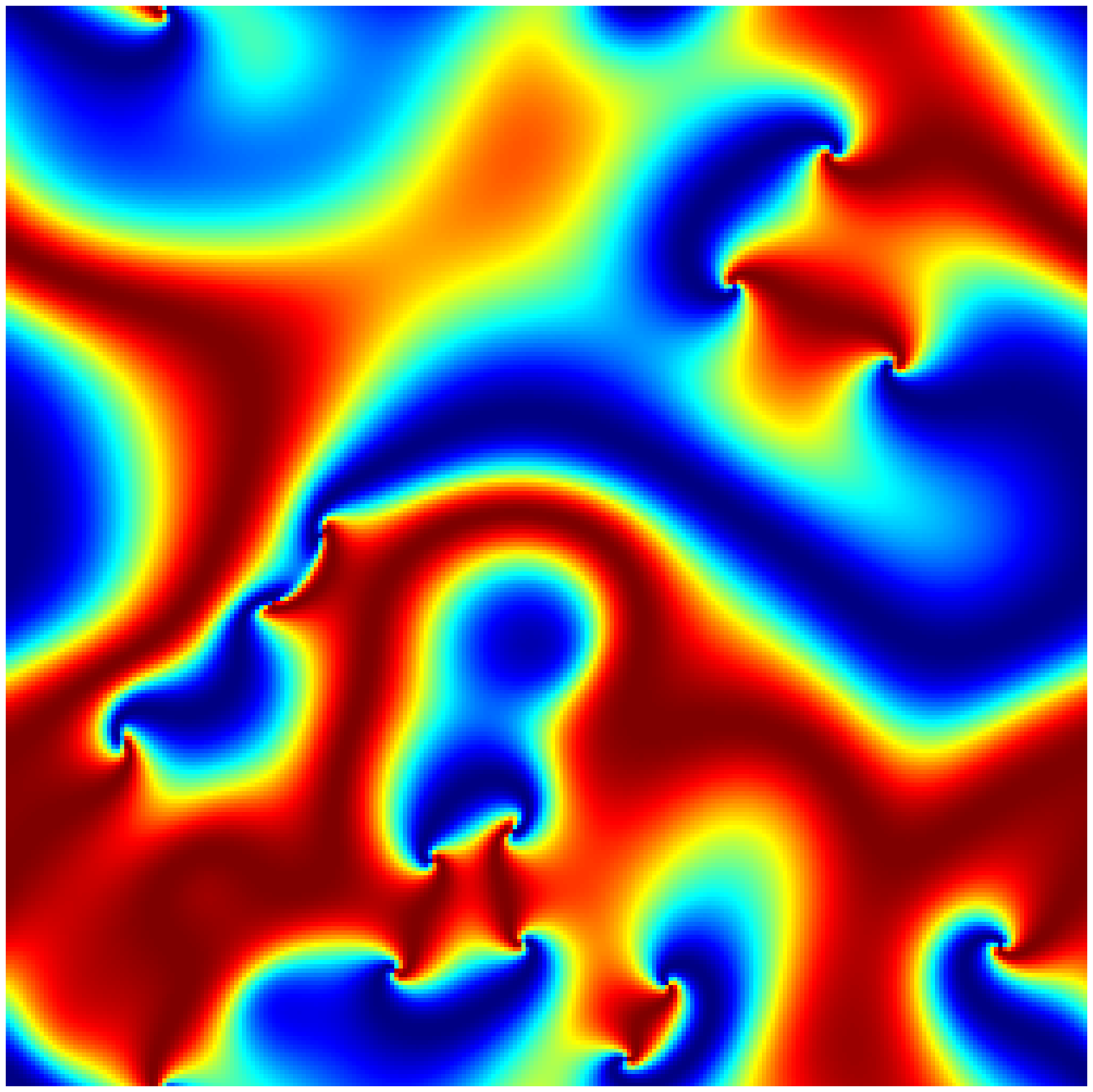}\label{fig:dyn_int_chg11b}}\\
    \subfigure [$t=259$]
    {\includegraphics[width=4.1cm]{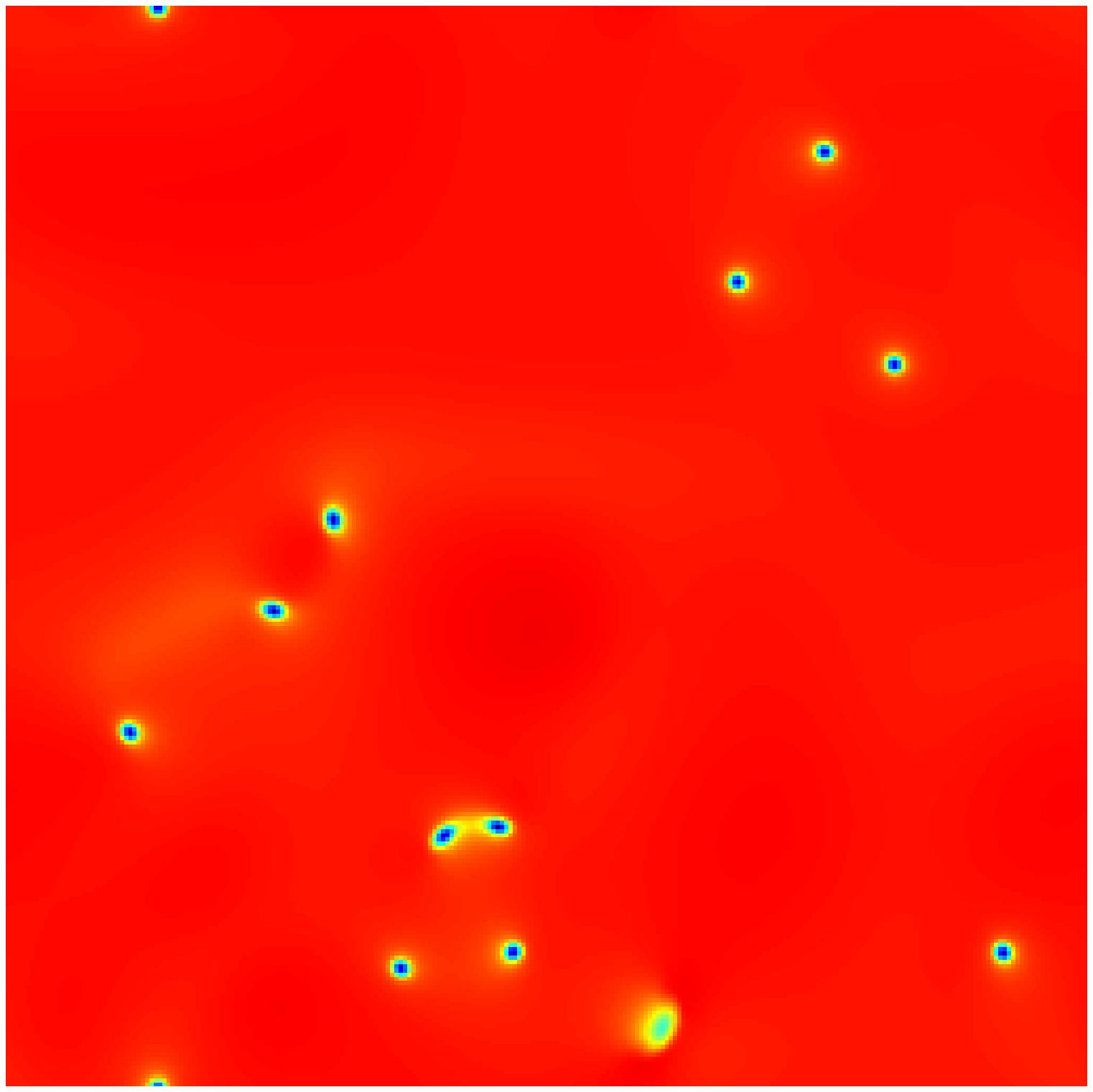}\label{fig:dyn_int_chg12a}}
    \subfigure [$t=259$]
    {\includegraphics[width=4.1cm]{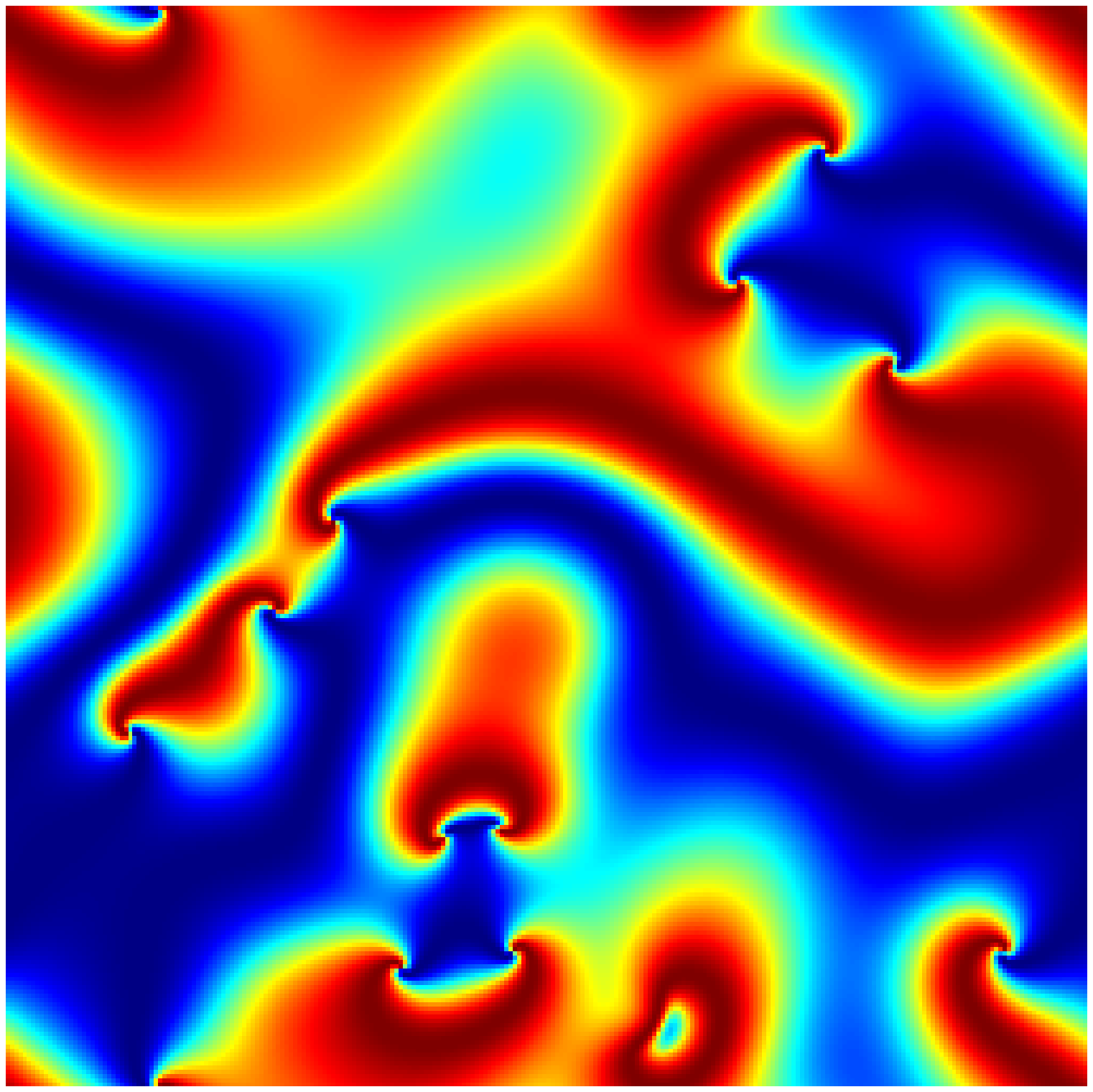}\label{fig:dyn_int_chg12b}}\\
    \subfigure [$t=263$]
    {\includegraphics[width=4.1cm]{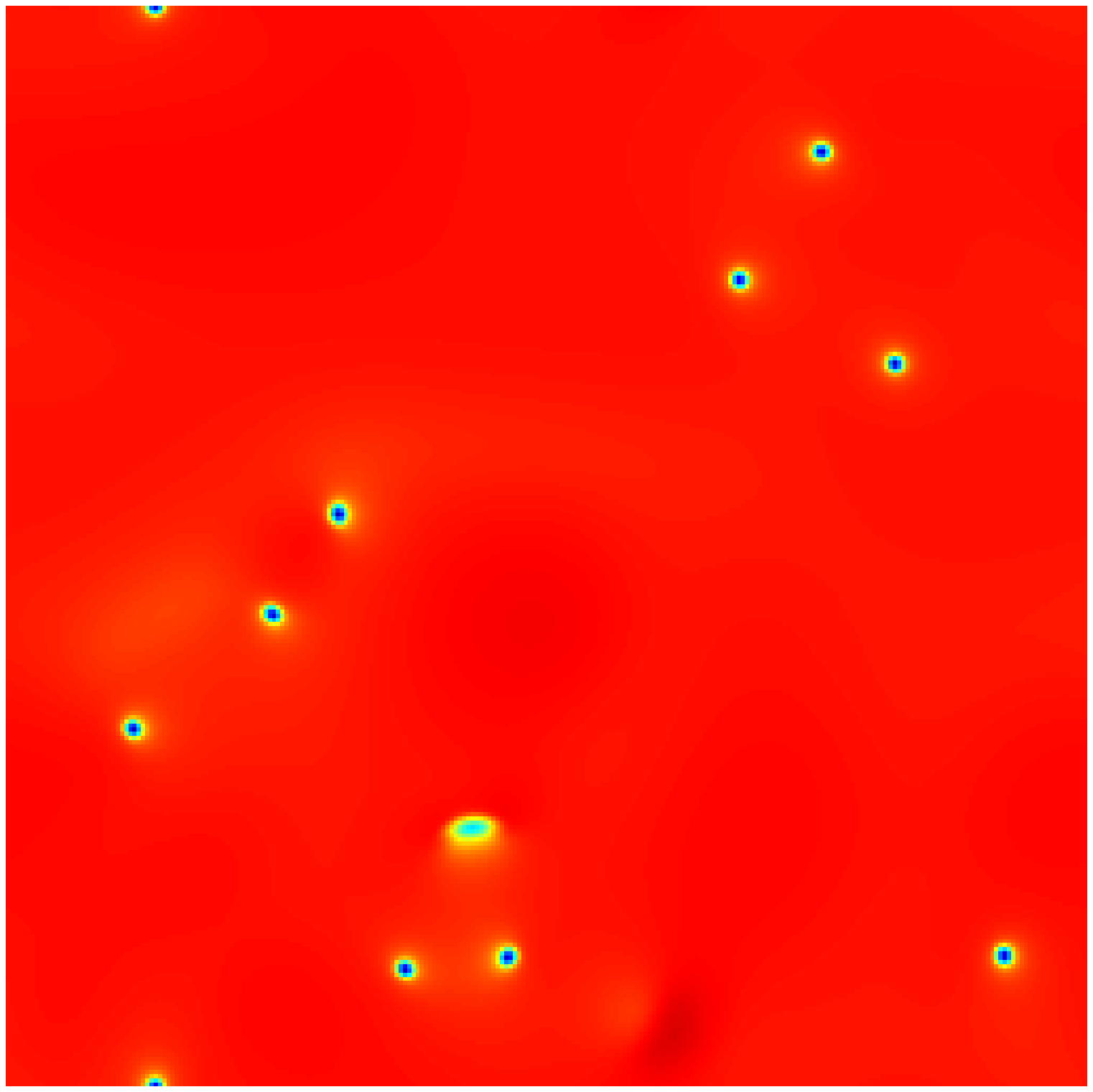}\label{fig:dyn_int_chg13a}}
    \subfigure [$t=263$]
    {\includegraphics[width=4.1cm]{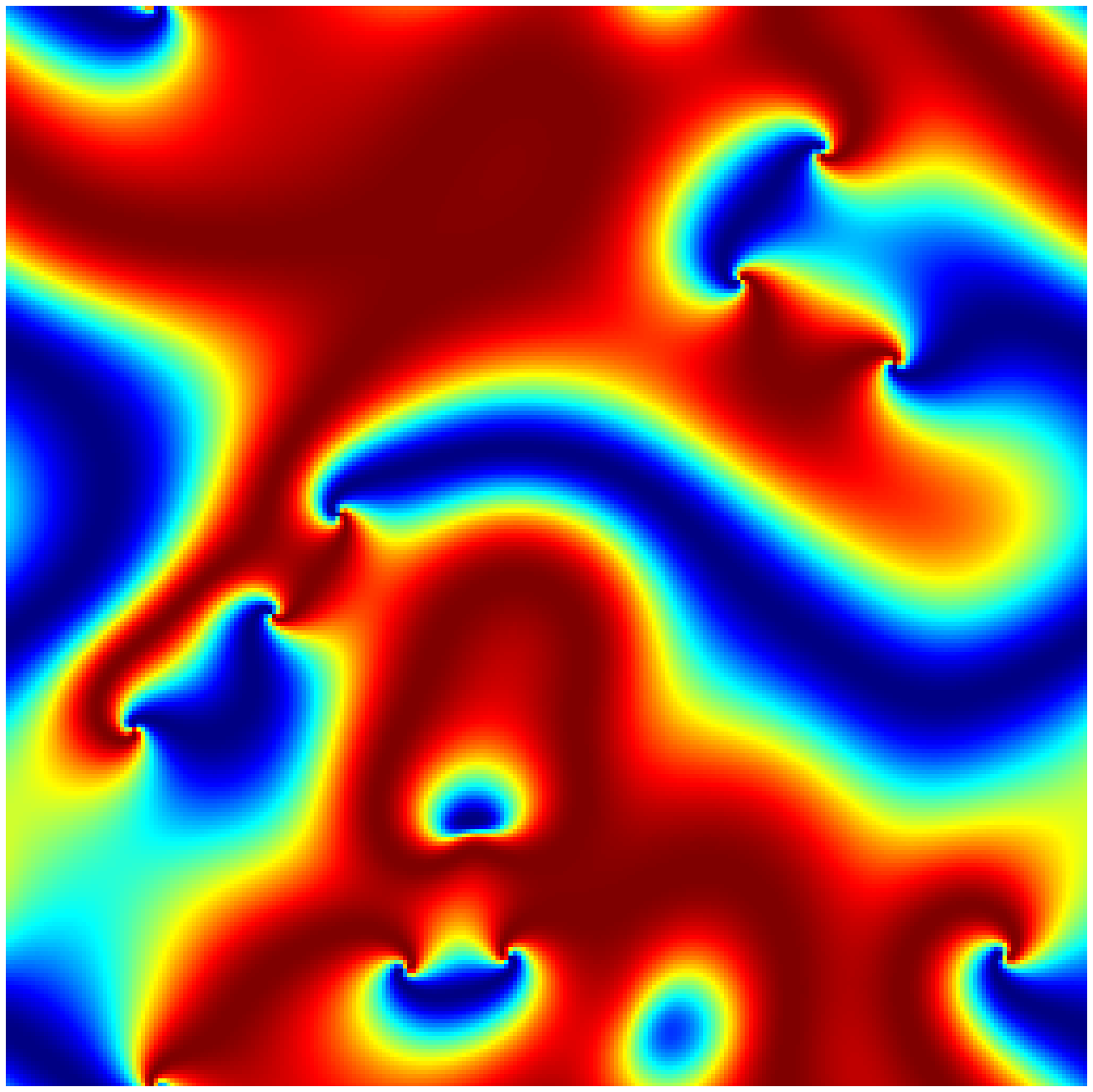}\label{fig:dyn_int_chg13b}}\\
    \subfigure [$t=267$]
    {\includegraphics[width=4.1cm]{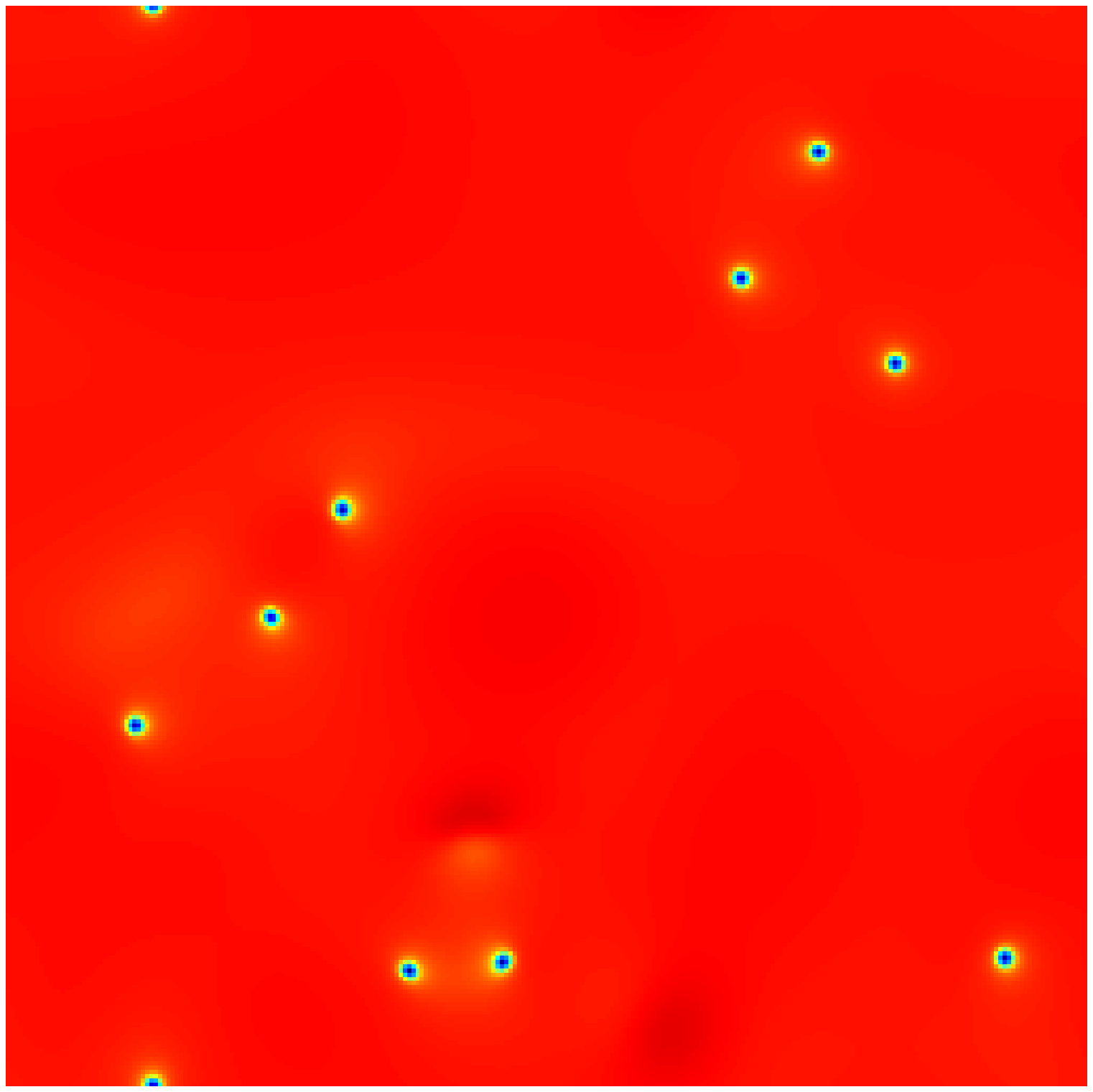}\label{fig:dyn_int_chg14a}}
    \subfigure [$t=267$]
    {\includegraphics[width=4.1cm]{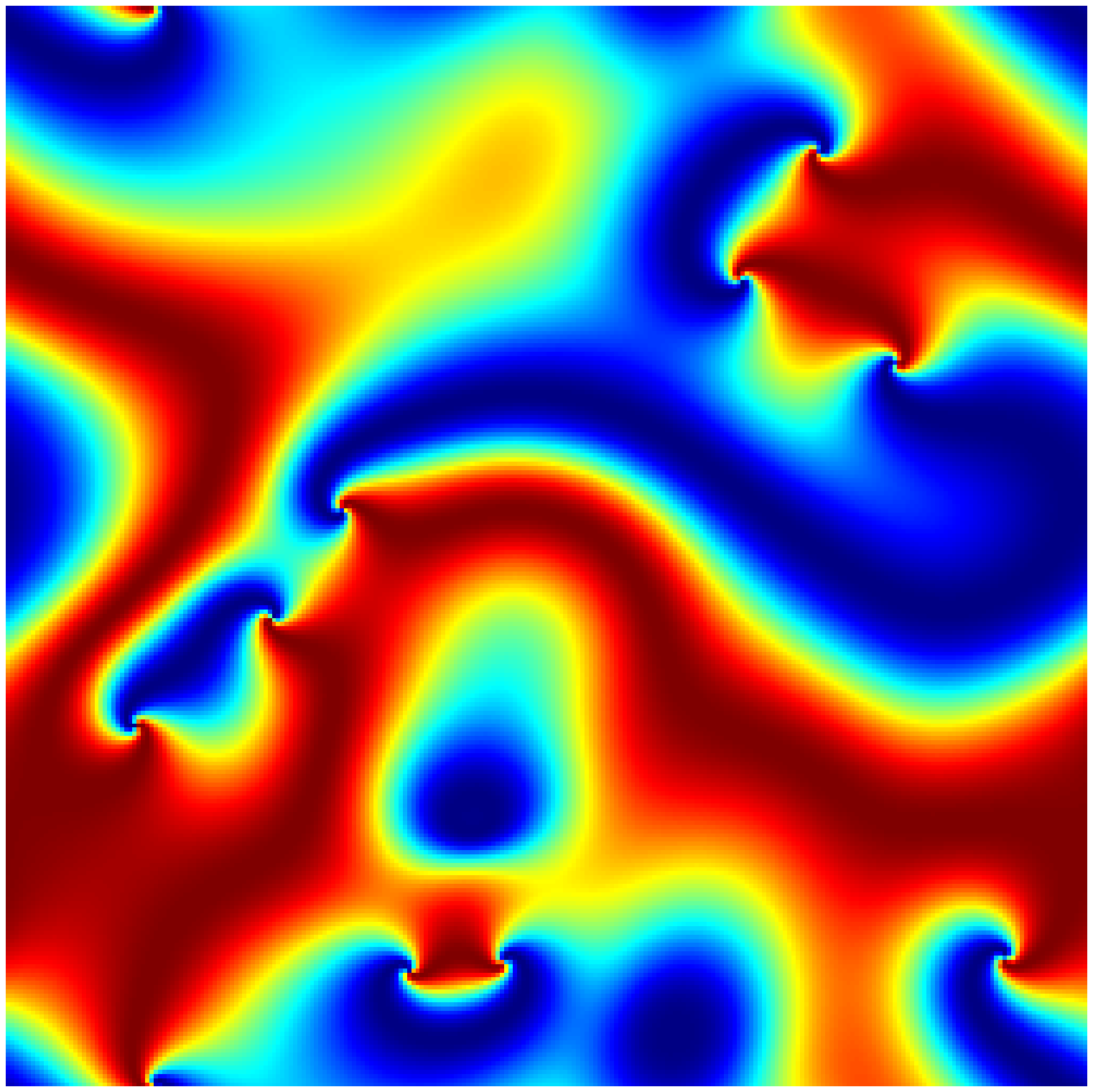}\label{fig:dyn_int_chg14b}}\\
    \end{center}
    \captcont*{Cont'd}
    \label{fig:int_chg}
\end{figure*}

\begin{figure*}[htbp]
    \begin{center}
    \subfigure [$t=281$]
    {\includegraphics[width=4.1cm]{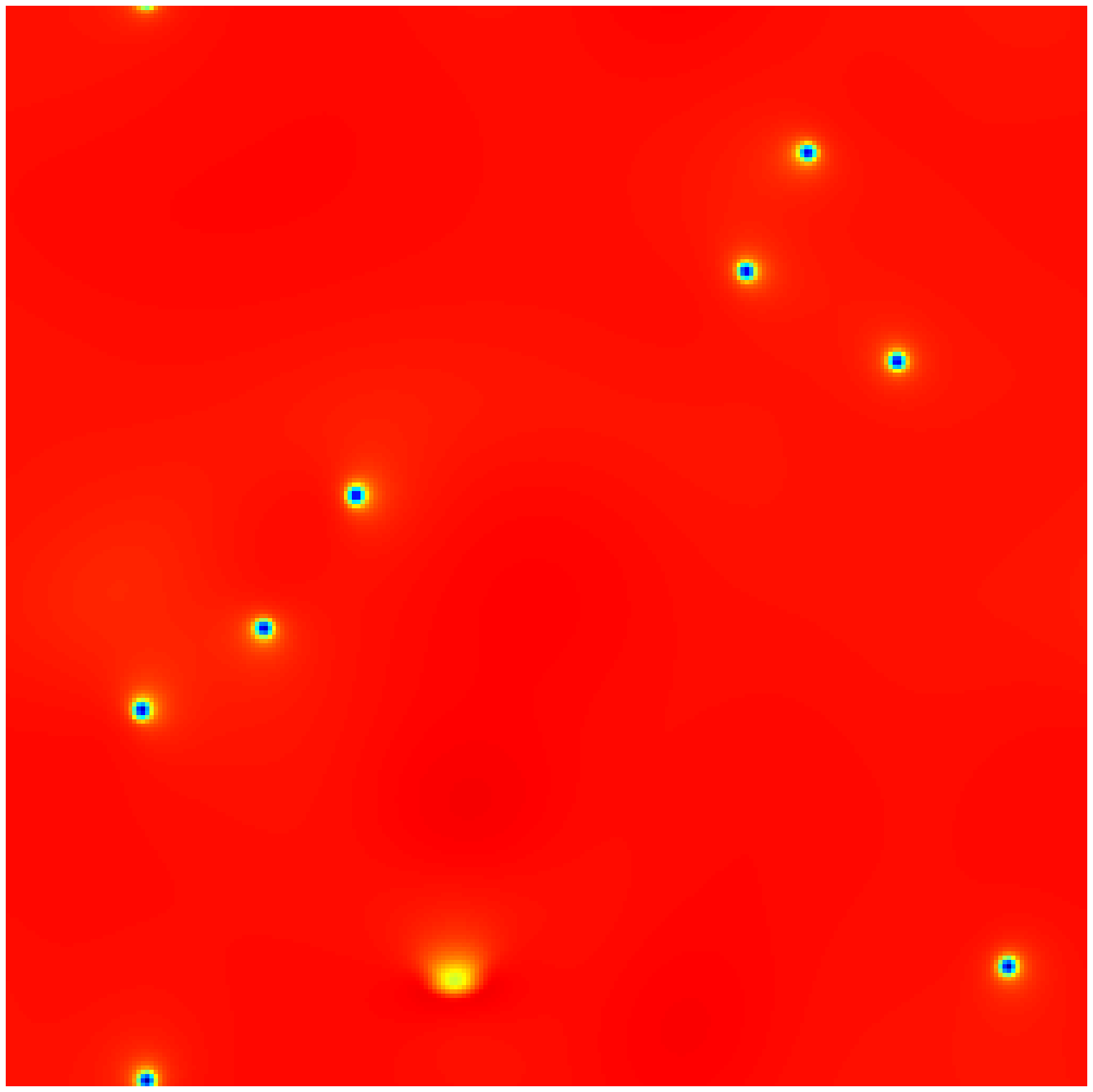}\label{fig:dyn_int_chg15a}}
    \subfigure [$t=281$]
    {\includegraphics[width=4.1cm]{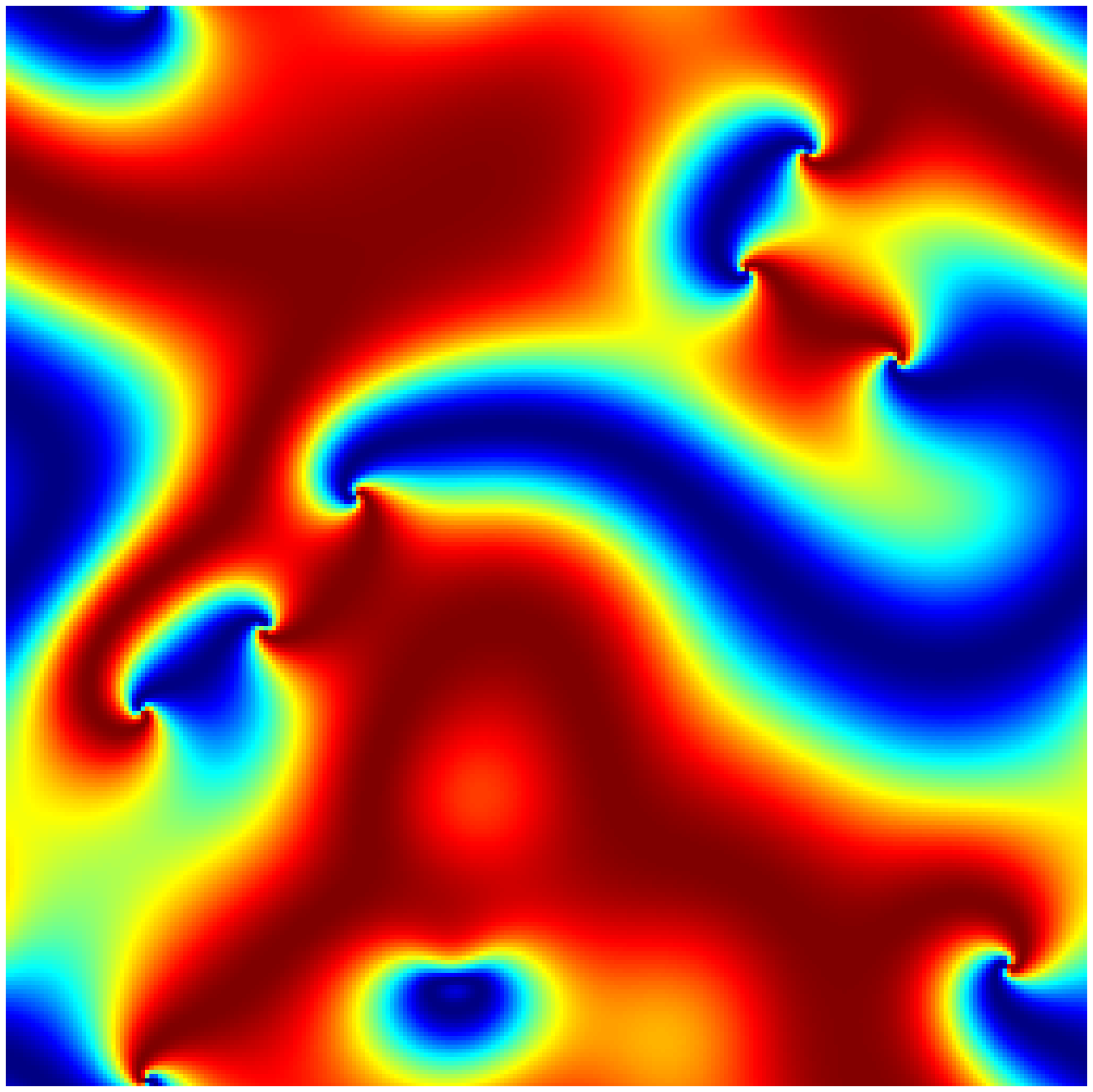}\label{fig:dyn_int_chg15b}}\\
    \subfigure [$t=293$]
    {\includegraphics[width=4.1cm]{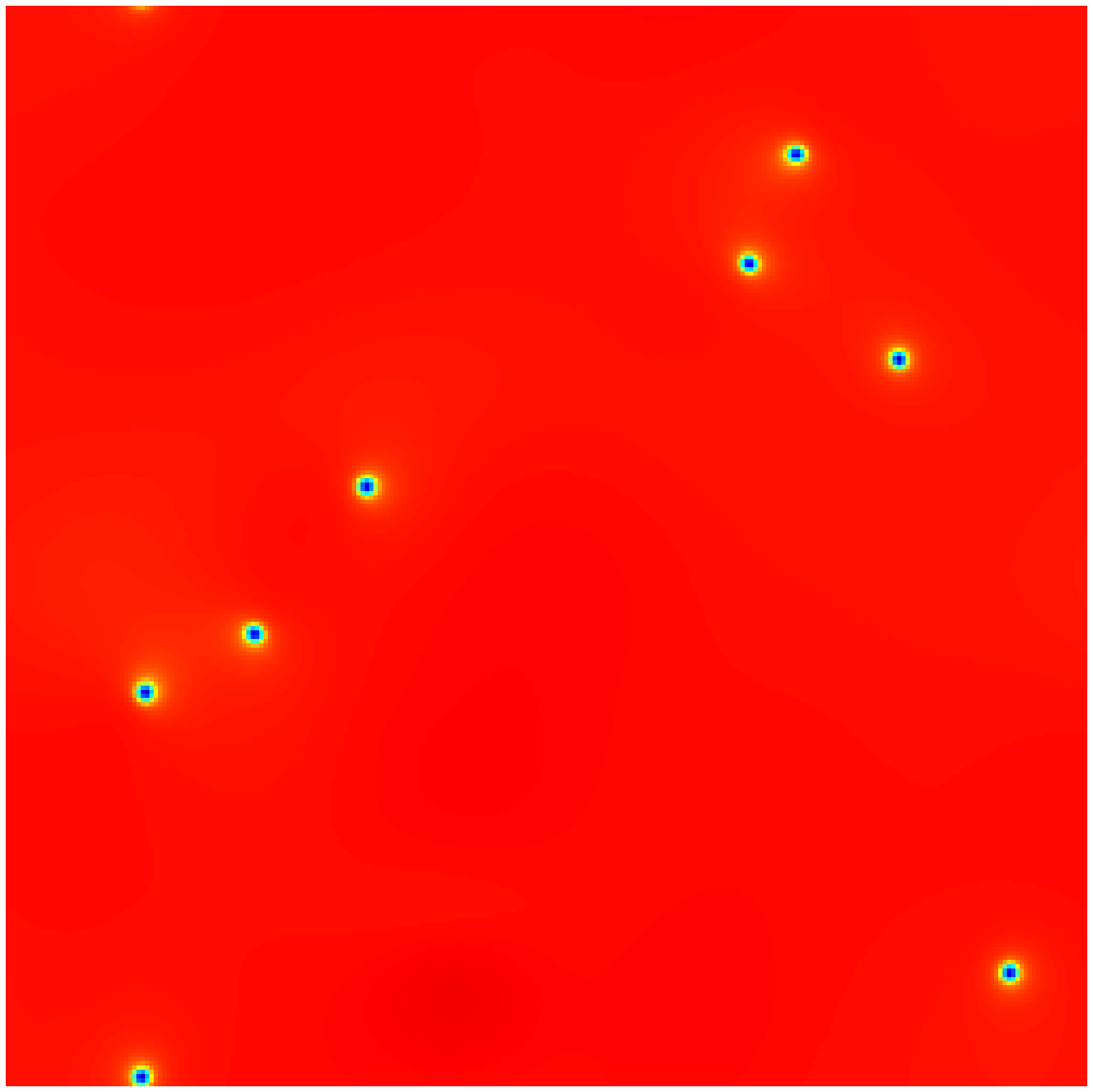}\label{fig:dyn_int_chg16a}}
    \subfigure [$t=293$]
    {\includegraphics[width=4.1cm]{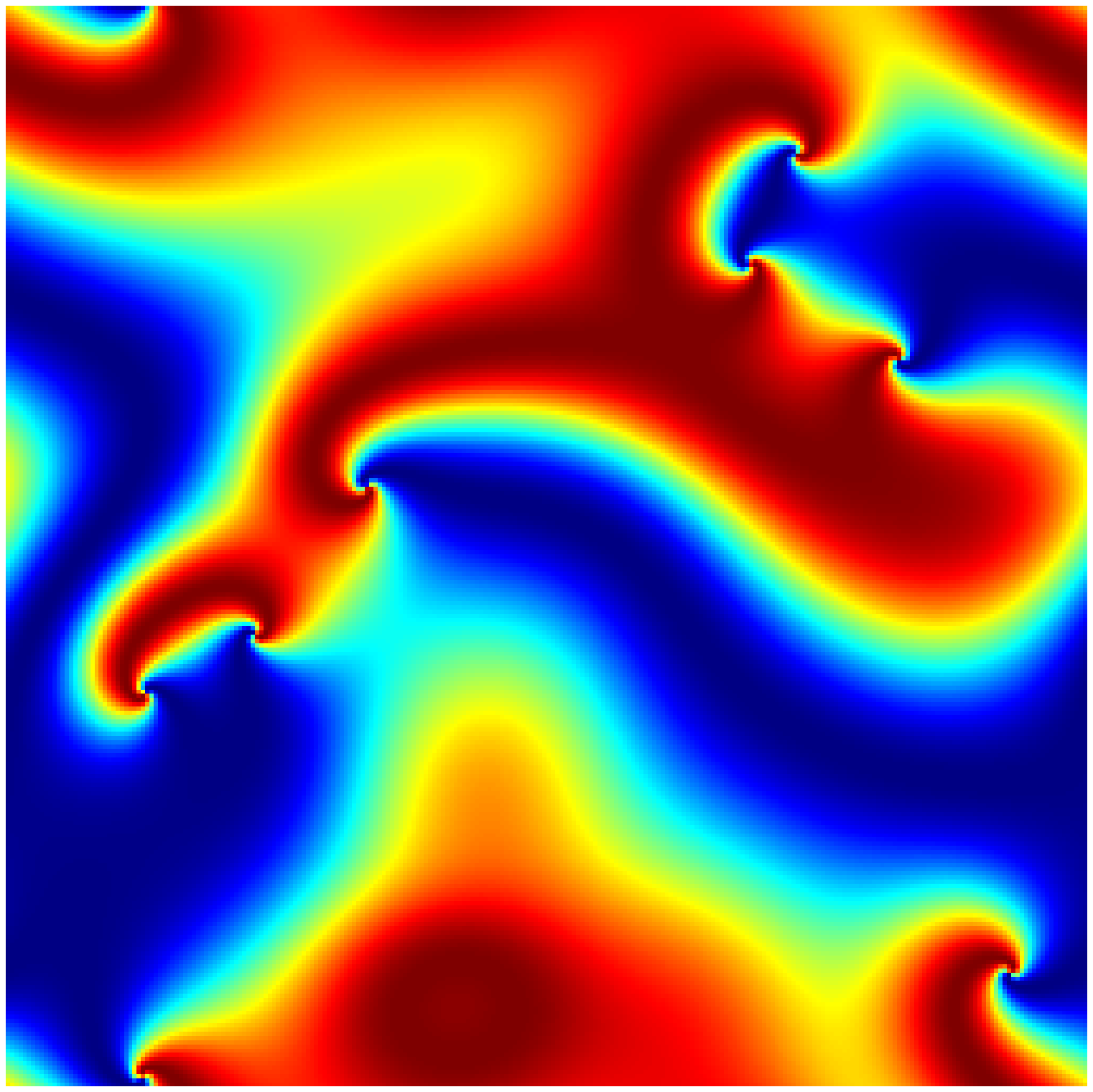}\label{fig:dyn_int_chg16b}}
    \end{center}
    \caption{Cont'd}
    \label{fig:int_chg}
\end{figure*}

\end{document}